\documentclass[a4paper,11pt,twopage]{article}

\usepackage{amsmath,amssymb}
\usepackage{maybemath}
\usepackage[top=1.5cm, bottom=1.8cm, inner=1.3cm, outer=1.3cm]{geometry}

\usepackage{multirow}
\usepackage{adjustbox}
\usepackage{imakeidx}
\makeindex[columns=1]
    \usepackage[range-units = single, separate-uncertainty = true]{siunitx}
    \DeclareSIUnit[]{\solarMasses}{\ensuremath{\mathrm{M}_\odot}}
    \DeclareSIUnit[]{\year}{yr}
    \DeclareSIUnit[]{\barn}{b}
    
\newcommand{\wgonetitle}{Fundamental Symmetries}
\newcommand{\wgonetonetitle}{Pauli Exclusion Principle violation experimental studies}
\newcommand{\wgonettwotitle}{Quantum gravity, CPT and Lorentz symmetries and the Pauli Exclusion Principle violation}
\newcommand{\wgonetthreetitle}{Quantum Collapse models and their experimental tests}

\newcommand{\wgtwotitle}{Direct measurements for Nuclear Astrophysics}
\newcommand{\wgtwotonetitle}{Open questions in astrophysics}
\newcommand{\wgtwottwotitle}{Challenges for nuclear physics: Underground and Recoil Mass Separator measurements}

\newcommand{\wgthreetitle}{Applied Nuclear Physics}
\newcommand{\wgthreetonetitle}{Mass spectrometry}
\newcommand{\wgthreettwotitle}{Diagnostics and modification of materials with ion beams}

    \newcommand{\nuclide}[2]{\ensuremath{{}^{#1}\mathrm{#2}}}
    
    \newcommand{\pp}{\ensuremath{\mathrm{p}}-\ensuremath{\mathrm{p}}}
    \newcommand{\pg}{\ensuremath{(\mathrm{p},\gamma)}}
    \newcommand{\pa}{\ensuremath{(\mathrm{p},\alpha)}}
    \newcommand{\PA}[1]{\ensuremath{(\mathrm{p},\alpha_{#1})}}
    \newcommand{\pag}{\ensuremath{(\mathrm{p},\alpha\gamma)}}
    \newcommand{\PAG}[1]{\ensuremath{(\mathrm{p},\alpha_{#1})}}

    \newcommand{\Dp}{\ensuremath{(\mathrm{d},\mathrm{p})}}
    
    \newcommand{\ngamma}{\ensuremath{(\mathrm{n},\gamma)}}
    
    \newcommand{\ag}{\ensuremath{(\alpha,\gamma)}}
    \newcommand{\an}{\ensuremath{(\alpha,\mathrm{n})}}
    \newcommand{\ad}{\ensuremath{(\alpha,\mathrm{d})}}
    \newcommand{\ap}{\ensuremath{(\alpha,\mathrm{p})}}
    
    \newcommand{\np}{\ensuremath{(\mathrm{n},\mathrm{p})}}
    \newcommand{\na}{\ensuremath{(\mathrm{n},\alpha)}}

    \newcommand{\reaction}[5]{\nuclide{#1}{#2}#3\nuclide{#4}{#5}}
        
    \newcommand{\Ex}{\ensuremath{E_x}}
    \newcommand{\Ecm}{\ensuremath{E_\mathrm{c.m.}}}
    
    \newcommand{\Ep}{\ensuremath{E_\mathrm{p}}}

    \newcommand{\eg}{\emph{e.\,g.}\@}
    \newcommand{\ie}{\emph{i.\,e.\@}}

\newcommand{\Hetag}{{^{3}\mathrm{He}(\alpha,\gamma)^{7}\mathrm{Be}}}

\newcommand{\cag}{^{12}\mathrm{C}(\alpha,\gamma)^{16}\mathrm{O}}
\newcommand{\CC}{^{12}\mathrm{C}$+${}^{12}\mathrm{C}}

\newcommand{\LUNAfourhundred}{LUNA 400\,kV}

\newcommand{\orgname}[1]{\emph{#1}}

\usepackage[colorlinks=false,anchorcolor=black,bookmarksopen=true,bookmarksopenlevel=\maxdimen]{hyperref}
\usepackage[
    backend=biber,	
    style=numeric-comp,
    giveninits=true,
    autocite=inline,
    sortcites=true,
    defernumbers=true,
    sorting=none,
    natbib=true,
    url=true, 
    doi=true,
    eprint=false
]{biblatex}
\bibliography{MidtermLNGS}
\usepackage{authblk}

\title{Nuclear Physics Midterm Plan at LNGS}
\date{}

\author[1,2]{R. Buompane} 
\author[3]{F. Cavanna} 
\author[4,5]{C. Curceanu} 
\author[1,2]{A. D'Onofrio} 
\author[*,6,2]{A. Di Leva}
\author[*,7]{A. Formicola}
\author[1,2]{L. Gialanella} 
\author[7]{C. Gustavino} 
\author[6,2]{G. Imbriani} 
\author[8]{M. Junker} 
\author[9,4]{A. Marcianò} 
\author[1,2]{F. Marzaioli} 
\author[10]{R. Nania} 
\author[4]{F. Napolitano} 
\author[{11},4]{K. Piscicchia} 
\author[{12},7]{O. Straniero} 

\author[13]{C. Abia} 
\author[14]{M. Aliotta} 
\author[15]{D. Bemmerer} 
\author[6,2]{A. Best} 
\author[15]{A. Boeltzig} 
\author[14]{C. Bruno} 
\author[16,17]{A. Caciolli} 
\author[18]{A. Chieffi} 
\author[19,20]{G. Ciani} 
\author[21]{G. D’Agata} 
\author[{22}]{R.J. DeBoer} 
\author[23,2]{M. De Cesare} 
\author[6,2]{D. Dell'Aquila} 
\author[24,25]{R. Depalo} 
\author[13]{I. Dominguez} 
\author[8]{F. Ferraro} 
\author[26]{J. Garcia Duarte} 
\author[24,25]{A. Guglielmetti} 
\author[27]{Gy. Gyürky} 
\author[28]{S. Hayakawa} 
\author[21]{M. La Cognata} 
\author[21,29]{L. Lamia} 
\author[30,31]{L.E. Marcucci} 
\author[15]{E. Masha} 
\author[16,17]{M. Mazzocco} 
\author[2,1]{E.L. Morales-Gallegos} 
\author[32,{33}]{S. Palmerini} 
\author[1,2]{I. Passariello} 
\author[1]{A. Petraglia} 
\author[16,17]{D. Piatti} 
\author[34,35]{M. Pignatari} 
\author[29,21]{R.G. Pizzone} 
\author[1,2]{G. Porzio} 
\author[6,2]{D. Rapagnani} 
\author[29,21]{G.G. Rapisarda} 
\author[29,21,36]{S. Romano} 
\author[1,2]{M. Rubino} 
\author[1,2]{C. Santonastaso} 
\author[29,21]{M.L. Sergi} 
\author[16,17]{J. Skowronski} 
\author[37,21]{R. Spartà} 
\author[1]{F. Terrasi} 
\author[37,21]{A. Tumino} 
\author[15]{S. Turkat} 
\author[22]{M. Wiescher} 
\author[38,39]{S. Zavatarelli} 

\affil[1]{\orgname{Università della Campania ``L. Vanvitelli''},
{Dipartimento di Matematica e Fisica},  
{
{I-81100}, {Caserta}, {Italy}}}
\affil[2]{\orgname{Istituto Nazionale di Fisica Nucleare},
{Sezione di Napoli},
{
{I-80126}, {Napoli}, {Italy}}}
\affil[3]{\orgname{Istituto Nazionale di Fisica Nucleare},
{Sezione di Torino}, 
{
{I-10125}, {Torino}, {Italy}}}
\affil[4]{\orgname{Istituto Nazionale di Fisica Nucleare},
{Laboratori Nazionali di Frascati},
{
{I-00044}, {Frascati}, {Italy}}}
\affil[5]{\orgname{Institutul National pentru Fizica si Inginerie Nucleara ``Horia Hulubei''},
{
{077125}, {Bucharest-Magurele}, {Romania}}}
\affil[6]{\orgname{Università di Napoli ``Federico II''},
{Dipartimento di Fisica ``E. Pancini"}, 
{
{I-80126}, {Napoli}, {Italy}}}
\affil[7]{\orgname{Istituto Nazionale di Fisica Nucleare},
{Sezione di Roma},
{
{I-00185}, {Roma}, {Italy}}}
\affil[8]{\orgname{Istituto Nazionale di Fisica Nucleare},
{Laboratori Nazionali del Gran Sasso},
{
{I-67100}, {Assergi L’Aquila}, {Italy}}}
\affil[9]{\orgname{Fudan University},
{Department of Physics (Jiangwan Campus)},
{
{200438}, {Shanghai}, {China}}}
\affil[10]{\orgname{Istituto Nazionale di Fisica Nucleare},
{Sezione di Bologna},
{
{I-40127}, {Bologna}, {Italy}}}
\affil[11]{\orgname{Centro Ricerche Enrico Fermi},
{Museo Storico della Fisica e Centro Studi e Ricerche ``Enrico Fermi''},
{
{I-00184}, {Roma}, {Italy}}}
\affil[12]{\orgname{Istituto Nazionale di Astrofisica},
{Osservatorio Astronomico d’Abruzzo},
{
{I-64100}, {Teramo}, {Italy}}}
\affil[13]{\orgname{Universidad de Granada},
{Dpto. Física Teórica y del Cosmos},
{
{18071}, {Granada}, {Spain}}}
\affil[14]{\orgname{University of Edinburgh},
{School of Physics and Astronomy},
{
{Edinburgh}, {United Kingdom}}}
\affil[15]{\orgname{Helmholtz-Zentrum Dresden-Rossendorf},
{
{D-01328}, {Dresden}, {Germany}}}
\affil[16]{\orgname{Università degli Studi di Padova},
{Dipartimento di Fisica e Astronomia},
{
{I-35131}, {Padova}, {Italy}}}
\affil[17]{\orgname{Istituto Nazionale di Fisica Nucleare},
{Sezione di Padova},
{
{I-35131}, {Padova}, {Italy}}}
\affil[18]{\orgname{Istituto Nazionale di Astrofisica},
{Osservatorio Astronomico di Roma},
{
{I-00078}, { Monte Porzio Catone}, {Italy}}}
\affil[19]{\orgname{Università degli Studi di Bari},
{Dipartimento di Fisica ``M. Merlin''},
{
{I-70125}, {Bari}, {Italy}}}
\affil[20]{\orgname{Istituto Nazionale di Fisica Nucleare},
{Sezione di Bari},
{
{I-70125}, {Bari}, {Italy}}}
\affil[21]{\orgname{Istituto Nazionale di Fisica Nucleare},
{Laboratori Nazionali del Sud},
{
{I-95123}, {Catania}, {Italy}}}
\affil[22]{\orgname{University of Notre Dame},
{Department of Physics and the Joint Institute for Nuclear Astrophysics},
{
{IN 46556}, {Notre Dame}, {USA}}}
\affil[23]{\orgname{Centro Italiano di Ricerche Aerospaziali},
{
{I-81043}, {Capua}, {Italy}}}
\affil[24]{\orgname{Università degli Studi di Milano},
{Dipartimento di Fisica},
{
{I-20133}, {Milano}, {Italy}}}
\affil[25]{\orgname{Istituto Nazionale di Fisica Nucleare},
{Sezione di Milano},
{
{I-20133}, {Milano}, {Italy}}}
\affil[26]{\orgname{Lawrence Livermore National Laboratory},
{Nuclear and Particle Physics Group},
{
{CA 94550}, {Livermore}, {USA}}}
\affil[27]{\orgname{Institute for Nuclear Research (ATOMKI)},
{
{H-4001}, {Debrecen}, {Hungary}}}
\affil[28]{\orgname{Center for Nuclear Study, University of Tokyo},
{RIKEN Campus},
{
{351-0198}, {Wakō, Saitama}, {Japan}}}
\affil[29]{\orgname{Università degli Studi di Catania},
{Dipartimento di Fisica e Astronomia ``E. Majorana''},
{
{I-95123}, {Catania}, {Italy}}}
\affil[30]{\orgname{Università degli Studi di Pisa},
{Dipartimento di Fisica},
{
{I-56127}, {Pisa}, {Italy}}}
\affil[31]{\orgname{Istituto Nazionale di Fisica Nucleare},
{Sezione di Pisa},
{
{I-56127}, {Pisa}, {Italy}}}
\affil[32]{\orgname{Università degli Studi di Perugia},
{Dipartimento di Fisica e Geologia},
{
{I-06123}, {Perugia}, {Italy}}}
\affil[33]{\orgname{Istituto Nazionale di Fisica Nucleare},
{Sezione di Perugia},
{
{I-06123}, {Perugia}, {Italy}}}
\affil[34]{\orgname{Konkoly Observatory},
{
{H-1121}, {Budapest}, {Hungary}}}
\affil[35]{\orgname{University of Hull},
{E. A. Milne Centre for Astrophysics},
{
{HU6 7RX}, {Hull}, {United Kingdom}}}
\affil[36]{\orgname{Università degli Studi di Enna ``Kore"},
{Facoltà di Ingegneria e Architettura},
{
{I-94100}, {Enna}, {Italy}}}
\affil[37]{\orgname{Centro Siciliano di Fisica Nucleare e Struttura della Materia},
{
{I-95123}, {Catania}, {Italy}}}
\affil[38]{\orgname{Istituto Nazionale di Fisica Nucleare},
{Sezione di Genova},
{
{I-16146}, {Genova}, {Italy}}}
\affil[39]{\orgname{Università degli Studi di Genova},
{Dipartimento di Fisica},
{
{I-16146}, {Genova}, {Italy}}}
\affil[*]{Correspondig author, \href{mailto:antonino.dileva@unina.it}{antonino.dileva@unina.it}, \href{mailto:alba.formicola@roma1.infn.it}{alba.formicola@roma1.infn.it}}

\begin{document}

\maketitle

\begin{abstract}
The Istituto Nazionale di Fisica Nucleare - Laboratori Nazionali del Gran Sasso (LNGS) is one the largest underground physics laboratory, a very peculiar environment suited for experiments in Astroparticle Physics, Nuclear Physics and Fundamental Symmetries. The newly established Bellotti Ion Beam facility represents a major advance in the possibilities of studying nuclear processes in an underground environment.
A workshop was organised at LNGS in the framework of the Nuclear Physics Mid Term Plan Italy, an initiative of the Nuclear Physics Division of the Instituto Nazionale di Fisica Nucleare to discuss the opportunities that will be possible to study in the near future by employing state-of-the-art detection systems. In this report a detailed discussion of the outcome of the workshop is presented.
\end{abstract}


\section{Executive summary}

The Laboratori Nazionali del Gran Sasso (LNGS) of the Istituto Nazionale di Fisica Nucleare (INFN) is one the largest underground physics laboratory devoted to fundamental research programs in Astroparticle Physics, Nuclear Physics and Fundamental Symmetries \cite{LNGSurl}.
The rock overburden (3800\,m of water equivalent) makes the LNGS a very peculiar environment suited for the study of “rare”  processes. The reduction of the cosmic ray flux is of great advantage for measurements where background suppression plays a crucial role \cite{Laubenstein2020,Ambrosio1995,Best2016}.
At present LNGS hosts activities in Fundamental Symmetries and Nuclear Astrophysics, funded and supervised by the Nuclear Physics Division (CSN3) of INFN. 

Among other laboratories having tight connections with LNGS, there is the Center for Isotopic Research on Cultural and Environmental heritage (CIRCE) of the Department of Mathematics and Physics, University of Campania, Caserta, Italy \cite{CIRCEurl}, where a wide range of applied nuclear physics and interdisciplinary activities are routinely conducted, together with INFN-CSN3 related researches mainly in nuclear astrophysics.

\bigskip 

In the framewok of CSN3 scientific activities there were already a few occasions to put together the large variety of expertises present, and are summarized in \cite{Broggini2019,Badala2022}. Following these successful initiatives the Nuclear Physics Mid Term Plan in Italy \cite{MidTermurl}, was organized by the CSN3 to allow the experimental and theoretical nuclear physics community to meet and plan a scientific vision for the near future in Italy, as presented in the introductory paper of this Focus issue \cite{Benzoni2023}. \\*
As for the Laboratori Nazionali di Legnaro (LNL) \cite{Ballan2023} and Laboratori Nazionali del Sud (LNS) \cite{Agodi2023}, and later the Laboratori Nazionali di Frascati (LNF) \cite{midtermLNF}, the LNGS Session \cite{MidTermLNGSurl} has given the occasion to survey the opportunities that the community gathering around the INFN-CSN3 funded projects can exploit in the next years. It also aims at strengthening and exploring further synergies between LNGS and other laboratories.

\subsection{\wgonetitle{}}

The fundamental symmetries play a crucial role in Quantum Mechanics (QM), arising from very few, general assumptions, and are incorporated directly in the theoretical framework of the Standard Model (SM). 
In this context, high precision experiments testing theories and models beyond the standard QM and SM, may pave the road towards a new theory, where the open questions may be fully clarified. 

At LNGS, fundamental symmetries and their relation to QM and SM are investigated from different directions and perspectives within the VIP Collaboration, searching for possible violations of the Pauli Exclusion Principle (PEP), deeply connected to CPT and Lorentz symmetry, but also quantum gravity. \\* 
As midterm plan, from a theoretical standpoint, a series of activities are foreseen, aiming to connect, and cross-fertilize various developments in non-commutative quantum gravity (NCQG), leading to the PEP violations. In this context, we also plan to seek new phenomenological connections between the quantization schemes of NCQG and the wave-function collapse and investigate their experimental signatures.
The experimental activity is divided in Open systems (where a new state is introduced from outside) and Closed systems (where the state was already present) experiments. 
In the former, the VIP-2 apparatus, currently in operation for direct searches of PEP violating atomic transitions, will be upgraded in the upcoming years, with focus on new, more efficient detectors and targets, and optimizations of the critical setup parameters to increase the sensitivity. In the latter, a dedicated data taking using a broad Energy Germanium (BEGe) detector will investigate new aspects and effects in PEP violation predicted by quantum gravity, such as signal anisotropy, and to explore the connection with CPT and Lorentz symmetries.
Finally, we plan to develop and run new dedicated apparatuses based on germanium detectors, to search for the patterns of the spontaneous radiation predicted by refined dissipative and non-Markovian collapse models and study their interplay with gravity. 

\subsection{\wgtwotitle{}}

Since more than three decades, accelerators are in use in the underground halls of LNGS. After the pioneering measurements with the 50\,kV accelerator \cite{Junker1998}, in 2000 the LUNA 400\,kV accelerator was installed and it is still in operation. More recently with the installation of the 3.5\,MV accelerator of the Bellotti Ion Beam Facility (BIBF) started its operation. The availability of these two accelerators provides great opportunities for the direct measurements of nuclear reactions of astrophysical interest, possibly directly reaching the energies corresponding to stellar environments. The most compelling cases will be presented in Section \ref{wg2:t1}. The experimental opportunities envisioned at LNGS and CIRCE, will be thoroughly discussed in Section \ref{wg2:t2}. A summary, including the needed main upgrades and a gross estimate of the time to deployment, is given in Tables \ref{wg2:t2:tab1} and \ref{wg2:t3:tab1} for LNGS and CIRCE, respectively. 

\subsection{\wgthreetitle{}}
The development of nuclear astrophysics experiments at CIRCE has led to a fruitful interchange between basic and applied research activities. Indeed, conventional and accelerator based mass spectrometry have been used for studies processes of astrophysical interest by means of isotopic ratio measurements in terrestrial as well as in meteoritic materials. Similarly, some technical developments needed in nuclear physics experiments, especially in ion beam production, detectors and related equipment, have led to applications in various fields. The very intense $^7\mathrm{Be}$ beam developed for the measurement of the $\reaction{7}{Be}{\pg}{8}{B}$ cross section, is now used in wear studies of materials for several applications, including aerospace. Mass spectrometry and other Ion Beam Analysis techniques are routinely applied for the characterisation of materials used in detector development, among which the study of water diffusion and related micro-leaks in the KM3NeT deep-sea neutrino telescope equipment.

\section{Introduction}
\label{wg0:intro}

The Laboratori Nazionali del Gran Sasso (LNGS) of the Istituto Nazionale di Fisica Nucleare (INFN) is one the largest underground physics laboratory devoted to fundamental research programs in Astroparticle Physics, Nuclear Physics and Fundamental Symmetries \cite{LNGSurl}.
At present, among activities funded and supervised by INFN-CSN3, LNGS hosts experiments in Fundamental Symmetries and Nuclear Astrophysics. \\*
Therefore, the Nuclear Physics Mid Term Plan in Italy - LNGS Session \cite{MidTermLNGSurl}, was the opportunity to discuss the present and near future prospects in these two main activities, with an outlook going beyond the state-of-the-art.

In the context of Fundamental Symmetries, a consolidated experimental activity, the VIolation of Pauli exclusion principle (VIP) and VIP-2 experiments, aim at improving the limit on the probability of violation of the Pauli Exclusion Principle (PEP), and in parallel conduct high sensitivity tests of quantum collapse models of the wave function.

LNGS has been for a long time a unique infrastructure hosting an accelerator devoted to Nuclear Astrophysics, the Laboratory for Underground Nuclear Astrophysics (LUNA). The \LUNAfourhundred{} accelerator, in almost two decades, produced a wealth of high precision cross sections, that significantly advanced our knowledge on nuclear processes occurring inside stars and other astrophysical objects. 
A new 3.5\,MV accelerator, the Bellotti Ion Beam Facility (BIBF), has been recently installed, that is able to deliver very intense proton, helium and carbon beams. It is for the Nuclear Astrophysics community the opportunity to extend the measurements to elusive reactions occurring in advanced stellar burning stages, and offers to a wider Nuclear Physics community a potential to be identified and exploited.

Underground measurements are often complemented with overground measurements, either to extend the energy range, for ancillary measurements or specific R\&D tailored towards the low activity environment. Among other laboratories having tight connections with LNGS, there is the Center for Isotopic Research on Cultural and Environmental heritage (CIRCE), of the Department of Mathematics and Physics, University of Campania, Caserta, Italy, where the European Recoil separator for Nuclear Astrophysics (ERNA) is located, and hosts also other INFN-CSN3 related research \cite{CIRCEurl}.

\subsection{LNGS}
\label{wg0:LNGS}
The rock overburden (3800\,m of water equivalent) makes the LNGS a very peculiar environment suited for the study of “rare”  processes. The reduction of the cosmic ray flux is of great advantage for measurements where background suppression plays a crucial role \cite{Arpesella1996,Ambrosio1995,Best2016}. 

\begin{figure*}[!hbt]
\begin{center}
\includegraphics[width=.5\textwidth]{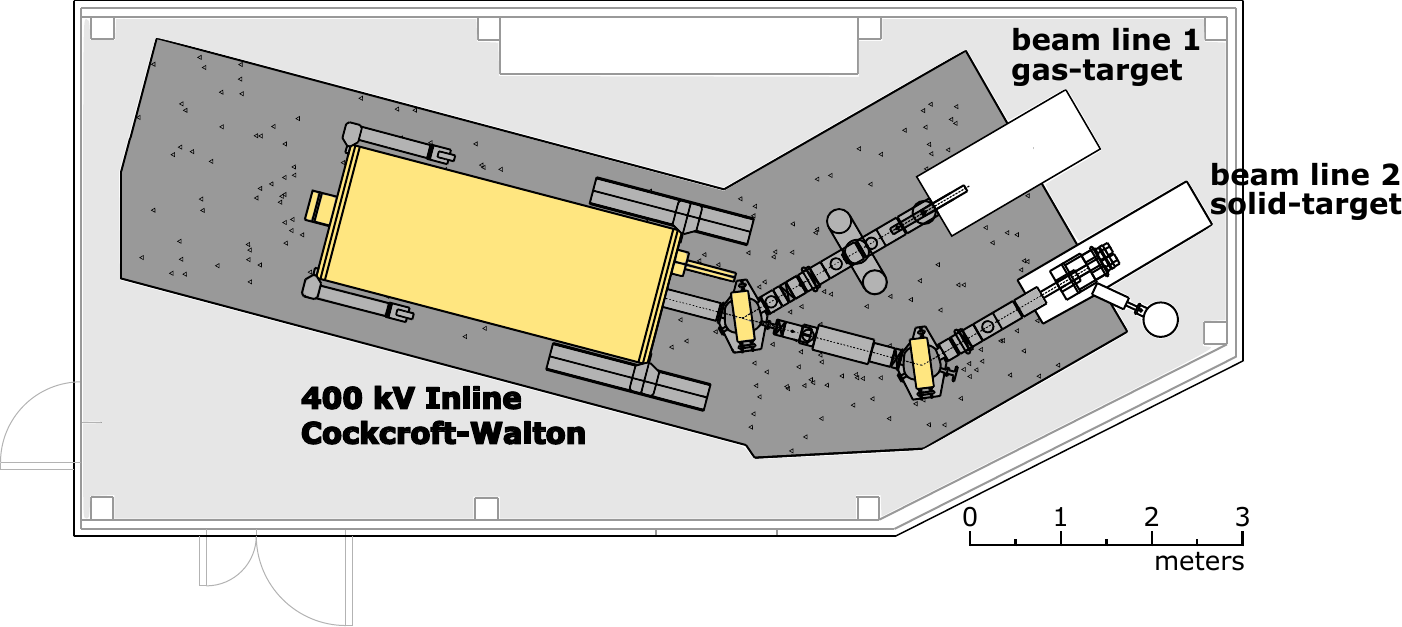}\\[5mm]
\includegraphics[width=.75\textwidth]{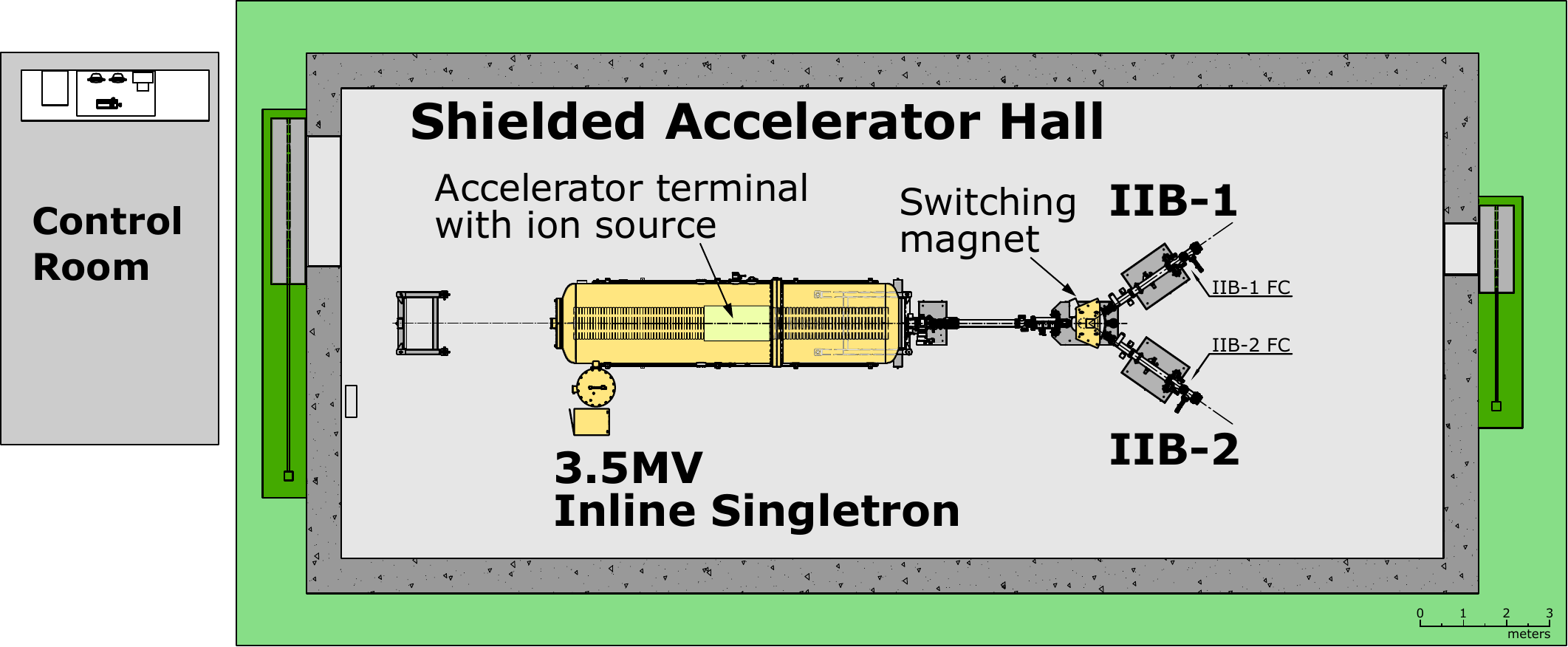}
\end{center}
	\caption{Layout of the \LUNAfourhundred{} (top) and the 3.5\,MV Inline Singletron (bottom) accelerators installed at the Laboratori Nazionali del Gran Sasso (drawings have different scales).
	\label{wg0:LNGS:fig:LUNAMV}}
\end{figure*}

Direct cross section measurements typically require accelerators providing high-intensity beams, ultra-pure stable targets, and the highest possible detection efficiency. Moreover, when approaching the astrophysically relevant energy range, the counting rate from a nuclear fusion reaction signal may become much smaller than the environmental background produced by cosmic radiation and naturally-occurring radioactive elements. 
To overcome this problem, the Laboratory for Underground Nuclear Astrophysics (LUNA) was built at LNGS in the early 90's, hosting a 400\,kV Inline-Cockcroft-Walton electrostatic accelerator produced by High Voltage Engineering Europa (HVEE), see Figure \ref{wg0:LNGS:fig:LUNAMV} top panel. The accelerator has a HV-ripple of 30\,Vpp and a reproducibility as well as long-term stability of the HV of 20\,Vpp at 400\,kV over several days. The radiofrequency ion source -mounted directly on the accelerator tube- provides ion beams of $1$\,mA hydrogen (75\% H$^+$) and $500\,\mu\mathrm{A}$ He$^+$ over a continuous operating time of about 40 days \cite{Formicola2003}. Alternatively, the ion beam can be guided through the $0^\circ$ port of the magnet, going to the first beam line equipped with a gas target, and a second $45^\circ$ magnet (identical to the first magnet) into an additional beam line with a solid target station.
Thanks to this accelerator several crucial reactions for hydrogen, helium burning and Big Bang Nucleosynthesis were studied.  

The LNGS has recently established the ``Bellotti Ion Beam Facility'' by building a new infrastructure in Hall B to host a 3.5\,MV Singletron machine designed and built by HVEE \cite{Sen2019}. This was possible also through a special funding of the Italian Ministry of Research (Progetto Premiale ``LUNAMV'') awarded to the LUNA Collaboration. The accelerator is equipped with two independent beam-lines, see Figure \ref{wg0:LNGS:fig:LUNAMV} bottom panel, and can deliver proton, helium, and carbon beams with maximumm intensity of 1, 0.5 and 0.15\,mA respectively. 
The start of scientific exploitation is foreseen by second half of 2023.

 \subsubsection{Resident instrumentation for nuclear physics}\label{wg0:LNGS:sub1}

\begin{itemize}
\item VIP-2 detector \cite{Napolitano2022SY}: based on the experience of the VIP experiment, the apparatus make use of cutting-edge radiation Silicon drift detectors (SDDs) developed in collaboration with the Stefan Meyer Institute (SMI) of the Austrian Academy of Sciences, Politecnico di Milano and the Fondazione Bruno Kessler. The SDD are employed as ideal X-ray detectors due to their excellent spectroscopic response and high resolution of approximately 190\,eV (FWHM) at 8\,keV. 
The SDD cells possess large geometrical acceptance, an overall active area of 0.64\,cm$^2$, a thickness of 450\,$\mu$m, and an efficiency of 99\% at 8\,keV, surpassing charge coupled devices used in VIP. The experiment setup includes a copper target consisting of copper strips for direct current circulation, with each strip measuring 71\,mm in length, 20\,mm in height, and 25\,$\mu$m in thickness, where a direct current up to 180\,A can be circulated. To maintain a low temperature, a cooling pad is placed between the strips and a closed chiller circuit is employed. Two parallel arrays of SDD cells, each containing 2$\times$4 cells, are positioned on both sides of the copper strips and operated at -90$^\circ$C. 
The vacuum chamber, maintained at a pressure of 10$^{-5}$ mbar, houses the detectors, their front-end electronics, and an in-situ calibration $^{55}$Fe radioactive source. Externally, a passive shield composed of copper and lead bricks provides additional protection against the residual environmental radiation present in the laboratory halls.

\item Solid target: vacuum chambers with different geometries are available to host different solid targets or serve different purposes.
Immediately prior to entering the target chamber the beam passes through a liquid nitrogen (LN$_2$) cooled copper tube that serves to reduce rest gas contaminants and prevents the build-up of carbon on the target surface. The target flange is isolated from all other beam-line components and acts as a Faraday cup to determine the total charge accumulated in the course of a measurement. The electron suppression is provided by a negative voltage applied to the copper tube. The target is typically water-cooled with de/ionized water.\\*
In addition, the target preparation laboratory is equipped with an apparatus to produce solid oxygen targets by anodic oxidation of tantalum backings in isotopically enriched water, a technique known to produce targets with highly uniform stoichiometry and homogeneous thicknesses \cite{Caciolli2012}.

\item Windowless gas target \cite{Ferraro2018a}: the \LUNAfourhundred{} accelerator is also equipped with a windowless gas target. The use of a gas target provides the advantage of isotopical purity and high stability over prolonged periods. The differentially pumped gas target consists of three pumping stages and the interaction chamber, separated from each other with flux collimators which also serve as beam collimators. The sequence of pumping stages generates a pressure gradient from $10^{-6}$\,mbar, the typical accelerator tube pressure, to the interaction chamber pressure in the order of a few mbar. The pumping system is remotely controlled through a Labview program. The gas coming out from the first and the second pumping stages can be collected and sent to an industrial purifier (MonoTorr II) to be cleaned and then recirculated. The target pressure is kept stable with an automated system controlling the inlet flux, using an active feedback from the pressure reading. Several gases have been used as targets, e.g. D, $\nuclide{3}{He}$, N, Ne. The beam stop in the interaction chamber serves as a calorimeter to measure beam intensity. 

\item HPGe detectors: High Purity Germanium (HPGe) detectors are used by the VIP collaboration for studies on the foundations of quantum mechanics \cite{Donadi2020NP}. The experimental setup consisted of a coaxial p-type high-purity germanium detector surrounded by a complex shielding structure, with the outer part made of pure lead and the inner part made of electrolytic copper. The germanium crystal is characterized by a diameter of $8.0\,\mathrm{cm}$ and a length of $8.0\,\mathrm{cm}$, with an inactive layer of $0.075\ \mathrm{mm}$ of lithium-doped germanium all around the crystal. The active germanium volume of the detector is $375\,\mathrm{cm}^3$. The outer part of the passive shielding of the high-purity germanium detector consists of lead ($30\,\mathrm{cm}$ from the bottom and $25\,\mathrm{cm}$ from the sides). The inner layer of the shielding ($5\,\mathrm{cm}$) is composed of electrolytic copper. The sample chamber has a volume of about $15\,\mathrm{l}$ ($250\,\mathrm{mm} \times 250\,\mathrm{mm} \times 240\,\mathrm{mm}$). The shield together with the cryostat are enclosed in an air-tight steel housing of $1\ \mathrm{mm}$ thickness, which is continuously flushed with boil-off nitrogen from a liquid nitrogen storage tank, in order to reduce the contact with external air (and thus radon) to a minimum.
Additionally, the VIP-GATOR collaboration employs the Gator~\cite{baudis2011JINST} facility and its 2.2\,kg HPGe with 100.5\% relative efficiency\footnote{The efficiency of HPGe detectors is usually reported relative to the efficiency of a $3\times3$ NaI scintillator at 25\,cm source to detector distance.} in the context of searches for PEP violating transitions at higher $Z$ target materials.\\*
The LUNA Collaboration utilizes four ultra low-background HPGe detectors mainly for prompt $\gamma$-ray measurements. Relative efficiencies range from 90\% to 140\%. An electro-mechanical cooling system is available and can be used with two of the smaller detectors, in order to avoid the use, and periodic filling, of LN$_{2}$ and also to exploit more compact measurement setup configurations. The detectors can be coupled with various massive lead shielding described below.

\item A segmented and highly efficient BGO detector \cite{Boeltzig2018, Skowronski2023}: it is made of an array of six prismatic crystals, each 28\,cm long arranged around a cylindric borehole of 6\,cm diameter each covering an azimuthal angle of 60 degrees. The radial thickness of the segments is at least 7\,cm. The crystals are optically insulated and coupled to Hamamatsu R1847-07 photomultipliers (PMTs). The target chamber fit in the central borehole and for the solid target at the centre of the detector, a close to 4$\pi$ solid angle coverage is achieved. The detector features an energy resolution of about 11$\%$ FWHM at 1.33\,MeV.

\item Neutron detectors \cite{Csedreki2021}: at LUNA a detector array containing eighteen \nuclide{3}{He} filled proportional counters with stainless steel housing is available. The \nuclide{3}{He} tubes are embedded in a high-density polyethylene (PE) for neutron moderation. The counters are arranged in two concentric rings around the target chamber: twelve counters of 40\,cm active length are located at a radius of 11\,cm, and six counters of 25\,cm active length are located at 6\,cm radius. This configuration allows for a nearly 4$\pi$ solid angle coverage around the target.

\item Massive shielding for $\gamma$-ray detection \cite{Boeltzig2018, Skowronski2023, Cavanna2014, Confortola2007}: to further reduce environmental backgrounds in experiments at the solid or gas target setup at \LUNAfourhundred{}, several shieldings have been designed for use with either the BGO or the HPGe detectors.\\*
A movable and compact lead shielding \cite{Boeltzig2018} used at the solid target beam line of the LUNA 400\,kV accelerator, allows for a minimum thickness of 10\,cm for the BGO detector and 15\,cm for the HPGe detector. The BGO detector can be placed at the centre of the shielding, while the HPGe detector can either be mounted at 0$^\circ$ or at a 55$^\circ$ angle with respect to beam direction, the latter is convenient to reduce the influence of unknown angular distributions.\\*
Since in a typical experiment using solid targets they have to be replaced frequently, the shielding is sectioned in several parts mounted on rails that can be easily slid for easy access, or for concurrent use of different detectors.\\*
In the case of the gas target beam line of the LUNA 400\,kV accelerator, is available a lead shielding composed of lead bricks for a total volume of 0.5\,m$^3$ ($80\times70$\,cm$^2$ of base and 95\,cm of height). It allows to surround up to two HPGe detectors with 25\,cm thickness of lead  \cite{Cavanna2014}. The shielding includes as innermost layer oxygen free copper of 4\,cm thickness with a very low intrinsic radioactivity. The achieved reduction of the environmental background is of a factor $10^{-5}$ for $\gamma$-rays below 2\,MeV. The shielding can be enclosed in a plastic box (radon box) that can be filled with N$_{2}$ gas to reduce at minimum the radon to get in contact with the detectors.
\end{itemize}

\subsubsection{New instrumentation for nuclear physics}\label{wg0:LNGS:sub2}
\begin{itemize}
\item VIP-3 experiment: the experimental effort of the VIP-2 experiment will be extended in depth and range to violations of the PEP in transitions on targets with higher $Z$ materials. To this aim, the upgraded experiment VIP-3 will make use of 64 SDD cells in a larger vacuum chamber, doubling the active area. The calibration system will be upgraded as well as the target cooling systems, which will allow the maximum direct current to reach 400\,A. Additionally, the VIP-3 setup will be equipped in a later stage with 1\,mm thick SDDs improving the reach to higher $Z$ material and enhancing the quantum efficiency of the detectors.
\item New massive shielding for $\gamma$-rays detection: to measure the $\CC$ fusion cross section, detecting the two low energy $\gamma$-rays from the de-excitation of \nuclide{20}{Ne}{} and \nuclide{23}{Na}{}, the LUNA Collaboration is developing a setup composed by a lead (\qty{25}{\cm}) and copper (\qty{1}{\cm}) shielding to suppress the environmental $\gamma$-ray background. The $\gamma$-rays detection setup comprises a HPGe detector (\qty{150}{\percent} relative efficiency) installed at \qty{0}{\degree} with respect to beam direction (in close geometry to the target to maximize the signal to noise ratio) and \num{8} NaI scintillator detectors surrounding both the target and the HPGe to reduce the Compton background and to enhance detection efficiency (see section \ref{wg2:t2:subsection:carbon}). The finalization of this setup is supported by the Italian Ministry of Research through the funding action PRIN2022.
\item The assembly of a recirculating gas target and neutron- as well as $\gamma$-ray detection systems is underway and has received funding from the ERC and the Italian Ministry of Research, respectively. The neutron detection system consists of a coupled array of liquid scintillators and $^3$He counters (see fig. \ref{wg2:t2:fig:shades} for the setup). Through capture-gating and coincidence/anticoincidence between the two detector types energy sensitivity is achieved, which allows to remove unwanted beam-induced background from $\an$ reactions on contaminants like boron or carbon.
\begin{figure}[htb]
\centering
\includegraphics[width=.7\columnwidth]{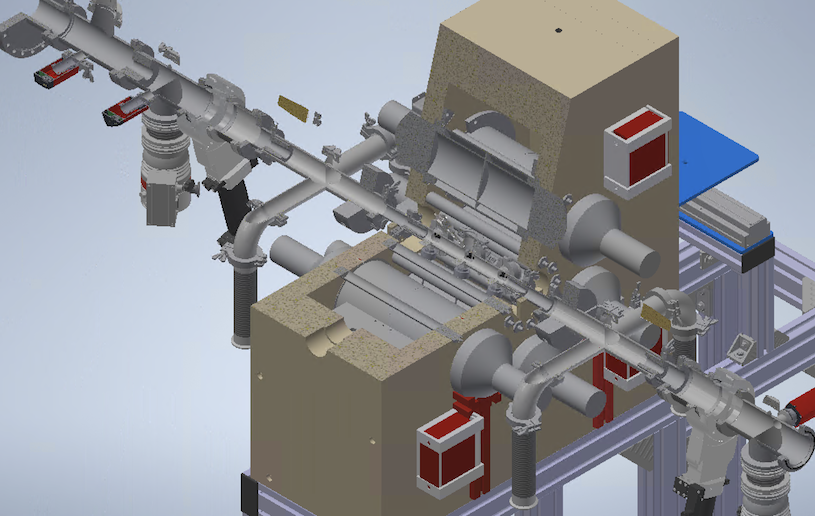}
\caption{Drawing of the gastarget and detector configuration for the \reaction{22}{Ne}{\an}{25}{Mg} measurement.}
\label{wg2:t2:fig:shades}
\end{figure}

\end{itemize}

\subsection{CIRCE}
\label{wg0:CIRCE}

\begin{figure*}[!h]
\begin{center}	
	\includegraphics[width=.85\textwidth]{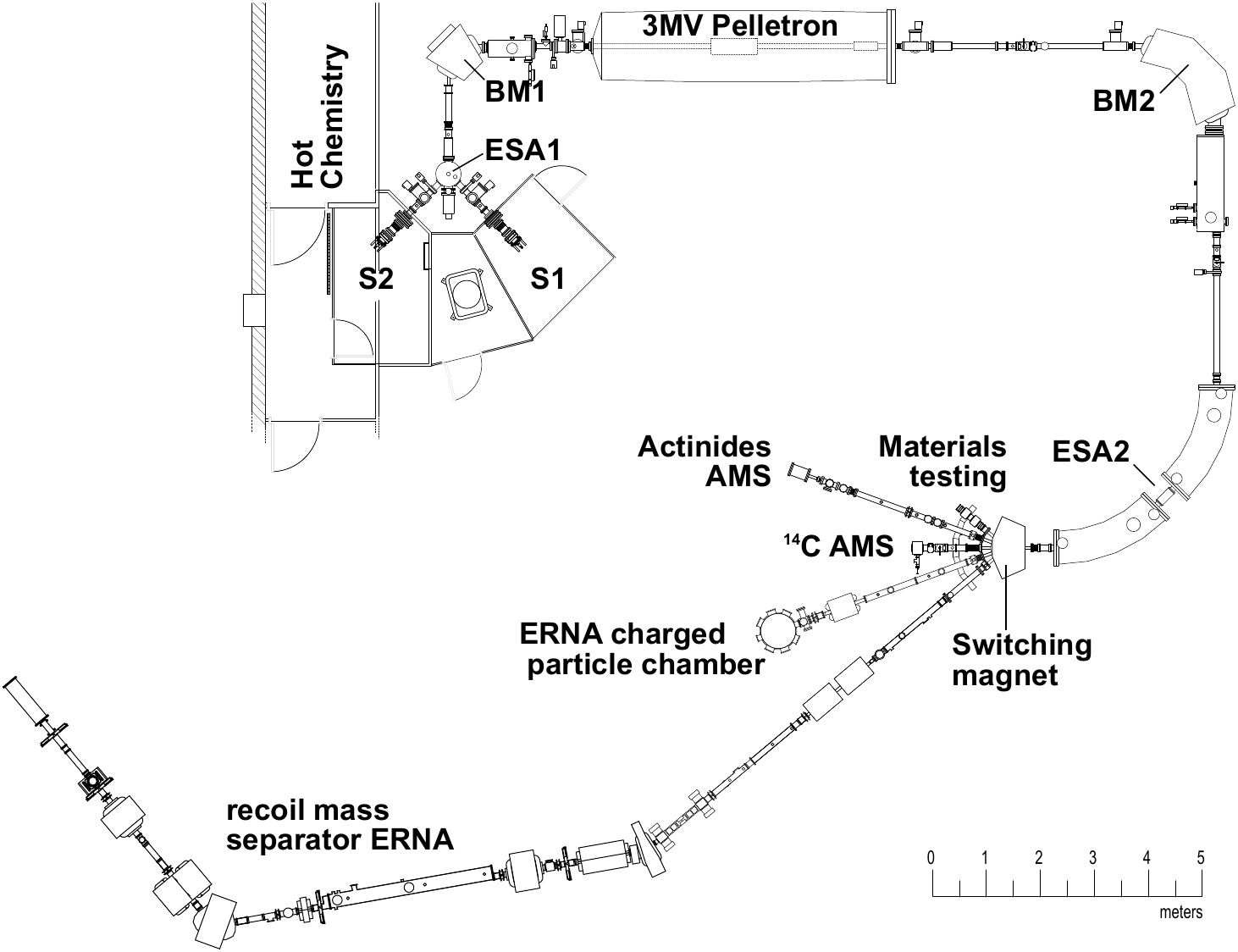}
\end{center}
	\caption{The layout of CIRCE Tandem Accelerator Laboratory. 
	\label{wg0:CIRCE:fig:CIRCE}}
\end{figure*}

The Center for Isotopic Research on Cultural and Environmental heritage, is a complex of laboratories of the University of Campania ``L. Vanvitelli'', located in San Nicola la Strada (Caserta), Italy. The Tandem Accelerator Laboratory, shown in Figure \ref{wg0:CIRCE:fig:CIRCE}, hosts a 3.0\,MV NEC 9SDH-2 Pelletron accelerator \cite{Terrasi2007}.\\*
The accelerator is equipped with two 40-sample MC-SNICS Cs-sputter negative ion sources. One is used for the production of stable ion beams (e.g. C up to 100\,$\mu$A, H up to 10\,$\mu$A, depending on energy) and Accelerator Mass Spectrometry (AMS) measurements. The second ion source is dedicated to the production of offline Radioactive Ion (RI) beams, in particular $\nuclide{7}{Be}$ \cite{Gialanella2002,Limata2008}.\\*
The addition of an RF Charge Exchange Ion Source (Alphatross) will allow the production of $\mathrm{He}^-$, as well as $\mathrm{H}^-$, $\mathrm{NH}^-$, and $\mathrm{O}^-$ beams with intensities up to several $\mu$A. \\*
These sources give the opportunity to investigate processes with stable and RI beams in both fundamental research and applications.

 \subsubsection{Resident instrumentation for nuclear physics}\label{wg0:CIRCE:sub1}

The main instrumentation present at CIRCE for fundamental Nuclear Physics studies is tailored for Nuclear Astrophysics. Most of the equipment is developed and maintained by the ERNA Collaboration:
\begin{itemize}
\item ERNA Recoil Mass Separator (RMS): is one of the few recoil mass separators devoted to the measurement of nuclear cross section of astrophysical interest. 
The ERNA RMS has two options for the target system: an extended windowless gas target \cite{Schuermann2013} and a gas jet target \cite{Rapagnani2017,Brandi2020}.
Different end detectors are available: a two stages ionization chamber used as $\Delta E$-$E_{\rm res}$ telescope \cite{Rogalla1999} and a position sensitive ToF-E detector capable of charge and mass identification, respectively, of the detected particles.
An array of 18 NaI detectors can surround the jet target in a geometry optimized to measure angular distributions. The coincidence condition with recoils in the end detector provides almost background free $\gamma$-ray spectra.

\item GASTLY: a dedicated beam-line for the study of nuclear reactions relevant for astrophysics with charged particle spectroscopy is active since 2014. It is equipped with a large chamber hosting an array of $\Delta E$-$E_{\rm res}$ detectors, named GASTLY, optimized for low energy proton and $\alpha$ particle discrimination \cite{Romoli2018}. 
Each GASTLY module houses in an aluminium pyramidal case, a detector telescope consisting of an Ionization Chamber (IC) for the $\Delta E$ signal, and immediately behind the IC, a large area silicon strip detector (SSD) acting as the $E_{\mathrm{res}}$. The array is presently mounted at backward angles, $\theta_\mathrm{Lab}\sim96^\circ$ to $163^\circ$, each SSD strip subtends about $1^\circ$. The array can be rotated as a whole in $90^\circ$ steps around the vertical axis. This detection setup has been successfully used for the measurement of the $\CC$ cross section \cite{Morales-Gallegos2017}.
\end{itemize}

\subsubsection{Infrastructures and instrumentation for applied physics}\label{wg0:CIRCE:sub3}
The CIRCE Laboratory offers different opportunities with accelerated ion beams and hosts ``State of the art'' mass spectrometers routinely devoted to the measurement of stable and radioactive isotope ratios:
\begin{itemize}
\item Ion beam analysis with RBS and NRA is presently performed at CIRCE with different beams and setups. The ion beam is produced by the tandem accelerator and is purified by the AMS system, then is focused on target by means of a quadrupole magnet triplets. Targets can be mounted on a multi-sample holder. A new dedicated scattering chamber with a motorised five-axis target holder is being installed, in order to expand the available analysis techniques to Channeling-RBS and ERDA. 
\item SIRMS: Stable Isotope Ratio Mass Spectrometry measurements for light elements (i.e. H, C, N, O and S) are performed by means of 2 Thermofisher Delta V mass Spectrometers equipped with one Dual Inlet injection system for measurement of gaseous samples and 2 Continuous Flow interfaces with different peripherals (Elemental Analyser (EA), Thermal Conversion -Elemental Analysis (TC/EA), Gas Chromatograh coupled with Combustion or TC and Gas Bench)) allowing for the measurement of different matrices (solid, liquids and gases) in automated mode. 
\item MC-ICP-MS: Stable Isotope Ratio Mass Spectrometry measurements for other than light elements are performed by Thermofisher Neptune Plus Multi Collector (MC) Inductively Coupled Plasma (ICP) Mass Spectrometer (MS) allowing the simultaneous measurement of different isotpe signals with interferent (up to 10 different movable faraday cups are present) also with the possibility of using a Single Electron Multiplier to expand the measurement dynamic range. 
\item Multi Source Evaporation Machine: this device allows the production of thin films, possibly multi-layered, to realize targets or coatings. It is equipped with an Electron Beam source and a Plasma Magnetron RF Sputtering source. The possibility of producing layers of controlled thickness deposited on several materials is open to new applications. At present the machine is used to produce metallic layers on polymeric material for the windows production for gas detector used in nuclear physics application like the ERNA final detector \cite{Rogalla1999}.

The RF can be moved from the magnetron sputtering source to the sample holder, by changing its polarization, in order to change the site of sputtering process and clean the backings' surface just before the evaporation. The liquid-cooled multi target holder can host up to 10 samples. 
A glovebox, with a gas purging system with inert gases, \nuclide{}{Ar} or \nuclide{}{N}, and a dedicated window for samples loading, is connected to the evaporation machine. This allows to avoid contamination by contact with ambient air and preserve purity of the evaporated samples. \\*
In the study of nuclear reactions the production of pure, stable and uniform target is of paramount importance. The target production with this evaporator allows, right at the end of the production process, for a complete characterization of thickness, composition and uniformity. These measurements are performed with the IBA techniques (NRA, RBS, ERDA). The target analysis can be also complemented by Energy-Dispersive X-ray spectroscopy using a Scanning Electron Microscope (EDX-SEM) that allow the analysis of surface uniformity and the damage of the target.
\end{itemize}

\section{\wgonetitle}
\label{wg1}

The experimental search for possible violations of the Pauli Exclusion Principle (PEP) and the investigation of the foundational aspects of quantum mechanics (QM), including the collapse of the wave-function, are fueling the VIP collaboration's activities. These topics are interwoven and characterized by theoretical and experimental cross-fertilizations.
On one side the VIP experiment challenges those phenomenological spin-statistics violation models, like the quon model, which require experimental verification to constrain the free parameter of the theory. VIP provided the strongest bounds on this class of models, constantly improving the PEP violation probability, fulfilling the Messiah-Greenberg superselection (MGS) rule, to which they are subject.
On the other side non-commutative space-time, as a class of universality to investigate quantum gravity models, provides the suitable framework for the investigation of both fundamental symmetries and quantum collapse. 
Space-time non-commutativity induces spin-statistics violations at the fundamental level, not subject to the MGS rule. At the same time it is connected to the possibility of recovering phenomenological models for the collapse of the wave-function, while unveiling the measurement problem in QM. Indeed, phenomenological models of (gravitational) collapse of the wave-functions can be derived from a theoretical top-down approach to the quantization of geometry. 
Moreover, recent phenomenological investigations by VIP offer, for the first time, the possibility to experimentally unveil the mechanism which is responsible for the collapse. 
At the same time, both fundamental symmetries and quantum foundations, can be experimentally falsified resorting to ultra-high sensitivity probes, and in particular to atomic transitions, by the VIP collaboration at LNGS.

VIP recently excluded the $\theta$-Poincar\'e model  \cite{Piscicchia2022PRL,Piscicchia2023PRD}, one of the most commonly adopted non-commutative quantum gravity (NCQG) models, for non-vanishing electric-like components of the theta tensor. The case in which the electric-like components are taken to be zero has been excluded up to almost one tenth of the Planck scale. Consequently, the collaboration has arrived, by running a high-precision experiment and developing a phenomenological and theoretical interpretation, to test quantum gravity, and other theories beyond the Standard Model (SM), with a very high precision that is still impossible to achieve at present day using accelerators in direct experiments.

On the other hand, the experimental studies carried out by the collaboration on wave function collapse models, contribute to shed light on the measurement problem in QM, and consequently on its foundational aspects. VIP demonstrated that high sensitivity X- and gamma-rays tests, looking for the collapse induced spontaneous radiation emission, are the most sensitive probes of these models. Experimental efforts and phenomenological analyses carried out by the VIP collaboration in this context enabled to falsify \eg{} the Diósi-Penrose model \cite{Donadi2020NP} in its original cutoff-less version, and set stringent limits in its present formulation. The analysis of a novel geometric method for the stochastic quantization of fields that has been useful to bridge top-down theories of quantum gravity to the gravitational collapse of the wave-function is now under scrutiny \cite{LulliInPreparation}.

\begin{figure*}
    \centering
    \includegraphics[width=.75\textwidth]{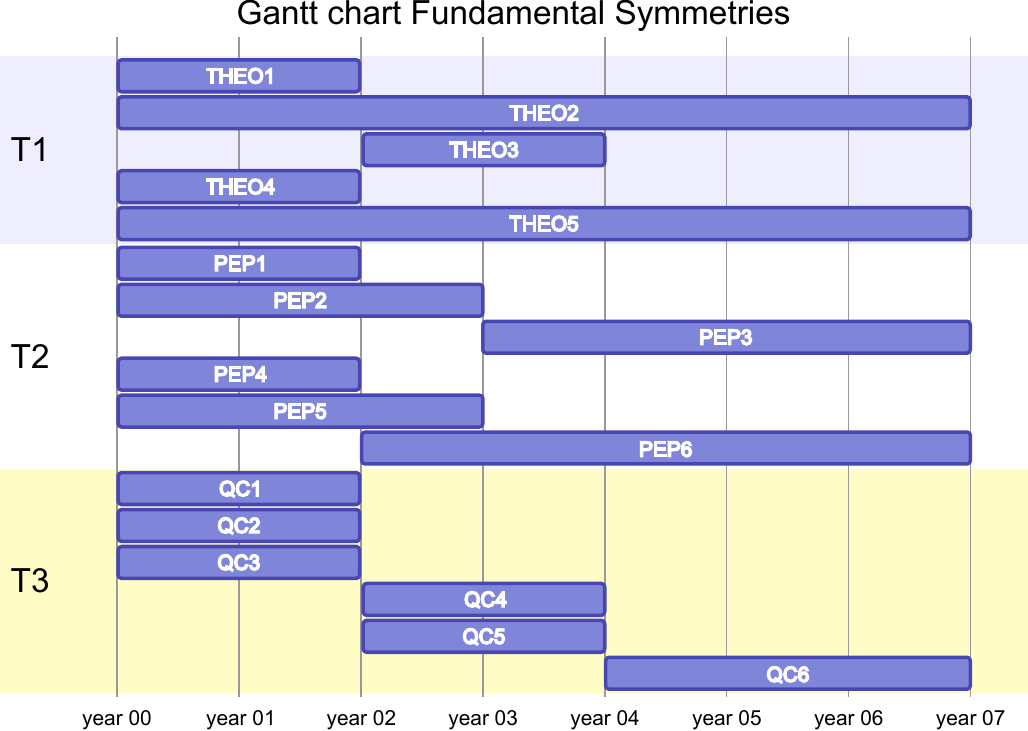}
    \caption{GANTT chart of the activities for the mid term future of WG1. T1 are the planned activities of Section \ref{wg1:t1}, T2 of Section \ref{wg1:t2}, T3 of Section \ref{wg1:t3}. In T1, THE01 refers to the task ``\textit{Explore of symmetries for PEP-violations on Open Systems}'', THEO2 to ``\textit{Inhomogeneity and anisotropy of the geometry ground state}'', THEO3 to ``\textit{Generalized Uncertainty Principle as a class of universality}'', THEO4 to ``\textit{Extend falsification of quantum gravity models}'', THEO5 to ``\textit{CPT violation and quantum decoherence, and the role of gravity}''. For T2 the PEP1 refers to ``\textit{Open-System data taking with the VIP-2 Experiment}'', PEP2 to ``\textit{Open-System experimental activity with the VIP-3 upgrade}'', PEP3 to ``\textit{Open-System activity with the 1\,mm layers Silicon Drift Detectors}'', PEP4 to ``\textit{Open-System activity with the GATOR Setup}'', PEP5 to ``\textit{Closed-System activity with the Broad Energy Germanium Detector}'', PEP6 to ``\textit{Closed-System setup with anisotropy detection}''. For T3, QC1 refers to the task ``\textit{Extension of the Collapse Models}'', QC2 to ``\textit{Study of cancellation effects}'', QC3 to ``\textit{Study of the dynamics of the spontaneous radiation at low energy}'', QC4 to ``\textit{Upgrade of the low-noise electronics}'', QC5 to ``\textit{Study of dependence on target atomic number}'', QC6 to ``\textit{Realization of a dedicated setup}''. }
    \label{wg1:fig:gantt}
\end{figure*}

The main objectives of the mid term planned activities at experimental, phenomenological and  theoretical level, are: 
\begin{enumerate}
    \item to develop and investigate possible violations of the PEP in atomic transitions, focusing on several theoretical approaches: phenomenological~\cite{Addazi2018CPC}, different classes of universality models of quantum gravity~\cite{Addazi2020EPJC}, considering various theoretical possibilities~\cite{Piscicchia2022PRL,Piscicchia2023PRD}, including the deformation/violation of fundamental symmetries~\cite{Amelino2007FPS}, the emergence of non-locality effects at high-energy scales~\cite{Arzano2008PRD}, claimed departures from unitarity~\cite{Addazi2020IJMPA} as well as the emergence of effects manifesting the existence of extra-dimensions;
    \item to fully exploit the ultra-sensitivity on a broad energy range, and directionality capability in the detection of electromagnetic radiation that will be probed by the collaboration,  hence making the difference, with respect to similar attempts in the literature, in constraining very tightly the parameter space of the examined models;
    \item to connect classes of universality of quantum gravity models that have been distinguished according to their possible implications to the fate of the PEP, as well as further classes of universality of quantum gravity models added to phenomenologically account for the collapse of the wave-function, to the possibility of testing foundational aspects of QM, through the experimental data available on the measurement problem;
    \item to recover a solid theoretical link between models of quantum gravity (and related universality classes) and their impact on the foundational aspects of QM that can be experimentally falsified, hence prospecting a new way to develop quantum gravity phenomenology.    
\end{enumerate}
In Section \ref{wg1:t1} a theoretical overview on the symmetries, theories and models which will be developed in the mid term future is provided (T1). Section \ref{wg1:t2} focuses on experimental activities planned to investigate the possible violation of the PEP (T2), while Section \ref{wg1:t3} provides a similar plan for experiments devoted to quantum collapse models (T3). Figure \ref{wg1:fig:gantt} reports the overall plan, see caption for details.


\subsection{\wgonetonetitle}
\label{wg1:t1}

The search for possible violations of the PEP has hitherto enabled the VIP collaboration to exclude several scenarios of non-commutative space-time \cite{Piscicchia2022PRL,Piscicchia2023PRD} that represent different classes of universality for models of quantum gravity. This result has been achieved by considering PEP-violating atomic transitions, probed by the VIP-2 experiment, to arise from the deformation of the space-time symmetries that eventually would induce deformations of the spin-statistics theorem.

From the experimental side, this scenario can be further investigated by improving the statistics of the data acquired within the next upgrades of the experiment, and searching for the presence of background directional effects arising from the eventual non-commutative structure of space-time.  From a theoretical perspective, either departures from unitarity or non-locality, proper to some attempts of quantum gravity models, or the possible imprinting of extra dimensions at ``our dimensions'', or the deformation of the symplectic structures of the theory that may result in a generalization of the Heisenberg uncertainty principle, might as well be responsible for PEP violating atomic transitions. 

Top-down approaches in high energy quantum field theory have been also developed to connect models for the gravitationally induced collapse of the wave function to quantum gravity theories, hence seeking to disentangle a longstanding issue in foundations of QM, such as the measurement problem. Among several models existing in the literature, interesting theoretical and phenomenological developments have been achieved in connection to the non-relativistic limit of the Stochastic Ricci Flow approach \cite{LulliInPreparation}. 
This is related to the notable model for the (gravitationally induced) collapse of the wave-function that goes under the names of Diosi and Penrose \cite{Diosi1987PRA,Diosi1989PRA,Penrose1996SPR,Penrose2014SPR}. The Stochastic Ricci Flow approach involves the breakdown of the fundamental symmetries of space-time and is consequently inextricably interwoven to the previous discussed topics of investigation of the fundamental space-time symmetries. Indeed, the decohering effect of the stochastic gravitational noise is at the base of both the collapse of the wave function, as driven by the Ricci drift term, as well as the breakdown of the space-time symmetries in out-of-equilibrium configurations.


The mid term aim of the collaboration is to further develop the phenomenology of fundamental symmetries, including discrete and continuous space-time symmetries, and to study the foundational aspects of QM, including the measurement problems and related studies on the collapse of the wave-function of quantum systems, as basis of future  experimental upgrades. These latter include electromagnetic radiation spectra in the keV-MeV energy range, as well as related measurements of atomic transitions. More specifically, we plan:
\begin{enumerate}
\item to enhance the understanding of physics at the Planck scale, while testing a fundamental principle of nature such as the PEP, with very high precision and at microscopic scales hitherto inaccessible;
\item to solve longstanding issues related to the interpretation of the measurement process in QM, from a statistical-mechanics and complex system perspective that is relevant to quantum many body systems and quantum information theory. This is related to decoherence of entangled quantum systems and to the appearance of stochastic noise of gravitational origin in the out-of-equilibrium dynamics.
\end{enumerate}

We then aim at unveiling the ultimate consequences of quantum space-time fuzziness, modeled as stochastic noise, to the out-of-equilibrium dynamics of quantum-many-body systems, then to the decohering effects and the collapse of the wave-functions of the systems experimentally under scrutiny.

The outcomes of these combined experiment, phenomenological analyses and theoretical investigations will then either unveil a first signal of physics beyond the SM, or rule out entire classes of quantum gravity models, with related consequences for the quantum measurement problem and models of collapse of the wave function, thus representing a milestone in the investigation of the fundamental symmetries and of the foundational aspects of QM.

Resorting to its consolidated expertise on this matter, the collaboration intends to develop detailed phenomenological analyses of the consequences of the constraints recovered by the future upgrades of the experimental apparatus on the atomic transitions that violate the PEP. These limits will be used to constrain the parameter spaces of several models of quantum gravity, using universality classes of non-commutative space-time, as well as stochastic models of quantization of gravity and concrete realizations of quantum gravity scenarios that either violate non-locality or deform the symplectic symmetry structure of the theory.

\begin{figure}
    \centering
    \includegraphics[width=.6\textwidth]{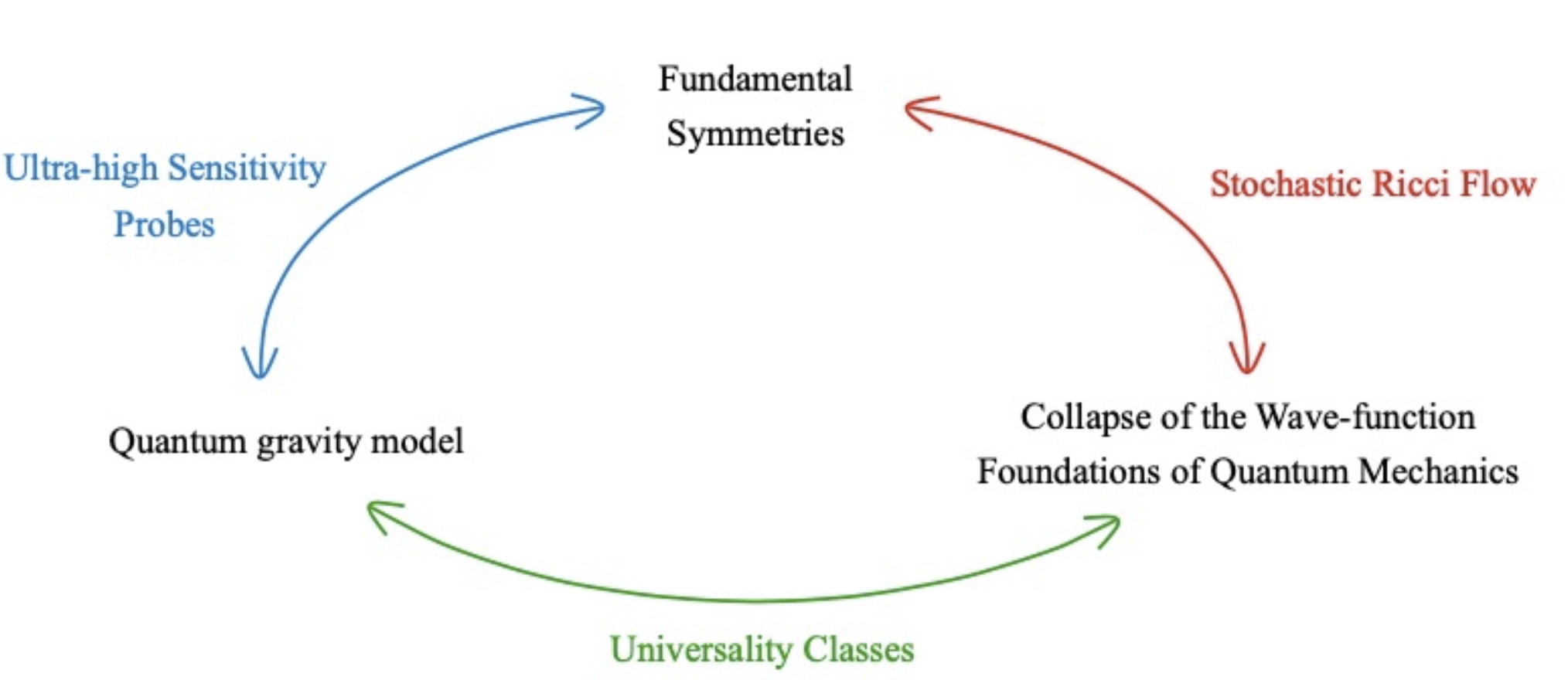}
    \caption{Main objective of the mid term plan in fundamental symmetries.}
    \label{wg1:t1:fig:fig1}
\end{figure}

\bigskip
Therefore, the main objectives inspiring the program of the mid term group activities  provide a circular feedback, intertwining experiment, phenomenology and theory, as shown in Figure \ref{wg1:t1:fig:fig1}.
In particular, the activities in the mid term plan for T1 are as follows.\\*
The THEO1 activity will focus on the exploration of symmetries for PEP-violations on Open Systems (year 1 to year 2). The collaboration will continue developing a detailed phenomenology of non-commutative space-time models, investigated as a class of universality of quantum gravity. \\*
Tighter constraints on these universality classes will be disentangled from the stronger limits experimentally recovered, while information about directionality of the electromagnetic radiation emitted will enable to test possible quantum non-commutative background anisotropies that have been predicted THEO2 (year 1 to year 7).

In this area, another class of universality worth of investigation is represented by those models that break/deform the symplectic symmetry, which will allow not only to connect PEP-violations to space-time non-commutativity, but also more in general to possible subsequent (THEO3, year3 to year 4) generalizations of the Heisenberg uncertainty principle, allowing to falsify numerous theoretical models yielding prediction on this.\\*
In the THEO4 activity (year 1 to year 2), an extension of falsification of quantum gravity models will also be done. The class of universality of non-local quantum field theories will be addressed in great detail. Specifically, a connection between non-local quantum field theory approaches to the quantization of gravity — in particular super-renormalizable theories of gravity, the non-locality of which is induced by the introduction in the action of the theory of form factors expressed by transcendental functions — and possible breakdown/deformation of the spin-statistic theorem, will be investigated in order to derive falsifications of this universality class of quantum gravity models, in which also specific realizations of string theory and causal triangulations fall.

The Stochastic Ricci Flow is a tool for the stochastic quantization of the gravitational field that is complemented with the conformal transformations generated by the classical Ricci (minus the matter Ricci target) drift term --- in other words, the Einstein equations reached classically at equilibrium. This general scheme was introduced over the last four years by a member of the collaboration, in light of the possible implications for quantum cosmology, in order to go beyond the frozen-time formalism, and more in general for the renormalization group flow of the gravitational field. Nonetheless, the same framework has been proved to have profound consequences for the very foundations of QM, for both the deep and intimate link of the stochastic quantization approach to Nelsonian models of QM  \cite{Nelson1967PUP,Nelson1966PR,Guerra1973PRL,Smolin2006Arxiv} and the phenomenological models for the gravitationally induced collapse of the wave-function of quantum systems. While the collaboration aims at clarifying the possible observational signatures of the dynamical relaxation toward the fundamental symmetries at equilibrium, as predicted by the Stochastic Ricci Flow scheme, it will be also relevant for this research line of our planned activities to achieve a solid establishment of the link between this geometric stochastic quantization method, seen as a class of universality of quantum gravity models, and the phenomenological models of wave-function collapse (THEO5, year 1 to year 7).\\*
This will open the pathway to a novel direction to test quantum gravity models through the tight constraints that will be derived experimentally on the models of wave-function collapse from the limits on the accompanying electromagnetic radiation that is supposed to be emitted.

\subsection{\wgonettwotitle}
\label{wg1:t2}

The experimental investigation of fundamental symmetries, through high sensitivity tests of Spin-Statistics violation/deformation, follows two parallel lines which are named VIP Open Systems and VIP Closed Systems, as a direct consequence of the two distinct classes of theoretical models which embed Spin-Statistics violation/deformation. 
For both VIP Open Systems and VIP Closed Systems the searched new-physics experimental signature is represented by PEP violating atomic transitions to the fundamental level, when this is eventually already filled by two electrons. The energy of this transition is expected at a different energy with respect to the standard one. The calculation of the Pauli-forbidden radiative-transition
energies is independent - at the first order - from the specific model and is performed by using the MCDFGME numerical
code \cite{Desclaux1975CPC}. This program solves the multiconfiguration Dirac-Fock equations self-consistently, taking into account relativistic
effects (see \cite[and references therein]{Shi2018EPJC,Okun1987JL}). 

Attempts of formulating theoretical models which violate the statistics of identical particles were pioneered by Fermi \cite{Fermi1934SC,Milotti2007AX}, who discussed the implications of an even tiny non-identity of electrons.  Gentile introduced an intermediate statistics \cite{Gentile1940INC}, while a parastatistics was developed by Green \cite{dellantonio1964NY}.  
Ignatiev and Kuzmin presented a model consisting in a deformation of the standard Fermi oscillator \cite{ignatiev1987QRK,ignatiev1987YF,Ignatiev2006RPC}, also discussed by Okun \cite{Okun1987JL}. In this approach, a three level Fermi oscillator is considered, in which the additional level  can be accessed with a probability $\beta^2/2$. $\beta$ is still currently used in this research field to represent the amplitude of a PEP violating transition. Rahal and Campa investigated the consequences of small PEP violations on the thermodynamic properties of matter \cite{Rahal1988PRA}. Greenberg and Mohapatra \cite{Greenberg1987PRL, Greenberg1990PRL} formulated a Local Quantum Field theory embedding PEP violation, named \emph{quon model}. We refer to Ref. \cite{Piscicchia2020EPJC} for a more detailed review. 
The experimental investigation of the above-mentioned theories is strongly demanded to constraint the free parameter of the models, but is complicated by a stringent condition known as the MGS rule \cite{Messiah1964PR}: the transition amplitude between two different symmetry states is zero.
VIP Open Systems sets the most stringent bounds on $\beta$ respecting MGS~\cite{Napolitano2022SY}, by checking the newly formed symmetry state, which follows the introduction of new fermions in a given system of identical fermions, according to a suggestion by Greenberg and Mohapatra \cite{Greenberg1987PRL}. The present status and results of the VIP-2 Open Systems experiment are outlined in Section \ref{wg1:t2:sec:vip2}.

More recent is the prediction of Spin-Statistics deformation related to the space-time non commutativity \cite{Addazi2018CPC,Addazi2020IJMPA,Balachandran2006IJMPA,Balachandran2007PRD}, common to several Quantum Gravity frameworks (the two main classes of NCQG are $\kappa$-Poincar\'e and $\theta$-Poincar\'e) or connected to CPT deformation \cite{Mavromatos2018EPJ}. This class of models predicts PEP violation, not constrained by the MGS. Depending on the specific model, the PEP violation probability $\delta^2$ is a function of the energies involved in the transition under study. It depends on the characteristic energy scale of new-physics emergence, and can be subject to not isotropic corrections.
The present status and results of the VIP Closed systems experiment, which is investigating this rich new phenomenology, are described in Section \ref{wg1:t2:sec:closedpresent}.   

\subsubsection{VIP-2 Open Systems, present status and results}\label{wg1:t2:sec:vip2}

The goal of the ongoing VIP-2 experiment is to improve by at least two orders of magnitude the result obtained by VIP ($\beta^2/2 < 4.7 \times 10^{-29}$ \cite{Curceanu2011JP,sperandio2008TV}). 

The main improvements of the experimental apparatus (see Refs. \cite{Shi2018EPJC,Curceanu2017EN} for a detailed description) consist in: $a)$ the replacement of the charged coupled devices for the X-rays detection, with state-of-the-art, 450\,$\mu$m thick, Silicon Drift Detectors (SDDs), characterized by a higher energy resolution (190\,eV (FWHM) at 8\,keV), large geometrical acceptance and an efficiency of 99$\%$ at 8\,keV, $b)$ a more compact and thinner target ensuring higher acceptance and efficiency, $c)$ a new target cooling system which allows an enhanced circulating current (with a peak value of 180\,A with respect to the 40\,A of VIP). After an exploratory data taking run (2016-2017) \cite{Curceanu2017EN,Shi2018EPJC} exploiting two arrays of 1$\times$3 SDDs, the fully upgraded setup was completed (2018-2019), by mounting four arrays of 2$\times$4 SDDs cells and the external passive shielding complex (an outer lead layer surrounding an internal copper layer), aimed to provide further suppression of the environmental background radioactivity. The VIP-2 experiment has been operating in its final configuration since May 2019, by alternating data taking periods with current flowing to periods without current, which provide the reference background spectrum.    

The sensitivity of the VIP-2 experiment was demonstrated by a progressive approach of the $\beta^2/2$ limit to the foreseen goal \cite{Curceanu2017EN,Shi2018EPJC,Piscicchia2020EN}. The strongest bound on the PEP violation probability, consistent with the MG superselection rule (see Ref. \cite{Napolitano2022SY}), was recently achieved by VIP-2, from the analysis of the data corresponding to about six months of experiment operation in its final configuration. Two analysis frameworks were followed, a Bayesian statistical model and a frequentist Confidence Levels (CL$_s$) approach, which share the same spectral shape description for the signal and the control spectra. The obtained limits are found to be well consistent within one sigma. The analyses yield:
\begin{equation}
    \beta^2/2 \leq 8.6 \times 10^{-31} \,\,  \textrm{(Bayesian)} \qquad
    \beta^2/2 \leq 8.9 \times 10^{-31} \,\,  \textrm{(CL}_s\textrm{)},
\end{equation}
when the electron propagation in the target is described by means of an electron diffusion model \cite{Ramberg1990PLB}, i.e. the number of electron-atom interactions is obtained from the ratio of the target length and the electrons scattering length in copper. According to more realistic diffusion models \cite{Milotti2018EN,Milotti2020SY}, on which we are recently working, electron-atom interactions in copper occur over a characteristic time $\tau=3.3\times 10^{-17}$\,s, therefore significantly increasing the number of independent PEP tests performed by each current electron, leading to the enhanced limits:
\begin{equation}
    \beta^2/2 \leq 6.8 \times 10^{-43} \,\,  \textrm{(Bayesian)} \qquad
    \beta^2/2 \leq 7.1 \times 10^{-43} \,\,  \textrm{(CL}{}_s\textrm{)}.
\end{equation}

\subsubsection{VIP Closed Systems, present status and results}
\label{wg1:t2:sec:closedpresent}

The analysis of the total data set collected by the VIP Closed System experiment recently provided the strongest (atomic-transitions) experimental test of the $\theta$-Poincar\'e NCQG model \cite{Piscicchia2022PRL,Piscicchia2023PRD}, looking for anomalous atomic transitions in the X-ray domain.

The experimental setup is based on a co-axial p-type HPGe detector, about 2\,kg in mass. The detector is surrounded by a target, consisting of three 5\,cm thick cylindrical sections of radio-pure Roman lead, for a total mass of about 22\,kg (see Ref. \cite{Piscicchia2020EPJC} for a detailed description of the apparatus and the acquisition system).

The first phenomenological analysis was performed, which accounts for the predicted energy dependence of the PEP violation probability.  
Upper limits are set on the expected signal of PEP violating K$_\alpha$ and K$_\beta$ transitions, generated in the target, by means of a Bayesian comparison of the measured spectrum with the violating K complex shape predicted by the $\theta$-Poincar\'e model. 

The analysis yields stringent bounds on the non-commutativity energy scale, which exclude $\theta$-Poincar\'e up to $2.6\times10^2$ Planck scales when the ``electric-like'' components of the $\theta_{\mu \nu}$ tensor are different from zero, and up to $6.9\times10^{-2}$ Planck scales if they vanish.

\subsubsection{VIP Open Systems future plans}\label{wg1:t2:sec:opensys}

The mid term plans for T2 on VIP Open System research line over year 1 to year 2 aims at concluding the data taking with the VIP-2 detector (PEP1) and preparing for the experimental upgrade.

The future of the research for MG allowed PEP violation, demands for a scan of the PEP violation probability - with sensitivity comparable to VIP-2 - as a function of the atomic number.
The importance of ``such tests, over the entire periodic table'' was outlined \eg{} in Ref. \cite{Okun1987JL}. 
The main technical challenge to be faced, when testing PEP-violating atomic transitions for metals characterized by a higher atomic number than copper, is represented by the SDD detectors quantum efficiency decrease, as a function of the increasing energy to be detected. The scan will be performed by means of the VIP-3 apparatus, using innovative 1\,mm thick SDD detectors, till atomic numbers as high as $Z\sim50$ as described in the following. Intermediate atomic number materials will be tested exploiting a layered architecture of 1mm thick SDDs (see Section \ref{wg1:t2:sec:layer}). The higher $Z$ range will be explored by the VIP-GATOR collaboration using the low-background GATOR germanium counting facility~\cite{baudis2011JINST}, with the implementation of dedicated ultra-pure target systems (see Section \ref{wg1:t2:sec:vipgator}). 

\paragraph{The VIP-3 experiment}\label{wg1:t2:sec:vip3}

\begin{figure}[h!]
\centering
\includegraphics[width=.5\textwidth,clip]{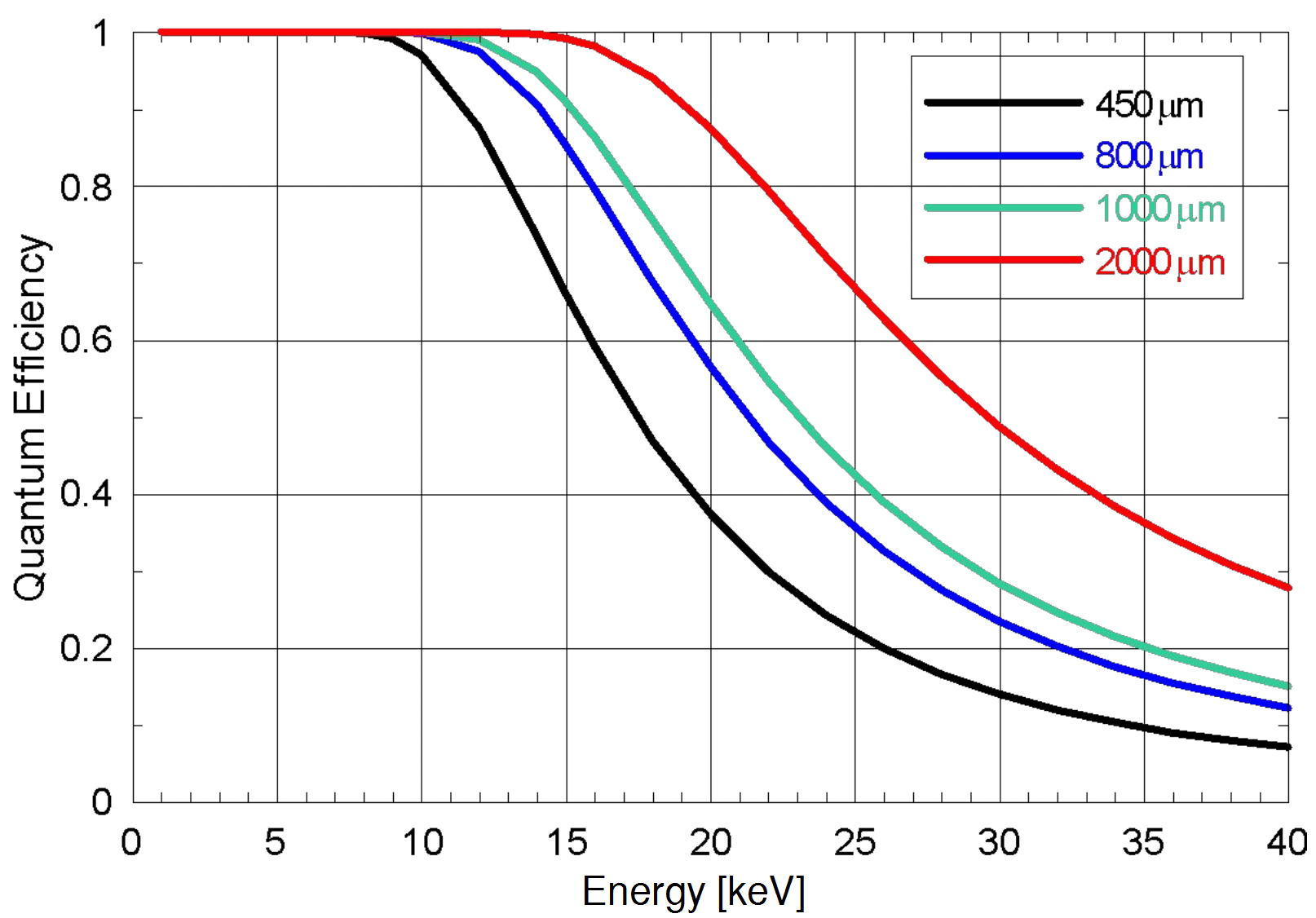}
\caption{The figure, from \cite{Piscicchia2022APPA}, shows the simulated quantum efficiency as a function of the energy, for SDD devices of various thicknesses. The black curve corresponds to the detectors currently used in VIP-2, the green curve shows the efficiency achievable with the new 1\,mm thick SDDs which we are presently developing for the VIP-3 experiment.}
\label{wg1:t2:fig:efficiency}
\end{figure}

The VIP-3 represents the WG effort on T2 for the year 1 to year 3 (PEP2).
Our group is presently developing new cutting edge SDD detectors in collaboration with Fondazione Bruno Kessler (FBK, Italy) and Politecnico di Milano (PoliMi, Italy). The new technology is characterized by double thickness, with respect to
the standard detectors used in the VIP-2 experiment
(1\,mm versus 0.45\,mm). We have already demonstrated (see Figure \ref{wg1:t2:fig:efficiency}) that the quantum efficiency of the new SDDs, in the energy range $(20 - 25)$\,keV, is roughly double with respect to the standard detectors, keeping constant the energy resolution. This will allow to investigate eventual PEP violation induced deviations from the standard K$_{\alpha}$ transitions, in palladium, silver and tin (see also Ref. \cite{Piscicchia2022APPA}).
As an example, in silver the PEP violating K$_{\alpha1}$ transition is shifted of 482\,eV with respect to the standard line, the corresponding shift for the K$_{\alpha2}$ is 478\,eV. Comparable shifts are found in palladium and tin.

\begin{figure}[h!]
\centering
\mbox{\includegraphics[width=.6\textwidth,clip]{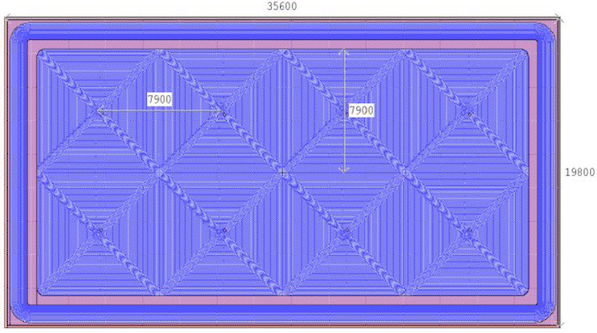}}
\caption{The figure, from \cite{Piscicchia2022APPA}, shows as seen from the anode side, the layout of the main SDDs array being produced for the VIP-3 experiment .}
\label{wg1:t2:fig:array}
\end{figure}

The production of the new SDD devices is presently under finalization. The new system is characterized by pixel dimensions of 7.9\,mm$\times$7.9\,mm, the width of the last ring was extended in order to improve collection at the border of the active area. The total chip dimensions are 35.6\,mm$\times$19.8\,mm \ie{} about 2\,mm wider than the previous chips. The geometry of the SDD arrays will consist of a 2$\times$4 matrix, whose anode side is shown in Figure \ref{wg1:t2:fig:array}. Among the main improvements characterizing the new detector system we want to mention the introduction of a layout solution on the window-side, to reduce the charge sharing effect, moreover the robustness of the bonding pads was enhanced \cite{Piscicchia2022APPA}.

During PEP2, the activity will proceed with the finalization of the setup. More in detail the layout of the compact targets-SDD detectors system is shown in Ref. \cite{Piscicchia2022APPA}. The planned configuration consists of 8 SDD arrays, facing two target strips, where the direct current will be circulated. With respect to the 4 SDD arrays presently arranged in the VIP-2 setup, VIP-3 will exploit a total of 64 SDD cells, for a double active area of about 41\,cm$^2$, in order to increase the geometrical efficiency. A new calibration system will be developed; moreover a new thermal contact between the cold-finger and the SDD detectors, and a new target cooling system will be built, both in OHFC ultra-pure copper, to minimize the natural copper radio-contamination. The improved thermal conductivity, with respect to the steel thermal contact and cooling system presently used in VIP-2, will reduce the detectors working temperature (improving the energy and timing resolution) and will also increase the applicable maximum current circulating in the target (from the 180\,A peak current of VIP-2 up to 400\,A). 
The design of the new vacuum chamber is also shown in Ref. \cite{Piscicchia2022APPA}.

This will be followed by assembly, testing, debugging and mounting of the setup and of an upgraded shielding, implementing a further inner polyethylene tier, at LNGS. 

The introduced improvements listed above compensate the quantum efficiency reduction for increasing the energy from 8\,keV up to 25\,keV, thus keeping at least constant the sensitivity of the experiment.  

The next 3 years period will be dedicated to data taking, alternating runs with and without current, and to the analysis of the collected data.

\paragraph{Beyond VIP-3: 1 mm thick SDDs}\label{wg1:t2:sec:layer}

After the experimental upgrade VIP-3 (PEP2), from year 4 to year 7, the Open System activities will focus on the 1mm thick SDDs (PEP3).
In fact, the search for PEP-violating atomic transitions for elements characterized by intermediate atomic numbers ($Z\sim60$), with sensitivity comparable to VIP-2, requires a further improvement in the Quantum Efficiency of the detectors. This will be obtained by means of the development of layered structures of 1\,mm thick SDDs. 
For higher $Z$ elements the shift of the violating K$_{\alpha}$ lines, with respect to the standard transition, becomes comparable or larger than the HPGe detectors energy resolution, hence an HPGe based setup becomes preferable (see Section \ref{wg1:t2:sec:vipgator}).  

Research activity on a layered architecture of 1\,mm thick SDDs (where two SDDs are placed one above the other) will be performed during year 3. The period year 6 to year 7 will be dedicated to the development of the layered structures of 1\,mm thick SDDs, to the design and realization of the improved experimental apparatus (new SDDs system cooling block, new vacuum flange connectors (more pins) to bring the signals out to the DAQ, etc.). We then plan a data taking period of three years, alternating periods with current circulating in the targets to periods without current, energy calibration runs, continuous maintenance of the VIP laboratory space and of the setup.

\paragraph{VIP-GATOR: Open Systems test for high Z elements}\label{wg1:t2:sec:vipgator}

During 2022 a joint-effort started between the VIP collaboration and the ETH Zurich group (lead by Prof. Laura Baudis) aimed at the realization of high sensitivity measurements of $\beta^2/2$, fulfilling the MGS, in the high $Z$ range of the periodic table. 
This will articulate an experimental effort from year 1 to year 2 (PEP4).

The experiment will use the GATOR facility~\cite{baudis2011JINST} at LNGS, with the implementation of a dedicated target system composed by: high radio-purity materials (Pb, Au, Pt, \dots) targets fed by a 400\,A direct current power supply, target cooling system and feed-through flange. GATOR is a low-background germanium counting facility with a core p-type coaxial HPGe detector with 2.2\,kg sensitive mass. The energy resolution of the detector is 1.1\,keV FWHM at 74.96\,keV, corresponding to the standard K$_{\alpha1}$ line in Pb. The energy shift among the standard and the PEP violating K$_{\alpha1}$ transitions is 1.25\,keV.   

An exploratory measurement was performed above ground, as reported in \cite{Elliott2012Foundations}, exploiting a point-contact Ge detector and a Pb target circulated by a 110\,A current, and the limit $\beta^2/2 < 1.5 \times 10^{-27}$ was established. 

\begin{figure}[h!]
\centering
\includegraphics[width=.6\textwidth,clip]{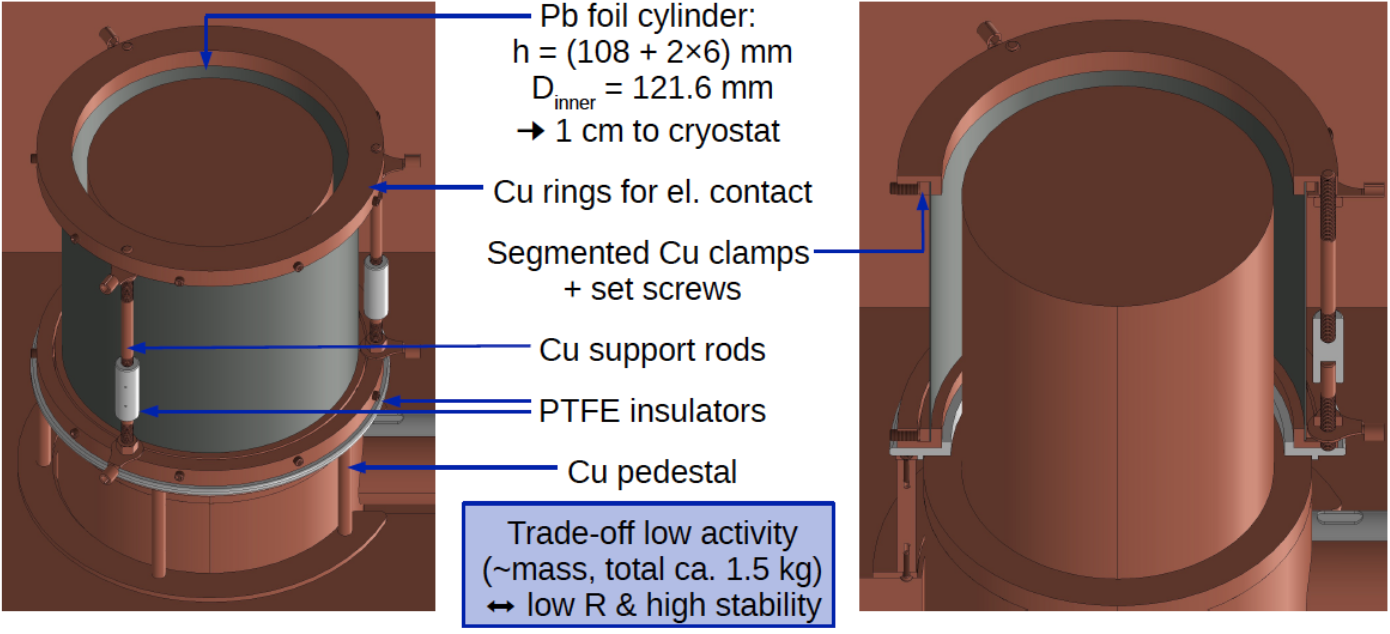}
\caption{Design of the optimized inner configuration of the VIP-GATOR experiment detector system, figure courtesy of Alexander Bismark.}
\label{gatortarget}
\end{figure}

Geant4 Monte Carlo simulations were performed with the aim to optimize the target geometry and thickness, in order to maximize the photons detection efficiency and the maximum circulating current which is compatible with power dissipation. A design of the optimized inner configuration of the detector system is shown in Figure \ref{gatortarget}; to give an example the thickness of the cylindrical Pb target is 25\,$\mu$m. 

The realization of the Pb target is presently ongoing. The design of the feed-through flange, of the cabling/connectors and of the target/cables cooling system are presently under finalization. Based on the performed MC studies a three months measurement of the VIP-GATOR experiment will improve the result in \cite{Elliott2012Foundations} of at least two orders of magnitude. 

The finalization of the setup realization and three/four months of data taking with Pb target are previewed for year 1. A scan over several high $Z$ targets will be performed in the year 2.

\subsubsection{VIP Closed Systems future plans}
\label{wg1:t2:sec:closed}

The mid term future experimental activity in the Closed Systems context is aimed to extend the falsification of Quantum Gravity models, to perform tests of the Generalized Uncertainty Principle, to investigate the interplay among CPT violation and quantum decoherence and the role of gravity. This will be accomplished by means of a Broad Energy Germanium detector (BEGe) based system, as described below, during year 1 to year 3 (PEP5). The long-term plan will be addressed to the search for signal of inhomogeneity and anisotropy of the geometry ground state in year 2 to year 7 (PEP6).

\paragraph{VIP Closed Systems - BEGe based setup}
\label{wg1:t2:sec:bege}

The falsification of the $\theta$-Poincar\'{e} NCQG model, for vanishing $\theta_{\mu\nu}$ ``electric-like'' components, requires the improvement of the current lower limit on the non commutativity scale $\Lambda>6.9 \times 10^{-2}$ \cite{Piscicchia2022PRL,Piscicchia2023PRD} of a factor 15. Considered that the PEP violation probability is given by:
\begin{equation} \label{wg1:t2:eq:cadu}
\delta^ 2 \simeq \frac{\bar{E}_1}{\Lambda} \frac{\bar{E}_2}{\Lambda}\enspace,
\end{equation}
where $\bar{E}_{1,2}$ are the energy levels occupied by the initial and the final electrons, the strategy is to use the Germanium as an active target, and to search for K$_{\alpha}$ violating transitions in the Ge crystal itself. Considered the energy of the levels in Ge and Pb, the sensitivity of the experiment has to be improved of a factor $s \sim 2\times 10^4$. This will be accomplished exploiting the incremented efficiency, which is $\epsilon_{Pb} = (5.39 \pm 0.11) \times 10^{-5}$ for the detection of a PEP violating K$_{\alpha_1}$ transition occurring in the Pb target of our previous experiment, and is $\epsilon_{Ge} \sim 1$ for the detection of a PEP violating K$_{\alpha_1}$ transition generated in the Ge crystal. The main difficulty is represented by the high background level, due to electronic noise,  which characterizes the HPGe detectors below 20\,keV, which makes impossible the measurement of the PEP violating Germanium K$_{\alpha}$ lines. The strategy is to exploit a BEGe based setup, which allows, by means of pulse shape discrimination techniques, to reject the electronic noise and to disentangle photons produced inside the crystal from photons impinging from outside the detector. 

\begin{figure}[h!]
\centering
\includegraphics[width=.6\textwidth,clip]{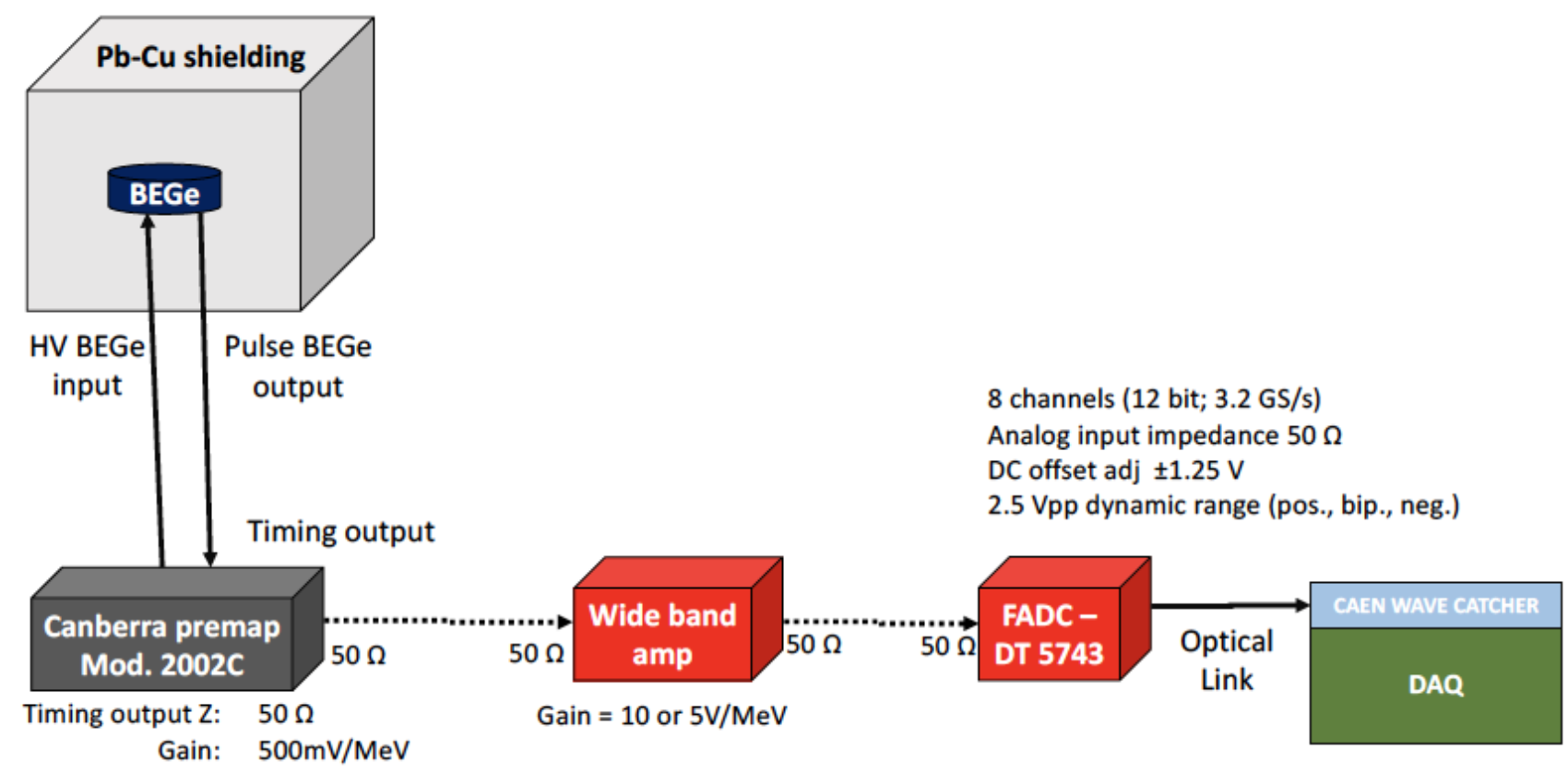}
\caption{Block diagram of the BEGe base system.}
\label{wg1:t2:fig:begefig}
\end{figure}

A test setup was already realized and two data taking runs were performed in the period 2021-2022, which served for the DAQ optimization. A block diagram of the BEGe base system is shown in Figure \ref{wg1:t2:fig:begefig}. The BEGe detector is cooled to 80\,K by means of LN$_2$, contained in a dedicated dewar, and is surrounded by two layers of Cu and Pb shielding. The CAEN FADC is characterized by a 12\,bit resolution and a 3.2Gs/s sampling rate, which allows to reconstruct the shape of the incoming signals. A wide band low noise amplifier, with a gain of a factor 10 in voltage, serves to increase the amplitude of the BEGe signals corresponding to photons of few keV. An extremely-low noise power supply was realized for the digitizer and the amplifier. 

\begin{figure}[h!]
\centering
\includegraphics[width=.9\textwidth,clip]{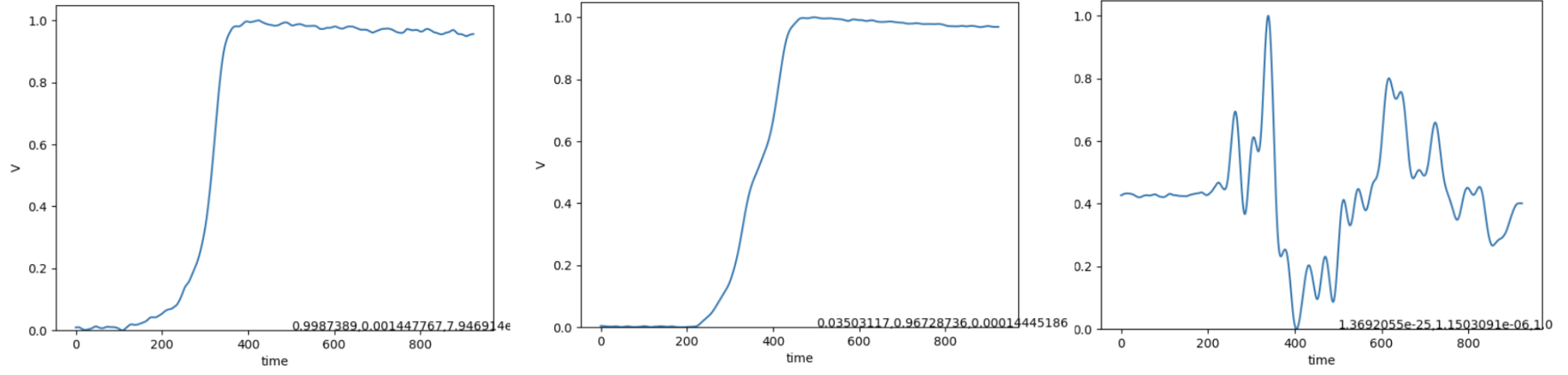}
\caption{Examples of pulse shapes from physics data from the BEGe detector, the vertical axis reports the normalized pulse amplitude, and the horizontal axis the time in AU. (Left) Single-site event, in which the energy is deposited in a single location within the detector, identified with 99\% accuracy. (Center) Multi-site event, in which the energy is distributed over multiple locations, identified with 97\% accuracy. (Right) Noise event, representing background radiation or other sources of interference, identified at 99\% accuracy.}
\label{wg1:t2:fig:cnn}
\end{figure}

We have developed and tested a machine learning based pulse shape discrimination (PSD) procedure. While dedicated algorithms for PSD require extensive fine-tuning  and calibrations, deep Learning techniques have already been pioneered for the GERDA and MAJORANA experiments \cite{Holl2019EPJC} showing the potential to match the state-of-the-art algorithms. We applied convolutional neural networks (CNN) for discriminating between single site events, in which the energy is deposited in a single location within the detector, and multi-site events, in which the energy is distributed over multiple locations. A CNN was developed and trained on a dataset of simulated pulse shapes from single site and multi-site events, representative samples are shown in Figure \ref{wg1:t2:fig:cnn}. The CNN performance was evaluated by using various metrics and our results show that is able to accurately discriminate between single site and multi-site events, with a classification accuracy of better than 95\%.

For PEP5, the period year 1 to year 2 will be dedicated to data taking and to the analysis of the collected data in the context of NCQG, CPT violation and Generalized Uncertainty Principle theories. 
Additionally, during year 3 to year 4 several data taking campaigns will be performed alternating various targets, to also explore the energy dependence of the $\delta^2$ which is predicted in NCQG. 

\paragraph{Test of Quantum Gravity induced anisotropy effects}\label{wg1:t2:sec:direction}

The longer term plan experimental activity on fundamental symmetries in Closed Systems will be focused on the first time ever investigation of Quantum Gravity induced anisotropy effects. The experimental setup will be capable of probing the directionality of the electromagnetic field background emission and will allow to recover tighter limits on space-time non-commutativity, testing possible tensor background effects that are peculiarly expected in string theory. These effects are predicted because of the non-vanishing vacuum expectation value of anti-symmetric background fields, which is at the very origin of space-time non-commutativity, and hence of the deformation of the fundamental symmetries.

The experiment will measure directionality of the electromagnetic radiation emitted in atomic transitions exploiting a 4$\pi$ detection geometry. 
The central targets will be surrounded by layered structures of 1\,mm thick SDDs. The targets will be embedded in magnetic fields which will introduce the directionality-dependence, necessary to unveil a characteristic and still unexplored signature of the quantum gravity predictions. The setup will be surrounded by a complex of pure lead, electrolytic copper and polyethylene shielding system and by a performant veto system against residual cosmic and environmental radioactivity background.

The previewed experimental program PEP6 will have a duration of five years. The first two years will be dedicated to the R\&D of a new setup, one year will serve for the realization and test of the setup, which will be followed by two years of data taking. 

\subsection{\wgonetthreetitle}
\label{wg1:t3}

The wave function collapse, or the measurement problem in QM refers to the conflict between the deterministic evolution of a quantum system described by the Schrödinger equation, and the stochastic collapse of the wave function that occurs as a consequence of the measurement. This issue was first brought to light by Schrödinger's famous thought experiment \cite{schrodinger1935NW}, known as the Schrödinger's cat paradox, which highlighted the contradiction between the microscopic and macroscopic realms in QM.

The Schrödinger equation describes the evolution of a quantum system in terms of a linear, deterministic differential equation. However, when a measurement is performed on a system in superposition of different states, the wave function abruptly collapses to a single eigenstate of the observable being measured. This process, known as wave function reduction, is non-linear and stochastic in nature and has been postulated as an ad-hoc addition to the Schrödinger equation to account for the lack of superposition in macroscopic objects.

At a fundamental level, several questions arise circa the foundation of QM. Why and how do we have a boundary between the two dynamics? Why the quantum properties of microscopic systems, \eg{} the possibility of being in a superposition of different states at once, do not carry over to larger objects? Fundamental question have a relevance also in light of the recent advancements in quantum technologies, which make use of quantum superposition. Will an isolated system manifest linear and deterministic Schrödinger evolution forever?

Spontaneous collapse models propose dynamics able to explain the breakdown of the linear, deterministic evolution into the collapse \cite{Leggett1980PTPS,Weinberg2014OKML,Bell2004CUP,Ghirardi1986PRD,Adler2004CUP,Weinberg2012PRA}. The most discussed in literature are the Diósi-Penrose (DP) and Continuous Spontaneous Localization (CSL) models.
In the DP model \cite{Penrose1996SPR,Penrose2014SPR,Howl2019NJP,Diosi1987PRA,Diosi1989PRA}, gravity plays a major role. As soon as a 'significant' amount of space-time curvature in introduced, the rules of quantum linear superposition must fail. If two states in superposition are displaced by a distance $\mathbf{d}$, with a mass density $\mu(\mathbf{r})$, the self potential difference $\Delta E_{\text{DP}}$ of the states in superposition can be written:
\begin{equation}
    \Delta E_{\text{DP}}({\bf d}) = - 8\pi G \int \text{d}\bf{r}\int \text{d}\bf{r'}\frac{\mu(\bf{r}) [\mu(\bf{r'}+\bf{d})-\mu(\bf{r'}) ] }{\mid \bf{r}-\bf{r'} \mid}\enspace.
\end{equation}
The wave function collapse takes place with a characteristic time $\tau_{\text{DP}}=\hbar/\Delta E_{\text{DP}}$. For microscopic systems, the superposition is preserved, and the characteristic time is very long. For a proton, for example, $\tau_{\text{DP}}\simeq 10^6$\,years, while for a dust grain $\tau_{\text{DP}}\simeq 10^{-8}$\,seconds. Superposition in macroscopic systems is almost immediately suppressed, recovering the reduction postulate. 

The CSL model \cite{Ghirardi1986PRD,Pearle1989PRA,Ghirardi1990PRA} is a stochastic and non-linear modification of the Schrödinger equation:
\begin{align*}
    d\mid\psi_{t}\rangle= & \left[ -\frac{i}{\hbar}Hdt+\frac{\sqrt{\lambda}}{m_{0}}\int d\boldsymbol{x}(\mu(\boldsymbol{x})-\langle\mu(\boldsymbol{x})\rangle_{t})dW_{t}(\boldsymbol{x})-\right. \\
    &\left.-\frac{\lambda}{2m_{0}^{2}}\int d\boldsymbol{x}\int d\boldsymbol{y}e^{-\frac{(\boldsymbol{x}-\boldsymbol{y})^{2}}{4r_{C}^{2}}}(\mu(\boldsymbol{x})-\langle\mu(\boldsymbol{x})\rangle_{t})(\mu(\boldsymbol{y})-\langle\mu(\boldsymbol{y})\rangle_{t}) \right]\mid\psi_{t}\rangle\enspace.
\end{align*}
In the first addend, one can recognize the system's Hamiltonian $H$ within the standard Schrödinger dynamics. The model depends on the parameters $\lambda$ and $r_C$, the collapse strength and correlation length, respectively. $\mu(\boldsymbol{x})$ is the mass density of the system. 
The non-linearity is introduced via the $\langle\mu(\boldsymbol{x})\rangle_t$ term. 
The stochastic noise field $W_t(t)$ depends on $r_C$, with the properties of a white noise in the original formulation of the model. $dW_{t}(\boldsymbol{x})$ are a set a Wiener increments with zero average and correlation $\mathbb{E}[dW_{t}(\boldsymbol{x})dW_{t}(\boldsymbol{y})]=dte^{-\frac{(\boldsymbol{x}-\boldsymbol{y})^{2}}{4r_{C}^{2}}}$.
The parameter can effectively tune the dynamics of collapse. For example, for $r_C\sim 10^{-5}$\,cm, $\lambda\sim 10^{-8\pm2}\,\text{s}^{-1}$, gives the boundary between the microscopic and mesoscopic world; $\lambda\sim 10^{-17}\,\text{s}^{-1}$ the boundary between the mesoscopic and macroscopic one.

In both models, if the particle undergoing the collapse is charged, the diffusive Brownian-like motion which takes place as a result of the spontaneous collapse implies the emission of radiation. This results in the radiation emission rates:
\begin{align} 
\label{wg1:t3:eqrate}
  \begin{split}
    \frac{\text{d}\Gamma}{\text{d}E}\Big{\rvert}_{\text{DP}} = \frac{2}{3} \frac{G e^2 N^2 N_a}{\pi^{3/2}\epsilon_0 c^3 R^3_0 E},\\
  \end{split}
\quad
  \begin{split}
    \frac{\text{d}\Gamma}{\text{d}E}\Big{\rvert}_{\text{CSL}} = (N^2+N)N_a \frac{\hbar\lambda}{4 \pi^2 \epsilon_0 m_0^2 r^2_C c^3 E}\\
  \end{split}\enspace.
\end{align}
Both models predict an emission rate which follows the $1/E$ dependence.
On the left, the variables $G$, $e$, $\epsilon_0$, and $c$ represent the gravitational constant, the charge of an electron, the vacuum permittivity, and the speed of light, respectively. The term $N_a$ stands for the total number of atoms being considered, while the factor $N^2$ takes into account the atomic number's quadratic effect on the outcome. $R_0$ is the size of the particle's mass density, as proposed by Penrose, to be similar to the size of the nucleus's wave function. However, $R_0$ can also be used as a free variable in the model \cite{Donadi2020NP}.
On the right, $r_C$ and $\lambda$ are the two main parameters: the spatial resolution of collapse and the strength of noise. The term $(N^2+N)N_a$ takes into account both the coherent emission from nuclei (accounted for by $N^2$) and the incoherent emission from electrons (accounted for by $N$); $m_0$ represents the reference mass of a nucleon (in mass-proportional CLS model).

\subsubsection{Quantum Collapse, Present Status and Results}
At LNGS, collapse models can be strongly constrained, searching for the spontaneous radiation. The extremely low background allows detectors to be sensitive to even the small signal predicted by the models, if it exists. 
\begin{figure}[h!]
    \centering
    \includegraphics[width=.5\textwidth]{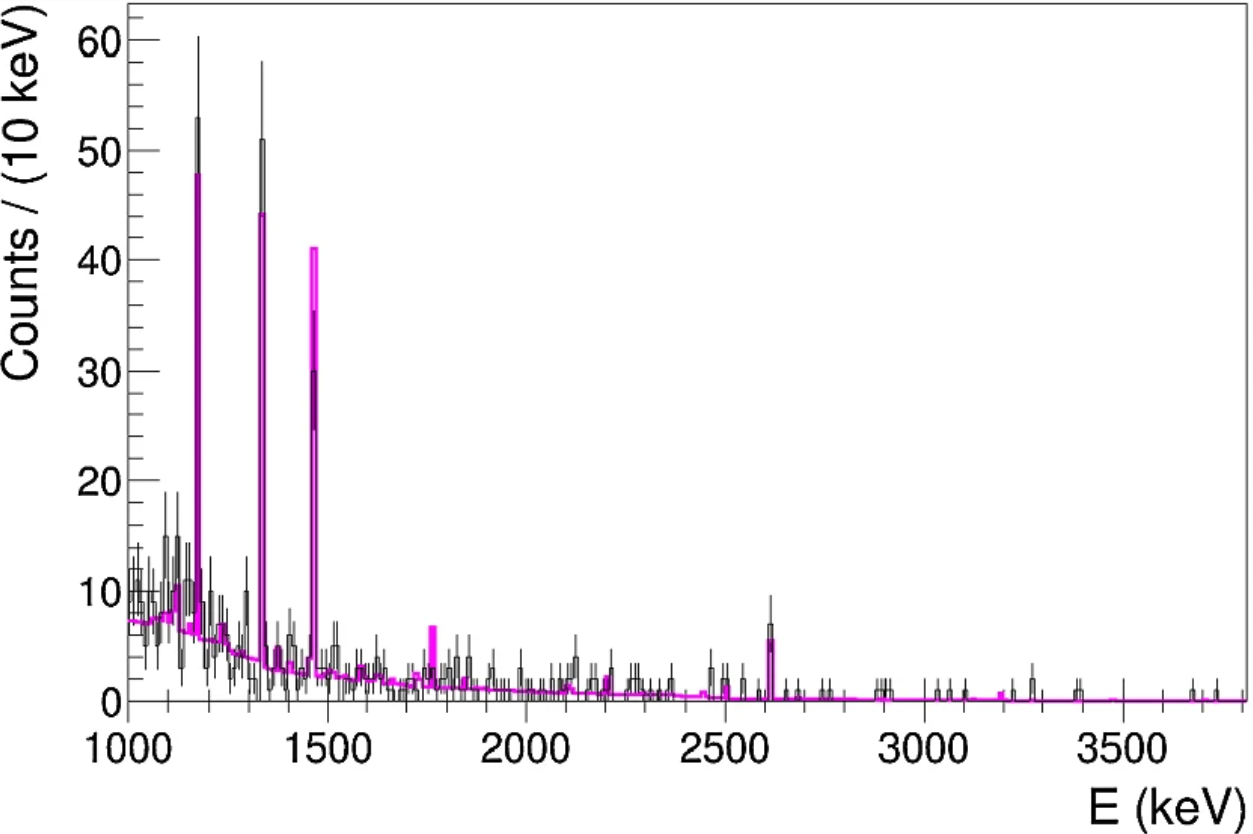}
    \caption{Measured X-rays spectrum (black histogram) and simulated background distribution (magenta histogram) in the energy range $\Delta E$ = (1000–3800)\,keV from the germanium detector described in \cite{Donadi2021EPJC}.}
    \label{wg1:t3:fig:donadi_spectrum}
\end{figure}

\begin{figure}[h!]
    \centering
    \includegraphics[width=.4\textwidth]{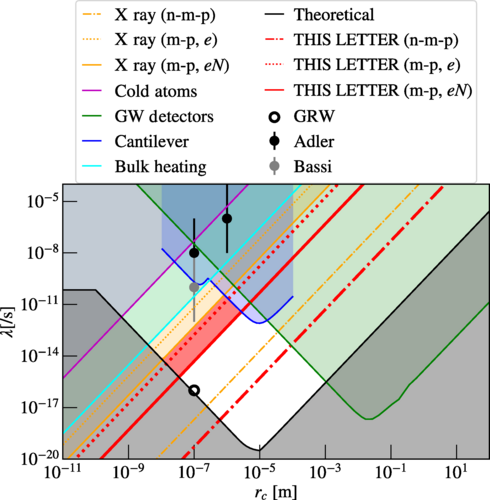}
    \caption{Updated exclusion plot of the ($r_C-\lambda$) plane from the MAJORANA DEMONSTRATOR \cite{Arnquist2022PRL}. Solid line and dotted represent the mass-proportional case for quasi free electrons (m-p,$e$) and coherent emission from the nuclei (m-p, $eN$). Dash-dotted line represent the non-mass-proportional (n-m-p) case. The limits come from cold atom experiments (in magenta) \cite{Bilardello2017PRA,Bilardello2016PASMA}, the interpretation of gravity wave experiments from LIGO and LISA Pathfinder (in green) \cite{Carlesso2016PRD}, measurements using cantilevers where an anomalous result was reported (in blue) \cite{Vinante2020PRL,Vinante2017PRL,vinante2016PRL}, and the interpretation of heat leaks in low temperature experiments in terms of the CSL heating effect (in cyan) \cite{Adler2018PRA,Bahrami2018PRA}. The limits from previous x-ray studies at LNGS \cite{Piscicchia2017E,Donadi2021EPJC} are represented by orange lines. Additionally, a theoretical lower limit is shown in black, which is based on the idea that a graphene disk with a radius of 10 micrometers (the minimum resolution of the human eye) should be localized in less than 1ms (the minimum time resolution of the human eye) \cite{Toro2018JPA}. Some proposed lower bounds by Adler \cite{Adler2007JPA}, Bassi \cite{Bassi2010EPL}, and Ghirardi, Rimini, and Weber \cite{Ghirardi1986PRD} are also shown as black vertical lines, a gray vertical line, and a black hollow circle, respectively.}
    \label{wg1:t3:fig:csl}
\end{figure}

Using germanium detectors at LNGS, the available parameter space of the DP and CLS models have been severely constrained \cite{Donadi2020NP,Donadi2021EPJC}. The data taken by a coaxial p-type HPGe detector, with active crystal volume of 375\,cm$^3$ and shielded by environmental radiation with a copper and lead casing, is shown in Figure \ref{wg1:t3:fig:donadi_spectrum}. Since no evidence of the signal, which has the shape of a falling distribution as a function of $E$, is found, it is possible to exclude at 90\% C.L. the available parameter space.
The ($r_C-\lambda$) plane of the CSL model is progressively reduced, as shown in Figure \ref{wg1:t3:fig:csl} in the most up-to-date plot.

\subsubsection{Quantum collapse models and their experimental tests future plans}
The activities planned for the mid term at LNGS regarding the quantum collapse model and their experimental tests will pave the way to a deeper understanding of the measurement problem, with implication ranging from fundamental physics to quantum technologies. The activities in this subgroup are outlined in Figure \ref{wg1:fig:gantt}, task  T3, with time horizon for the next 2, 5 and 8 years from the first quarter of 2023.

\paragraph{Generalized Models, Cancellation Effects and Energy dependence}

The first activities of T3 (year 1 to year 2) consist of theoretical and phenomenological work on the collapse models, with focus on: generalized models (QC1), cancellation effects (QC2) and energy dependence on the target atomic number (QC3).

The DP model is ruled out in the parameter-free version, however the model can be extended to more general cases. 
This will be realized in collaboration with the theoretical groups (Diósi, Penrose, Adler, Bassi and others).
The following extensions are foreseen:
\begin{itemize}
    \item addition of dissipation terms to the master equation and stochastic nonlinear Schrödinger equation of the DP theory. This addition will mitigate a known criticism on the model, which is the runaway energy increase.
    \item inclusion of non-Markovian correlation function.
\end{itemize}
The extensions of the model are expected to lead to stronger dependence on the energy of the spontaneous radiation, in relation to the atomic structure.\\*
With the present formulation of the models, the spontaneous radiation rate is expressed with the Eq. \ref{wg1:t3:eqrate}.
However, in the lower energy regime, the photon wavelength is comparable to the atomic orbits dimensions, and the coherent emission of electrons together with protons leads to cancellation effects; their strength and behavior strongly correlates with the atomic structure. Moreover, due to the different interplay of the models parameters, the DP and CSL models are expected to originate different behaviors at low-energy. This feature will effectively change the phenomenology, which will be dependent on the target atomic number and the model, allowing for a richer and more dedicated optimization of the searches. There will be great advantage with the use of ultra-pure Aluminum, Silver, Gold, Lead, each of them exhibiting a different energy dependence at lower energy for the different collapse models.

\begin{figure*}
    \centering
    \includegraphics[width=.6\textwidth]{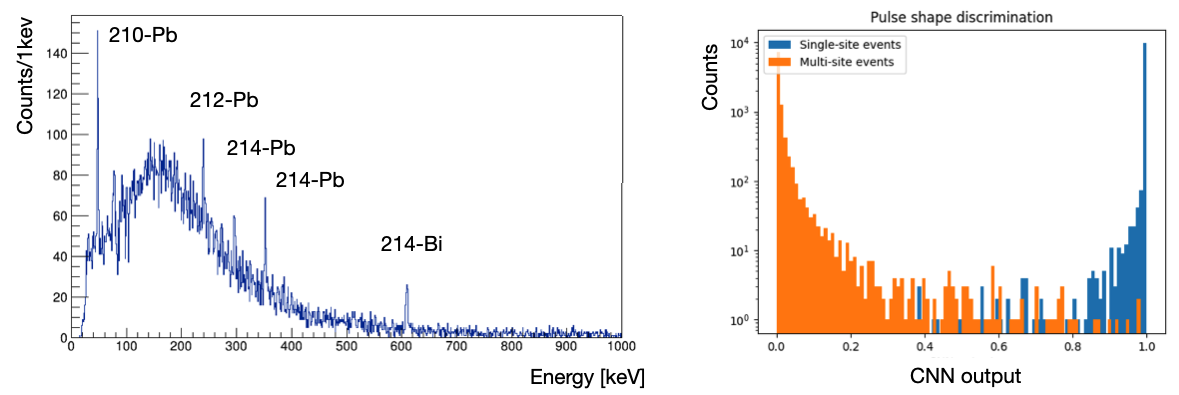}
    \caption{Preliminary data from a BEGe devoted to collapse studies. On the left, the data acquired in 74\,days in the region 0 - 1\,MeV, where the lines originating from the Lead isotopes are indicated. Lead is used both as target and shielding in this setup. On the right, the output of a machine learning approach trained to discriminate signals acquired from the germanium detector into two categories: single-site, and multi-site events.}
    \label{wg1:t3:fig:bege}
\end{figure*}

\paragraph{Preliminary data taking}
The activities from year 3 to year 4 focus on exploiting experimentally what found in the previous ones. The energy dependence will be targeted with a preliminary study of a BEGe detector of around 1\,kg of mass. This activity has already been pioneered with a BEGe of smaller mass. Preliminary results of the data taking of this detector are shown in Figure \ref{wg1:t3:fig:bege}, where the spectrum of 74 days of data taking is shown, together with a CNN output trained to discriminate single-site from multi-site events, the latter contributing to the background. Leveraging Machine Learning techniques is in fact expected to further enhance the physics capability of the detector.
In order to exploit a stronger energy dependence on the target material (QC5), lower energies are necessary to fully disentangle the model and atomic number dependence. This will be achieved with an upgrade of the front-end electronics, via a low-noise data acquisition system and a high insulation, low noise amplifier (QC4).

\paragraph{Dedicated setup}
Finally, from year 5 to year 7 (QC6) a dedicated setup with focus on quantum wave function collapse is planned. To this end, an array of p-type point contact Germanium detectors with the total active crystal mass of more than 10 kg is needed. The setup will have to be equipped with a compact shield and an active muon veto, in order to suppress the environmental and cosmic background. In order to exploit shape dependence, it will need a lower energy threshold, with the possibility of carrying out a data taking campaign with different targets. This last feature (not typically present at deep underground experiments searching for effects beyond the SM) will enable exploiting the atomic mass dependence, and potentially strongly constrain the dissipative and non-Markovian DP and CLS collapse models. Moreover, the setup will be at the same time suited for Dark Matter searches.

\section{\wgtwotitle{}}
\label{wg2}
An overview of the INFN-CSN3 activities in Nuclear Astrophysics is presented in \cite{Broggini2019}, and further discussed in the Nuclear Physics Mid Term Plan in Italy of LNL and LNS sessions \cite{Ballan2023,Agodi2023}. In the following sections a selection of challenging scientific cases debated during the LNGS session \cite{MidTermLNGSurl} are reviewed.

Nuclear Astrophysics is devoted to shed light on the nuclear processes active in stellar environments and Big Bang Nucleosynthesis (BBN).
Nuclear Astrophysics studies include: the measurement at very low energies of the cross section of nuclear reactions related to various stages of stellar evolution and nucleosynthesis; the study of the neutron induced reactions responsible for the production of the heavy elements in the Universe; the understanding of nuclear processes far from stability (associated with stellar explosions or extreme condition of matter) such as $r$-process, p-process and $r$p-process.

In the last decades huge experimental efforts were devoted to the study of a large number of reactions, using different approaches, that were developed and brought to their best performances. Several projects have been conducted in close cooperation between nuclear physicists and astrophysicists, stimulating further improvements both on the experimental nuclear research side, and on the interpretation of astronomical observations. 
The main scientific motivations are presented in Section \ref{wg2:t1}, while the experimental aspects on most relevant cases are discussed in Section \ref{wg2:t2}. The Tables \ref{wg2:t2:tab1} and \ref{wg2:t3:tab1} present a list of key reactions that might be investigated at LNGS and CIRCE laboratories, respectively, in the next 5 to 10 years.
\begin{figure*}[!htbp]
    \centering
    \includegraphics[width=0.9\textwidth]{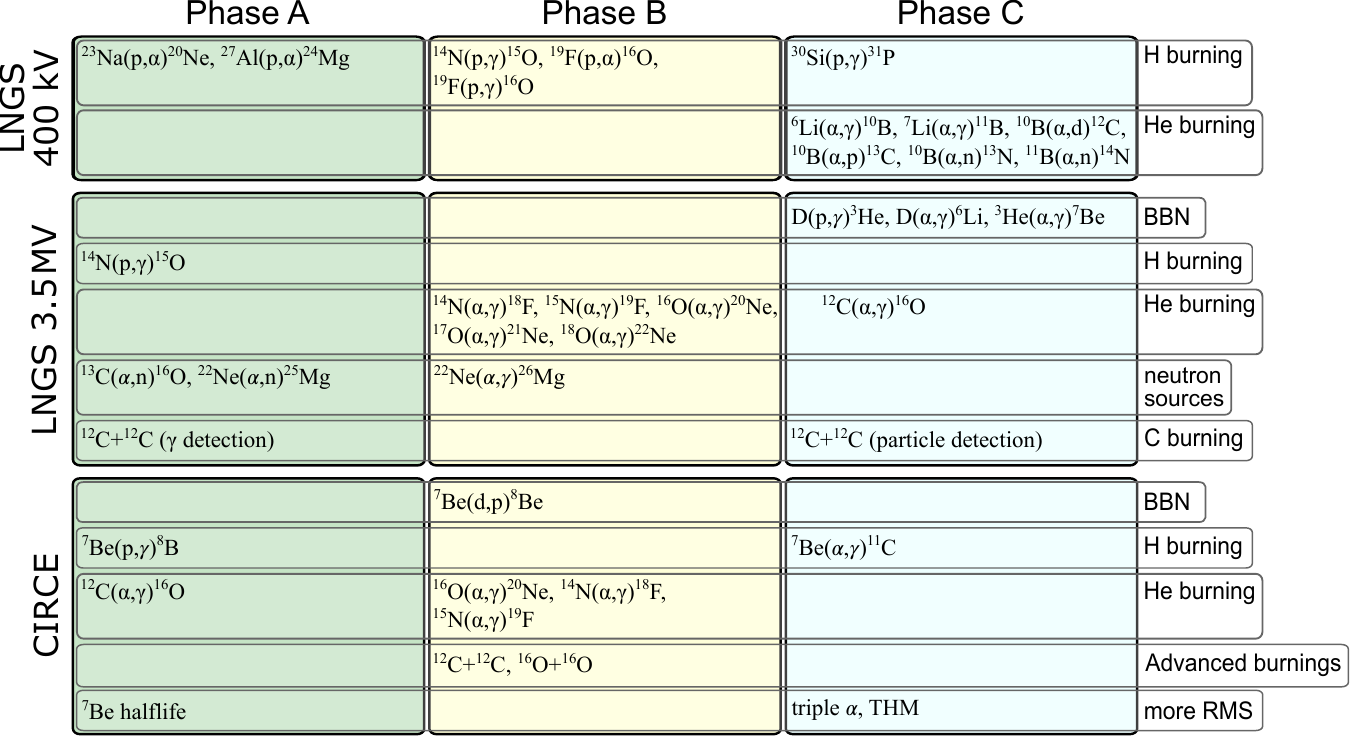}
    \caption{GANTT chart of the activities of WG2 for the mid term. Phase A corresponds to the next 2-3 years, phase B to 3-5 years, phase C to 5-7 years.}
    \label{wg2:fig:gantt}
\end{figure*}


\subsection{\wgtwotonetitle{}}
\label{wg2:t1}\label{intro_astro}

An impressive advance in the astronomical facilities, those already available, such as the Gaia mission and the new James Webb Space Telescope, and those under development, such as the next generation of large aperture ground-based telescopes like the Extremely Large Telescope, allow more accurate determinations of the physical and the chemical properties of single stars and, more in general, of stellar populations. The exploitation of this huge amount of data requires sophisticated models of stellar evolution and nucleosynthesis, based on extended nuclear networks and very accurate sets of nuclear physics inputs, such as reaction rates.

In this section a brief overview of different astrophysics scenarios, that are influenced by the reactions discussed in the LNGS Session of the Nuclear Physics Mid Term Plan in Italy, is presented.

\subsubsection{H burning}\label{J_astro}
Stars burn hydrogen via different sets of nuclear reactions: the proton-proton (\pp) chain and the CNO cycle. 
The latter is the dominant energy source, and thus of the nucleosynthesis, in stars having mass larger than about $1.2\,M_\odot$.
Unlike the \pp{} chain, the CNO cycle is a catalytic cycle, \ie{} it converts 4 protons into one helium nucleus but does so via reactions on the preexisting seed nuclei of carbon, nitrogen and oxygen. Depending on the temperature attained in stellar interiors, different branches of the CNO cycle are active. For instance, at \emph{low} temperatures ($T\sim 20$\,MK) only the CN cycle works, while at higher energies ($T\gtrsim 30$\,MK) the NO cycle is efficiently active as well.
The rate of the CN with respect to the NO cycle depends on the branching ratio of the proton capture on \nuclide{15}{N}, i.e. the \reaction{15}{N}{\pg}{16}{O} and \reaction{15}{N}{\pa}{12}{C} reaction cross sections. In all cases, the \reaction{14}{N}{\pg}{15}{O} is the slowest reaction of the cycle and, in turn, it determines the whole H-burning rate. 
Once central hydrogen has been converted into helium, proton burning continues in a shell surrounding the core. 
The synthesis of heavier elements, up to Si, can occur through the NeNa and the MgAl cycles, provided that large enough temperatures are attained. A detailed knowledge of the nuclear processes active in these cycles is mandatory to properly determine the nucleosynthesis during the star evolution, in particular in the Red Giant Branch and Asymptotic Giant Branch (AGB) phases, as well as in Novae explosions. The abundances of the synthesized elements critically depend on the rates of the nuclear processes involved, often through non-trivial nucleosynthesis reaction chains, combined with complex mixing mechanisms.

In addition, in extremely metal-poor stars, as the first generation of stars, owing to the scarcity of C, N and O, the H burning cannot occur through the CNO cycle. In this case, even massive stars burn H only through the \pp{} chain, but the released nuclear energy is not sufficient to balance the contraction of the core. As a result, the central temperature increases, until the triple-$\alpha$ process is efficiently activated, as will be further discussed in the next Subsection (see also Section \ref{metalpoor_astro}).

\subsubsection{He burning}\label{he_astro}

\paragraph{Triple-\texorpdfstring{\maybebm{\alpha}}{alpha} and the \maybebm{A=5} and \maybebm{A=8} gaps}
\label{triplea_astro}

The triple-$\alpha$ reaction, \ie{} the fusion of three $\alpha$ particles into a \nuclide{12}{C} nucleus, plays a role in many astrophysical processes. Firstly, it is the main mechanism behind the synthesis of carbon in stellar interiors, essential to the development of life on Earth (at least). Secondly, it is the mechanism by which the first generation of stars, created from the hydrogen and helium ashes of the Big Bang (the so called population III), were able to bridge the $A=5$ and $A=8$ mass gaps, thus allowing nucleosynthesis to proceed toward heavier elements. 
The triple-$\alpha$ reaction proceeds as a two-step process, strongly enhanced by two low-lying s-wave ($l=0$) resonances, $\alpha + \alpha \xrightarrow{}\nuclide{8}{Be}(0_1^+)$; $\alpha + \nuclide{8}{Be}(0_1^+) \xrightarrow{} \nuclide{12}{C}(0_2^+)$. In the intermediate-temperature domain ($0.1$-$2$\,GK), the rate of this important reaction is fully determined by the properties of the Hoyle state in \nuclide{12}{C} ($7.654$\,MeV, $0^+$) \cite{Rolfs}, and, in particular, by the competition between its $\alpha$ and \emph{radiative} decays. At the energies corresponding to this temperature range, the reaction cross section can not be directly measured, and the rate determination relies mainly on measurements of the radiative decay partial width of the Hoyle state \cite{Freer94}. This leads, in turn, to a relatively precise determination of the reaction rate ($10\%$), even though ambiguities still persist in the experimental findings related to the radiative decay width of the Hoyle state \cite{Kibedi2020}. The latter directly affects the reaction rate, and calls for new experimental investigations. 

The determination of the reaction rate is far more complicated in the low- and high-temperature domains, \ie{} $T<0.1$\,GK and $T>2$\,GK, respectively. In particular, complex three-body models are necessary to calculate the rate in the low-temperature regime \cite{Akahori15,Suno15}. However, the commonly used evaluation of the reaction rate (NACRE, \cite{NACRE}) assumes that the breakup of the \nuclide{12}{C} Hoyle state into three particles proceeds exclusively as a sequential two-step process via the ground state of \nuclide{8}{Be} (SD). In practice, direct decay (DD) into three $\alpha$-particles, associated with three-body processes that bypass the formation of the \nuclide{8}{Be} resonance, are neglected. The competition between SD and DD has been for a long time the subject of theoretical and experimental efforts. From a theoretical point of view, the Hoyle state is often described as a cluster state made by three weakly interacting $\alpha$-particles, but the peculiarities of this geometrical arrangement, which affect the observed DD branching ratio \cite{Ishikawa14}, are still quite debated \cite{Morinaga1956, VonOertzen97, Uegaki77, Kamimura81, Tohsaki01, Chernykh07, Epelbaum11, Ishikawa14}. In spite of the critical importance, the experimental determination of the DD partial width has received relatively little scrutiny until recently \cite{Freer94, Raduta11, Manfredi12, Kirsebom12, Rana13, Itoh14}. Improved high-precision experiments \cite{DellAquila17,Smith2017,Rana19,Bishop20} have finally allowed to prove the dominance of the SD decay over the DD one, reaching a level of sensitivity that now enables to experimentally test the predictions of microscopic nuclear structure calculations \cite{Ishikawa14}. The existence of a non-vanishing DD partial width for the Hoyle state has a relevance in astrophysics, as it may affect low temperature triple-$\alpha$ reaction rate and, in turn, the synthesis of C during the H-burning phase of primordial stars. The first stellar generation was probably made of peculiar massive stars. Indeed, owing to the lack of CNO isotopes, the H burning proceeded in the core of these stars through the \pp{} chain and, in order to generate enough energy to replace the radiative energy loss, the temperature rises up, until the triple-$\alpha$ reactions starts. Then, the C released by the triple-$\alpha$ activates the CNO cycle, causing the development of a quite extended convective core. A fundamental question arises: is the triple-$\alpha$ reaction the only process in nature that makes it past the gaps at $A=5$ and $8$? According to the current stellar nucleosynthesis models, the answer is yes. However, these scenarios include alternative nuclear chains involving reactions such as \reaction{6}{Li}{\ag}{10}{B}, \reaction{7}{Li}{\ag}{11}{B}, 
\reaction{7}{Be}{\ag}{11}{C}, 
\reaction{10}{B}{\ad}{12}{C}, \reaction{10}{B}{\ap}{13}{C}, \reaction{10}{B}{\an}{13}{N}, and \reaction{11}{B}{\ad}{14}{N}, whose cross sections are barely known at the stellar thermal energies. Their study, although interesting, is beyond the mid term discussed in this paper. 

The high temperature domain is instead interesting for the explosive nucleosynthesis, as it occurs in core-collapse supernovae (SNe) and neutron star mergers. At $T>4$\,GK, baryons attain a nuclear statistical equilibrium, in which matter photodisintegrates into protons, neutrons and $\alpha$ particles. Then, as this matter expands and the temperature drops below $3$-$4$\,GK, fusions restarts building up heavier nuclei. However, the lack of stable nuclei with $A=5$ and $8$ hamper this process that can be overcome through triple-$\alpha$ reactions. 
At these high temperatures, the triple-$\alpha$ rate is determined by higher energy states of \nuclide{12}{C} and by possible interferences among them. In this respect, the properties (spin, parity, partial widths) of roto-vibrational excitations of the Hoyle state, which are predicted by many cluster models \cite{Bijker00,Marin-Lambarri14,Cardella2023}, are expected to play a role. For example, the possible occurrence of a monopole breathing-mode excitation of the Hoyle state, recently conjectured in \cite{Li22}, is expected to modify the temperature dependence of the triple-$\alpha$ process, and the predicted nucleosynthesis in explosive environments and neutron star mergers. These theoretical findings are yet to be fully confirmed by experimental investigations. 

\paragraph{
\texorpdfstring{
\maybebm{\reaction{12}{C}{\ag}{16}{O}}}{12C(alpha,gamma)16O}, the Holy Grail}
\label{holygrail_astro}

The competition between the triple-$\alpha$ and the \reaction{12}{C}{\ag}{16}{O} reaction determines the relative abundances of C and O left by the He burning. The C/O ratio, which scales inversely with the rate of the \reaction{12}{C}{\ag}{16}{O}, has a profound impact on the more advanced phases of the stellar evolution. For instance, the cooling timescale of C-O white dwarfs is longer in case of higher C/O \cite{Prada2002} and the amount of \nuclide{56}{Ni} ejected by type Ia SNe is higher in case of a higher C/O ratio \cite{Dominguez2001}.
Concerning core-collapse SNe, the C/O ratio left after the He burning affects the chemical composition of the ejecta \cite{Imbriani2001, deBoer2017}. 
Hence, a robust determination of the \reaction{12}{C}{\ag}{16}{O} cross section around 300\,keV (with an uncertainty of 15\% or better) would boost our understanding of the final fate of stellar evolution. At such low energy, the cross section of this reaction is dominated by ground state captures through two sub-threshold resonances ($J^\pi =1^-$ and $2^+$). Interferences with higher energy states and the direct capture (DC) component can substantially affect the cross section. The most accurate estimation of the astrophysical $S$-factor, as obtained by means of R-matrix calculations, combines direct and indirect measurements \cite{deBoer2017} and found an uncertainty of $\sim 30$\% at 300\,keV. 


\subsubsection{Neutron sources}
\label{sct2_wg1}\label{nsources_astro}

The \reaction{22}{Ne}{\an}{25}{Mg} reaction is the main source of neutrons for the $s$-process nucleosynthesis taking place in massive AGB stars, $M>4-5\,M_\odot$ \cite{Karakas2014}, and massive stars, $M>10\,M_\odot$ \cite{Pignatari2010}. In particular, massive stars are contributing to the nucleosynthesis of the bulk of nuclei in the mass range $70<A<110$ via the {\it weak $s$-process} component in the solar system \cite{Kaeppeler2011}. This would account for most of the copper, gallium and germanium in the Solar System \cite{Pignatari2010}. In principle, the abundance of \nuclide{22}{Ne}, \ie{}, the fuel for the $s$-process activation, depends on the stellar metallicity. Indeed, the bulk of the CNO isotopes is converted into \nuclide{14}{N} during the H-burning phase, and then to \nuclide{22}{Ne} via the \reaction{14}{N}{\ag}{18}{F}$(\beta^+\nu)$\reaction{18}{O}{\ag}{22}{Ne} chain during He burning. Therefore, the amount of \nuclide{22}{Ne} available for the $s$-process in the stellar core should be more than the 60\% of the initial metallicity of the star \cite{Raiteri1991, Kaeppeler1994}. In this framework, the contribution of massive stars to the production of $s$-process elements is expected to decrease with decreasing the initial stellar metallicity \cite{Prantzos1990, Baraffe1992}. 
However, for fast-rotating massive stars the amount of \nuclide{22}{Ne} can be boosted by mixing induced by rotational instabilities, such as meridional circulation or shear (in case of differential rotation). 
In this case, the main source of \nuclide{22}{Ne} becomes the \nuclide{12}{C} made during He burning, by the triple-$\alpha$ reaction, and the resulting $s$-process production is significantly enhanced \cite{Pignatari2008}. The relevance of the $s$-process from fast rotating massive stars in the galactic chemical evolution is matter of current research \cite{Cescutti2013,Prantzos2020}. 
Recent nucleosynthesis calculation in rotating massive stars \cite{Limongi2018} have shown that they may contribute significantly to the weak $s$-process down to a metallicity [Fe/H]$\sim -1.5$. Anyway, a precise evaluation of the \reaction{22}{Ne}{\an}{25}{Mg} reaction rate is mandatory to understand the synthesis of various elements, such as Se, Ge, As, Sb and to distinguish between different scenarios for the $s$-process in massive stars. Furthermore, this reaction affects the magnesium isotopic ratio $\nuclide{24}{Mg}$/$\nuclide{25}{Mg}$/$\nuclide{26}{Mg}$ in the material ejected to the interstellar medium by massive AGB and/or massive stars.
Currently both kind of objects are the proposed polluters of the second generation of stars observed in many globular clusters, which show chemical anomalies in terms of O-Na and Al-Mg abundance anticorrelations \cite[see \eg{}][]{Yong2006}. In this context, an important player is the concurrent reaction \reaction{22}{Ne}{\ag}{26}{Mg}. Indeed, at high temperature most of the \nuclide{22}{Ne} is burned through the $\an$ channel, while the $\ag$ dominates at low temperatures, being the former hampered by the neutron threshold. Note that the minimum temperature for the $\an$ activation depends on many nuclear resonances in the \nuclide{26}{Mg} compound nucleus. The magnesium isotopic ratio is one of the few elemental isotopic ratios which can be measured directly in stars and the interstellar medium with modern astronomical facilities. Therefore, the feedback between Mg isotopic ratio measurements in stars and the interstellar medium and the improvement of the \reaction{22}{Ne}{\an}{25}{Mg} and \reaction{22}{Ne}{\ag}{26}{Mg} reaction rates certainly will enlighten the puzzle of the origin of abundance anomalies in globular cluster as well as the contribution of massive stars to the synthesis of beyond-iron-peak elements.

On the other hand, the {\it main $s$-process} component, which is mainly responsible for the nucleosynthesis of nuclei with mass $90<A<210$, is efficiently active in low-mass AGB stars ($1.5<M/M_\odot<3$) \cite{Kaeppeler2011}. In this case, the main neutron source is the $\nuclide{13}{C}\an)\nuclide{16}{O}$ reaction, while the \reaction{22}{Ne}{\an}{25}{Mg} plays a secondary role \cite{Straniero1995, Gallino1998, Straniero2006}. The former reaction is indeed active at rather low temperatures (about 90-100\,MK) and generates a low neutron density (about $10^7$ neutrons per cm$^{-3}$) but for a long time (up to \qty{e5}{\year}), thus providing a sufficiently high neutron exposure, as needed to produced the heavier $s$-process elements, from Sr to Pb. In practice, an AGB star undergoes recurrent He-shell flashes, called thermal pulses (TP), which are thermonuclear runaways whose power attains a few $10^8\,L_\odot$. Then, a dredge-up episode may occur after a TP, when the external layers expand and cool down, until the H-burning shell eventually dies, and the external convection can penetrate the He- and C-rich mantel. According to the standard paradigm, a partial mixing occurring at the bottom of the convective envelope at the time of the dredge-up leaves a thin pocket where the H mass fraction is $X_H<0.01$, while the carbon mass fraction is about 0.2. Then, at the H re-ignition, a substantial amount of \nuclide{13}{C} is produced by the \reaction{12}{C}{\pg}{13}{N} reaction followed by the \nuclide{13}{N} decay, and, later on, the $s$-process can start, due to the activation of the \nuclide{13}{C} neutron source. A second neutron burst, as due to the activation of the \reaction{22}{Ne}{\an}{25}{Mg} reaction, may eventually occur at the bottom of the convective shell powered by a TP, but only if the temperature exceeds $300$\,MK.


\subsubsection{Carbon and Oxygen burnings}\label{advanced_astro}
The C burning phase is triggered by the $\CC$ reaction, whose main channels release protons and $\alpha$ particles that react with various isotopes, from C to Si, so that a rich nucleosynthesis arises.
Below 2.3\,MeV, the cross section of this reaction is largely debated. Contrasting scenarios are possible: $\alpha$-clusters and molecular states in the \nuclide{24}{Mg} compound nucleus could substantially enhance the cross section \cite{Tumino2018}, while fusion hindrance could depress it \cite{Jiang2007}. Anyway, the cross section between 1.5 and 2\,MeV determines the astrophysical reaction rate and, in turn, the temperature of the C burning. This uncertainty has profound consequences in our understanding the late phase of massive stars evolution, \ie{}, the progenitors of core-collapse SNe \cite{Chieffi2021}, and the carbon-simmering phase preceding the dynamical breakout in type Ia SNe progenitors \cite{Bravo2011}.

In massive stars, the C burning rate affects the number, the extension and the duration of the C convective shells and, in turn, it determines the final mass-to-radius relation or, equivalently, the structural compactness at the onset of the core collapse. This parameter plays a key role in understanding the final fate of a massive star. In particular, it determines: i) if the initial implosion is converted into an explosion or a full collapse takes place, directly or after a fall-back; ii) if the remnant will be a neutron star or a black hole; iii) the ``islands'' of explodability, \ie{} the initial-mass ranges for which the explosion is possible; iv) the relative abundances of intermediate-mass nuclei in the SNe ejecta. 

The $\CC$ reaction rate also determines the minimum stellar mass that undergoes a C burning phase \cite{Straniero2019}. Stars whose initial mass is below this threshold skip the C burning phase and directly enter the AGB phase, ending their live as C-O white dwarfs. Stars with mass slightly larger than this threshold will ignite C in a degenerate core and, later on, enter the super-AGB phase. Their final fate may be the development of an O-Ne white dwarf or an electron-capture SN. Therefore, the knowledge of this mass limit is of paramount importance in modern astrophysics.

In type Ia SNe, a thermonuclear explosion follows a hydrostatic C burning phase, called simmering. The rate of the C burning during the simmering, which is controlled by the $\CC$ reaction, determines the physical and the chemical structure of the exploding white dwarfs, namely: the temperature and the density profiles, as well as the degree of neutronization (see \cite{Piersanti2022} and references therein).
Concerning the explosive phase, the anomalous Ca/S and Ar/S mass ratios observed in type Ia SNe remnants suggest a substantial reduction (by a factor of 10) of the current estimates of the \nuclide{12}{C}+\nuclide{16}{O} reaction rate or, alternatively, a combined variation of the rates of $\CC$, \nuclide{12}{C}+\nuclide{16}{O}, \nuclide{16}{O}+\nuclide{16}{O} and the \reaction{16}{O}{$(\gamma,\alpha)$}{12}{C} \cite{Bravo2019}. This occurrence has important implications for the whole nucleosynthesis of Chandrasekhar and sub-Chandrasekhar mass SNe Ia explosions. 
The relative mass fraction of Ca to S also hints to a metallicity dependent yield during explosive oxygen burning. This ratio depends on the quantity of $\alpha$ particles: O burning can proceed through an $\alpha$-poor or an $\alpha$-rich branch. The $\alpha$-rich explosive O burning enhances the production of Ca with respect to S. The \reaction{16}{O}{\pa}{13}{N} reaction, followed by $\reaction{13}{N}{(\gamma,\mathrm{p})}{12}{C}$, has been identified as the origin of the metallicity dependence. This reaction chain boosts $\alpha$-rich oxygen burning when the proton abundance is large, increasing the synthesis of Ar and Ca with respect to S and Si. At high $Z$ the presence of free neutrons leads to a drop of the proton abundance and the above chain is not efficient. Moreover, an increase of the rate of this reaction by a factor of approximately 7 with respect to the usual tabulated values (\eg{} REACLIB) would be enough to explain all the measured Ca/S mass ratios in the Milky Way and Large Magellanic Cloud SN remnants. Cross section measurements for the \reaction{16}{O}{\pa}{13}{N}, within the Gamow peak for temperatures in the range 3 to \qty{5e9}{\kelvin}, are encouraged, as it is critical to elucidate the impact of this reaction on the synthesis of Ca in thermonuclear SNe. 

\subsection{\wgtwottwotitle{}}
\label{wg2:t2}

Stellar nucleosynthesis is driven by charged particle interactions that, due to the relatively low temperatures in stellar environments, occur at very low energies. The structures of the compound nuclei are often characterized by the single particle and cluster configurations, as well as possible Coulomb and quantum effects near the threshold. In the absence of direct reaction measurements, complementary information from indirect transfer or other reaction techniques helps to understand these configurations. Extrapolations from high energy experimental data, by means of physical or phenomenological nuclear models, to energies of astrophysical interest have been developed reaching the so called Gamow Peak \cite{Rolfs}, but often result in large reaction rate uncertainties. 
There is, therefore, the need to push direct measurements to lower and lower energies. However, most cross sections are too small to be directly measured in the laboratory at these energies. This is because, in a stellar environment, the energy available to the nuclei is much lower than the Coulomb barrier and nuclear reactions proceed via quantum tunneling. The main challenge in direct measurements comes from the background signals, which, together with the low yields, set a limit to the energy range that can be investigated. Cosmic rays, environmental radioactivity, and beam-induced background reactions on target impurities, all represent major limitations to the measurement of cross sections down to the Gamow Peak. 
Different experimental approaches can be used to maximize signal to background in the measurement of cross sections at very low energies.

One approach is the exploitation of the extremely low background at LNGS, as was first proposed by the LUNA Collaboration. The LUNA experiments have established underground nuclear physics as a powerful tool for determining nuclear reaction rates at Gamow peak energies (which represents the energies where most of the reactions take place in the interior of a star at a given temperature), during thirty years of continuous refinement of the experimental procedures.
Over the last few years, several reactions belonging to the CNO, MgAl and NeNa cycles have been studied by the LUNA Collaboration, using the \LUNAfourhundred{} accelerator, a Singletron accelerator which has been installed in year 2000 \cite{Formicola2003}. These studies have contributed to unveil the origin of some meteoric stardust \cite{Lugaro2017} and to shed light on the speed at which the NeNa cycle operates \cite{Ferraro2018b, Boeltzig2019}. It was also measured, with high precision, the cross section of the most important reaction affecting the BBN, \ie{} \reaction{}{D}{\pg}{3}{He}. LUNA results have contributed to precisely establish one of most uncertain nuclear physics input to BBN calculations leading to an accurate determination of the density of baryonic matter \cite{Mossa2020}. More recently, thanks to the intense He beam available, it was measured for the first time the rate of the most important stellar neutron source, the \reaction{13}{C}{\an}{16}{O} reaction, directly inside the Gamow peak \cite{Ciani2021}, thus improving our understanding of the branch points along the $s$-process \cite{Ciani2021}.

A second possible approach is the use of more complex experimental apparatuses to maximize signal to background through high selectivity and larger detection efficiency. An example is provided by RMSs, that allow to measure the cross section of radiative capture reactions by means of the detection of the residual nucleus, without the need of $\gamma$-ray measurements, that can be optionally performed in order to gain additional information on the nuclear process. 
The ERNA RMS is one of the few recoil separators devoted to the measurement of nuclear cross section of astrophysical interest. In the early 2000s, the ERNA RMS was realized at the 4MV Dynamitron Tandem Laboratorium of the Ruhr-Universität Bochum (Germany) \cite{Rogalla1999,Gialanella2000,Rogalla2003,Gialanella2004,Schurmann2004}, with the aim of studying the $\cag$ reaction. In 2009, the ERNA RMS was moved to the Tandem Accelerator Laboratory of CIRCE, its updated layout is shown in Figure \ref{wg0:CIRCE:fig:CIRCE}.

The ERNA RMS so far has provided data on several processes of astrophysical relevance. The first results were obtained from the measurement of the total cross-section of the $\cag$ reaction \cite{Schurmann2005}. It was possible to get a new insight on the relevance of the cascade transitions that were later investigated by DRAGON \cite{Matei2006} and ERNA \cite{Schurmann2011}.
The use of ERNA RMS for the measurement of the $\Hetag{}$ reaction cross-section 
give an important contribution to solve the then long-standing issue of the difference between extrapolation based on experiments selection according to the measurement method \cite{DiLeva2008}.

The ERNA RMS was used to measure the resonances in the \reaction{15}{N}{\ag}{19}{F} reaction at \mbox{$\Ecm=1323$} and 1487\,keV. The reaction was studied in inverse kinematics, using a \nuclide{15}{N} beam \cite{DiLeva2012} on a windowless extended \nuclide{4}{He} target \cite{DiLeva2017}.
 More recently, the ERNA Collaboration completed the measurement of the \reaction{7}{Be}{\pg}{8}{B} total cross-section in the energy range $\Ecm=367.2$\,keV to 812.2\,keV \cite{Buompane2018,Buompane2021} using a windowless gas target \cite{Schuermann2013} and the radioactive \nuclide{7}{Be} beam available at CIRCE with intensities up to \qty{e9}{pps} \cite{Limata2008}. 

The ERNA Collaboration also performed at CIRCE measurements of the $\CC$ cross section using the array of GASTLY detectors \cite{Romoli2018} at backward angles. The GASTLY detectors allowed for unambiguous identification of both protons and alfa particles of $\CC$ \cite{Morales2018}

\subsubsection{Big Bang Nucleosynthesis}

\paragraph{\texorpdfstring{\maybebm{\reaction{}{D}{\pg}{3}{He}}}{D(p,gamma)3He} and \texorpdfstring{\maybebm{\reaction{}{D}{\ag}{6}{Li}}}{D(alpha,gamma)6Li}}
\label{wg2:t2:subsection:2H-p-g-3He}

Deuterium (D) is an excellent indicator of cosmological parameters in the early Universe because its primordial abundance is the most sensitive to the baryon density of the Universe.
Although astronomical observations of primordial deuterium abundance reached percent accuracy \cite{Cooke2018}, theoretical predictions \cite{Pitrou2018,Coc2015} based on BBN were hampered by large uncertainties on the cross section of the deuterium burning reaction \reaction{}{D}{\pg}{3}{He}.

A recent measurement at the \LUNAfourhundred{} accelerator reached \qty{3}{\percent} accuracy in the energy range $\Ecm = \qtyrange{33}{263}{keV}$ \cite{Mossa2020}, that provided remarkable boost to these studies: a BBN estimates of the baryon density at the \qty{1.6}{\percent} level has become possible, now in excellent agreement with cosmic microwave background analyses.
However, the maximum achievable beam energy (\qty{400}{keV}) did not allow to completely explore the energy range relevant for BBN, that extends up to $\Ecm = \qty{369}{keV}$ for \qty{95}{\percent} coverage. 
The BIBF is perfectly suited to extend precision measurements in the energy range of interest up to $E = \qty{1.5}{MeV}$ using an approach similar to the one used in \cite{Mossa2020}, with a windowless gas target \cite{Mossa2020-EPJA}. Given the high $Q$ value of the reaction (\qty{5.5}{MeV}) the photons are emitted with an energy where the background suppression in $\gamma$-ray spectra is most effective in an underground laboratory. In addition to the important cosmological implications, precise measurements of the angular distribution and of the cross section over a broad energy range are relevant to test nuclear physics models. In fact, the \reaction{}{D}{\pg}{3}{He} reaction cross section has been calculated within a microscopic \emph{ab-initio} approach, \ie{} using realistic models for the nuclear interactions and currents, and applying numerical techniques able to calculate the bound- and scattering-state wave functions, including the Coulomb interaction without approximation (see refs. \cite{Viviani2000,Marcucci2005,Marcucci2016}) and a comparison with precise experimental data would be very helpful. 

Another important reaction contributing to BBN is the \reaction{}{D}{\ag}{6}{Li} responsible for the production of the \nuclide{6}{Li} isotope. Its cross section has been measured around \qty{130}{keV} with the \LUNAfourhundred{} accelerator \cite{Anders2014,Trezzi2017}. It was the first direct measurement in the BBN energy range. Coulomb dissociation experiments provided only upper limits \cite{Kiener1991,Hammache2010}. The BIBF 3.5\,MV accelerator offers the possibility to extend the measurements up to $\Ecm = \qty{1.2}{MeV}$, by using an $\alpha$ beam in the \qtyrange{0.2}{3.5}{MeV} range and a deuterium gas target to fully cover the BBN energy range.

\paragraph{\texorpdfstring{\maybebm{\reaction{3}{He}{\ag}{7}{Be}}}{3He(4He,gamma)7Be}}
\label{wg2:t2:subsection:3He-4He-g-7Be}

The \reaction{3}{He}{\ag}{7}{Be} reaction plays an important role in the production of \nuclide{7}{Li} during BBN ($E = \qtyrange{160}{380}{keV}$) and is also crucial for the hydrogen-burning in low mass stars.

Due to its very low cross section, the \reaction{3}{He}{\ag}{7}{Be} reaction cannot be measured directly at solar energies and the data relies on high energy data extrapolations using different theoretical calculations. 
A recent theoretical work \cite{Zhang2020} showed that the angular distribution of the \reaction{3}{He}{\ag}{7}{Be} $S$-factor is directly correlated to the extrapolated astrophysical $S$-factor at solar energies.
To better constrain the extrapolations down to solar energies, experimental data which connect the BBN range data \cite{Confortola2007} with higher energy data \cite{DiLeva2009} is still missing. A new high energy measurement might be performed at the BIBF 3.5\,MV accelerator using a gas target setup similar to the one used in \cite{Cavanna2015} in combination with a large (\qty{140}{\percent}) HPGe. 

\paragraph{\texorpdfstring{\maybebm{^7\mathrm{Be}\Dp^8\mathrm{Be}}}{7Be(d,p)8Be} reaction via THM at BBN energies}
\label{wg2:t3:paragraphBB_7Be+d}

The quest of a nuclear solution to the long-standing cosmological lithium-problem is matter of debate and research in nuclear astrophysics \cite{Coc2004,Fields2011,Cyburt2016}. Indeed, while BBN nicely describes \nuclide{2}{H}, \nuclide{3}{He} and \nuclide{4}{He} primordial abundances in connection with the most recent astronomical observations, it fails in predicting the \nuclide{7}{Li} ones being these a factor $\sim$2-3 larger with respect the observed ones in halo stars, as discussed in \cite{Sbordone2010}. Besides the solutions provided by cosmology and stellar physics, nuclear physics solutions have been also investigated in recent years in order to constrain the processes affecting the nucleosynthesis of \nuclide{7}{Li}. For instance, the Trojan Horse Method (THM, see \eg{} \cite[and references therein]{Tumino2021}) was applied to study the \reaction{7}{Be}{+}{}{\mathrm{n}} induced reactions, opening the new frontier of applying THM on reactions in which unstable nuclei and neutrons are involved. In detail, the \reaction{7}{Be}{\na}{4}{He} THM measurement \cite{Lamia2017, Lamia2019} was performed at the EXOTIC facility at LNL \cite{Farinon2008,Ballan2023} while a subsequent measurement \cite{Hayakawa2021} allowed to investigate also the competing \reaction{7}{Be}{\np}{7}{Li} channel via a devoted experiment performed at CRIB-Riken in which both p$_0$ and p$_1$ exit channels have been measured.\\*
In order to further investigate the impact of nuclear reactions involving the unstable \nuclide{7}{Be}-isotope, a further channel to be investigated is the \reaction{7}{Be}{\Dp}{8}{Be} because of its role in depleting \nuclide{7}{Be} nuclei during BBN by populating several \nuclide{8}{Be} exited states in the exit channel as given in \cite{Ali2022}. 
 
However, direct cross section measurements partially cover the BBN energy region (\ie{} energies $\approx$ 100-300\,keV) thus requiring additional investigations via indirect techniques.

For such a reason, THM could be applied to the quasi-free $\reaction{6}{Li}{(\nuclide{7}{Be},\mathrm{p}\,\nuclide{8}{Be})}{4}{He}$ break-up reaction by using the LNS THM group know-how and a 7-9\,MeV \nuclide{7}{Be} beam, available at the CIRCE laboratory, impinging on a \nuclide{6}{LiF} target, evaporated on a C-backing. Indeed, \nuclide{6}{Li} represents a suitable TH-nucleus because of its marked $\alpha$+d structure, its relative low binding energy, $\sim$1.47\,MeV, and its well known momentum distribution for the $\alpha$-d intercluster motion occurring in s-wave \cite{Tumino2021}. 


\subsubsection{Hydrogen burning}

\paragraph{\texorpdfstring{\maybebm{^{7}\mathrm{Be}(\mathrm{p},\gamma)^{8}\mathrm{B}}}{7Be(p,gamma)8B}}
\label{wg2:t3:paragraphHB_7Be+p}

The reaction \reaction{7}{Be}{\pg}{8}{B} plays an important role in the Sun, where it determines the high energy component of the solar neutrino spectrum. The importance of this reaction triggered several experiments over the last decades. Recently, the ERNA Collaboration measured for the first time, with adequate precision the cross section of \reaction{7}{Be}{\pg}{8}{B} using the a radioactive \nuclide{7}{Be} beam and a RMS in the energy range from $\Ecm$=367\,keV to 812\,keV \cite{Buompane2018,Buompane2021}.
A combined analysis of their results produces an overall consistent picture for the energy dependence of the cross section, while an inflation of the quoted uncertainties is needed to accommodate the observed discrepancy in the absolute scale of the different data sets \cite{Adelberger2011, Buompane2021}.
It is important to note that in the energy range above $\Ecm=1$\,MeV the different data sets and theoretical models show the greatest discrepancy. The available direct measurements data, in this energy range, are all obtained in direct kinematics, with a proton beam on a \nuclide{7}{Be} target \cite{Adelberger2011,Cyburt2004}. A completely new measurement, performed in inverse kinematics, with a \nuclide{7}{Be} beam impinging on a proton target, using the ERNA RMS, will help shed light on the origin of the discrepancy and the selection of a more accurate theoretical extrapolation model.
Recently, the accelerator terminal has been equipped with a solid stripper system, that leads to an increase in the probability of higher charge states and then more intense \nuclide{7}{Be} beams at high energy. This opens the possibility to measure with the RMS the cross section up to $\Ecm=1.3$\,MeV. A new extended hydrogen gas target, with reduced dimension respect the one used in the lower energy range, is necessary to optimize the energy and angular acceptance of the ERNA RMS. Considering an average \nuclide{7}{Be} beam current of $\approx 10^8$pps, to perform the measurements in the energy range range $\Ecm=0.8$\,MeV to 1.3\,MeV will be needed about 80\,GBq of \nuclide{7}{Be} to produce the cathodes and 8\,weeks measurement time.


\paragraph{\texorpdfstring
{\maybebm{\reaction{14}{N}{\pg}{15}{O}}}
{14N(p,gamma)15O}
}
\label{wg2:t2:subsection:14N-p-g-15O}

As already mentioned, the reaction \reaction{14}{N}{\pg}{15}{O} plays a fundamental role in the CNO cycle of stellar hydrogen burning, and has strong impact in various other astrophysical phenomena \cite{Wiescher2010}, being the bottle-neck of the cycle. An extrapolation to the astrophysical relevant energies requires its cross section to be known with high precision over a wider energy range. Recent experiments exhibit some contradictions \cite{Li2016,Wagner2018,Gyurky2022,Frentz2022}, this emphasizes the need for a new, high-precision measurement of the \reaction{14}{N}{\pg}{15}{O} cross section. The energy range of BIBF 3.5\,MV accelerator, is perfectly suited to  address these emerging inconsistencies.

Preparations for the experiments at the BIBF are already ongoing. A first setup will use a single $\gamma$-ray detector, readily available at LUNA, and can then be extended to a multi-detector setup for angular distribution experiments. Different options of target preparation are investigated, including implanted and sputtered solid state targets. For the high proton energies and intensities, beam-induced background and target stability are crucial issues. 

Additionally, a further study of the \reaction{14}{N}{\pg}{15}{O} at the \LUNAfourhundred{} accelerator is highly desirable, to measure the cross section at energies above and below the well known resonance at $\Ecm = \qty{259}{keV}$ \cite{Marta2010}. In particular, cross section data with branching ratios for the decay of \nuclide{15}{O} at energies below the resonance have large uncertainties \cite{Formicola2004}. The use of a new scintillation detector with high efficiency and granularity, currently under planning at LUNA, would allow to reach the ambitious goal of measuring the \reaction{14}{N}{\pg}{15}{O} total and partial cross sections down to very low energies. The principle of using a segmented summing detector to reconstruct the branching ratios has been already studied by the LUNA Collaboration\cite{Ferraro2018a,Ferraro2018b}.

\paragraph{\texorpdfstring
{\maybebm{\reaction{19}{F}{\PA{0,1}}{16}{O}}, \maybebm{\reaction{19}{F}{\PAG{2,3,4}}{16}{O}}, \maybebm{\reaction{19}{F}{\pg}{16}{O}}}
{19F(p,a\{0,1\})16O, 19F(p,alpha\{2,3,4\}+gamma)16O, 19F(p,gamma)20Ne}
}
\label{wg2:t2:subsection:19F-p-a-16O}
\label{wg2:t2:subsection:19F-p-g-20Ne}

The galactic abundance of \nuclide{19}{F} is a long standing puzzle in nuclear astrophysics. 
The dominant \reaction{19}{F}{\PA{0}}{16}{O} channel has been studied at energies below \qty{400}{keV} (of interest for AGB star nucleosynthesis) via direct and indirect measurements \cite{Lombardo2015, Indelicato2017} with results in qualitative agreement. However, a recent review paper \cite{deBoer2021} reported a disagreement between aforementioned data and old results \cite{Herndl1991}.
The \reaction{19}{F}{\PA{1}}{16}{O} reaction is expected to have a negligible contribution to the total reaction rate \cite{deBoer2021}. In a recent work \cite{Lombardo2019}, however, the crucial impact on the \reaction{19}{F}{\PA{1}}{16}{O} reaction $S(E)$ of the elusive resonance at $\Ecm \sim \qty{206}{keV}$ (corresponding to $\Ex = \qty{13095}{keV}$) was clearly pointed out, demanding experimental efforts, especially at unexplored astrophysical energies. At the Jinping Underground Nuclear Astrophysics (JUNA) laboratory recently was measured for the first time the \reaction{19}{F}{\PAG{2,3,4}}{16}{O} reaction $S(E)$ down to $\Ecm = \qty{72.4}{keV}$ \cite{Zhang2021}, with a significant difference with respect to a previous measurement performed down to $\Ep = \qty{200}{keV}$, that was leading to a quite different extrapolation, thus new independent data would help in settling the \dots.\\*
The \pg{} channel investigation is severely hampered by the huge background produced by the \pag{} reaction. Another measurement performed at JUNA, however, recently reported new $S$-factor data down to $E\sim$180~keV, resulting in an increase of the rate of this reaction by almost an order of magnitude \cite{Zhang2022}. Given the significant impact of such results on our understanding of observed abundances in population III stars \cite{Clarkson2021} together with the fact that we still need to rely on extrapolation at Gamow energies, a new measurement is required to confirm and possibly improve the JUNA results. 

Given the high intensity beam of \LUNAfourhundred{} it is crucial to investigate the optimal target among the many options available in literature.
For the $\alpha$ detection the setup designed for the \reaction{23}{Na}{\pa}{20}{Ne} reaction might be used with minor changes, see below. 
The particle detection system will be coupled with \num{2} CeBr$_3$ scintillators and a HPGe detector in forward geometry to measure the prompt $\gamma$-rays produced by the \PAG{} channel. Coincidence measurements between $\alpha$ particles and $\gamma$-rays will improve the discrimination between the different $\alpha$ groups.
A second campaign will focus on the \PA{1} and the \pg{} channels and it would be performed with a high efficiency segmented $4\pi$ scintillator detector, placed all around the target chamber. The \reaction{19}{F}{\PA{1}}{16}{O} reaction, indeed, populates the $0^+$ state at $\Ex = \qty{6049}{keV}$ in \nuclide{16}{O}, which de-excites via $e^-e^+$ pair production. The emitted positron quickly annihilates producing two photons of $\qty{511}{keV}$ energy each in opposite directions, which will be detected in coincidences by two opposite crystals of the clover scintillator. In the study of the of the \reaction{19}{F}{\pg}{20}{Ne} reaction, one of the major challenges is the background due to the concurrent \PAG{} channel, its identification would benefit of a highly segmented detector.

\paragraph{\texorpdfstring
{\maybebm{\reaction{23}{Na}{\pa}{20}{Ne}}, \maybebm{\reaction{27}{Al}{\pa}{24}{Mg}}}
{23Na(p,alpha)20Ne, 27Al(p,alpha)24Mg}
}
\label{wg2:t2:subsection:23Na-p-a-20Ne}
\label{wg2:t2:subsection:27Al-p-a-24Mg}

The \reaction{23}{Na}{\pa}{20}{Ne} and \reaction{27}{Al}{\pa}{24}{Mg} reactions belong to the NeNa and MgAl cycles, respectively, active during H-burning when the temperature exceeds $\sim \qty{50}{MK}$. In particular, over the temperature range relevant for Hot-Bottom Burning in the final stages of intermediate-mass stars ($M \approx \qtyrange{4}{8}{\solarMasses}$), and for core and shell-hydrogen burning in massive stars ($M \ge \qty{8}{\solarMasses}$), the \reaction{23}{Na}{\pa}{20}{Ne} appears \cite{Cesaratto2013} to totally dominate the \nuclide{23}{Na} destruction due to proton capture. As such, its rate is key to evaluate how much Na is synthesized in stars and subsequently ejected in the interstellar medium by means of \emph{winds} and/or SNe explosions.

One of the most debated issues in modern astrophysics is the origin of the peculiar chemical patterns observed in Globular Clusters, gravitationally-bound groups of very old stars ($\approx \qtyrange{12}{13}{\giga\year}$), whose ages set a lower limit to the age of our Universe. Understanding the chemical characteristics of Globular Clusters can provide us with the key to understanding the formation of our Galaxy from the epochs of the early Universe until today. Of particular interest in recent years has been the so-called Oxygen-Sodium (O-Na) anti-correlation \cite{Carretta2009}, \ie{} a strong sodium enhancement when oxygen is heavily depleted. Therefore, having a precise measurement of the \reaction{23}{Na}{\pa}{20}{Ne} rate is important for testing the plausibility of the various hypotheses suggested so far \cite{Slemer2017,DAntona2016, Renzini2015}. 

The \reaction{27}{Al}{\pa}{24}{Mg} reaction is also active in stellar interiors during the hydrogen burning stages and significantly affects the abundances of Mg and Al isotopes. Current uncertainties do not allow an unambiguous conclusion about the existence of a Mg-Al cycle, for example according to present knowledge at temperatures $T < \qty{50}{\mega \kelvin}$ the ratio of the cross section of the $\pg$ and the $\pa$ channels ranges from about \num{0.01} to \num{1000} \cite{Iliadis2001} due to a number of key nuclear resonances for which only upper limits exist \cite{LaCognata2022}.

In the framework of the ELDAR ERC starting grant\footnote{UKRI ERC-StG grant EP/X019381/1}, a new measurements of the \reaction{23}{Na}{\pa}{20}{Ne} and \reaction{27}{Al}{\pa}{24}{Mg} reactions are already planned. A new charged particle detection setup specifically designed and built for underground in beam charged-particle spectroscopy will be mounted at the \LUNAfourhundred{} accelerator.

For the $^{23}$Na(p,$\alpha$) reaction, the solid target will likely be a tantalum-backed target of the type used and tested in \cite{Boeltzig2019} at LUNA. Alpha particles will be detected at backward angles by an array of silicon detectors, shielded by thin aluminized Mylar foils from the flux of recoiling beam protons. Tests are underway to employ neutron-transmutation doped (nTD) silicon detectors to identify and discriminate charged particles via pulse shape discrimination without the need of a $\Delta E$-$E$ telescope or similar arrangements that would cause a larger energy degradation. For light particles this type of detectors has already been shown to be capable of particle identification at higher energies compared to those expected for this experiment \cite{Assie2015}. Particle identification would further reduce the background and help identifying the alpha signal of interest. 

\paragraph{\texorpdfstring
{\maybebm{\reaction{30}{Si}{\pg}{31}{P}}}
{30Si(p,gamma)31P}
}
\label{wg2:t2:subsection:30Si-p-g-31P}

An underground measurement of the $\reaction{30}{Si}{\pg}{31}{P}$ reaction would represent a fundamental piece of knowledge to solve the Globular Cluster abundance anomalies puzzle.
Knowledge on the cross section of this reaction for astrophysical scenarios is sorely lacking and a measurement of the $\reaction{30}{Si}{\pg}{31}{P}$ reaction ($Q = \qty{7297}{keV}$) would be possible at the LUNA 400\,kV accelerator.
The high intensity and the possibility of beam-induced background issues requires a dedicated study to find the best target and eventually the optimal experimental strategy. 

Because of the low rate expected, particularly at low energies, a high efficiency $4\pi$ scintillator detector should be used, combined with light (Al) target chamber and holder, and with a properly designed shielding to further reduce the residual background.

\subsubsection{Hydrogen burning in extremely metal poor stars}\label{metalpoor_astro}

\paragraph{\texorpdfstring
{\maybebm{\reaction{6}{Li}{\ag}{10}{B}}, \maybebm{\reaction{7}{Li}{\ag}{11}{B}}, \maybebm{\reaction{10}{B}{\ad}{12}{C}}, \maybebm{\reaction{10}{B}{\ap}{13}{C}}, \maybebm{\reaction{10}{B}{\an}{13}{N}}}
{6Li(alpha,gamma)10B, 7Li(alpha,gamma)11B, 10B(alpha,d)12C, 10B(alpha,p)13C, 10B(alpha,n)13N}
}
\label{wg2:t2:subsection:6Li-a-g-10B}
\label{wg2:t2:subsection:7Li-a-g-11B}
\label{wg2:t2:subsection:10B-a-d-12C}
\label{wg2:t2:subsection:10B-a-p-13C}
\label{wg2:t2:subsection:10B-a-n-13N}

The importance of these reactions has been discussed in subsection \ref{triplea_astro}.
The strong background reduction in underground and the intense $\alpha$-beam available at the LUNA 400\,kV accelerator allow to study the \nuclide{10,11}{B}+$\alpha$ and \nuclide{6,7}{Li}+$\alpha$ reactions at the most relevant energies, \ie{} $\Ecm\gtrsim100$\,keV. 
Specifically, we plan to investigate the \nuclide{10}{B}\ap{}, \ad{} and \an{} using a combined detection system for charged-particles and neutrons.
Separate measurements will be done for the \reaction{6,7}{Li}{\ag}{10,11}{B} radiative captures.
These studies will serve to address the contribution and location of broad and interfering resonances in the compound systems, as well as DC transitions with the intent of reduce reaction rate uncertainties.

Count-rate estimates on the basis of present knowledge suggest a data taking time of about \qty{2}{weeks} per reaction. 
Estimates are calculated on the basis of $R$-matrix extrapolations of existing $S$-factor data, assuming \qty{200}{\mu{}A} $\alpha$-beam current, \qty{2e8}{atoms/\cm\squared} active nuclei in the target, a \qty{10}{\percent} (\qty{20}{\percent}) detection efficiency for charged-particles (neutrons).

\paragraph{\texorpdfstring{\maybebm{\reaction{7}{Be}{\ag}{11}{C}}}{7Be(alpha,gamma)11C}}
\label{wg2:t3:paragraphHeB_7Be+alpha}

The \reaction{7}{Be}{\ag}{11}{C} reaction rate remains very uncertain because of the quite limited knowledge on its cross section. Until recent only the strength of the two resonances at $\Ecm=884$ and 1376\,keV were known \cite{hardie1984}. This triggered new experimental efforts and recently the strength of two other resonances at $\Ecm=1110$ and 1155\,keV were measured in inverse kinematics using the DRAGON RMS \cite{Psaltis2022a,Psaltis2022b}. Thus, the NACRE II compilation \cite{Xu2013} to evaluate the cross section uses a potential model analysis, and the strengths of higher energy resonances are estimated on the basis of the corresponding states in the mirror nucleus \nuclide{11}{B}. Likewise, DC contribution is calculated from the mirror reaction \reaction{7}{Li}{\ag}{11}{B}. As a consequence the the \reaction{7}{Be}{\ag}{11}{C} $S$-factor is loosely constrained, and the rate below $T=0.2$\,GK has an estimated uncertainty of about an order of magnitude. Further, the known resonance strengths rest on just a single experimental investigation. Measurement at higher energies, would be interesting also to better determine the \reaction{10}{B}{\pg}{11}{C} and \reaction{10}{B}{\pa}{7}{Be} cross sections, since they populate the same compound nucleus \cite{VandeKolk2022}. 
At these higher energies, it would be very useful to measure the resonance that corresponds to the threshold state in the $^{10}$B+p reaction as well as broader higher energy resonances where their level assignments are still somewhat uncertain.

\subsubsection{Helium burning}

\paragraph{\texorpdfstring{\maybebm{^{12}\mathrm{C}\ag^{16}\mathrm{O}}, \maybebm{^{16}\mathrm{O}\ag^{20}\mathrm{Ne}}}{12C(a,gamma)16O, 16O(a,gamma)20Ne}}
\label{wg2:t3:paragraphHeB_12C+alpha}
\label{wg2:t3:paragraphHeB_16O+alpha}

The $\cag$ is challenging, being the cross-section in the Gamow window ($\approx$\qty{300}{keV}) the result of the composition of overlapping resonances and non resonant components.
The estimated cross section at $\Ecm = 300$\,keV is about \qty{2e-17}{\barn} \cite{deBoer2017} which makes a direct measure unfeasible. The estimation is done by extrapolating higher energy measurements to the stellar energies through a phenomenological model such as R-matrix.
The \reaction{12}{C}{\ag}{16}{O} reaction has been studied for more than 60 years now, see e.g. \cite{deBoer2017}. Its astrophysical interest was known since 1957 \cite{Burbidge1957} but experiments specifically for this purpose started only 13 years later with \cite{Jaszczak1970}. Given its central role in the astrophysical scenario, many research groups faced the goal of measuring the \reaction{12}{C}{\ag}{16}{O} cross section from that moment on. The demand for more accurate cross section extrapolations, pushed the design of increasingly sophisticated apparatuses and analysis techniques to solve the experimental and theoretical difficulties arising.
Despite all of the work mentioned above, the uncertainty on the extrapolated $S$-factor sits around \qty{15}{\percent}. Fits are now mostly limited by inconsistencies of data sets available in literature especially for the $E2$ ground state transition. In synthesis, the following focus on future measurements are desired \cite{deBoer2017}:
\begin{itemize}
    \item differential cross section,
    \item interference behavior,
    \item accurate cascade transitions,
    \item lower energy measurements with RMSs.
\end{itemize}
However, beam induced background, mainly from \ngamma{} reactions induced by neutrons from the \reaction{13}{C}{\an}{16}{O} reaction, will still remain one of the major challenges to be afforded.
Past measurements have shown that this background can be strongly suppressed using time-of-flight (\eg{} \cite{Makii2005, Makii2009, Plag2012}), and that this outweighs the loss of beam intensity because of the gain in signal to background. The $E1$ and $E2$ multipole components of the ground state transition are the dominant components of the total cross section at low energy.
The $E1$ is most pronounced at \qty{90}{\degree} \cite{Dyer1974}, while the $E2$ and the interference term distribute the $\gamma$ intensity more evenly over a larger angular inteval. Therefore a high efficiency detection setup able to cover forward, \qty{90}{\degree}, and backward angles will be required. \\*
It is very important to anchor any low energy measurements with a scan of the well known broad $1^-$ level at a center of mass energy of about \qty{2.3}{MeV}. 
Measurements with improved accuracy are needed to better constrain the shape of the tails of the $1^-$ resonance thereby improving the extrapolation to low energy because the resonance shape is distorted due to interference with the sub-threshold state.
These measurements are desirable both above and below the $1^-$ resonance, as demonstrated by the greatly improved constraint provided by the results reported in \cite{Schurmann2012}.
$E2$ measurements over this region are also highly desirable where there is a great deal of discrepancy in past data \cite{Smith2021, deBoer2017}.
Finally, very low energy $E2$ measurements, which will be extremely challenging, can constrain the relative contributions of the $E2$ sub-threshold state and the $E2$ DC component \cite{Shen2020}.

Another possible approach to measure the \reaction{12}{C}{\ag}{16}{O} is to use a RMS where the $^{16}$O produced in the reaction is detected in coincidence with $\gamma$-rays, thus measuring at the same time the total cross section and the angular distribution of the $\gamma$-rays. The environmental background is negligible on the charged particle detection and is strongly suppressed in the $\gamma$-ray spectra by applying coincidence between recoils and photons. This has been done in the past with the ERNA separator \cite{Schurmann2005}. The total cross section data are still, after almost 20 years, the most stringent constrain to the R-matrix fit in the energy range covered with RMS measurements. The ERNA collaboration has upgraded the apparatus with a new beam optics layout and a recirculating helium jet gas target combined with a $2\pi$ scintillator array to extend its capability both in terms of energy range and angular resolution of the $\gamma$-ray detection. The ongoing campaign will asses the cross section down to $\Ecm\sim1$\,MeV with a focus on the $E1$ and $E2$ contributions. \\
The \reaction{16}{O}{\ag}{20}{Ne} reaction represents the endpoint of the main reaction sequence $^{4}$He(2$\alpha$,$\gamma$)$\nuclide{12}{C}$($\alpha$,$\gamma$)$\nuclide{16}{O}$($\alpha$,$\gamma$)$^{20}$Ne and determines, together with the rate of $\nuclide{12}{C}$($\alpha$,$\gamma$)$\nuclide{16}{O}$, the $\nuclide{16}{O}$ abundance at the ignition of the carbon burning phase in late stellar evolution.
As for \reaction{12}{C}{\ag}{16}{O}, the ERNA RMS at CIRCE can be used for the measurement of the \reaction{16}{O}{\ag}{20}{Ne} total cross section using the already available gas target and the same final detector with a few upgrades in the setup.

\paragraph{\texorpdfstring{\maybebm{^{14}\mathrm{N}\ag^{18}\mathrm{F}, ^{15}\mathrm{N}\ag^{19}\mathrm{F}}}{14N(alpha,gamma)18F, 15N(alpha,gamma)19F}}
\label{wg2:t3:paragraphHeB_14N+alpha}
\label{wg2:t3:paragraphHeB_15N+alpha}

Thanks to their high $Q$-value both the reactions are good candidates for measurements at the BIBF 3.5\,MV accelerator through $\gamma$-ray spectroscopy.

At temperatures from \qtyrange{0.1}{0.5}{GK} the rate of the \reaction{14}{N}{\ag}{18}{F} reaction rate is dominated by the contribution of a $J_\pi = 1^-$ resonance at $\Ecm = \qty{572}{keV}$.
Below \qty{0.1}{GK} additional contributions are possible from the low-energy tail of the \qty{445}{keV} resonance, the DC component, and the $J_\pi = 4^+$ resonance at \qty{237}{keV}.
The lowest directly measured resonance is the one at \qty{445}{keV} \cite{Iliadis2010}, whose cross section is sufficiently high to allow a new measurement.
The strength of the $\Ecm = \qty{237}{keV}$ resonance has never been measured and is far below present experimental measurement possibilities.
The DC contribution for this reaction is only theoretically estimated from spectroscopic factors, using some crude approximations, to be about \qtyrange{0.7}{0.8}{\kilo\electronvolt\barn} \cite{Goerres2000}. An underground measurement is an opportunity to determine it.

The rate of the \reaction{15}{N}{\ag}{19}{F} reaction is dominated by the resonant contribution of several low-lying states in \nuclide{19}{F}.
At astrophysical relevant energies, the main contributions to the thermonuclear reaction rate are from the $\Ecm = \qty{364}{keV}$, \qty{536}{keV} and \qty{542}{keV} resonances.
The strength of the \qty{364}{keV} resonance, directly influencing the reaction rate at temperatures lower than \qty{200}{MK}, has been measured only indirectly \cite{deOliveira1996}.
The counting rate at this energy is about \num{100} reactions per day considering an $\alpha$ beam of \qty{100}{\mu\ampere}, allowing for a direct measurement in the underground facility. Also in this case, the high beam intensity might allow to perform a measurement of the DC component. 

However, such measurements require a new and high efficiency $4\pi$ segmented detector and a HPGe detector in close geometry, while monitoring target degradation for a proper cross section normalization. An optimization of the setup (\eg{} low absorption aluminum vacuum pipes, lead and polyethylene shielding around the detectors) will also concur to maximize the signal to noise ratio to achieve the best possible precision.

The \reaction{15}{N}{\ag}{19}{F} reaction has been recently studied also using the ERNA RMS \cite{DiLeva2017}. Similarly the $\reaction{14}{N}{\ag}{18}{F}$ might be investigated. The measurement of the total cross section through the detection of the recoils produced in the reactions can provide complementary information about the processes and experimental systematic effects.

\paragraph{\texorpdfstring
{\maybebm{\reaction{17}{O}{\ag}{21}{Ne}, \reaction{18}{O}{\ag}{22}{Ne}}}
{17O(alpha,gamma)21Ne, 18O(alpha,gamma)22Ne}
}
\label{wg2:t2:subsection:17O-a-g-21Ne}
\label{wg2:t2:subsection:18O-a-g-22Ne}

Both the \reaction{17}{O}{\ag}{21}{Ne} and \reaction{18}{O}{\ag}{22}{Ne} reactions have been scarcely studied in the past and no DC data exist for these reactions. The strengths of some of the resonances are known, although with large uncertainties \cite{Goerres2000,Dombos2022}.
Both nitrogen and oxygen targets are very well known and were studied many times during the last years.
Additionally, the experimental technique could involve the activation method \cite{DiLeva2014} for the first reaction and the total absorption spectroscopy \cite{Ferraro2018b} for the second one.
Furthermore, both the techniques would require the use of a highly segmented and efficient scintillator clover detector \cite{Boeltzig2018}, thus both the reactions could be studied during the same beam time.


\subsubsection{Neutron sources}
\paragraph{\texorpdfstring{\maybebm{\reaction{22}{Ne}{\an}{25}{Mg}, \reaction{22}{Ne}{\ag}{26}{Mg}}}{22Ne(alpha,n)25Mg, 22Ne(alpha,gamma)26Mg}}
\label{wg2:t2:subsection:22Ne-a-n-25Mg}
\label{wg2:t2:subsection:22Ne-a-g-26Mg}

The energy range of interest for both reactions is around 460 - 750\,keV (for a stellar temperature of 300\,MK), and below the neutron production threshold at $E_\alpha = 561.9$\,keV only the $\gamma$-ray channel is active. Both below and above the threshold a number of natural-parity compound states have been identified, but their resonance strengths are not known (mostly due to undetermined $\alpha$ widths) \cite{Adsley2021}. Only a single resonance in the stellar energy window (at $E_\alpha=830$\,keV) has been directly measured in both channels, leading to large uncertainties in nucleosynthesis calculations.

Using dedicated detector and target systems (see section \ref{wg0:LNGS:sub2}), the high beam intensity of the BIBF 3.5\,MV accelerator and the ultra-low background in the underground an increase in sensitivity for both reactions by at least three orders of magnitude is expected.

The measurement of the neutron channel is funded by the SHADES ERC starting grant\footnote{SHADES ERC-StG grant n. 852016}. Instead of traditionally used energy-blind moderating neutron counters it relies on a coupled array of liquid scintillators and \nuclide{3}{He} counters, allowing for a certain energy sensitivity. This latter point is of great importance to be able to identify possible beam-induced backgrounds, which have been an issue in past experiments \cite{Jaeger2001}. In addition, it allows to suppress a part of the remaining higher-energy natural neutron background, improving sensitivity even more.

The measurement of the \ag channel has strong synergies with the neutron setup, utilizing the same target and electronics system. Design of a high-efficiency, low background $\gamma$-ray detection system is underway.

The program for both measurements is similar: re-measuring the 830\,keV resonance and performing a high-sensitivity search for resonances in the astrophysical energy range.

\paragraph{\texorpdfstring{\maybebm{\reaction{13}{C}{\an}{16}{O}}}{13C(alpha,n)16O}}
\label{wg2:t2:subsection:13C-a-n-16O}

As discussed in Section \ref{wg2:t1}, this reaction is the main neutron source in AGB stars, at Gamow energies of \qtyrange{140}{230}{keV}. 
Due to the nuclear structure of the \nuclide{17}{O} compound nucleus, the reaction rate in this region of interest is dominated by two resonances: the near-threshold resonance at $\Ex = \qty{-3(8)}{keV}$ and a wide resonance at $\Ecm = \qty{880}{keV}$ ($\Ex = \qty{7215}{keV}$, $\Gamma = \qty{300}{keV}$).
As for direct measurements, two recent measurements were performed in underground laboratories by the LUNA \cite{Ciani2021} and JUNA collaborations \cite{Gao2022}, entering the Gamow window and improving the extrapolation towards lower energies. In particular, at LUNA was used an $\alpha$ beam impinging on a solid highly enriched \nuclide{13}{C} target, produced by evaporation. Neutrons were detected using the \nuclide{3}{He} counters detection array \cite{Csedreki2021}. The cross section was determined in the $\Ecm = \qtyrange{230}{305}{keV}$ energy range, with an overall uncertainty lower than \qty{20}{\percent} for all data points, thanks to the setup high efficiency and the low intrinsic background in combination with an innovative method for target monitoring \cite{Ciani2020}.
It has been shown, in particular, that this reaction rate is about \qty{20}{\percent} lower than previously estimated, with interesting consequences for the production of some $s$-process isotopes. This result was confirmed by the independent measurement performed at the JUNA underground laboratory \cite{Gao2022}. 
However, at high energy, the new measurements present a systematic difference with respect to the results of previous experiments. For this reason, a further study at higher energy would clarify the origin of this discrepancy, thus allowing a better R-matrix fit of the astrophysical $S$-factor down to the energies relevant for $s$-process nucleosynthesis.
Independently, a crucial open point is a \qty{30}{\percent} discrepancy between different data sets on the \qty{880}{keV} resonance, \ie{} the results reported in \cite{Heil2008} and in \cite{Harissopulos2005}. According to the sensitivity study presented in \cite{deBoer2020}, the inconsistency in the overall normalization of the \reaction{13}{C}{\an}{16}{O} experimental data remains a significant source of uncertainty.

Thanks to the BIBF 3.5\,MV accelerator, the LUNA measurement will be extended towards higher energies, providing a complete data set over a wide energy range.


\subsubsection{Carbon burning}
\label{wg2:t2:subsection:carbon}

\paragraph{\texorpdfstring{\maybebm{^{12}\mathrm{C}+{}^{12}\mathrm{C}, {}^{12}\mathrm{C}+{}^{16}\mathrm{O}}}
{12C+12C, 12C+16O}}

\label{wg2:t2:subsection:12C-12C}
\label{wg2:t3:paragraphCB_12C+12C}
\label{wg2:t3:paragraphCB_12C+16O}
\label{wg2:t3:subsectionCB}

Fusion reactions involving carbon nuclei are among the most important in stellar evolution since they determine the fate of massive stars. In particular, $\CC$ and $\nuclide{12}{C}$+$\nuclide{16}{O}$ reactions play a crucial role in the C and O burning phase of stellar evolution. Existing data do not provide an estimate of the stellar rate of these reactions with the precision required by stellar models. Recent indirect measurements indicated possible low energy resonances, dramatically changing the predicted rate, calling for improved direct investigations. 
In the past, some experiments provided hints of an anisotropic angular distribution of the $\CC$ reactions \cite{Becker1981}.

The $\CC$ will be the flagship reaction of the first \qty{5}{year} scientific program of the LUNA Collaboration with the BIBF 3.5\,MV accelerator. Energies in the center of mass system down to about \qty{1.5}{MeV} will be investigated. 
The reaction can proceed through different channels corresponding to the emission of a photon, a neutron, a proton, an $\alpha$ particle or even two $\alpha$ particles or a \nuclide{8}{Be} nucleus. Of these channels, the two more relevant are the emission of protons and alpha particles. The proton and alpha channels can be measured either by detecting the charged particles or by revealing the $\gamma$ decay of the first excited state to the ground state of the \nuclide{23}{Na} or \nuclide{20}{Ne} residual nuclei, respectively. 
A significant challenge in direct measurements is the beam induced background, produced by the interaction of the impinging C ions with \nuclide{1}{H} and \nuclide{2}{H} contaminations present in graphite targets. Different approaches have been attempted to mitigate this issue \cite{Zickefoose2018,Morales2018}, also at LNL the dedicated experiment HEAT (Hydrogen dEsorption from cArbon Targets) was started \cite{Ballan2023} to develop a reproducible technique for hydrogen desorption in different types of graphite targets through heating up to \qty{1200}{\degreeCelsius}, while assessing hydrogen content through different IBA techniques.
\begin{figure}[hbt!]
\centering
\includegraphics[width=.6\textwidth]{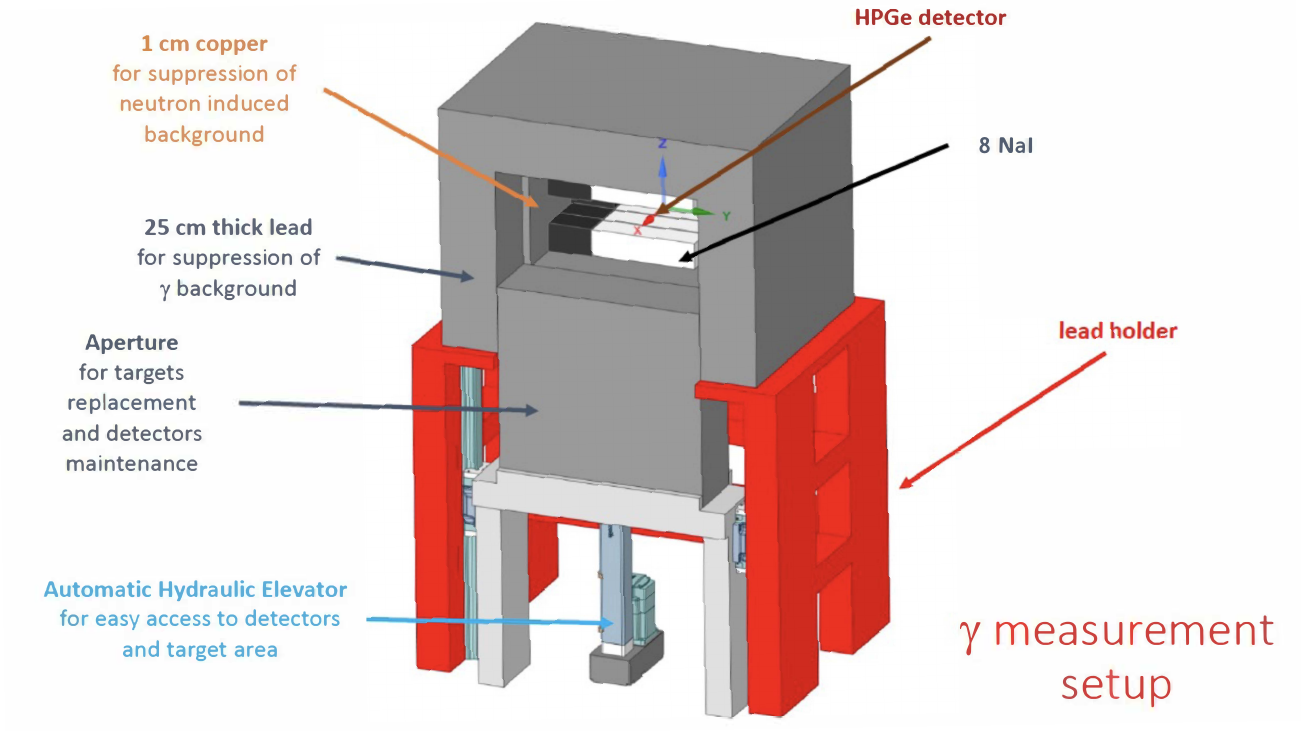}
\caption{Drawing of the shielding for the $\CC$ measurement which will be installed at LNGS.}
\label{wg2:t2:fig:shielding_12C}
\end{figure}

To measure the reaction cross section through the detection of $\gamma$-rays, the LUNA Collaboration is developing a setup composed by a lead (\qty{25}{\cm} thick) and copper (\qty{1}{\cm} thick) shielding to suppress the environmental $\gamma$-ray background. The detection setup will feature a HPGe detector (\qty{150}{\percent} relative efficiency) installed at \qty{0}{\degree} in close geometry to the target to maximize the signal to noise ratio and \num{8} NaI scintillator detectors surrounding both the target and the HPGe for Compton suppression, see Figure \ref{wg2:t2:fig:shielding_12C}.

To measure the reaction cross section detecting charged particles, their identification is needed, and different options are already under evaluation.\\* 
A promising candidate is the GASTLY (Gas-Silicon Two-Layer sYstem) detector \cite{Romoli2018} consisting of an ionization chamber (IC) and a Silicon Strip Detector (SSD) already used to measure this reaction at higher energies \cite{Morales-Gallegos2017}.
The intrinsic background of the GASTLY detectors expected from $\alpha$ emitter contaminations in the detector's materials, was studied within a joint effort of the ERNA and LUNA collaborations. Background measurements performed underground in different conditions, \eg{} several gas pressures, show background levels compatible with the expected yield for the $\CC$ reaction at 1.5\,MeV of a few reactions per day. Additional material characterization studies and Monte Carlo simulations are undergoing in order to identify most critical components and thus further reduce the intrinsic background with appropriate choice of materials or coatings.
Another option is given by nTD detectors using pulse shape discrimination analysis techniques, as already described in Section \ref{wg2:t2:subsection:22Ne-a-g-26Mg}.

Using the intense beam of the CIRCE laboratory accelerator, the ERNA Collaboration performed measurements of the $\CC$ down to $\Ecm\approx2.0$\,MeV \cite{Morales-Gallegos2017} and plans to measure $\nuclide{12}{C}$+$\nuclide{16}{O}$ down to $\Ecm=3.0$\,MeV, respectively. Such low energies have already a strong influence at the astrophysical region. Regions of energies where discrepancies still remain will be covered and data will overlap with the results obtained by indirect methods \cite{Tumino2018}. These data will clarify discrepancies, conceal results obtained with direct and indirect methods and minimize uncertainties.

\subsubsection{ERNA RMS innovative applications}
\label{wg2:t3:subsectionIA}

\paragraph{Half life Measurements}
\label{wg2:t3:paragraphIA_7BeHL}

Under terrestrial conditions, the short range of the strong interaction and its strength make the environmental influence on the atomic nucleus wave function negligible. Consequently, in most cases nuclear decay is not affected by environmental conditions. However, there are exceptions, like electron capture decay (EC). This process depends on the electron density in the nucleus, which can be modified by environmental conditions. \nuclide{7}{Be} is a nucleus decaying purely by EC to \nuclide{7}{Li} (Q=0.861\,MeV): the decay populates either the \nuclide{7}{Li} ground state (90$\%$) or the first excited state at 478\,keV (10$\%$) decaying to the \nuclide{7}{Li} ground state. The simple nuclear and atomic configurations of \nuclide{7}{Be} allow influencing the electron density at the nucleus rather easily, thus making \nuclide{7}{Be} ideal to constrain the decay nuclear matrix elements (NMEs). Moreover, the comparison of the so obtained experimental NMEs with calculated ones can provide a valuable contribution to the solution of the quenched gA puzzle. This latter refers to the need for quenching the axial coupling constant gA in nuclear many-body calculations employing truncated Hilbert spaces to reproduce beta-decay data \cite{Suhonen2017}.
In addition, it is worth noting that the change of the \nuclide{7}{Be} decay rate has been demonstrated to have a significant impact in solar physics and neutrino physics \cite{Das2005,Vescovi2019}.

A possible, yet unexplored, way to measure the change in half life for different \nuclide{7}{Be} ionization, is to observe its in-flight decay.  
This possibility is based on the fact that \nuclide{7}{Be} EC decay produces \nuclide{7}{Li} ions in the same charge state, being Auger emission from ionized \nuclide{}{Li} not allowed. Therefore, the measurement of the number ratio \nuclide{7}{Be}-to-\nuclide{7}{Li} ions allows determining the half life, provided:\\*
\phantom{xx}a. The \nuclide{7}{Be} ion beam is sufficiently purified from \nuclide{7}{Li}.\\*
\phantom{xx}b. Both \nuclide{7}{Be} and \nuclide{7}{Li} ions from the in-flight decay are fully collected.\\*
A full feasibility study has been performed at CIRCE by the ERNA Collaboration \cite{Santonastaso2019,Santonastaso2021}. Based on this work the measurement is briefly described below.\\*
Ions are accelerated toward the terminal, which is equipped with a gas stripper system, where molecules break and \nuclide{7}{Be} ions in different charge states $q_i$ are formed emerging from the accelerator with different energies. 
The \nuclide{7}{Li} contamination is eliminated populating the 4+ charge state, not accessible to \nuclide{}{Li}, in a post stripper at the accelerator exit. The pure $\nuclide{7}{Be}^{4+}$ beam is transported to the ERNA gas target through the AMS system, a dipole magnet-electrostatic analyzer-dipole magnet combination, that cleans up contamination with mass different from $A=7$. The gas target system can be operated with heavy or light gases to optimize the population of high or low charge states, respectively. Finally, the pure \nuclide{7}{Be} beam in the selected charge state enters the ERNA RMS and is transported to the end detector, a $\Delta$E-E telescope able of charged particle identification.

The in flight half life measurements described here is part of the project ASBeST (A 7-Beryllium electron capture STudy for nuclear and solid state physics), supported by the Italian Ministry of Research through the funding action PRIN2020.

\paragraph{THM with RMS}
\label{wg2:t3:paragraphIA_THM}

The measurement of the cross section of an $A+x\xrightarrow{}c+C$ two-body reaction of astrophysical interest using the THM (see \cite[\eg{}]{Tumino2021}) requires the selection of the quasi-free contribution of the three-body reaction $a+A\xrightarrow{}c+C+s$ measured in laboratory. 
For measurements where the TH nucleus $a$ is used as projectile the quasi-free reaction mechanism gives the maximum contribution to the three-body process for the spectator nucleus $s$ emitted at zero degree.
Although in a typical TH measurement nuclei $c$ and $C$ (ejectiles of the two-body process of interest) are detected in coincidence, in some cases spectator detection becomes necessary. In detail, it becomes crucial when one among $c$ or $C$ is a neutron as for the \reaction{13}{C}{\an}{16}{O} or \reaction{22}{Ne}{\an}{25}{Mg} reactions.
In such cases the use of a RMS, allowing the measurement of the spectator at zero degree, improves the coverage of the kinematical region where the quasi-free mechanism is favored, reducing the contribution due to other unwanted reaction mechanisms. This would allow for a different and complementary way to apply the THM. The excitation function of the two-body reaction of astrophysical interest will be measured in the best quasi-free kinematic conditions, that is momentum of the spectator equal to zero, with a change of the beam energy in order to span the energy range of astrophysical interest. 
Moreover, the possibility to measure at zero degree can improve the THM studies when one of the two ejectiles is emitted at very forward angles, allowing for a better covering of the quasi-free phase-space region. An actual physical case is represented by the $\reaction{23}{Na}{(\mathrm{d},\alpha\,\nuclide{20}{Ne})}{}{\mathrm{n}}$ THM reaction measured to study the \reaction{23}{Na}{\pa}{20}{Ne} (preliminary results obtained with standard THM are reported in \cite{Dagata2018}).
The ERNA RMS could allow to perform such kind of zero-degree measurements even if the available beam energies represent an important limit for the application of the THM where beam energies of tens of MeV might be required (even up to 100\,MeV).
Nevertheless, in order to investigate the coupling of the THM measurement with RMS and the possible experimental advantages, we propose a test consisting in the study of the \reaction{13}{C}{\an}{16}{O} applying the THM to the $\reaction{13}{C}{(\nuclide{6}{Li},\mathrm{n}\,\nuclide{16}{O})}{}{\mathrm{d}}$ reaction where the $\nuclide{6}{Li} = (\alpha \oplus\mathrm{d})$ is the TH nucleus. Such reaction was already measured with \nuclide{6}{Li} as target nucleus \cite{Lacognata2013,Trippella2017}. Now the idea is to use a \nuclide{6}{Li} beam delivered on a \nuclide{13}{C} target. Taking advantage of ERNA RMS the deuteron (spectator) will be detected at zero degree in coincidence with the \nuclide{16}{O} (measured using a position sensitive detector at target position) with a better selection of the quasi-free kinematical region and a reduced background.
To perform such measurement a scattering chamber to place a solid target and an array of DSSSD is required.

\paragraph{Triple-\texorpdfstring{\maybebm{\alpha}}{alpha} with RMS}
\label{wg2:t3:paragraphIA_Triple_a}

As stated before, the triple-$\alpha$ process is crucial to bridge the $A=5$ and $8$ mass gaps in the first generation of stars. Its rate is determined by the decay widths of the Hoyle state in \nuclide{12}{C}. The latter is the main channel in the intermediate-temperature domain. A large part of the uncertainty in the reaction rate estimation is currently related to the knowledge of the radiative decay width of the Hoyle state, \eg{} the recent estimation in \cite{Kibedi2020} is in disagreement with the former state-of-the-art \cite{Freer2014}. Unfortunately, the detailed measurement of the radiative decay of the Hoyle state is not trivial, as it requires precision $\gamma$-ray measurements. An alternative way is to probe the radiative decay branching ratio by exploiting particle-particle coincidence techniques, as previously done for example in Ref. \cite{Mak1975}. To this end, the Hoyle state in \nuclide{12}{C} can be formed as the final state of a nuclear reaction, such as $\nuclide{13}{C}(\nuclide{3}{He},\nuclide{4}{He})\nuclide{12}{C}$, and the radiative decay branching ratio of the Hoyle state can be measured from the coincidence rate between the two ions in the final state, given that the reaction ejectile tags the residual \nuclide{12}{C} in its Hoyle state. Usually, a pair of detectors can be used to detect the two ions in kinematical coincidence. However, the most favorable kinematical conditions are obtained when the light fragment is emitted backward, such that the recoiling \nuclide{12}{C} has the maximum energy. This poses challenges due to the high-rate measured at forward angle. An RMS such as ERNA, could be successfully exploited to overcome these limitations and crucially enhance the rate in the most favorable kinematical conditions.


\begin{table}[!htb]
\caption{List of the reactions described in previous sections that can be studied at LNGS, along with the machine required the measurement. The necessary upgrades to the experimental setup and time schedule are indicated: phase A corresponds to the next 2-3 years, phase B to 3-5 years, phase C to 5-7 years.}
\label{wg2:t2:tab1}
\centering
\small
\begin{tabular}{llll}

\hline \hline
Reaction & Machine & Upgrade & Phase \\ 
\hline
\reaction{}{D}{\pg}{3}{He}\phantom{\Large{0}} & 3.5\,MV & none &  C\\
\reaction{}{D}{\ag}{6}{Li} & 3.5\,MV & none &  C\\
\reaction{3}{He}{($^4$He,$\gamma$)}{7}{Be} & 3.5\,MV & targets &  C\\
\reaction{14}{N}{\pg}{15}{O} & 400\,kV & detectors &  B\\
\reaction{14}{N}{\pg}{15}{O} & 3.5\,MV & none &  A\\
\reaction{23}{Na}{\pa{}}{20}{Ne} & 400\,kV & detectors &  A\\
\reaction{27}{Al}{\pa{}}{24}{Mg} & 400\,kV & detectors &  A\\
\reaction{30}{Si}{\pg}{31}{P} & 400\,kV & detectors &  C\\
\reaction{19}{F}{\PA{0,1}}{16}{O} & 400\,kV & detectors &  B\\
\reaction{19}{F}{\PAG{2,3,4}}{16}{O} & 400\,kV & detectors &  B\\
\reaction{19}{F}{\pg}{20}{Ne} & 400\,kV & detectors &  B\\
\reaction{6}{Li}{\ag}{10}{B} & 400\,kV & targets &  C \\
\reaction{7}{Li}{\ag}{11}{B} & 400\,kV & targets &  C\\
\reaction{10}{B}{\ad}{12}{C} & 400\,kV & detectors &  C\\
\reaction{10}{B}{\ap}{13}{C} & 400\,kV & detectors &  C\\
\reaction{10}{B}{\an}{13}{N} & 400\,kV & detectors &  C\\
\reaction{11}{B}{\an}{14}{N} & 400\,kV & detectors &  C\\
\reaction{18}{O}{\ag}{22}{Ne} & 3.5\,MV & detectors &  B \\
\reaction{17}{O}{\ag}{21}{Ne} & 3.5\,MV & detectors &  B \\
\reaction{15}{N}{\ag}{19}{F} & 3.5\,MV & detectors &  B \\
\reaction{14}{N}{\ag}{18}{F} & 3.5\,MV & detectors &  B \\
\reaction{12}{C}{\ag}{16}{O} & 3.5\,MV & target and detectors &  C \\
\reaction{22}{Ne}{\an}{25}{Mg} & 3.5\,MV & gas target &  A \\
\reaction{22}{Ne}{\ag}{26}{Mg} & 3.5\,MV & gas target &  B \\
\reaction{13}{C}{\an}{16}{O} & 3.5\,MV & none &  A \\
$\CC$ & 3.5\,MV & target and detectors &  A (gamma)\\
$\CC$ & 3.5\,MV & target and detectors &  C (particles)\\
\hline \hline
\end{tabular}
\end{table}

\begin{table}[!htbp]
\caption{List of the reactions described in previous sections that can be studied at CIRCE. The necessary upgrades to the experimental setup, main needs and time schedule are indicated: phase A corresponds to the next 2-3 years, phase B to 3-5 years, phase C to 5-7 years.}
\label{wg2:t3:tab1}
\centering
\small
\begin{tabular}{lll}
\hline \hline
Reaction & Upgrade/Materials & Phase \\ 
\hline
\reaction{7}{Be}{\pg}{8}{B}\phantom{\Large{0}} & $^{1}$H Target/80\,GBq \nuclide{7}{Be} & A/B \\
\reaction{12}{C}{\ag}{16}{O} & none & A \\
\reaction{16}{O}{\ag}{20}{Ne} & none & A/B \\
\reaction{14}{N}{\ag}{18}{F} & none & A/B \\
\reaction{15}{N}{\ag}{19}{F} & none & A/B \\
$\reaction{7}{Be}{\ag}{11}{C}$ & RMS optics/60\,GBq \nuclide{7}{Be} & C \\
$\nuclide{7}{Be}\Dp\nuclide{8}{Be}$ & none/20\,GBq \nuclide{7}{Be}  & B \\
\nuclide{7}{Be} Half life & Windowless post-stripper/ & A \\
& 80\,GBq \nuclide{7}{Be} & \\
\nuclide{12}{C}+\nuclide{12}{C} & none  & B \\
\nuclide{12}{C}+\nuclide{16}{O} & none & B \\
Triple-$\alpha$ & \nuclide{3}{He} gas target & C \\
THM with RMS & Scattering Chamber, & C \\
& Array of DSSSD & \\
\hline \hline
\end{tabular}
\end{table}

\section{\wgthreetitle{}}
\label{wg3}
The development of nuclear astrophysics experiments at CIRCE has led to a fruitful interchange between basic and applied research activities. Indeed, analytical tools routinely used in applied nuclear physics can be devoted to fundamental research, and  the techniques and instrumentation developed for nuclear physics experiments contributes to an important know-how for further uses in applications. \\*
In Section \ref{wg3:t1} will be described the use of conventional and accelerator based mass spectrometry to study processes of astrophysical interest by means of isotopic ratio measurements. Similarly, some technical developments needed for nuclear physics experiments (ion beams, detectors and related equipment) have led to applications in various fields, as described in Section \ref{wg3:t2}. At present similar applications are performed in few specific research centers, however often involve only one analytical technique (mostly AMS). In this framework CIRCE, with the potential of multi-analytical isotopic techniques, represents an opportunity for nuclear physics research but also for the development and characterization of complex detectors. 

\subsection{\wgthreetonetitle}
\label{wg3:t1}

Isotope Ratio Mass Spectrometry (\ie{} the measurement of a sample isotope ratio based on high resolution magnetic selection and ion counting) is a specialized analytical technique usually utilized to address different scientific issues over several research fields such as Geology, Environmental Sciences, Archaeology, Food and Material Sciences \cite{White2023}. Here is given a short overview of the current state of the art of the Isotope Ratio Mass Spectrometric techniques applied to different field of research related to CSN3 activities at CIRCE, also highlighting possible developments over short to mid term time ranges.\\*
For the simplicity of illustration Isotope Ratio Mass Spectrometry techniques applications, either if historically originating from the same root \cite{Becker2008}, are introduced and discussed according to the specific isotope  ratio, rare vs stable, to be measured.

\subsubsection{Accelerator Mass Spectrometry for the measurement of rare isotope ratios}

The development of Accelerator Mass Spectrometry (AMS) allows the determination of the relative isotopic abundance of rare isotopes (long lived radioisotopes) with respect to stable ones. Such isotopic ratios spans from over \num{e-12} (or higher) to \num{e-16} \cite{Wallner2005}. AMS history is mostly related to the use and development of accelerators for nuclear physics experiments \cite{Kutschera2005} and its definitive birth is related to experimental activities related to the measurements of \nuclide{14}{C} (radiocarbon) isotope ratios for dating applications by cyclotron \cite{Muller1977} or tandem accelerators equipped with Cs negative ion sputter source \cite{Purser1977}. That is why, along its lifespan, AMS has always been tightly connected to nuclear physics research studies, sharing with them most of the technology utilized. AMS spectrometers allow the measurement of rare vs abundant isotope ratios being also characterized by a wide dynamic range, and guaranteeing the best precision performances, \ie{} from 0.3\% to c.a. 10\% relative error depending of the measured isotope at typical detection abundances through the use of fast cycling technologies such as Magnet Bouncing Systems \cite{Sie2000}.

\noindent AMS applications at CIRCE Laboratory described in this paper falls in\\*
\phantom{xx}i) the ultra-sensitive determination of \nuclide{14}{C} isotope ratios for archaeometry, forensics and environmental sciences;\\*
\phantom{xx}ii) \nuclide{14}{C} AMS measurements on tree rings to search for past astrophysical events in proximity of the Solar System;\\*
\phantom{xx}iii) the ultra-sensitive determination of actinides, U and Pu, for environmental studies like radionuclides release monitoring during Nuclear Power Plant operation and decommissioning;\\*
\phantom{xx}iv) the determination of \nuclide{244}{Pu} content in environmental matrices (carbonatic sediments), to search for natural occurrence of this isotope;\\*
\phantom{xx}v) ultra-sensitive determination of material composition for rare events physics studies
and\\* 
\phantom{xx}vi) the direct determination of nuclear reaction strengths such as \reaction{25}{Mg}{\pg}{26}{Al}.\\*
Measurements for points i) and iii) are reported here to highlight typical AMS application performed at CIRCE laboratory, but also to stress the point that methods and knowledge developed starting from nuclear physics technologies, after a period of development and specialization, may come back to their origins in order to be used for specific of analytical requests for nuclear physics studies. As examples of such applications of \nuclide{14}{C} AMS, we like to mention the dating of the “Acerenza portrait” in archaeometry, the determination of the year of death and age at death for forensics \cite{Marzaioli2011a} and the determination of colonization chronology for the Caribbean for environmental sciences and paleoanthropology \cite{Fernandes2021}. Also the development of new procedures allowing to increase the bouquet of datable materials such as Mortars \cite{Marzaioli2011b,Ricci2022} should be mentioned here. For example, attained sample preparation methodologies performances (\ie{} lower detection limits) can play a key role for the ultra-sensitive determination of material composition for rare events physics studies (vi).
 The developments needed in such researches,  leading on the average to practical feed-backs such as pushing toward the limits the minimum analyzable mass while guaranteeing  high precision (\ie{} specific compound radiocarbon analysis), might increase the possibilities of analyses for fundamental nuclear physics measurements applications drastically reducing the necessity of long time demanding chemical preparation. For Actinides it is the case to mention the characterization of environmental matrices collected around the Garigliano Nuclear Power Plant,  for the purpose of environmental monitoring during plant decommissioning, through the measurement of \nuclide{236}{U}/\nuclide{238}{U}, and \nuclide{239}{Pu}/\nuclide{240}{Pu} ratios \cite{Petraglia2022, Buompane2023}.

A bit higher degree of detail will be spent in discussing applications of \nuclide{14}{C} AMS developed within the ERNA experiment to identify in geological and environmental archives traces of past near Earth astrophysical events (point iii). Over the last decade a strong research effort was spent by several AMS laboratories aiming to look at fast \nuclide{14}{C} concentration increase by working onto annually resolved tree rings series. The combination of the ultra-precise measurements with a chronological proxy allow to infer about solar activity, namely Solar Proton Event (SPE), and several events have been identified and confirmed also by C biogeochemical modeling \cite{Miyake2013,Miyake2012}. However, also  SNe events and Gamma Ray Bursts (GRB) might have a role in producing such spikes. This aspect was sometimes discussed, but its role was neglected also because no known SNEs were reported for the period of interest. On the contrary, at CIRCE annual resolved tree ring series in the 1030-1080 AD interval from a Finnish Lapland Tree \cite{Uusitalo2018} was analyzed aiming to evaluate possible SNe influence on the period of the SN1054 \cite{Terrasi2020}. \ie{} the event that generated the Crab Nebula, that is sufficiently close to Earth to contemplate this possibility. While a clear oscillation during 1050–1063\,CE was found, having different characteristics with respect to other identified SPEs, its connection to the SN1054 remains unclear, and other measurements using \nuclide{14}{C} and/or different proxies, \eg{} \nuclide{10}{B} in polar ices, are required.

The search for astronomical events in the local Universe also involves the AMS of actinide nuclei. In particular, the determination of \nuclide{244}{Pu} abundances can be performed on environmental matrices (carbonatic sediments) to search for natural production of this isotope. Several studies attribute the production of this nuclide to explosive stellar nucleosynthesis environments, such as some types of SNe and neutron star mergers \cite{Wang2021}. Reported typical concentrations of \nuclide{244}{Pu} for environmental samples are in the range of tens to hundreds of attograms per gram of sediment (\num{e-18}) with an isotopic ratio of \nuclide{244}{Pu}/\nuclide{239}{Pu} c.a. \num{e-4}, therefore AMS is the  tool of election for such measurements. At CIRCE Laboratory, AMS performances have been assessed by measuring 2 reference samples, IAEA 410 and 412, for the validation of the analytical procedures (sample preparation and measurement) to be applied to sedimentary rock samples, with the final goal to infer about $r$-process captures that might lead to natural occurrence of \nuclide{244}{Pu}. 

AMS applications in Nuclear Astrophysics fall also into the direct cross section measurements through detection of isotope ratios. An example of this kind of measurements is the determination of the resonance strengths in the \reaction{25}{Mg}{\pg}{26}{Al} fusion process, that plays a key role in the Mg-Al cycle. The study of this reaction through $\gamma$-ray spectroscopy was complemented by a measurement of the \nuclide {27}{Al}/\nuclide{26}{Al} ratio in \nuclide {25}{Mg} samples, doped with a known amount of \nuclide{27}{Al}, that were irradiated at the LUNA 400 kV accelerator at LNGS at the energy corresponding to the resonance at $\Ecm= 304$\,keV \cite{Limata2010}.

\subsubsection{SIRMS and MC-ICP-MS or the measurement of stable isotope ratios}

The sensitivity prerequisite for the determination of stable isotope ratios in a sample, asks for an Isotope Ratio Mass Spectrometer capable to measure isotope ratios ranging from $10^{-2}$ till $10^{-4}$ \cite{Soti2019}. To reach such a ratio sensitivity it is possible to operate at lower energies and thus apparatuses of minor complexity are used. The precision that can be achieved range typically around 0.01\% to 0.001\% (relative error) depending on the measured isotopes and on the spectrometer specifications. For such purposes 2 different approaches can be applied: 1) using an adequate ion source (Electron Impact or Thermal Ionization) producing a quasi-monoenergetic beam (\ie{} $\Delta E$ 10 eV) only a magnetic selection can be applied (SIRMS or TIMS); ii) ion sources producing higher energy spread (Ion Coupled Plasma) requires double focusing using an ElectroStatic Analyzer (ESA) to reduce energy spread produced during ionization (SF-ICP-MS). Nowadays double-focusing mass spectrometers with forward or reverse Nier–Johnson \cite{Johnson1953} for isotope ratio measurements. Precision can be drastically improved using an array of position adjustable collectors (Faraday cups plus Single Electrons Multipliers) guaranteeing the desired dynamic range. These methodologies are routinely conducted for environment and cultural heritage studies and applications. 

Among the applications of stable isotope ratio mass spectrometry of interest for the INFN CSN3 activities performed at CIRCE  it is worth to mention the development and application of MC-ICP-MS to search for nucleosynthesis signatures in meteoritic materials. This project was initiated as an extension of a project of the University of Perugia \cite{RicciLisa2022} where the study of such materials was started on samples of the local collection. This kind of researches are highly interdisciplinary, since the meteorites are quite complex objects, having specific phases characterized by different geochemistry that need to be carefully assessed, \eg{} the metallic matrix and the olivine in metallic meteorites such as Pallasites \cite{Zucchini2018,Mugnaioli2022}, to understand the contribution due to the various processes that play a role to their isotopic enrichment. At CIRCE have been developed and characterized the procedures aimed at the measurement of the Ni isotopes ratio, since \nuclide{60}{Ni} is the stable product of the decay of \nuclide{60}{Fe}, that is one of the main nucleosynthesis products of core collapse SNe. The sample preparation and measurement procedures have been tested and validated onto the reference material NIST SRM 986 doped with Fe. Actual measurements can be performed provided the availability of the meteorite samples.

\subsection{\wgthreettwotitle}
\label{wg3:t2}

Several technical developments made at CIRCE to meet the requirements of the experiments described above turned out to have interesting applications in material and devices characterization. These applications are briefly presented in  the following sections.

\subsubsection{Wear measurements}

Radioactive ion implantation allows non‐contacting, online wear measurements with sub micrometric sensitivity to be performed, by monitoring the removed and/or residual activity on parts subject to wear. Comparative studies of different materials, including those who exhibit a low resistance to radiation damage, can be easily performed by means of this technique.
\nuclide{7}{Be} is a nuclide particularly interesting in this kind of applications because its intermediate half life, 53\,d, is long enough to perform experiments, but not too long to make its handling complicated, its decay mode (EC followed by a single low energy $\gamma$-ray) reduces its radio-toxicity, its low atomic number allows larger implantation depths while minimizing the damage to the tested materials. The development of a \nuclide{7}{Be} ion beam at CIRCE \cite{Limata2008} made these applications possible in fundamental and in applied industrial research. A dedicated implantation beam line is present at CIRCE, see Figure \ref{wg0:CIRCE:fig:CIRCE}. It allows for the realization of tailored radionuclide depth distributions on a variety of samples (e.g. bushings, joints, bearings) through a combination of different beam energies and passive absorbers. \\* In collaboration with the CIRA (Italian Aerospace Research Centre) the study of durability tests of Thermal Protection Systems (TPS) used in space science to protect space vessel in atmospheric reentry are performed. The \nuclide{7}{Be} ion implantation has been proposed as an innovative tool to study the deterioration of ablative TPS \cite{DeCesare2018,DeCesare2020,Rapagnani2021} and the IBA techniques are used for the composition estimation of the sample \cite{Rapagnani2020}. The collaboration in this field is still open, in the next years the \nuclide{7}{Be} ion implantation and the IBA methods will be applied to ceramic TPS to study, respectively, their recession rate and their surface oxidation.

In addition, among other equipment developed for fundamental nuclear physics research the ERNA collaboration has built and used different windowless gas targets, both extended and as supersonic jet. Supersonic jet gas targets are very similar, besides dimensions, to plasma wind tunnels used to study the durability of materials of interest for space missions. Due to its quite compact size a jet gas target with a high degree of ionization, \ie{} a supersonic plasma jet, is an interesting opportunity for material testing because of the major reduction of the complexity and both the building and running costs. Such a system would be also very interesting for fundamental nuclear research, in particular for the study of electron screening effects in nuclear reactions \cite{Assenbaum1987}.
The development of a supersonic plasma jet has been investigated in a joint effort with CIRA \cite{Rapagnani2023}. The use of such a system in nuclear physics is discussed in the LNF Session of the Nuclear Physics Mid Term Plan \cite{midtermLNF}.

\subsubsection{Materials characterization}

Another activity that benefits of the intense and well characterized \nuclide{7}{Be} ion beam is the ASBeST project. This project aims at measuring the effect of electron density variation on \nuclide{7}{Be} EC decay rate and possibly its applications. \\*
To reach this goal \nuclide{7}{Be} implantations in Silicon Carbide (SiC) devices are conducted in order to study the possible variation in the half life of the beryllium isotope under intense electrostatic field, of the order of MV\,cm$^{-1}$. For this study, is of paramount importance to understand how the implanted \nuclide{7}{Be} ions are located in the SiC lattice and the nature of this impurity. In fact, the \nuclide{7}{Be} can create a substitutional or interstitial impurity, that corresponds to a completely different behavior with respect to the electrostatic field. For this purpose, the channeling-RBS is a powerful tool \cite{Feldman2012}, therefore a scattering chamber with a five-axis in-vacuum sample manipulator is under commissioning phase, and will allow to perform both Channeling-RBS and Elastic Recoil Detection Analysis (ERDA) measurements.

\bigskip
Since 2018, adjacent to the CIRCE is present the CAPACITY (Campania AstroPArtiCle InfrastrucTure facilitY) laboratory, whose main goal is to build the detection units and related equipment for the KM3NeT deep-sea neutrino telescope \cite{KM3url}. The close contact between the two laboratories has generated a positive feedback. A program aimed at the investigation of the performances of materials and devices to be deployed at high depths, was started exploiting the high sensitivity of IBA and Mass Spectrometry techniques.
In fact, water penetration is expected to be one of the main factors limiting the lifetime of the KM3NeT underwater equipment. A first series of measurements showed that through these techniques it is possible to identify micro-leaks on a scale capable to produce a critical malfunctioning in twenty years from deep sea deployment. In detail, NRA and RBS can be used to trace penetration depth of deuterium in materials, such as glass or plastics, exposed to deuterated water (D\textsubscript{2}O) at high pressure. Deuterium can be also detected with SIRMS on microtome cut plastic samples of such exposed materials. Using SIRMS can be also quantified the amount of \nuclide{14}{C}, introduced as a spike in the inorganic carbon dissolved in water, that leaks into immersed air filled assembled parts (\eg{} detection modules) or liquid filled parts (e.g oil filled cables). 
In order to push the attainable mass sensitivity by such isotopic methodologies, an innovative high pressure test chamber with isotopic markers is under construction and an ISO4 clean room were installed. This equipment will allow sample tests to be carried out throughout the entire construction phase of the undersea telescope.



\section*{List of symbols, abbreviations and acronyms}\label{acro}
$L_\odot$ Solar luminosity \\
$M_\odot$ Solar mass \\
AGB Asymptotic Giant Branch \\ 
AMS Accelerator Mass Spectrometry \\
BEGe Broad Energy Germanium detector \\
BBN Big Bang Nucleosynthesis \\
CNN Convolutional Neural Network \\
CNO Carbon-Nitrogen-Oxigen \\
CSL Continuous Spontaneous Localization \\
DD Direct Decay \\
DP Diósi-Penrose \\
DSSSD Double Sided Silicon Strip Detector \\
EC Electron Capture \\
ERDA Elastic Recoil Detection Analisys \\
HPGe High Purity Germanium detector\\ 
IBA Ion Beam Analysis \\
LUNA Laboratory Underground for Nuclear Astrophysics \\
MC-ICP-MS Multi Collector-Inductively Coupled Plasma Mass Spectrometry \\
MGS Messiah-Greenberg Superselection \\
NCQG Non-Commutative Quantum Gravity \\
NME Nuclear Matrix Element \\
NRA Nuclear Reaction Analysis \\
PEP Pauli Exclusion Principle \\
PSD Pulse Shape Discrimination \\
QM Quantum Mechanics \\
RBS Rutherford Back Scattering \\
RI Radioactive Ion \\
RMS Recoil Mass Separator \\
SD Sequential Decay \\
SDD Silicon Drift Detectors \\ 
SF-ICP-MS Sector Field-Inductively Coupled Plasma Mass Spectrometry \\
SIRMS Stable Isotope Ratio Mass Spectrometry \\
SM Standard Model \\
SN Supernova \\
SPE Solar Proton Events\\ 
SSD Silicon Strip Detector \\
THM Trojan Horse Method \\
TIMS Thermal Ionization Mass Spectrometry \\
TP Thermal Pulse \\
TPS Thermal Protection System

\section*{Acknowledgments}
\addcontentsline{toc}{section}{Acknowledgments}

The Authors express their gratitude to LNGS for having hosted the the Nuclear Physics Mid Term Plan in Italy - LNGS Session, in particular the Scientific Secretary of the Research Division for their support.

The Authors in WG1 wish to acknowledge the support of Grant No. 62099 from the John Templeton Foundation (the opinions expressed in this publication are those of the authors and do not necessarily reflect the views of the John Templeton Foundation). The support from the Foundational Questions Institute and Fetzer Franklin Fund, a donor advised fund of Silicon Valley Community Foundation (Grants No. FQXi-RFP-CPW-2008 and No. FQXi-MGB-2011), and from the H2020 FET TEQ (Grant No. 766900) is also acknowledged. The VIP2 project is supported by the Austrian Science Foundation (FWF) through the Grant No. P25529-N20, Projects No. P30635-N36, and No. W1252-N27 (doctoral college particles and interactions). K. P. acknowledges support from the Centro Ricerche Enrico Fermi—Museo Storico della Fisica e Centro Studi e Ricerche ``Enrico Fermi'' (Open Problems in Quantum Mechanics project). A. M. wishes to acknowledge support by the Shanghai Municipality, through the Grant No. KBH1512299, by Fudan University, through the Grant No. JJH1512105, the Natural Science Foundation of China, through the Grant No. 11875113, and by the Department of Physics at Fudan University, through the Grant No. IDH1512092/001. 
The support from the STELLA facility at LNGS is gratefully acknowledged. A. M. thanks Rita Bernabei and Pierluigi Belli for useful discussions on the subject.
 
The Authors in WG2 and WG3 express their gratitude to the Accelerator Services of LNGS and CIRCE.
Support by MIUR through the ``Dipartimenti di eccellenza'' project ``Physics of the Universe'' program, the EU grants ERC-StG \emph{SHADES} no. 852016, UKRI ERC-StG \emph{ELDAR} no. EP/X019381/, \emph{ChETEC-INFRA}, no. 101008324, the DFG BE~4100-4/1, the Helm\-holtz Association ERC-RA-0016, the NKFIH K134197, PD129060, the COST Action ChETEC CA16117, is gratefully acknowledged.

\section*{Funding} Open access funding provided by Università degli Studi di Napoli ``Federico II'' within the CRUI-CARE Agreement.

\section*{Data availability} No data associated with this manuscript.

\section*{Rights and permissions} 
\addcontentsline{toc}{section}{Rights and permissions}
This work has been originally published in "The European Physical Journal Plus 139.3, 224 (Mar. 2024), p. 224. doi: \href{https://doi.org/10.1140/epjp/s13360-023-04840-2}{10.1140/epjp/s13360-023-04840-2} in Open Access under a Creative Commons Attribution 4.0 International License, which permits use, sharing, adaptation, distribution and reproduction in any medium or format, as long as appropriate credit to the original author(s) and the source is given, a link to the Creative Commons licence is provided, and indicated if changes were made.\\*
In this version only a change to reference \cite{midtermLNF} was made.\\*
To view a copy of the licence, visit \href{http://creativecommons.org/licenses/by/4.0/}{http://creativecommons.org/licenses/by/4.0/}.

\printbibliography[heading=bibintoc]

@String{APPA        = "Acta Phys. Pol. A"}

@String{AC          = "Anal. Chem."}

@String{ARNPS       = "Annu. Rev. Nucl. Part. Sci."}

@String{APPSC       = "Appl. Sci."}

@String{AAP         = "Astron. Astrophys."}

@String{APARTP      = "Astropart. Phys."}

@String{APJ         = "Astrophys. J."}

@String{APJL        = "Astrophys. J. Lett."}

@String{APJSS       = "Astrophys. J. Suppl. Ser."}

@String{ARI         = "Appl. Rad. Isotop."}

@String{ARNP        = "Annu. Rev. Nucl. Part. Sci."}

@String{CPC         = "Comput. Phys. Commun."}

@String{EPLETT      = "Europhys. Lett."}

@String{EPJA        = "Eur. Phys. J. A"}

@String{EPJC        = "Eur. Phys. J. C"}

@String{EPJN        = "Eur. Phys. J. Nucl. Sci. Technol."}

@String{EPJP        = "Eur. Phys. J. Plus"}

@String{EPJWC       = "Eur. Phys. J. Web Conf."}

@String{FP          = "Front. Phys."}

@String{FOUNDP      = "Found. Phys."}

@String{GRG         = "Gen. Relativ. Gravit."}

@String{IJMS        = "Int. J. Mass Spectrom."}

@String{IJMPA       = "Int. J. Mod. Phys. A"}

@String{JI          = "J. Instrum."}

@String{JPhysCS     = "J. Phys.: Conf. Ser."}

@String{JPhysA      = "J. Phys. A: Math. Theor."}

@String{JPhysD      = "J. Phys. D: Appl. Phys."}

@String{JPhysG      = "J. Phys. G: Nucl. Part. Phys."}

@String{MNRAS       = "Mon. Not. R. Astron. Soc."}

@String{MPS         = "Meteorit. Planet. Sci."}

@String{NIMA        = "Nucl. Instrum. Methods Phys. Res. A"}

@String{NIMB        = "Nucl. Instrum. Methods Phys. Res. B"}

@String{Nature      = "Nature"}

@String{NatureCom   = "Nat. Commun."}

@String{NatureAstro = "Nat. Astron."}

@String{NaturePhys  = "Nat. Phys."}

@String{NEWJP       = "New J. Phys."}

@String{NuclPhysA   = "Nucl. Phys. A"}

@String{PhysA       = "Physica A"}

@String{PLA         = "Phys. Lett. A"}

@String{PLB         = "Phys. Lett. B"}

@String{PRep        = "Phys. Rep."}

@String{PR          = "Phys. Rev."}

@String{PRA         = "Phys. Rev. A"}

@String{PRC         = "Phys. Rev. C"}

@String{PRD         = "Phys. Rev. D"}

@String{PRL         = "Phys. Rev. Lett."}

@String{PPNP        = "Prog. Part. Nucl. Phys."}

@String{PTP         = "Prog. Theor. Phys."}

@String{PTPS        = "Prog. Theor. Phys. Supp."}

@String{RMP         = "Rev. Mod. Phys."}

@String{RNC         = "Riv. Nuovo Cimento"}

@String{RPA         = "Rev. Phys. Appl. (Paris)"}

@String{RPC         = "Radiat. Phys. Chem."}

@String{Science     = "Science"}

@String{SciRep      = "Sci. Rep."}

@String{JETPL       = "J. Exp. Theor. Phys. Lett."}

@String{ZFPA        = "Z.~Phys.~A"}

@String{Elsevier = "Elsevier"}

@String{pasa = "Publications of the Astronomical Society of Australia"}

@misc{LNGSurl,
  url = 	 {https://www.lngs.infn.it},
}

@misc{CIRCEurl,
  url = 	 {https://www.circe.unicampania.it},
}

@misc{MidTermurl,
  title =  {Nuclear Physics Midterm Plan in Italy},
  url = 	 {https://web.infn.it/nucphys-plan-italy},
}

@misc{MidTermLNGSurl,
  title =	 {Nuclear Physics Midterm Plan in Italy - LNGS Session agenda},
  url =		 {https://agenda.infn.it/event/31580},
}

@article{midtermLNF,
  author =	 {P. Cardarelli and others},
  title =	 {Nuclear Physics Mid Term Plan in Italy: detector developments},
  journal=       EPJP,
  note =	 {Forthcoming}
}

@misc{KM3url,
  url =		 {https://www.km3net.org},
}

@article{Addazi2018CPC,
  doi =		 {10.1088/1674-1137/42/9/094001},
  year =	 2018,
  month =	 8,
  volume =	 42,
  number =	 9,
  pages =	 094001,
  author =	 {Andrea Addazi and Pierluigi Belli and Rita Bernabei and Antonino Marcian{\`{o}}},
  title =	 {Testing noncommutative spacetimes and violations of the {Pauli} Exclusion Principle through underground experiments},
  journal =	 CPC
}

@article{Addazi2020IJMPA,
  author =	 {{Addazi}, Andrea and {Marcian{\`o}}, Antonino},
  title =	 {A modern guide to $\theta$-{Poincar{\'e}}},
  journal =	 IJMPA,
  year =	 2020,
  month =	 11,
  volume =	 35,
  number =	 32,
  eid =		 {2042003-385},
  pages =	 {2042003-385},
  doi =		 {10.1142/S0217751X20420038}
}

@article{Addazi2020EPJC,
  title =	 {Phenomenology of the {Pauli} exclusion principle violations due to the non-perturbative generalized uncertainty principle},
  author =	 {Addazi, Andrea and Belli, Pierluigi and Bernabei, Rita and Marcian{\`o}, Antonino and Shababi, Homa},
  journal =	 EPJC,
  volume =	 {80},
  pages =	 {1--7},
  year =	 {2020},
  doi =		 {10.1140/epjc/s10052-020-8401-0}
}

@article{Adelberger2011,
  author =	 {E.G. Adelberger and others},
  OPTauthor =	 {E.G. Adelberger and {Garc{\'{\i}}a}, A. and {Robertson}, R.~G.~H. and {Snover}, K.~A. and {Balantekin}, A.~B. and {Heeger}, K. and {Ramsey-Musolf}, M.~J. and {Bemmerer}, D. and {Junghans}, A. and {Bertulani}, C.~A. and {Chen}, J.-W. and {Costantini}, H. and {Prati}, P. and {Couder}, M. and {Uberseder}, E. and {Wiescher}, M. and {Cyburt}, R. and {Davids}, B. and {Freedman}, S.~J. and {Gai}, M. and {Gazit}, D. and {Gialanella}, L. and {Imbriani}, G. and {Greife}, U. and {Hass}, M. and {Haxton}, W.~C. and {Itahashi}, T. and {Kubodera}, K. and {Langanke}, K. and {Leitner}, D. and {Leitner}, M. and {Vetter}, P. and {Winslow}, L. and {Marcucci}, L.~E. and {Motobayashi}, T. and {Mukhamedzhanov}, A. and {Tribble}, R.~E. and {Nollett}, K.~M. and {Nunes}, F.~M. and {Park}, T.-S. and {Parker}, P.~D. and {Schiavilla}, R. and {Simpson}, E.~C. and {Spitaleri}, C. and {Strieder}, F. and {Trautvetter}, H.-P. and {Suemmerer}, K. and {Typel}, S.},
  journal =	 RMP,
  pages =	 {195-246},
  title =	 {{Solar fusion cross sections. II. The pp chain and CNO cycles}},
  volume =	 {83},
  year =	 {2011},
  doi =		 {10.1103/RevModPhys.83.195},
}

@book{Adler2004CUP,
  doi =		 {10.1017/cbo9780511535277},
  year =	 2004,
  month =	 8,
  publisher =	 {Cambridge University Press},
  address =	 {UK},
  author =	 {Stephen L. Adler},
  title =	 {Quantum Theory as an Emergent Phenomenon}
}

@article{Adler2007JPA,
  doi =		 {10.1088/1751-8113/40/12/s03},
  year =	 2007,
  month =	 3,
  volume =	 40,
  number =	 12,
  pages =	 {2935--2957},
  author =	 {Stephen L Adler},
  title =	 {Lower and upper bounds on {CSL} parameters from latent image formation and {IGM} heating},
  journal =	 JPhysA
}

@article{Adler2018PRA,
  title =	 {Bulk heating effects as tests for collapse models},
  author =	 {Adler, Stephen L. and Vinante, Andrea},
  journal =	 PRA,
  volume =	 97,
  issue =	 5,
  pages =	 052119,
  numpages =	 6,
  year =	 2018,
  month =	 5,
  doi =		 {10.1103/PhysRevA.97.052119},
}

@ARTICLE{Adsley2021,
  OPTauthor =	 {{Adsley}, Philip and {Battino}, Umberto and {Best}, Andreas and {Caciolli}, Antonio and {Guglielmetti}, Alessandra and {Imbriani}, Gianluca and {Jayatissa}, Heshani and {La Cognata}, Marco and {Lamia}, Livio and {Masha}, Eliana and {Massimi}, Cristian and {Palmerini}, Sara and {Tattersall}, Ashley and {Hirschi}, Raphael},
  author =	 {{Adsley}, Philip and others},
  title =	 {Reevaluation of the {$^{22}\mathrm{Ne}\ap{}^{26}\mathrm{Mg}$} and {$^{22}\mathrm{Ne}\an{}^{25}\mathrm{Mg}$} reaction rates},
  journal =	 PRC,
  year =	 2021,
  month =	 1,
  volume =	 103,
  number =	 1,
  eid =		 015805,
  pages =	 015805,
  doi =		 {10.1103/PhysRevC.103.015805}
}

@ARTICLE{Agodi2023,
  author =	 {{Agodi}, C. and others},
  OPTauthor =	 {Agodi, C. and Cappuzzello, F. and Cardella, G. and Cirrone, G. A. P. and {De Filippo}, E. and {Di Pietro}, A. and Gargano, A. and La Cognata, M. and Mascali, D. and Milluzzo, G. and Nania, R. and Petringa, G. and Pidatella, A. and Pirrone, S. and Pizzone, R. G. and Rapisarda, G. G. and Sergi, M. L. and Tudisco, S. and {Valiente-Dob{\'o}n}, J. J. and Vardaci, E. and Abramczyk, H. and Acosta, L. and Adsley, P. and Amaducci, S. and Banerjee, T. and Batani, D. and Bellone, J. and Bertulani, C. and Biri, S. and Bogachev, A. and Bonanno, A. and Bonasera, A. and Borcea, C. and Borghesi, M. and Bortolussi, S. and Boscolo, D. and Brischetto, G. A. and Burrello, S. and Busso, M. and Calabrese, S. and Calinescu, S. and Calvo, D. and Capirossi, V. and Carbone, D. and Cardinali, A. and Casini, G. and Catalano, R. and Cavallaro, M. and Ceccuzzi, S. and Celona, L. and Cherubini, S. and Chieffi, A. and Ciraldo, I. and Ciullo, G. and Colonna, M. and Cosentino, L. and Cuttone, G. and {D'Agata}, G. and {De Gregorio}, G. and {Degl'Innocenti}, S. and Delaunay, F. and {Di Donato}, L. and {Di Nitto}, A. and Dickel, T. and Doria, D. and Ducret, J. E. and Durante, M. and Esposito, J. and Farrokhi, F. and {Fernandez Garcia}, J. P. and Figuera, P. and Fisichella, M. and F{\"u}l{\"o}p, Z. and Galat{\'a}, A. and {Galaviz Redondo}, D. and Gambacurta, D. and Gammino, S. and Geraci, E. and Gizzi, L. and Gnoffo, B. and Groppi, F. and Guardo, G. L. and Guarrera, M. and Hayakawa, S. and Horst, F. and Hou, S. Q. and Jarota, A. and Jos{\'e}, J. and Kar, S. and Karpov, A. and {Kierzkowska-Pawlak}, H. and Kiss, G. G. and Knyazheva, G. and Koivisto, H. and Koop, B. and Kozulin, E. and Kumar, D. and Kurmanova, A. and La Rana, G. and Labate, L. and Lamia, L. and Lanza, E. G. and Lay, J. A. and Lattuada, D. and Lenske, H. and Limongi, M. and Lipoglavsek, M. and Lombardo, I. and Mairani, A. and Manetti, S. and Marafini, M. and Marcucci, L. and Margarone, D. and Martorana, N. S. and Maunoury, L. and Mauro, G. S. and Mazzaglia, M. and Mein, S. and Mengoni, A. and Milin, M. and Mishra, B. and Mou, L. and Mrazek, J. and Nadtochy, P. and Naselli, E. and Nicolai, P. and Novikov, K. and Oliva, A. A. and Pagano, A. and Pagano, E. V. and Palmerini, S. and Papa, M. and Parodi, K. and Patera, V. and Pellumaj, J. and Petrone, C. and Piantelli, S. and Pierroutsakou, D. and Pinna, F. and Politi, G. and Postuma, I. and Prajapati, P. and {Prada Moroni}, P. G. and Pupillo, G. and Raffestin, D. and Racz, R. and Reidel, C. -A. and Rifuggiato, D. and Risitano, F. and Rizzo, F. and {Roca Maza}, X. and Romano, S. and Roso, L. and Rotaru, F. and Russo, A. D. and Russotto, P. and Saiko, V. and Santonocito, D. and Santopinto, E. and Sarri, G. and Sartirana, D. and Schuy, C. and Sgouros, O. and Simonucci, S. and Sorbello, G. and Soukeras, V. and Spart{\'a}, R. and Spatafora, A. and Stanoiu, M. and Taioli, S. and Tessonnier, T. and Thirolf, P. and Tognelli, E. and Torresi, D. and Torrisi, G. and Trache, L. and Traini, G. and Trimarchi, M. and Tsikata, S. and Tumino, A. and Tyczkowski, J. and Yamaguchi, H. and Vercesi, V. and Vidana, I. and Volpe, L. and Weber, U.},
  number =	 {11},
  pages =	 {1038},
  title =	 {Nuclear physics midterm plan at {LNS}},
  volume =	 {138},
  journal =	 EPJP,
  year =	 2023,
  doi =		 {10.1140/epjp/s13360-023-04358-7}
}

@article{Akahori15,
  title =	 {Imaginary-time formalism for triple-$\alpha$ reaction rates},
  author =	 {Akahori, T. and Funaki, Y. and Yabana, K.},
  journal =	 PRC,
  volume =	 92,
  issue =	 2,
  pages =	 022801,
  numpages =	 5,
  year =	 2015,
  month =	 8,
  doi =		 {10.1103/PhysRevC.92.022801}
}

@ARTICLE{Ali2022,
  OPTauthor =	 {{Ali}, Sk M. and {Gupta}, D. and {Kundalia}, K. and {Saha}, Swapan K. and {Tengblad}, O. and {Ovejas}, J.~D. and {Perea}, A. and {Martel}, I. and {Cederkall}, J. and {Park}, J. and {Szwec}, S.},
  author =	 {{Ali}, Sk M. and others},
  title =	 {Resonance Excitations in {$^{7}\mathrm{Be}(\mathrm{d,p})^{8}\mathrm{Be}^*$} to Address the Cosmological Lithium Problem},
  journal =	 PRL,
  year =	 2022,
  month =	 6,
  volume =	 128,
  number =	 25,
  eid =		 252701,
  pages =	 252701,
  doi =		 {10.1103/PhysRevLett.128.252701}
}

@article{Ambrosio1995,
  title =	 {Vertical muon intensity measured with {[MACRO]} at the {Gran Sasso} laboratory},
  author =	 {Ambrosio, M. and others},
  OPTauthor =	 {Ambrosio, M. and Antolini, R. and Auriemma, G. and Baker, R. and Baldini, A. and Barbarino, G. C. and Barish, B. C. and Battistoni, G. and Bellotti, R. and Bemporad, C. and Bernardini, P. and Bilokon, H. and Bisi, V. and Bloise, C. and Bower, C. and Bussino, S. and Cafagna, F. and Calicchio, M. and Campana, D. and Carboni, M. and Castellano, M. and Cecchini, S. and Cei, F. and Celio, P. and Chiarella, V. and Corona, A. and Coutu, S. and De Cataldo, G. and Dekhissi, H. and De Marzo, C. and De Mitri, I. and De Vincenzi, M. and Di Credico, A. and Erriquez, O. and Favuzzi, C. and Forti, C. and Fusco, P. and Giacomelli, G. and Giannini, G. and Giglietto, N. and Grassi, M. and Grillo, A. and Guarino, F. and Guarnaccia, P. and Gustavino, C. and Habig, A. and Hanson, K. and Hawthorne, A. and Heinz, R. and Hong, J. T. and Iarocci, E. and Katsavounidis, E. and Kearns, E. and Kyriazopoulou, S. and Lamanna, E. and Lane, C. and Levin, D. S. and Lipari, P. and Liu, R. and Longley, N. P. and Longo, M. J. and Lu, Y. and Ludlam, G. and Mancarella, G. and Mandrioli, G. and Margiotta-Neri, A. and Marini, A. and Martello, D. and Marzari-Chiesa, A. and Mazziotta, M. N. and Michael, D. G. and Mikheyev, S. and Miller, L. and Mittelbrunn, M. and Monacelli, P. and Montaruli, T. and Monteno, M. and Mufson, S. and Musser, J. and Nicol\'o, D. and Nolty, R. and Okada, C. and Orth, C. and Osteria, G. and Palamara, O. and Parlati, S. and Patera, V. and Patrizii, L. and Pazzi, R. and Peck,
                  C. W. and Petrera, S. and Pignatano, N. D. and Pistilli, P. and Popa, V. and Rain\'o, A. and Reynoldson, J. and Ronga, F. and Sanzgiri, A. and Sartogo, F. and Satriano, C. and Satta, L. and Scapparone, E. and Scholberg, K. and Sciubba, A. and Serra-Lugaresi, P. and Severi, M. and Sitta, M. and Spinelli, P. and Spinetti, M. and Spurio, M. and Steinberg, R. and Stone, J. L. and Sulak, L. R. and Surdo, A. and Tarl\'e, G. and Tassoni, F. and Togo, V. and Valente, V. and Walter, C. W. and Webb, R.},
  journal =	 PRD,
  volume =	 {52},
  issue =	 {7},
  pages =	 {3793--3802},
  numpages =	 {0},
  year =	 {1995},
  month =	 10,
  doi =		 {10.1103/PhysRevD.52.3793}
}

@article{Amelino2007FPS,
  title =	 {On the quantum-gravity phenomenology of multiparticle states},
  author =	 {Amelino-Camelia, Giovanni and Arzano, Michele and Marciano, Antonino},
  journal =	 {Frascati Phys. Ser.},
  volume =	 {43},
  pages =	 {155},
  year =	 {2007},
  url =		 {http://www.lnf.infn.it/sis/frascatiseries/Volume43/volume43.pdf}
}

@article{Anders2014,
  title =	 {First Direct Measurement of the $^{2}\mathrm{H}(\ensuremath{\alpha},\ensuremath{\gamma})^{6}\mathrm{Li}$ Cross Section at {Big Bang} Energies and the Primordial Lithium Problem},
  OPTauthor =	 {Anders, M. and Trezzi, D. and Menegazzo, R. and Aliotta, M. and Bellini, A. and Bemmerer, D. and Broggini, C. and Caciolli, A. and Corvisiero, P. and Costantini, H. and Davinson, T. and Elekes, Z. and Erhard, M. and Formicola, A. and F\"ul\"op, Zs. and Gervino, G. and Guglielmetti, A. and Gustavino, C. and Gy\"urky, Gy. and Junker, M. and Lemut, A. and Marta, M. and Mazzocchi, C. and Prati, P. and Rossi Alvarez, C. and Scott, D. A. and Somorjai, E. and Straniero, O. and Sz\"ucs, T.},
  author =	 {Anders, M. and others},
  collaboration ={LUNA Collaboration},
  journal =	 PRL,
  volume =	 113,
  issue =	 4,
  pages =	 042501,
  numpages =	 5,
  year =	 2014,
  month =	 7,
  doi =		 {10.1103/PhysRevLett.113.042501}
}

@article{NACRE,
  OPTauthor =	 {C. Angulo and M. Arnould and M. Rayet and P. Descouvemont and D. Baye and C. Leclercq-Willain and A. Coc and S. Barhoumi and P. Aguer and C. Rolfs and R. Kunz and J.W. Hammer and A. Mayer and T. Paradellis and S. Kossionides and C. Chronidou and K. Spyrou and S. Degl'Innocenti and G. Fiorentini and B. Ricci and S. Zavatarelli and C. Providencia and H. Wolters and J. Soares and C. Grama and J. Rahighi and A. Shotter and M. Lamehi Rachti},
  author =	 {C. Angulo and others},
  doi =		 {10.1016/S0375-9474(99)00030-5},
  issn =	 {0375-9474},
  journal =	 NuclPhysA,
  number =	 1,
  pages =	 {3 - 183},
  title =	 {A compilation of charged-particle induced thermonuclear reaction rates},
  volume =	 656,
  year =	 1999
}

@article{Arnquist2022PRL,
  title =	 {Search for Spontaneous Radiation from Wave Function Collapse in the Majorana Demonstrator},
  OPTauthor =	 {Arnquist, I. J. and Avignone, F. T. and Barabash, A. S. and Barton, C. J. and Bhimani, K. H. and Blalock, E. and Bos, B. and Busch, M. and Buuck, M. and Caldwell, T. S. and Chan, Y-D. and Christofferson, C. D. and Chu, P.-H. and Clark, M. L. and Cuesta, C. and Detwiler, J. A. and Efremenko, Yu. and Ejiri, H. and Elliott, S. R. and Giovanetti, G. K. and Green, M. P. and Gruszko, J. and Guinn, I. S. and Guiseppe, V. E. and Haufe, C. R. and Henning, R. and Hervas Aguilar, D. and Hoppe, E. W. and Hostiuc, A. and Kim, I. and Kouzes, R. T. and Lannen V., T. E. and Li, A. and Lopez, A. M. and L\'opez-Casta\~no, J. M. and Martin, E. L. and Martin, R. D. and Massarczyk, R. and Meijer, S. J. and Oli, T. K. and Othman, G. and Paudel, L. S. and Pettus, W. and Poon, A. W. P. and Radford, D. C. and Reine, A. L. and Rielage, K. and Ruof, N. W. and Tedeschi, D. and Varner, R. L. and Vasilyev, S. and Wilkerson, J. F. and Wiseman, C. and Xu, W. and Yu, C.-H. and Zhu, B. X.},
  author =	 {Arnquist, I. J. and others},
  collaboration ={Majorana Collaboration},
  journal =	 PRL,
  volume =	 129,
  issue =	 8,
  pages =	 080401,
  numpages =	 7,
  year =	 2022,
  month =	 8,
  doi =		 {10.1103/PhysRevLett.129.080401}
}

@article{Arpesella1996,
  title =	 {A low background counting facility at {Laboratori Nazionali del Gran Sasso}},
  journal =	 ARI,
  volume =	 47,
  number =	 9,
  pages =	 {991-996},
  year =	 1996,
  note =	 {Proceedings of the International Committee for Radionuclide Metrology Conference on Low-level Measurement Techniques},
  issn =	 {0969-8043},
  doi =		 {10.1016/S0969-8043(96)00097-8},
  author =	 {C. Arpesella}
}

@article{Arzano2008PRD,
  title =	 {Quantum fields, nonlocality and quantum group symmetries},
  author =	 {Arzano, Michele},
  journal =	 PRD,
  volume =	 {77},
  number =	 {2},
  pages =	 {025013},
  year =	 {2008},
  doi =		 {10.1103/PhysRevD.77.025013}
}

@ARTICLE{Assenbaum1987,
  author =	 {{Assenbaum}, H.~J. and {Langanke}, K. and {Rolfs}, C.},
  title =	 "{Effects of electron screening on low-energy fusion cross sections.}",
  journal =	 ZFPA,
  year =	 1987,
  month =	 1,
  volume =	 {327},
  number =	 {4},
  pages =	 {461-468},
  doi =		 {10.1007/BF01289572}
}

@ARTICLE{Assie2015,
  OPTauthor =	 {{Assi{\'e}}, M. and {Le Crom}, B. and {Genolini}, B. and {Chabot}, M. and {Mengoni}, D. and {Due{\~n}as}, J.~A. and {Ancelin}, S. and {Beaumel}, D. and {Blumenfeld}, Y. and {de S{\'e}r{\'e}ville}, N. and {Dormard}, J. -J. and {Faul}, T. and {Guillot}, J. and {Jallat}, A. and {Le Ven}, V. and {Martel}, I. and {Rauly}, E. and {Suzuki}, D. and {Torrento}, A. -S.},
  author =	 {{Assi{\'e}}, M. and others},
  title =	 {Characterization of light particles ({$Z\leq2$}) discrimination performances by pulse shape analysis techniques with high-granularity silicon detector},
  journal =	 EPJA,
  year =	 2015,
  month =	 1,
  volume =	 {51},
  eid =		 {11},
  pages =	 {11},
  doi =		 {10.1140/epja/i2015-15011-6}
}

@article{Badala2022,
  author =	 {Badal{\`a}, A. and others},
  OPTauthor =	 {Badal{\`a}, A. and La Cognata, M. and Nania, R. and Osipenko, M. and Piantelli, S. and Turrisi, R. and Barion, L. and Capra, S. and Carbone, D. and Carnesecchi, F. and Casula, E. A. R. and Chatterjee, C. and Ciani, G. F. and Depalo, R. and Di Nitto, A. and Fantini, A. and Goasduff, A. and Guardo, G. L. and Kraan, A. C. and Manna, A. and Marsicano, L. and Martorana, N. S. and Morales-Gallegos, L. and Naselli, E. and Scordo, A. and Valdr{\'e}, S. and Volpe, G.},
  journal =	 {La Rivista del Nuovo Cimento},
  month =	 3,
  number =	 {3},
  pages =	 {189--277},
  title =	 {Trends in particle and nuclei identification techniques in nuclear physics experiments},
  volume =	 {45},
  year =	 {2022},
  doi =		 {10.1007/s40766-021-00028-5}
}

@article{Bahrami2018PRA,
  title =	 {Testing collapse models by a thermometer},
  author =	 {Bahrami, M.},
  journal =	 PRA,
  volume =	 97,
  issue =	 5,
  pages =	 052118,
  numpages =	 8,
  year =	 2018,
  month =	 5,
  doi =		 {10.1103/PhysRevA.97.052118}
}

@article{Balachandran2006IJMPA,
  author =	 {{Balachandran}, A.~P. and {Mangano}, G. and {Pinzul}, A. and {Vaidya}, S.},
  title =	 {Spin and Statistics on the {Groenewold-Moyal} Plane: {Pauli}-Forbidden Levels and Transitions},
  journal =	 IJMPA,
  year =	 2006,
  month =	 1,
  volume =	 {21},
  number =	 {15},
  pages =	 {3111-3126},
  doi =		 {10.1142/S0217751X06031764}
}

@article{Balachandran2007PRD,
  OPTauthor =	 {{Balachandran}, A.~P. and {Govindarajan}, T.~R. and {Mangano}, G. and {Pinzul}, A. and {Qureshi}, B.~A. and {Vaidya}, S.},
  author =	 {{Balachandran}, A.~P. and others},
  title =	 {Statistics and {UV-IR} mixing with twisted {Poincar{\'e}} invariance},
  journal =	 PRD,
  year =	 2007,
  month =	 2,
  volume =	 {75},
  number =	 {4},
  eid =		 {045009},
  pages =	 {045009},
  doi =		 {10.1103/PhysRevD.75.045009}
}

@ARTICLE{Ballan2023,
  author =	 {{Ballan}, M. and others},
  OPTauthor =	 {{Ballan}, M. and {Bottoni}, S. and {Caama{\~n}o}, M. and {Caciolli}, A. and {Campostrini}, M. and {Cicerchia}, M. and {Crespi}, F.~C.~L. and {Cristallo}, S. and {Dell'Aquila}, D. and {Depalo}, R. and {Fioretto}, E. and {Galtarossa}, F. and {Gasques}, L.~R. and {Gottardo}, A. and {Gramegna}, F. and {Gulminelli}, F. and {Kurtukian-Nieto}, T. and {La Cognata}, M. and {Lenzi}, S.~M. and {Marchi}, T. and {Mazurek}, K. and {Mengoni}, D. and {Mou}, L. and {Nania}, R. and {Pupillo}, G. and {Valiente-Dob{\'o}n}, J.~J. and {Zanon}, I. and {Acosta}, L. and {Alvarez}, M.~A.~G. and {Andrighetto}, A. and {Arazi}, A. and {Arzenton}, A. and {Assi{\'e}}, M. and {Bagatin}, M. and {Barbaro}, F. and {Barbieri}, C. and {Barlini}, S. and {Basiric{\`o}}, L. and {Battistoni}, G. and {Beaumel}, D. and {Bentley}, M.~A. and {Benzoni}, G. and {Bertoldo}, S. and {Bertulani}, C. and {Bonasera}, A. and {Camaiani}, A. and {Canton}, L. and {Capirossi}, V. and {Carante}, M.~P. and {Carraro}, C. and {Carturan}, S.~M. and {Casini}, G. and {Cavanna}, F. and {Centofante}, L. and {Ch{\'a}vez}, E.~R. and {Chbihi}, A. and {Ciema{\l}a}, M. and {Cisternino}, S. and {Colombi}, A. and {Colucci}, M. and {Compagnucci}, A. and {Corradetti}, S. and {Corradi}, L. and {D'Agata}, G. and {de Angelis}, G. and {De Dominicis}, L. and {De Salvador}, D. and {DeFilippo}, E. and {Del Fabbro}, M. and {Di Nitto}, A. and {Ditalia Tchernij}, S. and {Donzella}, A. and {Duguet}, T. and {Esposito}, J. and {Favela}, F. and
                  {Fern{\'a}ndez-Garc{\'\i}a}, J.~P. and {Flavigny}, F. and {Fontana}, A. and {Fornal}, B. and {Forneris}, J. and {Fraboni}, B. and {Frankland}, J. and {Gamba}, E. and {Geraci}, E. and {Gerardin}, S. and {Giuliani}, S.~A. and {Gnoffo}, B. and {Groppi}, F. and {Gruyer}, D. and {Haddad}, F. and {Isaak}, J. and {Kmiecik}, M. and {Koning}, A. and {Lamia}, L. and {Le Neindre}, N. and {Leoni}, S. and {L{\'e}pine-Szily}, A. and {Lilli}, G. and {Lombardo}, I. and {Loriggiola}, M. and {Loriggiola}, L. and {Lunardon}, M. and {Maggioni}, G. and {Maj}, A. and {Manenti}, S. and {Manzolaro}, M. and {Marcucci}, L.~E. and {Mar{\'\i}n-L{\'a}mbarri}, D.~J. and {Mariotti}, E. and {Martin Hernandez}, G. and {Massimi}, C. and {Mastinu}, P. and {Mazzocco}, M. and {Mazzolari}, A. and {Mijatovi{\'c}}, T. and {Mishenina}, T. and {Mizuyama}, K. and {Monetti}, A. and {Montagnoli}, G. and {Morselli}, L. and {Moschini}, L. and {Musacchio Gonzalez}, E. and {Nannini}, A. and {Niu}, Y.~F. and {Ota}, S. and {Paccagnella}, A. and {Palmerini}, S. and {Pellegri}, L. and {Perego}, A. and {Piantelli}, S. and {Piatti}, D. and {Picollo}, F. and {Pignatari}, M. and {Pinna}, F. and {Pirrone}, S. and {Pizzone}, R.~G. and {Polettini}, M. and {Politi}, G. and {Popescu}, L. and {Prete}, G. and {Quaranta}, A. and {Raabe}, R. and {Ramos}, J.~P. and {Raniero}, W. and {Rapisarda}, G.~G. and {Recchia}, F. and {Rigato}, V. and {Roca Maza}, X. and {Rocchini}, M. and {Rodriguez}, T. and {Roncolato}, C. and
                  {Rudolph}, D. and {Russotto}, P. and {S{\'a}nchez-Ben{\'\i}tez}, {\'A}. M. and {Savran}, D. and {Scarpa}, D. and {Scheck}, M. and {Sekizawa}, K. and {Sergi}, M.~L. and {Sgarbossa}, F. and {Silvestrin}, L. and {Singh Khwairakpam}, O. and {Skowronski}, J. and {Som{\`a}}, V. and {Spart{\`a}}, R. and {Spieker}, M. and {Stefanini}, A.~M. and {Steiger}, H. and {Stevanato}, L. and {Stock}, M.~R. and {Vardaci}, E. and {Verney}, D. and {Vescovi}, D. and {Vittone}, E. and {Werner}, V. and {Wheldon}, C. and {Wieland}, O. and {Wimmer}, K. and {Wyss}, J. and {Zago}, L. and {Zenoni}, A.},
  title =	 "{Nuclear physics midterm plan at Legnaro National Laboratories (LNL)}",
  journal =	 EPJP,
  year =	 2023,
  month =	 8,
  volume =	 {138},
  number =	 {8},
  eid =		 {709},
  pages =	 {709},
  doi =		 {10.1140/epjp/s13360-023-04249-x}
}

@ARTICLE{Baraffe1992,
  author =	 {{Baraffe}, I. and {El Eid}, M.~F. and {Prantzos}, N.},
  title =	 {The {$s$}-process in massive stars of variable composition},
  journal =	 AAP,
  year =	 1992,
  month =	 5,
  volume =	 258,
  number =	 2,
  pages =	 {357-367}
}

@article{Bassi2010EPL,
  doi =		 {10.1209/0295-5075/92/50006},
  year =	 2010,
  month =	 12,
  volume =	 92,
  number =	 5,
  pages =	 50006,
  author =	 {A. Bassi and D.-A. Deckert and L. Ferialdi},
  title =	 {Breaking quantum linearity: Constraints from human perception and cosmological implications},
  journal =	 EPLETT
}

@article{baudis2011JINST,
  title =	 {{Gator: a low-background counting facility at the Gran Sasso Underground Laboratory}},
  author =	 {Baudis, L and others},
  OPTauthor =	 {Baudis, L and Ferella, AD and Askin, A and Angle, J and Aprile, E and Bruch, T and Kish, A and Laubenstein, M and Manalaysay, A and Undagoitia, T Marrodan and others},
  journal =	 JI,
  volume =	 {6},
  number =	 {08},
  pages =	 {P08010},
  year =	 {2011},
  doi =		 {10.1088/1748-0221/6/08/P08010}
}

@article{Becker1981,
  title =	 {The {$^{12}\mathrm{C}+^{12}\mathrm{C}$} Reaction at Subcoulomb Energies {(II)*}},
  author =	 {Becker, H.W. and Kettner, K.U. and Rolfs, C. and Trautvetter, H.P.},
  journal =	 ZFPA,
  volume =	 303,
  pages =	 {305-312},
  year =	 1981,
  doi =		 {10.1007/BF01421528}
}

@book{Becker2008,
  year =	 2008,
  publisher =	 {John Wiley \& Sons},
  address =	 {US},
  author =	 {Becker, Sabine},
  title =	 {Inorganic Mass Spectrometry},
  subtitle =	 {Principles and Applications}
}

@book{Bell2004CUP,
  doi =		 {10.1017/cbo9780511815676},
  year =	 2004,
  month =	 6,
  publisher =	 {Cambridge University Press},
  address =	 {UK},
  author =	 {J. S. Bell and Alain Aspect},
  title =	 {Speakable and Unspeakable in Quantum Mechanics}
}

@article{Benzoni2023,
  author =	 {Benzoni, G. and Bettoni, D. and Bossi, F. and Colonna, M. and Di Leva, A. and Fioretto, E. and Formicola, A. and Fortunato, L. and Gammino, S. and Gramegna, F. and Gustavino, C. and Junker, M. and La Cognata, M. and Lombardo, I. and Nania, R. and Pisano, S. and Previtali, E. and Romano, S. and Russotto, P. and Soramel, F. and Valiente-Dob{\'o}n, J. J.},
  journal =	 epjp,
  number =	 6,
  pages =	 526,
  title =	 {Nuclear physics midterm plan in {Italy: introduction} to the series},
  volume =	 138,
  year =	 2023,
  doi =		 {10.1140/epjp/s13360-023-04108-9}
}

@article{Best2016,
  title =	 {Low energy neutron background in deep underground laboratories},
  journal =	 NIMA,
  volume =	 {812},
  pages =	 {1-6},
  year =	 {2016},
  issn =	 {0168-9002},
  doi =		 {10.1016/j.nima.2015.12.034},
  OPTauthor =	 {Andreas Best and Joachim Görres and Matthias Junker and Karl-Ludwig Kratz and Matthias Laubenstein and Alexander Long and Stefano Nisi and Karl Smith and Michael Wiescher},
  author =	 {Andreas Best and others}
}

@article{Bijker00,
  author =	 {{Bijker}, R. and {Iachello}, F.},
  title =	 {Cluster states in nuclei as representations of a {U({\ensuremath{\nu}}+1)} group},
  journal =	 PRC,
  year =	 2000,
  month =	 6,
  volume =	 {61},
  number =	 {6},
  eid =		 {067305},
  pages =	 {067305},
  doi =		 {10.1103/PhysRevC.61.067305}
}

@article{Bilardello2016PASMA,
  title =	 {Bounds on collapse models from cold-atom experiments},
  journal =	 PhysA,
  volume =	 462,
  pages =	 {764-782},
  year =	 2016,
  issn =	 {0378-4371},
  doi =		 {10.1016/j.physa.2016.06.134},
  author =	 {Marco Bilardello and Sandro Donadi and Andrea Vinante and Angelo Bassi}
}

@article{Bilardello2017PRA,
  title =	 {Collapse in ultracold {Bose Josephson} junctions},
  author =	 {Bilardello, M. and Trombettoni, A. and Bassi, A.},
  journal =	 PRA,
  volume =	 95,
  issue =	 3,
  pages =	 032134,
  numpages =	 13,
  year =	 2017,
  month =	 3,
  doi =		 {10.1103/PhysRevA.95.032134}
}

@article{Bishop20,
  OPTauthor =	 {{Bishop}, J. and {Rogachev}, G.~V. and {Ahn}, S. and {Aboud}, E. and {Barbui}, M. and {Bosh}, A. and {Hunt}, C. and {Jayatissa}, H. and {Koshchiy}, E. and {Malecek}, R. and {Marley}, S.~T. and {Pollacco}, E.~C. and {Pruitt}, C.~D. and {Roeder}, B.~T. and {Saastamoinen}, A. and {Sobotka}, L.~G. and {Upadhyayula}, S.},
  author =	 {{Bishop}, J. and others},
  title =	 {Almost medium-free measurement of the {Hoyle} state direct-decay component with a {TPC}},
  journal =	 PRC,
  year =	 2020,
  month =	 10,
  volume =	 102,
  number =	 4,
  eid =		 041303,
  pages =	 041303,
  doi =		 {10.1103/PhysRevC.102.041303}
}

@article{Boeltzig2018,
  OPTauthor =	 {{Boeltzig}, A. and {Best}, A. and {Imbriani}, G. and {Junker}, M. and {Aliotta}, M. and {Bemmerer}, D. and {Broggini}, C. and {Bruno}, C.~G. and {Buompane}, R. and {Caciolli}, A. and {Cavanna}, F. and {Chillery}, T. and {Ciani}, G.~F. and {Corvisiero}, P. and {Csedreki}, L. and {Davinson}, T. and {deBoer}, R.~J. and {Depalo}, R. and {Di Leva}, A. and {Elekes}, Z. and {Ferraro}, F. and {Fiore}, E.~M. and {Formicola}, A. and {F{\"u}l{\"o}p}, Z. and {Gervino}, G. and {Guglielmetti}, A. and {Gustavino}, C. and {Gy{\"u}rky}, G. and {Kochanek}, I. and {Menegazzo}, R. and {Mossa}, V. and {Pantaleo}, F.~R. and {Paticchio}, V. and {Perrino}, R. and {Piatti}, D. and {Prati}, P. and {Schiavulli}, L. and {St{\"o}ckel}, K. and {Straniero}, O. and {Strieder}, F. and {Sz{\"u}cs}, T. and {Tak{\'a}cs}, M.~P. and {Trezzi}, D. and {Wiescher}, M. and {Zavatarelli}, S.},
  author =	 {{Boeltzig}, A. and others},
  title =	 {Improved background suppression for radiative capture reactions at {LUNA} with {HPGe} and {BGO} detectors},
  journal =	 JPhysG,
  year =	 2018,
  month =	 2,
  volume =	 45,
  number =	 2,
  pages =	 025203,
  doi =		 {10.1088/1361-6471/aaa163}
}

@ARTICLE{Boeltzig2019,
  OPTauthor =	 {{Boeltzig}, A. and {Best}, A. and {Pantaleo}, F.~R. and {Imbriani}, G. and {Junker}, M. and {Aliotta}, M. and {Balibrea-Correa}, J. and {Bemmerer}, D. and {Broggini}, C. and {Bruno}, C.~G. and {Buompane}, R. and {Caciolli}, A. and {Cavanna}, F. and {Chillery}, T. and {Ciani}, G.~F. and {Corvisiero}, P. and {Csedreki}, L. and {Davinson}, T. and {deBoer}, R.~J. and {Depalo}, R. and {Di Leva}, A. and {Elekes}, Z. and {Ferraro}, F. and {Fiore}, E.~M. and {Formicola}, A. and {F{\"u}l{\"o}p}, Zs. and {Gervino}, G. and {Guglielmetti}, A. and {Gustavino}, C. and {Gy{\"u}rky}, Gy. and {Kochanek}, I. and {Lugaro}, M. and {Marigo}, P. and {Menegazzo}, R. and {Mossa}, V. and {Munnik}, F. and {Paticchio}, V. and {Perrino}, R. and {Piatti}, D. and {Prati}, P. and {Schiavulli}, L. and {St{\"o}ckel}, K. and {Straniero}, O. and {Strieder}, F. and {Sz{\"u}cs}, T. and {Tak{\'a}cs}, M.~P. and {Trezzi}, D. and {Wiescher}, M. and {Zavatarelli}, S.},
  author =	 {{Boeltzig}, A. and others},
  title =	 {Direct measurements of low-energy resonance strengths of the {$^{23}\mathrm{Na}\pg^{24}\mathrm{Mg}$} reaction for astrophysics},
  journal =	 PLB,
  year =	 2019,
  month =	 8,
  volume =	 795,
  pages =	 {122-128},
  doi =		 {10.1016/j.physletb.2019.05.044}
}

@ARTICLE{Brandi2020,
  author =	 {{Brandi}, F. and others},
  OPTauthor =	 {{Brandi}, F. and {Labate}, L. and {Rapagnani}, D. and {Buompane}, R. and {Leva}, A. di and {Gialanella}, L. and {Gizzi}, L.~A.},
  title =	 "{Optical and spectroscopic study of a supersonic flowing helium plasma: energy transport in the afterglow}",
  journal =	 SciRep,
  year =	 2020,
  month =	 3,
  volume =	 {10},
  eid =		 {5087},
  pages =	 {5087},
  doi =		 {10.1038/s41598-020-61988-y}
}

@ARTICLE{Bravo2011,
  OPTauthor =	 {{Bravo}, E. and {Piersanti}, L. and {Dom{\'\i}nguez}, I. and {Straniero}, O. and {Isern}, J. and {Escartin}, J.~A.},
  author =	 {{Bravo}, E. and others},
  title =	 {{Type Ia} supernovae and the {$^{12}\mathrm{C}$+$^{12}\mathrm{C}$} reaction rate},
  journal =	 AAP,
  year =	 2011,
  month =	 11,
  volume =	 535,
  eid =		 {A114},
  pages =	 {A114},
  doi =		 {10.1051/0004-6361/201117814}
}

@ARTICLE{Bravo2019,
  author =	 {{Bravo}, E.},
  title =	 "{$^{16}$O(p,{\ensuremath{\alpha}})$^{13}$N makes explosive oxygen burning sensitive to the metallicity of the progenitors of type Ia supernovae}",
  journal =	 AAP,
  year =	 2019,
  month =	 7,
  volume =	 {627},
  eid =		 {A146},
  pages =	 {A146},
  doi =		 {10.1051/0004-6361/201936024}
}

@article{Broggini2019,
  author =	 {{Broggini}, C. and others},
  OPTauthor =	 {{Broggini}, C. and {Straniero}, O. and {Taiuti}, M.~G.~F. and {de Angelis}, G. and {Benzoni}, G. and {Bruno}, G.~E. and {Bufalino}, S. and {Cardella}, G. and {Colonna}, N. and {Contalbrigo}, M. and {Cosentino}, G. and {Cristallo}, S. and {Curceanu}, C. and {De Filippo}, E. and {Depalo}, R. and {Di Leva}, A. and {Feliciello}, A. and {Gammino}, S. and {Galat{\`a}}, A. and {La Cognata}, M. and {Lea}, R. and {Leoni}, S. and {Lombardo}, I. and {Manzari}, V. and {Mascali}, D. and {Massimi}, C. and {Mengoni}, A. and {Mengoni}, D. and {Napoli}, D.~R. and {Palmerini}, S. and {Piano}, S. and {Pirrone}, S. and {Pizzone}, R.~G. and {Politi}, G. and {Prati}, P. and {Prete}, G. and {Russotto}, P. and {Tagliente}, G. and {Urciuoli}, G.~M.},
  journal =	 RNC,
  month =	 3,
  pages =	 {103},
  title =	 {{Experimental nuclear astrophysics in Italy}},
  volume =	 {42},
  year =	 2019,
  doi =		 {10.1393/ncr/i2019-10157-1}
}

@article{Buompane2018,
  OPTauthor =	 {Buompane, R. and De Cesare, N. and Di Leva, A. and D'Onofrio, A. and Gialanella, L. and Romano, M. and De Cesare, M. and Duarte, J. G. and F\"ul\"op, Zs. and Morales-Gallegos, L. and Gy\"urky, Gy. and Gasques, L. R. and Marzaioli, F. and Palumbo, G. and Porzio, G. and Rapagnani, D. and Roca, V. and Rogalla, D. and Romoli, M. and Sabbarese, C. and Sch\"urmann, D. and Terrasi, F.},
  author =	 {Buompane, R. and others},
  title =	 {Test measurement of {$^7\mathrm{Be}\pg^8\mathrm{B}$} with the recoil mass separator {ERNA}},
  DOI =		 {10.1140/epja/i2018-12522-6},
  journal =	 EPJA,
  year =	 2018,
  volume =	 54,
  number =	 6,
  pages =	 92
}

@article{Buompane2021,
  OPTauthor =	 {Buompane, R. and A. {Di Leva} and L. Gialanella and A. {D'Onofrio} and M. {De Cesare} and J. G. Duarte and Z. F\"ul\"op and L. R. Gasques and Gy. Gy\"urky and L. Morales-Gallegos and F. Marzaioli and G. Palumbo and G. Porzio and D. Rapagnani and V. Roca and D. Rogalla and M. Romoli and C. Santonastaso and D. Sch\"urmann},
  author =	 {Buompane, R. and others},
  title =	 {Determination of the $^7\mathrm{Be}\pg^8 \mathrm{B}$ cross section at astrophysical energies using a radioactive $^7\mathrm{Be}$ ion beam},
  journal =	 PLB,
  year =	 2021,
  pages =	 136819,
  doi =		 {10.1016/j.physletb.2021.136819}
}

@ARTICLE{Buompane2023,
  author =	 {Buompane, R. and others},
  OPTauthor =	 {Buompane, R. and D'Onofrio, A. and Gialanella, L. and Marzaioli, F. and Petraglia, A. and Porzio, G. and Romoli, M. and Sabbarese, C. and Terrasi, F.},
  title =	 {Isotopic techniques for environmental monitoring and nuclear waste management at {CIRCE}},
  journal =	 {Il Nuovo Cimento C},
  year =	 {2023},
  volume =	 {46},
  number =	 {2},
  doi =		 {10.1393/ncc/i2023-23049-2}
}

@article{Burbidge1957,
  title =	 {Synthesis of the Elements in Stars},
  author =	 {Burbidge, E. Margaret and Burbidge, G. R. and Fowler, William A. and Hoyle, F.},
  journal =	 RMP,
  volume =	 {29},
  issue =	 {4},
  pages =	 {547--650},
  numpages =	 {0},
  year =	 {1957},
  month =	 10,
  doi =		 {10.1103/RevModPhys.29.547}
}

@ARTICLE{Caciolli2012,
  OPTauthor =	 {{Caciolli}, A. and {Scott}, D.~A. and {Di Leva}, A. and {Formicola}, A. and {Aliotta}, M. and {Anders}, M. and {Bellini}, A. and {Bemmerer}, D. and {Broggini}, C. and {Campeggio}, M. and {Corvisiero}, P. and {Depalo}, R. and {Elekes}, Z. and {F{\"u}l{\"o}p}, Z. and {Gervino}, G. and {Guglielmetti}, A. and {Gustavino}, C. and {Gy{\"u}rky}, G. and {Imbriani}, G. and {Junker}, M. and {Marta}, M. and {Menegazzo}, R. and {Napolitani}, E. and {Prati}, P. and {Rigato}, V. and {Roca}, V. and {Rolfs}, C. and {Rossi Alvarez}, C. and {Somorjai}, E. and {Salvo}, C. and {Straniero}, O. and {Strieder}, F. and {Sz{\"u}cs}, T. and {Terrasi}, F. and {Trautvetter}, H.~P. and {Trezzi}, D.},
  author =	 {{Caciolli}, A. and others},
  title =	 {Preparation and characterisation of isotopically enriched {$\mathrm{Ta}_2\mathrm{O}_5$} targets for nuclear astrophysics studies},
  journal =	 epja,
  year =	 2012,
  month =	 10,
  volume =	 48,
  pages =	 {144},
  doi =		 {10.1140/epja/i2012-12144-0}
}

@article{Cardella2023,
  author =	 {{Cardella}, Giuseppe and others},
  OPTauthor =	 {{Cardella}, Giuseppe and {Bonasera}, Aldo and {Martorana}, Nunzia, Simona and {Acosta}, Luis and {De Filippo}, Enrico and {Geraci}, Elena and {Gnoffo}, Brunilde and {Guazzoni}, Chiara and {Maiolino}, Concettina and {Pagano}, Angelo and {Pagano}, Emanuele, Vincenzo and {Papa}, Massimo and {Pirrone}, Sara and {Politi}, Giuseppe and {Risitano}, Fabio and {Rizzo}, Francesca and {Russotto}, Paolo and {Trimarchi}, Marina},
  title =	 "{Potential experimental evidence of an Efimov state in $^{12}$C and its influence on astrophysical carbon creation}",
  booktitle =	 EPJWC,
  year =	 2023,
  series =	 {European Physical Journal Web of Conferences},
  volume =	 {279},
  month =	 9,
  eid =		 {03001},
  pages =	 {03001},
  doi =		 {10.1051/epjconf/202327903001}
}

@article{Carlesso2016PRD,
  title =	 {Experimental bounds on collapse models from gravitational wave detectors},
  author =	 {Carlesso, Matteo and Bassi, Angelo and Falferi, Paolo and Vinante, Andrea},
  journal =	 PRD,
  volume =	 94,
  issue =	 12,
  pages =	 124036,
  numpages =	 7,
  year =	 2016,
  month =	 12,
  doi =		 {10.1103/PhysRevD.94.124036}
}

@ARTICLE{Carretta2009,
  author =	 {{Carretta}, E. and {Bragaglia}, A. and {Gratton}, R. and {Lucatello}, S.},
  title =	 {{Na-O} anticorrelation and {HB. VIII}. {Proton}-capture elements and metallicities in 17 globular clusters from {UVES} spectra},
  journal =	 AAP,
  year =	 2009,
  month =	 10,
  volume =	 505,
  number =	 1,
  pages =	 {139-155},
  doi =		 {10.1051/0004-6361/200912097}
}

@ARTICLE{Cavanna2014,
  author =	 {{Cavanna}, F. and others},
  OPTauthor =	 {{Cavanna}, F. and {Depalo}, R. and {Menzel}, M. -L. and {Aliotta}, M. and {Anders}, M. and {Bemmerer}, D. and {Broggini}, C. and {Bruno}, C.~G. and {Caciolli}, A. and {Corvisiero}, P. and {Davinson}, T. and {di Leva}, A. and {Elekes}, Z. and {Ferraro}, F. and {Formicola}, A. and {F{\"u}l{\"o}p}, Zs. and {Gervino}, G. and {Guglielmetti}, A. and {Gustavino}, C. and {Gy{\"u}rky}, Gy. and {Imbriani}, G. and {Junker}, M. and {Menegazzo}, R. and {Prati}, P. and {Rossi Alvarez}, C. and {Scott}, D.~A. and {Somorjai}, E. and {Straniero}, O. and {Strieder}, F. and {Sz{\"u}cs}, T. and {Trezzi}, D.},
  title =	 {A new study of the {${}^{22}\mathrm{Ne}(\mathrm{p},\gamma){}^{23}\mathrm{Na}$} reaction deep underground: Feasibility, setup and first observation of the 186\,{keV} resonance},
  journal =	 EPJA,
  year =	 2014,
  volume =	 {50},
  pages =	 {179},
  doi =		 {10.1140/epja/i2014-14179-5}
}

@article{Cavanna2015,
  title =	 {Three New Low-Energy Resonances in the ${}^{22}\mathrm{Ne}(\mathrm{p},\gamma){}^{23}\mathrm{Na}$ Reaction},
  author =	 {Cavanna, F. and others},
  OPTauthor =	 {Cavanna, F. and Depalo, R. and Aliotta, M. and Anders, M. and Bemmerer, D. and Best, A. and Boeltzig, A. and Broggini, C. and Bruno, C. G. and Caciolli, A. and Corvisiero, P. and Davinson, T. and di Leva, A. and Elekes, Z. and Ferraro, F. and Formicola, A. and F\"ul\"op, Zs. and Gervino, G. and Guglielmetti, A. and Gustavino, C. and Gy\"urky, Gy. and Imbriani, G. and Junker, M. and Menegazzo, R. and Mossa, V. and Pantaleo, F. R. and Prati, P. and Scott, D. A. and Somorjai, E. and Straniero, O. and Strieder, F. and Sz\"ucs, T. and Tak\'acs, M. P. and Trezzi, D.},
  collaboration ={{LUNA} {Collaboration}},
  journal =	 PRL,
  volume =	 {115},
  pages =	 {252501},
  numpages =	 {6},
  year =	 {2015},
  doi =		 {10.1103/PhysRevLett.115.252501}
}

@ARTICLE{Cesaratto2013,
  OPTauthor =	 {{Cesaratto}, J.~M. and {Champagne}, A.~E. and {Buckner}, M.~Q. and {Clegg}, T.~B. and {Daigle}, S. and {Howard}, C. and {Iliadis}, C. and {Longland}, R. and {Newton}, J.~R. and {Oginni}, B.~M.},
  author =	 {{Cesaratto}, J.~M. and others},
  title =	 {Measurement of the {$E_{r}^\mathrm{c.m.} = 138$\,keV} resonance in the {${}^{23}\mathrm{Na}(\mathrm{p},\gamma){}^{24}\mathrm{Mg}$} reaction and the abundance of sodium in {AGB} stars},
  journal =	 PRC,
  year =	 2013,
  month =	 12,
  volume =	 88,
  number =	 6,
  eid =		 065806,
  pages =	 065806,
  doi =		 {10.1103/PhysRevC.88.065806}
}

@ARTICLE{Cescutti2013,
  OPTauthor =	 {{Cescutti}, G. and {Chiappini}, C. and {Hirschi}, R. and {Meynet}, G. and {Frischknecht}, U.},
  author =	 {{Cescutti}, G. and others},
  title =	 {The {$s$}-process in the Galactic halo: the fifth signature of spinstars in the early {Universe}?},
  journal =	 AAP,
  year =	 2013,
  month =	 5,
  volume =	 553,
  eid =		 {A51},
  pages =	 {A51},
  doi =		 {10.1051/0004-6361/201220809}
}

@article{Chernykh07,
  title =	 {Structure of the Hoyle State in $^{12}\mathrm{C}$},
  OPTauthor =	 {Chernykh, M. and Feldmeier, H. and Neff, T. and von Neumann-Cosel, P. and Richter, A.},
  author =	 {Chernykh, M. and others},
  journal =	 PRL,
  volume =	 98,
  issue =	 3,
  pages =	 032501,
  numpages =	 4,
  year =	 2007,
  month =	 1,
  doi =		 {10.1103/PhysRevLett.98.032501}
}

@ARTICLE{Chieffi2021,
  OPTauthor =	 {{Chieffi}, Alessandro and {Roberti}, Lorenzo and {Limongi}, Marco and {La Cognata}, Marco and {Lamia}, Livio and {Palmerini}, Sara and {Pizzone}, Rosario Gianluca and {Spart{\`a}}, Roberta and {Tumino}, Aurora},
  author =	 {{Chieffi}, Alessandro and others},
  title =	 {Impact of the New Measurement of the {$^{12}\mathrm{C}+^{12}\mathrm{C}$} Fusion Cross Section on the Final Compactness of Massive Stars},
  journal =	 APJ,
  year =	 2021,
  month =	 8,
  volume =	 916,
  number =	 2,
  eid =		 79,
  pages =	 79,
  doi =		 {10.3847/1538-4357/ac06ca}
}

@article{Ciani2020,
  OPTauthor =	 {{Ciani, G. F.} and {Csedreki, L.} and {Balibrea-Correa, J.} and {Best, A.} and {Aliotta, M.} and {Barile, F.} and {Bemmerer, D.} and {Boeltzig, A.} and {Broggini, C.} and {Bruno, C. G.} and {Caciolli, A.} and {Cavanna, F.} and {Chillery, T.} and {Colombetti, P.} and {Corvisiero, P.} and {Davinson, T.} and {Depalo, R.} and {Di Leva, A.} and {Di Paolo, L.} and {Elekes, Z.} and {Ferraro, F.} and {Fiore, E. M.} and {Formicola, A.} and {F\"ul\"op, Zs.} and {Gervino, G.} and {Guglielmetti, A.} and {Gustavino, C.} and {Gy\"urky, Gy.} and {Imbriani, G.} and {Junker, M.} and {Kochanek, I.} and {Lugaro, M.} and {Marigo, P.} and {Masha, E.} and {Menegazzo, R.} and {Mossa, V.} and {Pantaleo, F. R.} and {Paticchio, V.} and {Perrino, R.} and {Piatti, D.} and {Prati, P.} and {Schiavulli, L.} and {St\"ockel, K.} and {Straniero, O.} and {Sz\"ucs, T.} and {Tak\'acs, M. P.} and {Terrasi, F.} and {Trezzi, D.} and {Zavatarelli, S.}},
  author =	 {{Ciani, G. F.} and others},
  title =	 {A new approach to monitor ${}^{13}\mathrm{C}$-targets degradation in situ for ${}^{13}\mathrm{C}(\alpha,\mathrm{n}){}^{16}\mathrm{O}$ cross-section measurements at {LUNA}},
  DOI =		 "10.1140/epja/s10050-020-00077-0",
  journal =	 EPJA,
  year =	 2020,
  volume =	 56,
  number =	 3,
  pages =	 75
}

@article{Ciani2021,
  title =	 {{Direct Measurement of the ${}^{13}\mathrm{C}(\alpha,\mathrm{n}){}^{16}\mathrm{O}$ Cross Section into the $s$-Process Gamow Peak}},
  OPTauthor =	 {Ciani, G. F. and Csedreki, L. and Rapagnani, D. and Aliotta, M. and Balibrea-Correa, J. and Barile, F. and Bemmerer, D. and Best, A. and Boeltzig, A. and Broggini, C. and Bruno, C. G. and Caciolli, A. and Cavanna, F. and Chillery, T. and Colombetti, P. and Corvisiero, P. and Cristallo, S. and Davinson, T. and Depalo, R. and Di Leva, A. and Elekes, Z. and Ferraro, F. and Fiore, E. and Formicola, A. and F\"ul\"op, Zs. and Gervino, G. and Guglielmetti, A. and Gustavino, C. and Gy\"urky, Gy. and Imbriani, G. and Junker, M. and Lugaro, M. and Marigo, P. and Masha, E. and Menegazzo, R. and Mossa, V. and Pantaleo, F. R. and Paticchio, V. and Perrino, R. and Piatti, D. and Prati, P. and Schiavulli, L. and St\"ockel, K. and Straniero, O. and Sz\"ucs, T. and Tak\'acs, M. P. and Terrasi, F. and Vescovi, D. and Zavatarelli, S.},
  author =	 {Ciani, G. F. and others},
  collaboration ={LUNA Collaboration},
  journal =	 PRL,
  volume =	 127,
  issue =	 15,
  pages =	 152701,
  numpages =	 7,
  year =	 2021,
  month =	 10,
  doi =		 {10.1103/PhysRevLett.127.152701}
}

@ARTICLE{Clarkson2021,
  author =	 {{Clarkson}, O. and {Herwig}, F.},
  title =	 "{Convective H-He interactions in massive population III stellar evolution models}",
  journal =	 MNRAS,
  year =	 2021,
  month =	 1,
  volume =	 {500},
  number =	 {2},
  pages =	 {2685-2703},
  doi =		 {10.1093/mnras/staa3328}
}

@article{Coc2004,
  doi =		 {10.1086/380121},
  year =	 {2004},
  month =	 1,
  volume =	 {600},
  number =	 {2},
  pages =	 {544},
  author =	 {Alain Coc and Elisabeth Vangioni-Flam and Pierre Descouvemont and Abderrahim Adahchour and Carmen Angulo},
  title =	 {Updated {Big Bang Nucleosynthesis} Compared with {Wilkinson Microwave Anisotropy Probe} Observations and the Abundance of Light Elements},
  journal =	 APJ
}

@article{Coc2015,
  title =	 {New reaction rates for improved primordial $\mathrm{D}/\mathrm{H}$ calculation and the cosmic evolution of deuterium},
  OPTauthor =	 {Coc, Alain and Petitjean, Patrick and Uzan, Jean-Philippe and Vangioni, Elisabeth and Descouvemont, Pierre and Iliadis, Christian and Longland, Richard},
  author =	 {Coc, Alain and others},
  journal =	 PRD,
  volume =	 92,
  issue =	 12,
  pages =	 123526,
  numpages =	 20,
  year =	 2015,
  month =	 12,
  doi =		 {10.1103/PhysRevD.92.123526}
}

@ARTICLE{Confortola2007,
  OPTauthor =	 {{Confortola}, F. and {Bemmerer}, D. and {Costantini}, H. and {Formicola}, A. and {Gy{\"u}rky}, Gy. and {Bezzon}, P. and {Bonetti}, R. and {Broggini}, C. and {Corvisiero}, P. and {Elekes}, Z. and {F{\"u}l{\"o}p}, Zs. and {Gervino}, G. and {Guglielmetti}, A. and {Gustavino}, C. and {Imbriani}, G. and {Junker}, M. and {Laubenstein}, M. and {Lemut}, A. and {Limata}, B. and {Lozza}, V. and {Marta}, M. and {Menegazzo}, R. and {Prati}, P. and {Roca}, V. and {Rolfs}, C. and {Alvarez}, C. Rossi and {Somorjai}, E. and {Straniero}, O. and {Strieder}, F. and {Terrasi}, F. and {Trautvetter}, H.~P.},
  author =	 {{Confortola}, F. and others},
  title =	 {Astrophysical {$S$} factor of the {$^3\mathrm{He}(\alpha,\gamma)^7\mathrm{Be}$} reaction measured at low energy via detection of prompt and delayed {\ensuremath{\gamma}} rays},
  journal =	 PRC,
  year =	 2007,
  month =	 6,
  volume =	 {75},
  number =	 {6},
  eid =		 {065803},
  pages =	 {065803},
  doi =		 {10.1103/PhysRevC.75.065803}
}

@article{Cooke2018,
  doi =		 {10.3847/1538-4357/aaab53},
  year =	 2018,
  month =	 3,
  volume =	 855,
  number =	 2,
  pages =	 102,
  author =	 {Ryan J. Cooke and Max Pettini and Charles C. Steidel},
  title =	 {One Percent Determination of the Primordial Deuterium Abundance},
  journal =	 APJ
}

@article{Csedreki2021,
  title =	 {Characterization of the {LUNA} neutron detector array for the measurement of the ${}^{13}\mathrm{C}(\alpha,\mathrm{n}){}^{16}\mathrm{O}$ reaction},
  journal =	 NIMA,
  volume =	 994,
  pages =	 165081,
  year =	 2021,
  issn =	 {0168-9002},
  doi =		 {10.1016/j.nima.2021.165081},
  OPTauthor =	 {L. Csedreki and G.F. Ciani and J. Balibrea-Correa and A. Best and M. Aliotta and F. Barile and D. Bemmerer and A. Boeltzig and C. Broggini and C.G. Bruno and A. Caciolli and F. Cavanna and T. Chillery and P. Colombetti and P. Corvisiero and T. Davinson and R. Depalo and A. {Di Leva} and Z. Elekes and F. Ferraro and E.M. Fiore and A. Formicola and Zs. Fülöp and G. Gervino and A. Guglielmetti and C. Gustavino and Gy. Gyürky and G. Imbriani and Z. Janas and M. Junker and I. Kochanek and M. Lugaro and P. Marigo and E. Masha and C. Mazzocchi and R. Menegazzo and V. Mossa and F.R. Pantaleo and V. Paticchio and R. Perrino and D. Piatti and P. Prati and L. Schiavulli and K. Stöckel and O. Straniero and T. Szücs and M.P. Takács and F. Terrasi and S. Zavatarelli},
  author =	 {L. Csedreki and others}
}

@article{Curceanu2011JP,
  doi =		 {10.1088/1742-6596/306/1/012036},
  year =	 2011,
  month =	 7,
  volume =	 306,
  pages =	 012036,
  OPTauthor =	 {C Curceanu and S Bartalucci and S Bertolucci and M Bragadireanu and M Cargnelli and S Di Matteo and J -P Egger and C Guaraldo and M Iliescu and T Ishiwatari and M Laubenstein and J Marton and E Milotti and D Pietreanu and T Ponta and A Rizzo and A Romero Vidal and A Scordo and D L Sirghi and F Sirghi and L Sperandio and O Vazquez Doce and E Widmann and J Zmeskal},
  author =	 {C Curceanu and others},
  title =	 {Experimental tests of quantum mechanics {\textendash} {Pauli} exclusion principle violation (the {VIP} experiment) and future perspective},
  journal =	 JPhysCS
}

@article{Curceanu2017EN,
  doi =		 {10.3390/e19070300},
  year =	 2017,
  month =	 6,
  volume =	 19,
  number =	 7,
  pages =	 300,
  OPTauthor =	 {Catalina Curceanu and Hexi Shi and Sergio Bartalucci and Sergio Bertolucci and Massimiliano Bazzi and Carolina Berucci and Mario Bragadireanu and Michael Cargnelli and Alberto Clozza and Luca De Paolis and Sergio Di Matteo and Jean-Pierre Egger and Carlo Guaraldo and Mihail Iliescu and Johann Marton and Matthias Laubenstein and Edoardo Milotti and Marco Miliucci and Andreas Pichler and Dorel Pietreanu and Kristian Piscicchia and Alessandro Scordo and Diana Sirghi and Florin Sirghi and Laura Sperandio and Oton Vazquez Doce and Eberhard Widmann and Johann Zmeskal},
  author =	 {Catalina Curceanu and others},
  title =	 {Test of the {Pauli} Exclusion Principle in the {VIP}-2 Underground Experiment},
  journal =	 {Entropy}
}

@article{Cyburt2004,
  title =	 {Determination of ${S}_{17}(0)$ from published data},
  author =	 {Cyburt, R. H. and Davids, B. and Jennings, B. K.},
  journal =	 PRC,
  volume =	 {70},
  issue =	 {4},
  pages =	 {045801},
  numpages =	 {9},
  year =	 {2004},
  month =	 10,
  doi =		 {10.1103/PhysRevC.70.045801}
}

@article{Cyburt2016,
  title =	 {Big bang nucleosynthesis: Present status},
  author =	 {Cyburt, Richard H. and Fields, Brian D. and Olive, Keith A. and Yeh, Tsung-Han},
  journal =	 RMP,
  volume =	 {88},
  issue =	 {1},
  pages =	 {015004},
  numpages =	 {22},
  year =	 {2016},
  month =	 2,
  doi =		 {10.1103/RevModPhys.88.015004}
}

@ARTICLE {Dagata2018,
  author =	 {{D'Agata}, G. and others},
  OPTauthor =	 {{D'Agata}, G. and {Pizzone}, R.~G. and {La Cognata}, M. and {Indelicato}, I. and {Spitaleri}, C. and {Burjan}, V. and {Cherubini}, S. and {Di Pietro}, A. and {Guardo}, G.~L. and {Gulino}, M. and {La Commara}, M. and {Lamia}, L. and {Lattuada}, M. and {Mazzocco}, M. and {Mrazek}, J. and {Milin}, M. and {Palmerini}, S. and {Parascandolo}, C. and {Pierroutsakou}, D. and {Rapisarda}, G.~G. and {Romano}, S. and {Sergi}, M.~L. and {Soi{\'c}}, N. and {Spart{\'a}}, R. and {Trippella}, O. and {Tumino}, A.},
  title =	 {The {$^{19}\mathrm{F}(\alpha,\mathrm{p})^{22}\mathrm{Ne}$} and {$^{23}\mathrm{Na}(\mathrm{p},\alpha)^{20}\mathrm{Ne}$} reaction in {AGB} nucleosynthesis via {THM}},
  journal =	 EPJWC,
  year =	 2018,
  series =	 {European Physical Journal Web of Conferences},
  volume =	 {184},
  pages =	 {02003},
  doi =		 {10.1051/epjconf/2018184h02003},
}

@ARTICLE{DAntona2016,
  OPTauthor =	 {{D'Antona}, F. and {Vesperini}, E. and {D'Ercole}, A. and {Ventura}, P. and {Milone}, A.~P. and {Marino}, A.~F. and {Tailo}, M.},
  author =	 {{D'Antona}, F. and others},
  title =	 {A single model for the variety of multiple-population formation(s) in globular clusters: a temporal sequence},
  journal =	 MNRAS,
  year =	 2016,
  month =	 5,
  volume =	 458,
  number =	 2,
  pages =	 {2122-2139},
  doi =		 {10.1093/mnras/stw387}
}

@article{Das2005,
  title =	 {Terrestrial {$^{7}\mathrm{Be}$} decay rate and {$^{8}\mathrm{B}$} solar neutrino flux},
  author =	 {Das, P. and Ray, A.},
  journal =	 PRC,
  volume =	 71,
  issue =	 2,
  pages =	 025801,
  numpages =	 7,
  year =	 2005,
  month =	 2,
  doi =		 {10.1103/PhysRevC.71.025801}
}

@ARTICLE{deBoer2017,
  OPTauthor =	 {{deBoer}, R.~J. and {G{\"o}rres}, J. and {Wiescher}, M. and {Azuma}, R.~E. and {Best}, A. and {Brune}, C.~R. and {Fields}, C.~E. and {Jones}, S. and {Pignatari}, M. and {Sayre}, D. and {Smith}, K. and {Timmes}, F.~X. and {Uberseder}, E.},
  author =	 {{deBoer}, R.~J. and others},
  title =	 {The {$^{12}\mathrm{C}(\alpha,\gamma){}^{16}\mathrm{O}$} reaction and its implications for stellar helium burning},
  journal =	 RMP,
  year =	 2017,
  month =	 7,
  volume =	 89,
  number =	 3,
  eid =		 035007,
  pages =	 035007,
  doi =		 {10.1103/RevModPhys.89.035007}
}

@article{deBoer2020,
  title =	 {{Sensitivity of the $^{13}\mathrm{C}(\alpha,\mathrm{n}){}^{16}\mathrm{O}$ $S$ factor to the uncertainty in the level parameters of the near-threshold state}},
  OPTauthor =	 {deBoer, R. J. and Brune, C. R. and Febrarro, M. and G\"orres, J. and Thompson, I. J. and Wiescher, M.},
  author =	 {deBoer, R. J. and others},
  journal =	 PRC,
  volume =	 101,
  issue =	 4,
  pages =	 045802,
  numpages =	 9,
  year =	 2020,
  month =	 4,
  doi =		 {10.1103/PhysRevC.101.045802}
}

@article{deBoer2021,
  title =	 {{${}^{19}\mathrm{F}(\mathrm{p},\gamma){}^{20}\mathrm{Ne}$} and {${}^{19}\mathrm{F}(\mathrm{p},\alpha){}^{16}\mathrm{O}$} reaction rates and their effect on calcium production in {Population III} stars from hot {CNO} breakout},
  OPTauthor =	 {deBoer, R. J. and Clarkson, O. and Couture, A. J. and G\"orres, J. and Herwig, F. and Lombardo, I. and Scholz, P. and Wiescher, M.},
  author =	 {deBoer, R. J. and others},
  journal =	 PRC,
  volume =	 103,
  issue =	 5,
  pages =	 055815,
  numpages =	 21,
  year =	 2021,
  month =	 5,
  doi =		 {10.1103/PhysRevC.103.055815}
}

@ARTICLE{DeCesare2018,
  author =	 {{De Cesare}, M. and {Di Leva}, A. and {Del Vecchio}, A. and {Gialanella}, L.},
  title =	 "{A novel recession rate physics methodology for space applications at CIRA by means of CIRCE radioactive beam tracers}",
  journal =	 JPhysD,
  year =	 2018,
  month =	 3,
  volume =	 {51},
  number =	 {9},
  eid =		 {09LT01},
  pages =	 {09LT01},
  doi =		 {10.1088/1361-6463/aaa834}
}

@article{DeCesare2020,
  title =	 {Gamma and infrared novel methodologies in Aerospace re-entry: {$\gamma$}-rays crystal efficiency by {GEANT4} for {TPS} material recession assessment and simultaneous dual color infrared temperature determination},
  journal =	 NIMB,
  volume =	 {479},
  pages =	 {264-271},
  year =	 {2020},
  issn =	 {0168-583X},
  doi =		 {10.1016/j.nimb.2020.02.005},
  author =	 {M. {De Cesare} and others},
  OPTauthor =	 {M. {De Cesare} and L. Savino and A. {Di Leva} and D. Rapagnani and A. {Del Vecchio} and A. D'Onofrio and L. Gialanella}
}

@book{dellantonio1964NY,
  author =	 {{Dell'Antonio}, G. and Greenberg, O. and Sudarshan, O.},
  title =	 {Group theoretical concepts and methods in elementary particle physics. Lectures at the {Istanbul} Summer School of Theoretical Physics, 1962},
  year =	 1964,
  editor =	 {F. Gursey},
  address =      {New York},
  publisher =	 {Gordon and Breach}
}

@article{DellAquila17,
  OPTauthor =	 {{Dell'Aquila}, D. and {Lombardo}, I. and {Verde}, G. and {Vigilante}, M. and {Acosta}, L. and {Agodi}, C. and {Cappuzzello}, F. and {Carbone}, D. and {Cavallaro}, M. and {Cherubini}, S. and {Cvetinovic}, A. and {D'Agata}, G. and {Francalanza}, L. and {Guardo}, G.~L. and {Gulino}, M. and {Indelicato}, I. and {La Cognata}, M. and {Lamia}, L. and {Ordine}, A. and {Pizzone}, R.~G. and {Puglia}, S.~M.~R. and {Rapisarda}, G.~G. and {Romano}, S. and {Santagati}, G. and {Spart{\`a}}, R. and {Spadaccini}, G. and {Spitaleri}, C. and {Tumino}, A.},
  author =	 {{Dell'Aquila}, D. and others},
  title =	 {High-Precision Probe of the Fully Sequential Decay Width of the {Hoyle} State in $^{12}\mathrm{C}$},
  journal =	 PRL,
  year =	 2017,
  month =	 9,
  volume =	 119,
  number =	 13,
  eid =		 132501,
  pages =	 132501,
  doi =		 {10.1103/PhysRevLett.119.132501}
}

@ARTICLE{deOliveira1996,
  OPTauthor =	 {{de Oliveira}, F. and {Coc}, A. and {Aguer}, P. and {Angulo}, C. and {Bogaert}, G. and {Kiener}, J. and {Lefebvre}, A. and {Tatischeff}, V. and {Thibaud}, J. -P. and {Fortier}, S. and {Maison}, J.~M. and {Rosier}, L. and {Rotbard}, G. and {Vernotte}, J. and {Arnould}, M. and {Jorissen}, A. and {Mowlavi}, N.},
  author =	 {{de Oliveira}, F. and others},
  title =	 {Determination of {$\alpha$}-widths in {${}^{19}\mathrm{F}$} relevant to fluorine nucleosynthesis},
  journal =	 NuclPhysA,
  year =	 1996,
  month =	 2,
  volume =	 597,
  number =	 2,
  pages =	 {231-252},
  doi =		 {10.1016/0375-9474(95)00455-6}
}

@article{Desclaux1975CPC,
  doi =		 {10.1016/0010-4655(75)90054-5},
  year =	 1975,
  month =	 1,
  volume =	 9,
  number =	 1,
  pages =	 {31--45},
  author =	 {J.P. Desclaux},
  title =	 {A multiconfiguration relativistic {DIRAC}-{FOCK} program},
  journal =	 CPC
}

@article{DiLeva2008,
  title =	 {Recoil separator {ERNA}: Measurement of {$\Hetag$}},
  journal =	 NIMA,
  volume =	 595,
  number =	 2,
  pages =	 {381-390},
  year =	 2008,
  issn =	 {0168-9002},
  doi =		 {10.1016/j.nima.2008.07.082},
  OPTauthor =	 {A. {Di Leva} and M. {De Cesare} and D. Schürmann and N. {De Cesare} and A. D’Onofrio and L. Gialanella and R. Kunz and G. Imbriani and A. Ordine and V. Roca and D. Rogalla and C. Rolfs and M. Romano and E. Somorjai and F. Strieder and F. Terrasi},
  author =	 {A. {Di Leva} and others}
}

@article{DiLeva2009,
  author =	 {A. {Di Leva} and others},
  OPTauthor =	 {A. {Di Leva} and L. Gialanella and R. Kunz and D. Rogalla and D. Sch\"{u}rmann and F. Strieder and M. {De Cesare} and N. {De Cesare} and A. {D'Onofrio} and Z. F{\"u}l{\"o}p and G. Gy{\"u}rky and G. Imbriani and G. Mangano and A. Ordine and V. Roca and C. Rolfs and M. Romano and E. Somorjai and F. Terrasi},
  doi =		 {10.1103/PhysRevLett.102.232502},
  journal =	 prl,
  note =	 {Erratum: [Phys. Rev. Lett. 103, 159903 (2009)]},
  number =	 {23},
  numpages =	 {4},
  pages =	 {232502},
  title =	 {Stellar and Primordial Nucleosynthesis of {$^7\mathrm{Be}$}: Measurement of {$^3\mathrm{He}(\alpha, \gamma)^7\mathrm{Be}$}},
  volume =	 {102},
  year =	 {2009}
}

@ARTICLE{DiLeva2012,
  OPTauthor =	 {{Di Leva}, A. and {Pezzella}, A. and {De Cesare}, N. and {D'Onofrio}, A. and {Gialanella}, L. and {Romano}, M. and {Romoli}, M. and {Sch\"urmann}, D. and {Terrasi}, F. and {Imbriani}, G.},
  author =	 {{Di Leva}, A. and others},
  title =	 {{$^{14,15}\mathrm{N}$} beam from cyanide compounds},
  journal =	 NIMA,
  year =	 2012,
  month =	 10,
  volume =	 689,
  pages =	 {98-101},
  doi =		 {10.1016/j.nima.2012.06.037}
}

@article{DiLeva2014,
  title =	 {Underground study of the ${}^{17}\mathrm{O}(\mathrm{p},\gamma){}^{18}\mathrm{F}$ reaction relevant for explosive hydrogen burning},
  OPTauthor =	 {{Di Leva}, A. and Scott, D. A. and Caciolli, A. and Formicola, A. and Strieder, F. and Aliotta, M. and Anders, M. and Bemmerer, D. and Broggini, C. and Corvisiero, P. and Elekes, Z. and F\"ul\"op, Zs. and Gervino, G. and Guglielmetti, A. and Gustavino, C. and Gy\"urky, Gy. and Imbriani, G. and Jos\'e, J. and Junker, M. and Laubenstein, M. and Menegazzo, R. and Napolitani, E. and Prati, P. and Rigato, V. and Roca, V. and Somorjai, E. and Salvo, C. and Straniero, O. and Sz\"ucs, T. and Terrasi, F. and Trezzi, D.},
  author =	 {{Di Leva}, A. and others},
  collaboration ={LUNA Collaboration},
  journal =	 PRC,
  volume =	 89,
  issue =	 1,
  pages =	 015803,
  numpages =	 15,
  year =	 2014,
  month =	 1,
  doi =		 {10.1103/PhysRevC.89.015803}
}

@ARTICLE{DiLeva2017,
  OPTauthor =	 {{Di Leva}, A. and {Imbriani}, G. and {Buompane}, R. and {Gialanella}, L. and {Best}, A. and {Cristallo}, S. and {De Cesare}, M. and {D'Onofrio}, A. and {Duarte}, J.~G. and {Gasques}, L.~R. and {Morales-Gallegos}, L. and {Pezzella}, A. and {Porzio}, G. and {Rapagnani}, D. and {Roca}, V. and {Romoli}, M. and {Sch{\"u}rmann}, D. and {Straniero}, O. and {Terrasi}, F. and {ERNA Collaboration}},
  author =	 {{Di Leva}, A. and others},
  title =	 {Measurement of 1323 and 1487\,{keV} resonances in {${}^{15}\mathrm{N}(\alpha,\gamma){}^{19}\mathrm{F}$} with the recoil separator {ERNA}},
  journal =	 PRC,
  year =	 2017,
  month =	 4,
  volume =	 95,
  number =	 4,
  eid =		 045803,
  pages =	 045803,
  doi =		 {10.1103/PhysRevC.95.045803}
}

@article{Diosi1987PRA,
  author =	 {{Di{\'o}si}, Lajos},
  title =	 {A universal master equation for the gravitational violation of quantum mechanics},
  journal =	 PLA,
  year =	 1987,
  month =	 3,
  volume =	 {120},
  number =	 {8},
  pages =	 {377-381},
  doi =		 {10.1016/0375-9601(87)90681-5}
}

@article{Diosi1989PRA,
  author =	 {{Di{\'o}si}, Lajos},
  title =	 {Models for universal reduction of macroscopic quantum fluctuations},
  journal =	 PRA,
  year =	 1989,
  month =	 8,
  volume =	 {40},
  number =	 {3},
  pages =	 {1165-1174},
  doi =		 {10.1103/PhysRevA.40.1165}
}

@article{Dombos2022,
  title =	 {Measurement of Low-Energy Resonance Strengths in the $^{18}\mathrm{O}(\ensuremath{\alpha},\ensuremath{\gamma})^{22}\mathrm{Ne}$ Reaction},
  OPTauthor =	 {Dombos, A. C. and Robertson, D. and Simon, A. and Kadlecek, T. and Hanhardt, M. and G\"orres, J. and Couder, M. and Kelmar, R. and Olivas-Gomez, O. and Stech, E. and Strieder, F. and Wiescher, M.},
  author =	 {Dombos, A. C. and others},
  journal =	 PRL,
  volume =	 128,
  issue =	 16,
  pages =	 162701,
  numpages =	 6,
  year =	 2022,
  month =	 4,
  doi =		 {10.1103/PhysRevLett.128.162701}
}

@ARTICLE{Dominguez2001,
  author =	 {{Dom{\'\i}nguez}, Inma and {H{\"o}flich}, Peter and {Straniero}, Oscar},
  title =	 {Constraints on the Progenitors of {Type Ia} Supernovae and Implications for the Cosmological Equation of State},
  journal =	 APJ,
  year =	 2001,
  month =	 8,
  volume =	 557,
  number =	 1,
  pages =	 {279-291},
  doi =		 {10.1086/321661}
}

@article{Donadi2020NP,
  doi =		 {10.1038/s41567-020-1008-4},
  year =	 2020,
  month =	 9,
  volume =	 17,
  number =	 1,
  pages =	 {74--78},
  OPTauthor =	 {Sandro Donadi and Kristian Piscicchia and Catalina Curceanu and Lajos Di{\'{o}}si and Matthias Laubenstein and Angelo Bassi},
  author =	 {Sandro Donadi and others},
  title =	 {Underground test of gravity-related wave function collapse},
  journal =	 NaturePhys
}

@article{Donadi2021EPJC,
  doi =		 {10.1140/epjc/s10052-021-09556-0},
  year =	 2021,
  month =	 8,
  volume =	 81,
  number =	 8,
  OPTauthor =	 {Sandro Donadi and Kristian Piscicchia and Raffaele Del Grande and Catalina Curceanu and Matthias Laubenstein and Angelo Bassi},
  author =	 {Sandro Donadi and others},
  title =	 {Novel {CSL} bounds from the noise-induced radiation emission from atoms},
  journal =	 EPJC
}

@article{Dyer1974,
  Author =	 {P. Dyer and C.A. Barnes},
  Doi =		 {10.1016/0375-9474(74)90470-9},
  Journal =	 NuclPhysA,
  Number =	 2,
  Pages =	 {495-520},
  Title =	 {The {$^{12}\mathrm{C}(\alpha,\gamma){}^{16}\mathrm{O}$} reaction and stellar helium burning},
  Volume =	 233,
  Year =	 1974
}

@article{Elliott2012Foundations,
  OPTauthor =	 {{Elliott}, S.~R. and {LaRoque}, B.~H. and {Gehman}, V.~M. and {Kidd}, M.~F. and {Chen}, M.},
  author =	 {{Elliott}, S.~R. and others},
  title =	 {An Improved Limit on {Pauli-Exclusion-Principle} Forbidden Atomic Transitions},
  journal =	 FOUNDP,
  year =	 2012,
  month =	 8,
  volume =	 42,
  number =	 8,
  pages =	 {1015-1030},
  doi =		 {10.1007/s10701-012-9643-y}
}

@article{Epelbaum11,
  author =	 {{Epelbaum}, Evgeny and {Krebs}, Hermann and {Lee}, Dean and {Mei{\ss}ner}, Ulf-G.},
  title =	 {Ab Initio Calculation of the {Hoyle} State},
  journal =	 PRL,
  year =	 2011,
  month =	 5,
  volume =	 106,
  number =	 19,
  eid =		 192501,
  pages =	 192501,
  doi =		 {10.1103/PhysRevLett.106.192501}
}

@ARTICLE{Farinon2008,
  OPTauthor =	 {{Farinon}, F. and {Glodariu}, T. and {Mazzocco}, M. and {Battistella}, A. and {Bonetti}, R. and {Costa}, L. and {De Rosa}, A. and {Guglielmetti}, A. and {Inglima}, G. and {La Commara}, M. and {Maidikov}, V.~Z. and {Martin}, B. and {Mazzocchi}, C. and {Pierroutsakou}, D. and {Romoli}, M. and {Sandoli}, M. and {Signorini}, C. and {Soramel}, F. and {Stroe}, L. and {Vardaci}, E.},
  author =	 {{Farinon}, F. and others},
  title =	 {Commissioning of the {EXOTIC} beam line},
  journal =	 NIMB,
  year =	 2008,
  month =	 10,
  volume =	 266,
  number =	 {19-20},
  pages =	 {4097-4102},
  doi =		 {10.1016/j.nimb.2008.05.128}
}

@book{Feldman2012,
  title =	 {Materials Analysis by Ion Channeling: Submicron Crystallography},
  author =	 {Leonard C. Feldman and J. W. Mayer and Samuel Thomas Picraux},
  year =	 {2012},
  publisher =	 {Academic Press},
  address =	 {San Diego},
  isbn =	 {978-0-12-252680-0},
  doi =		 {10.1016/B978-0-12-252680-0.50003-6}
}

@article{Fermi1934SC,
  journal =	 {Scientia},
  author =	 {E. Fermi},
  year =	 1934,
  title =	 {Le Ultime Particelle Costitutive Della Materia},
  OPTnumber =	 55,
  volume =	 55,
  publisher =	 {Engelmann},
  pages =	 {21-28}
}

@article{Fernandes2021,
  OPTauthor =	 {Fernandes, Daniel M. and Sirak, Kendra A. and Ringbauer, Harald and Sedig, Jakob and Rohland, Nadin and Cheronet, Olivia and Mah, Matthew and Mallick, Swapan and Olalde, I{\~n}igo and Culleton, Brendan J. and Adamski, Nicole and Bernardos, Rebecca and Bravo, Guillermo and Broomandkhoshbacht, Nasreen and Callan, Kimberly and Candilio, Francesca and Demetz, Lea and Carlson, Kellie Sara Duffett and Eccles, Laurie and Freilich, Suzanne and George, Richard J. and Lawson, Ann Marie and Mandl, Kirsten and Marzaioli, Fabio and McCool, Weston C. and Oppenheimer, Jonas and {\"O}zdogan, Kadir T. and Schattke, Constanze and Schmidt, Ryan and Stewardson, Kristin and Terrasi, Filippo and Zalzala, Fatma and Ant{\'u}nez, Carlos Arredondo and Canosa, Ercilio Vento and Colten, Roger and Cucina, Andrea and Genchi, Francesco and Kraan, Claudia and La Pastina, Francesco and Lucci, Michaela and Maggiolo, Marcio Veloz and Marcheco-Teruel, Beatriz and Maria, Clenis Tavarez and Mart{\'\i}nez, Christian and Par{\'\i}s, Ingeborg and Pateman, Michael and Simms, Tanya M. and Sivoli, Carlos Garcia and Vilar, Miguel and Kennett, Douglas J. and Keegan, William F. and Coppa, Alfredo and Lipson, Mark and Pinhasi, Ron and Reich, David},
  author =	 {Fernandes, Daniel M. and others},
  date =	 {2021/02/01},
  doi =		 {10.1038/s41586-020-03053-2},
  isbn =	 {1476-4687},
  journal =	 Nature,
  number =	 7844,
  pages =	 {103--110},
  title =	 {A genetic history of the pre-contact Caribbean},
  volume =	 590,
  year =	 2021
}

@article{Ferraro2018a,
  OPTauthor =	 {Ferraro, F. and Tak{\'a}cs, M. P. and Piatti, D. and Mossa, V. and Aliotta, M. and Bemmerer, D. and Best, A. and Boeltzig, A. and Broggini, C. and Bruno, C. G. and Caciolli, A. and Cavanna, F. and Chillery, T. and Ciani, G. F. and Corvisiero, P. and Csedreki, L. and Davinson, T. and Depalo, R. and D'Erasmo, G. and Di Leva, A. and Elekes, Z. and Fiore, E. M. and Formicola, A. and F{\"u}l{\"o}p, Zs. and Gervino, G. and Guglielmetti, A. and Gustavino, C. and Gy{\"u}rky, Gy. and Imbriani, G. and Junker, M. and Kochanek, I. and Lugaro, M. and Marcucci, L. E. and Marigo, P. and Menegazzo, R. and Pantaleo, F. R. and Paticchio, V. and Perrino, R. and Prati, P. and Schiavulli, L. and St{\"o}ckel, K. and Straniero, O. and Sz{\"u}cs, T. and Trezzi, D. and Zavatarelli, S.},
  author =	 {Ferraro, F. and others},
  day =		 15,
  doi =		 {10.1140/epja/i2018-12476-7},
  issn =	 {1434-601X},
  journal =	 EPJA,
  month =	 3,
  number =	 3,
  pages =	 44,
  title =	 {A high-efficiency gas target setup for underground experiments, and redetermination of the branching ratio of the 189.5\,{keV} ${}^{22}\mathrm{Ne}(\mathrm{p},\gamma){}^{23}\mathrm{Na}$ resonance},
  volume =	 54,
  year =	 2018
}

@article{Ferraro2018b,
  OPTauthor =	 {Ferraro, F. and Tak\'acs, M. P. and Piatti, D. and Cavanna, F. and Depalo, R. and Aliotta, M. and Bemmerer, D. and Best, A. and Boeltzig, A. and Broggini, C. and Bruno, C. G. and Caciolli, A. and Chillery, T. and Ciani, G. F. and Corvisiero, P. and Davinson, T. and D'Erasmo, G. and Di Leva, A. and Elekes, Z. and Fiore, E. M. and Formicola, A. and F\"ul\"op, Zs. and Gervino, G. and Guglielmetti, A. and Gustavino, C. and Gy\"urky, Gy. and Imbriani, G. and Junker, M. and Karakas, A. and Kochanek, I. and Lugaro, M. and Marigo, P. and Menegazzo, R. and Mossa, V. and Pantaleo, F. R. and Paticchio, V. and Perrino, R. and Prati, P. and Schiavulli, L. and St\"ockel, K. and Straniero, O. and Sz\"ucs, T. and Trezzi, D. and Zavatarelli, S.},
  author =	 {Ferraro, F. and others},
  collaboration ={LUNA Collaboration},
  doi =		 {10.1103/PhysRevLett.121.172701},
  issue =	 17,
  journal =	 PRL,
  month =	 10,
  numpages =	 6,
  pages =	 172701,
  title =	 {Direct Capture Cross Section and the ${E}_\mathrm{p}=71$ and 105\,keV Resonances in the $^{22}\mathrm{Ne}(\mathrm{p},\gamma){}^{23}\mathrm{Na}$ Reaction},
  volume =	 121,
  year =	 2018
}

@article{Fields2011,
  author =	 {Fields, Brian D.},
  title =	 {The Primordial Lithium Problem},
  journal =	 ARNPS,
  volume =	 {61},
  number =	 {1},
  pages =	 {47-68},
  year =	 {2011},
  doi =		 {10.1146/annurev-nucl-102010-130445},
}

@ARTICLE{Formicola2003,
  OPTauthor =	 {{Formicola}, A. and {Imbriani}, G. and {Junker}, M. and {Bemmerer}, D. and {Bonetti}, R. and {Broggini}, C. and {Casella}, C. and {Corvisiero}, P. and {Costantini}, H. and {Gervino}, G. and {Gustavino}, C. and {Lemut}, A. and {Prati}, P. and {Roca}, V. and {Rolfs}, C. and {Romano}, M. and {Sch{\"u}rmann}, D. and {Strieder}, F. and {Terrasi}, F. and {Trautvetter}, H. -P. and {Zavatarelli}, S.},
  author =	 {{Formicola}, A. and others},
  title =	 {The {LUNA II} 400\,{kV} accelerator},
  journal =	 NIMA,
  year =	 2003,
  month =	 7,
  volume =	 507,
  number =	 3,
  pages =	 {609-616},
  doi =		 {10.1016/S0168-9002(03)01435-9}
}

@article{Formicola2004,
  OPTauthor =	 {{Formicola}, A. and {Imbriani}, G. and {Costantini}, H. and {Angulo}, C. and {Bemmerer}, D. and {Bonetti}, R. and {Broggini}, C. and {Corvisiero}, P. and {Cruz}, J. and {Descouvemont}, P. and {F{\"u}l{\"o}p}, Z. and {Gervino}, G. and {Guglielmetti}, A. and {Gustavino}, C. and {Gy{\"u}rky}, G. and {Jesus}, A.~P. and {Junker}, M. and {Lemut}, A. and {Menegazzo}, R. and {Prati}, P. and {Roca}, V. and {Rolfs}, C. and {Romano}, M. and {Rossi Alvarez}, C. and {Sch{\"u}mann}, F. and {Somorjai}, E. and {Straniero}, O. and {Strieder}, F. and {Terrasi}, F. and {Trautvetter}, H.~P. and {Vomiero}, A. and {Zavatarelli}, S.},
  author =	 {{Formicola}, A. and others},
  doi =		 {10.1016/j.physletb.2004.03.092},
  journal =	 PLB,
  pages =	 {61-68},
  title =	 {{Astrophysical S-factor of $^{14}$N(p,$\gamma$)$^{15}$O}},
  volume =	 591,
  year =	 2004
}

@article{Freer94,
  OPTauthor =	 {{Freer}, M. and {Wuosmaa}, A.~H. and {Betts}, R.~R. and {Henderson}, D.~J. and {Wilt}, P. and {Zurm{\"u}hle}, R.~W. and {Balamuth}, D.~P. and {Barrow}, S. and {Benton}, D. and {Li}, Q. and {Liu}, Z. and {Miao}, Y.},
  author =	 {{Freer}, M. and others},
  title =	 {Limits for the 3{\ensuremath{\alpha}} branching ratio of the decay of the {7.65\,MeV, $0^{+}_{2}$} state in {$^{12}\mathrm{C}$}},
  journal =	 PRC,
  year =	 1994,
  month =	 4,
  volume =	 49,
  number =	 4,
  pages =	 {R1751-R1754},
  doi =		 {10.1103/PhysRevC.49.R1751}
}

@article{Freer2014,
  author =	 {{Freer}, M. and {Fynbo}, H.~O.~U.},
  title =	 {The {Hoyle} state in $^{12}\mathrm{C}$},
  journal =	 PPNP,
  year =	 2014,
  month =	 9,
  volume =	 {78},
  pages =	 {1-23},
  doi =		 {10.1016/j.ppnp.2014.06.001}
}

@ARTICLE{Frentz2022,
  author =	 {{Frentz}, B. and others},
  OPTauthor =	 {{Frentz}, B. and {Aprahamian}, A. and {Boeltzig}, A. and {Borgwardt}, T. and {Clark}, A.~M. and {deBoer}, R.~J. and {Gilardy}, G. and {G{\"o}rres}, J. and {Hanhardt}, M. and {Henderson}, S.~L. and {Howard}, K.~B. and {Kadlecek}, T. and {Liu}, Q. and {Macon}, K.~T. and {Moylan}, S. and {Reingold}, C.~S. and {Robertson}, D. and {Seymour}, C. and {Strauss}, S.~Y. and {Strieder}, F. and {Vande Kolk}, B. and {Wiescher}, M.},
  title =	 {Investigation of the {$^{14}\mathrm{N}\pg^{15}\mathrm{O}$} reaction and its impact on the {CNO} cycle},
  journal =	 PRC,
  year =	 2022,
  month =	 12,
  volume =	 {106},
  number =	 {6},
  eid =		 {065803},
  pages =	 {065803},
  doi =		 {10.1103/PhysRevC.106.065803}
}

@ARTICLE{Gallino1998,
  OPTauthor =	 {{Gallino}, Roberto and {Arlandini}, Claudio and {Busso}, Maurizio and {Lugaro}, Maria and {Travaglio}, Claudia and {Straniero}, Oscar and {Chieffi}, Alessandro and {Limongi}, Marco},
  author =	 {{Gallino}, Roberto and others},
  title =	 {Evolution and Nucleosynthesis in Low-Mass {Asymptotic Giant Branch} Stars. {II}. {Neutron} Capture and the $s$-Process},
  journal =	 APJ,
  year =	 1998,
  month =	 4,
  volume =	 497,
  number =	 1,
  pages =	 {388-403},
  doi =		 {10.1086/305437}
}

@article{Gao2022,
  title =	 {{Deep Underground Laboratory Measurement of $^{13}\mathrm{C}\an{}^{16}\mathrm{O}$ in the Gamow Windows of the $s$ and $i$ Processes}},
  OPTauthor =	 {Gao, B. and Jiao, T. Y. and Li, Y. T. and Chen, H. and Lin, W. P. and An, Z. and Ru, L. H. and Zhang, Z. C. and Tang, X. D. and Wang, X. Y. and Zhang, N. T. and Fang, X. and Xie, D. H. and Fan, Y. H. and Ma, L. and Zhang, X. and Bai, F. and Wang, P. and Fan, Y. X. and Liu, G. and Huang, H. X. and Wu, Q. and Zhu, Y. B. and Chai, J. L. and Li, J. Q. and Sun, L. T. and Wang, S. and Cai, J. W. and Li, Y. Z. and Su, J. and Zhang, H. and Li, Z. H. and Li, Y. J. and Li, E. T. and Chen, C. and Shen, Y. P. and Lian, G. and Guo, B. and Li, X. Y. and Zhang, L. Y. and He, J. J. and Sheng, Y. D. and Chen, Y. J. and Wang, L. H. and Zhang, L. and Cao, F. Q. and Nan, W. and Nan, W. K. and Li, G. X. and Song, N. and Cui, B. Q. and Chen, L. H. and Ma, R. G. and Zhang, Z. C. and Yan, S. Q. and Liao, J. H. and Wang, Y. B. and Zeng, S. and Nan, D. and Fan, Q. W. and Qi, N. C. and Sun, W. L. and Guo, X. Y. and Zhang, P. and Chen, Y. H. and Zhou, Y. and Zhou, J. F. and He, J. R. and Shang, C. S. and Li, M. C. and Kubono, S. and Liu, W. P. and deBoer, R. J. and Wiescher, M. and Pignatari, M.},
  author =	 {Gao, B. and others},
  collaboration ={JUNA Collaboration},
  journal =	 PRL,
  volume =	 129,
  issue =	 13,
  pages =	 132701,
  numpages =	 7,
  year =	 2022,
  month =	 9,
  doi =		 {10.1103/PhysRevLett.129.132701}
}

@article{Gentile1940INC,
  doi =		 {10.1007/bf02960187},
  year =	 1940,
  month =	 12,
  volume =	 17,
  number =	 10,
  pages =	 {493--497},
  author =	 {Gentile, G. j.},
  title =	 {{\emph{Osservazioni}} sopra le statistiche intermedie},
  journal =	 {Il Nuovo Cimento}
}

@article{Ghirardi1986PRD,
  doi =		 {10.1103/physrevd.34.470},
  year =	 1986,
  month =	 7,
  volume =	 34,
  OPTnumber =	 2,
  pages =	 {470--491},
  author =	 {G. C. Ghirardi and A. Rimini and T. Weber},
  title =	 {Unified dynamics for microscopic and macroscopic systems},
  journal =	 PRD
}

@article{Ghirardi1990PRA,
  doi =		 {10.1103/physreva.42.78},
  year =	 1990,
  month =	 7,
  volume =	 42,
  OPTnumber =	 1,
  pages =	 {78--89},
  author =	 {Gian Carlo Ghirardi and Philip Pearle and Alberto Rimini},
  title =	 {Markov processes in {Hilbert} space and continuous spontaneous localization of systems of identical particles},
  journal =	 PRA
}

@Article{Gialanella2000,
  OPTauthor =	 {Gialanella, L. and Strieder, F. and Campajola, L. and D'Onofrio, A. and Greife, U. and Gyurky, G. and Imbriani, G. and Oliviero, G. and Ordine, A. and Roca, V. and Rolfs, C. and Romano, M. and Rogalla, D. and Sabbarese, C. and Somorjai, E. and Terrasi, F. and Trautvetter, H. P.},
  author =	 {Gialanella, L. and others},
  title =	 {Absolute cross section of {$\mathrm{p}(^7\mathrm{Be},\gamma)^8\mathrm{B}$} using a novel approach},
  journal =	 EPJA,
  year =	 2000,
  month =	 3,
  day =		 01,
  volume =	 7,
  number =	 3,
  pages =	 {303-305},
  issn =	 {1434-601X},
  doi =		 {10.1007/PL00013599}
}

@article{Gialanella2002,
  title =	 {Off-line production of a {$^7\mathrm{Be}$} radioactive ion beam},
  journal =	 NIMB,
  volume =	 197,
  number =	 1,
  pages =	 {150-154},
  year =	 2002,
  issn =	 {0168-583X},
  doi =		 {10.1016/S0168-583X(02)01386-1},
  OPTauthor =	 {L Gialanella and U Greife and N {De Cesare} and A D’Onofrio and M Romano and L Campajola and A Formicola and Z Fulop and G Gyurky and G Imbriani and C Lubritto and A Ordine and V Roca and D Rogalla and C Rolfs and M Russo and C Sabbarese and E Somorjai and F Strieder and F Terrasi and H.P Trautvetter},
  author =	 {L Gialanella and others}
}

@article{Gialanella2004,
  title =	 {Recoil separator {ERNA}: gas target and beam suppression},
  journal =	 NIMA,
  volume =	 522,
  number =	 3,
  pages =	 {432-438},
  year =	 2004,
  issn =	 {0168-9002},
  doi =		 {10.1016/j.nima.2003.11.386},
  OPTauthor =	 {L Gialanella and D Schürmann and F Strieder and A {Di Leva} and N {De Cesare} and A D’Onofrio and G Imbriani and J Klug and C Lubritto and A Ordine and V Roca and H Röcken and C Rolfs and D Rogalla and M Romano and F Schümann and F Terrasi and H.P Trautvetter},
  author =	 {L Gialanella and others}
}

@article{Goerres2000,
  title =	 {Low-energy resonances in {${}^{14}\mathrm{N}(\alpha,\gamma){}^{18}\mathrm{F}$} and their astrophysical implications},
  OPTauthor =	 {G\"orres, J. and Arlandini, C. and Giesen, U. and Heil, M. and K\"appeler, F. and Leiste, H. and Stech, E. and Wiescher, M.},
  author =	 {G\"orres, J. and others},
  journal =	 PRC,
  volume =	 62,
  issue =	 5,
  pages =	 055801,
  numpages =	 7,
  year =	 2000,
  month =	 10,
  doi =		 {10.1103/PhysRevC.62.055801}
}

@article{Greenberg1987PRL,
  doi =		 {10.1103/physrevlett.59.2507},
  year =	 1987,
  month =	 11,
  volume =	 59,
  number =	 22,
  pages =	 {2507--2510},
  author =	 {O. W. Greenberg and R. N. Mohapatra},
  title =	 {Local Quantum Field Theory of Possible Violation of the {Pauli} Principle},
  journal =	 PRL
}

@article{Greenberg1990PRL,
  doi =		 {10.1103/physrevlett.64.705},
  year =	 1990,
  month =	 2,
  volume =	 64,
  number =	 7,
  pages =	 {705--708},
  author =	 {O. W. Greenberg},
  title =	 {Example of infinite statistics},
  journal =	 PRL
}

@article{Guerra1973PRL,
  title =	 {New Interpretation of the Euclidean-{Markov} Field in the Framework of Physical {Minkowski} Space-Time},
  author =	 {Guerra, Franesco and Ruggiero, Patrizia},
  journal =	 PRL,
  volume =	 31,
  issue =	 16,
  pages =	 {1022--1025},
  numpages =	 0,
  year =	 1973,
  month =	 10,
  doi =		 {10.1103/PhysRevLett.31.1022}
}

@article{Gyurky2022,
  title =	 {{Activation cross section measurement of the $^{14}\mathrm{N}\pg^{15}\mathrm{O}$ astrophysical key reaction}},
  OPTauthor =	 {Gy{\"u}rky, Gy. and Hal{\'a}sz, Z. and Kiss, G. G. and Sz{\"u}cs, T. and F{\"u}l{\"o}p, Zs.},
  author =	 {Gy{\"u}rky, Gy. and others},
  journal =	 PRC,
  volume =	 105,
  issue =	 2,
  pages =	 {L022801},
  numpages =	 5,
  year =	 2022,
  month =	 2,
  doi =		 {10.1103/PhysRevC.105.L022801}
}

@article{Hammache2010,
  title =	 {High-energy breakup of $^{6}\mathrm{Li}$ as a tool to study the Big Bang nucleosynthesis reaction $^{2}\mathrm{H}$($\ensuremath{\alpha}$,$\ensuremath{\gamma}$)$^{6}\mathrm{Li}$},
  author =	 {Hammache, F. and others},
  OPTauthor =	 {Hammache, F. and Heil, M. and Typel, S. and Galaviz, D. and S\"ummerer, K. and Coc, A. and Uhlig, F. and Attallah, F. and Caamano, M. and Cortina, D. and Geissel, H. and Hellstr\"om, M. and Iwasa, N. and Kiener, J. and Koczon, P. and Kohlmeyer, B. and Mohr, P. and Schwab, E. and Schwarz, K. and Sch\"umann, F. and Senger, P. and Sorlin, O. and Tatischeff, V. and Thibaud, J. P. and Vangioni, E. and Wagner, A. and Walus, W.},
  journal =	 PRC,
  volume =	 {82},
  issue =	 {6},
  pages =	 {065803},
  numpages =	 {11},
  year =	 {2010},
  month =	 12,
  doi =		 {10.1103/PhysRevC.82.065803},
}

@article{hardie1984,
  title =	 {Resonant alpha capture by $^{7}\mathrm{Be}$ and $^{7}\mathrm{Li}$},
  OPTauthor =	 {Hardie, G. and Filippone, B. W. and Elwyn, A. J. and Wiescher, M. and Segel, R. E.},
    author =	 {Hardie, G. and others},
  journal =	 PRC,
  volume =	 29,
  issue =	 4,
  pages =	 {1199--1206},
  numpages =	 0,
  year =	 1984,
  month =	 4,
  doi =		 {10.1103/PhysRevC.29.1199},
}

@article{Harissopulos2005,
  title =	 {Cross section of the $^{13}\mathrm{C}(\alpha, \mathrm{n}){}^{16}\mathrm{O}$ reaction: A background for the measurement of geo-neutrinos},
  OPTauthor =	 {Harissopulos, S. and Becker, H. W. and Hammer, J. W. and Lagoyannis, A. and Rolfs, C. and Strieder, F.},
  author =	 {Harissopulos, S. and others},
  journal =	 PRC,
  volume =	 72,
  issue =	 6,
  pages =	 062801,
  numpages =	 5,
  year =	 2005,
  month =	 12,
  doi =		 {10.1103/PhysRevC.72.062801}
}

@ARTICLE{Hayakawa2021,
  OPTauthor =	 {{Hayakawa}, S. and {La Cognata}, M. and {Lamia}, L. and {Yamaguchi}, H. and {Kahl}, D. and {Abe}, K. and {Shimizu}, H. and {Yang}, L. and {Beliuskina}, O. and {Cha}, S.~M. and {Chae}, K.~Y. and {Cherubini}, S. and {Figuera}, P. and {Ge}, Z. and {Gulino}, M. and {Hu}, J. and {Inoue}, A. and {Iwasa}, N. and {Kim}, A. and {Kim}, D. and {Kiss}, G. and {Kubono}, S. and {La Commara}, M. and {Lattuada}, M. and {Lee}, E.~J. and {Moon}, J.~Y. and {Palmerini}, S. and {Parascandolo}, C. and {Park}, S.~Y. and {Phong}, V.~H. and {Pierroutsakou}, D. and {Pizzone}, R.~G. and {Rapisarda}, G.~G. and {Romano}, S. and {Spitaleri}, C. and {Tang}, X.~D. and {Trippella}, O. and {Tumino}, A. and {Zhang}, N.~T.},
  author =	 {{Hayakawa}, S. and others},
  title =	 {Constraining the Primordial Lithium Abundance: New Cross Section Measurement of the {$^{7}\mathrm{Be}$+$\mathrm{n}$} Reactions Updates the Total {$^{7}\mathrm{Be}$} Destruction Rate},
  journal =	 APJL,
  year =	 2021,
  month =	 7,
  volume =	 915,
  number =	 1,
  eid =		 {L13},
  pages =	 {L13},
  doi =		 {10.3847/2041-8213/ac061f}
}

@article{Heil2008,
  title =	 {{The ${}^{13}\mathrm{C}(\alpha,\mathrm{n})$ reaction and its role as a neutron source for the $s$ process}},
  OPTauthor =	 {Heil, M. and Detwiler, R. and Azuma, R. E. and Couture, A. and Daly, J. and G\"orres, J. and K\"appeler, F. and Reifarth, R. and Tischhauser, P. and Ugalde, C. and Wiescher, M.},
  author =	 {Heil, M. and others},
  journal =	 PRC,
  volume =	 78,
  issue =	 2,
  pages =	 025803,
  numpages =	 17,
  year =	 2008,
  month =	 8,
  doi =		 {10.1103/PhysRevC.78.025803}
}

@ARTICLE{Herndl1991,
  OPTauthor =	 {{Herndl}, H. and {Abele}, H. and {Staudt}, G. and {Bach}, B. and {Gr{\"u}n}, K. and {Scsribany}, H. and {Oberhummer}, H. and {Raimann}, G.},
  author =	 {{Herndl}, H. and others},
  title =	 {Direct reaction analysis of {$^{19}\mathrm{F}(\mathrm{p},\alpha)^{16}\mathrm{O}$} below the {Coulomb} barrier},
  journal =	 PRC,
  year =	 1991,
  month =	 9,
  volume =	 44,
  number =	 3,
  pages =	 {R952-R955},
  doi =		 {10.1103/PhysRevC.44.R952}
}

@article{Holl2019EPJC,
  doi =		 {10.1140/epjc/s10052-019-6869-2},
  year =	 2019,
  month =	 5,
  volume =	 79,
  number =	 6,
  OPTauthor =	 {P. Holl and L. Hauertmann and B. Majorovits and O. Schulz and M. Schuster and A. J. Zsigmond},
  author =	 {P. Holl and others},
  title =	 {Deep learning based pulse shape discrimination for germanium detectors},
  journal =	 EPJC
}

@article{Howl2019NJP,
  author =	 {{Howl}, Richard and {Penrose}, Roger and {Fuentes}, Ivette},
  title =	 {Exploring the unification of quantum theory and general relativity with a {Bose-Einstein} condensate},
  journal =	 NEWJP,
  year =	 2019,
  month =	 4,
  volume =	 21,
  number =	 4,
  eid =		 043047,
  pages =	 043047,
  doi =		 {10.1088/1367-2630/ab104a}
}

@article{ignatiev1987YF,
  Author =	 {Ignatiev, A. Yu. and Kuzmin, V.A.},
  Journal =	 {Yad. Fiz.},
  volume =	 46,
  pages =	 786,
  Title =	 {Search for slight violation of the {Pauli} principle},
  Year =	 1987
}

@incollection{ignatiev1987QRK,
  title =	 {Is small violation of the {Pauli} principle possible?},
  author =	 {Ignatiev, A. Yu. and Kuzmin, V.A.},
  booktitle =    {Quarks '86},
  editor =       {Tavkhelidze, A. N. and Matveev, V. A. and Pivovarov, A. A.},
  pages =	 263,
  year =	 1987,
  isbn =         {90-6764-097-2},
  publisher =    {VNU Science Press},
  address =      {Utrecht, The Netherlands}
}

@article{Ignatiev2006RPC,
  doi =		 {10.1016/j.radphyschem.2005.10.040},
  year =	 2006,
  month =	 11,
  publisher =	 {Elsevier {BV}},
  volume =	 75,
  number =	 11,
  pages =	 {2090--2096},
  author =	 {A. Yu. Ignatiev},
  title =	 {X rays test the {Pauli} exclusion principle},
  journal =	 RPC
}

@ARTICLE{Iliadis2001,
  OPTauthor =	 {{Iliadis}, C. and {D'Auria}, John M. and {Starrfield}, Sumner and {Thompson}, William J. and {Wiescher}, Michael},
  author =	 {{Iliadis}, C. and others},
  title =	 {Proton-induced Thermonuclear Reaction Rates for {$A=20-40$} Nuclei},
  journal =	 APJSS,
  year =	 2001,
  month =	 5,
  volume =	 134,
  number =	 1,
  pages =	 {151-171},
  doi =		 {10.1086/320364}
}

@ARTICLE{Iliadis2010,
  author =	 {{Iliadis}, C. and {Longland}, R. and {Champagne}, A.~E. and {Coc}, A. and {Fitzgerald}, R.},
  title =	 "{Charged-particle thermonuclear reaction rates: II. Tables and graphs of reaction rates and probability density functions}",
  journal =	 NuclPhysA,
  year =	 2010,
  month =	 10,
  volume =	 {841},
  number =	 {1-4},
  pages =	 {31-250},
  doi =		 {10.1016/j.nuclphysa.2010.04.009}
}

@ARTICLE{Imbriani2001,
  OPTauthor =	 {{Imbriani}, Gianluca and {Limongi}, Marco and {Gialanella}, Lucio and {Terrasi}, Filippo and {Straniero}, Oscar and {Chieffi}, Alessandro},
  author =	 {{Imbriani}, Gianluca and others},
  title =	 {The {$^{12}\mathrm{C}(\alpha,\gamma)^{16}\mathrm{O}$} Reaction Rate and the Evolution of Stars in the Mass Range {$0.8\leq M/M_{\odot}\leq25$}},
  journal =	 APJ,
  year =	 2001,
  month =	 9,
  volume =	 558,
  number =	 2,
  pages =	 {903-915},
  doi =		 {10.1086/322288}
}

@ARTICLE{Indelicato2017,
  OPTauthor =	 {{Indelicato}, I. and {La Cognata}, M. and {Spitaleri}, C. and {Burjan}, V. and {Cherubini}, S. and {Gulino}, M. and {Hayakawa}, S. and {Hons}, Z. and {Kroha}, V. and {Lamia}, L. and {Mazzocco}, M. and {Mrazek}, J. and {Pizzone}, R.~G. and {Romano}, S. and {Strano}, E. and {Torresi}, D. and {Tumino}, A.},
  author =	 {{Indelicato}, I. and others},
  title =	 {New Improved Indirect Measurement of the {$^{19}$F(p, {\ensuremath{\alpha}})$^{16}$O} Reaction at Energies of Astrophysical Relevance},
  journal =	 APJ,
  year =	 2017,
  month =	 8,
  volume =	 845,
  number =	 1,
  eid =		 19,
  pages =	 19,
  doi =		 {10.3847/1538-4357/aa7de7}
}

@article{Ishikawa14,
  author =	 {{Ishikawa}, S.},
  title =	 {Decay and structure of the {Hoyle} state},
  journal =	 PRC,
  year =	 2014,
  month =	 12,
  volume =	 90,
  number =	 6,
  eid =		 061604,
  pages =	 061604,
  doi =		 {10.1103/PhysRevC.90.061604}
}

@article{Itoh14,
  OPTauthor =	 {{Itoh}, M. and {Ando}, S. and {Aoki}, T. and {Arikawa}, H. and {Ezure}, S. and {Harada}, K. and {Hayamizu}, T. and {Inoue}, T. and {Ishikawa}, T. and {Kato}, K. and {Kawamura}, H. and {Sakemi}, Y. and {Uchiyama}, A.},
  author =	 {{Itoh}, M. and others},
  title =	 {Further Improvement of the Upper Limit on the Direct 3{\ensuremath{\alpha}} Decay from the {Hoyle} State in $^{12}\mathrm{C}$},
  journal =	 PRL,
  year =	 2014,
  month =	 9,
  volume =	 113,
  number =	 10,
  eid =		 102501,
  pages =	 102501,
  doi =		 {10.1103/PhysRevLett.113.102501}
}

@ARTICLE{Jaeger2001,
  author =	 {Jaeger, M. and Kunz, R. and Mayer, A. and Hammer, J. W. and Staudt, G. and Kratz, K. L. and Pfeiffer, B.},
  volume =	 87,
  journal =	 PRL,
  month =	 10,
  numpages =	 4,
  title =	 {$^{22}\mathrm{Ne}\an{}^{25}\mathrm{Mg}$: The Key Neutron Source in Massive Stars},
  year =	 2001,
  doi =		 {10.1103/PhysRevLett.87.202501},
  issue =	 20,
  pages =	 202501
}

@article{Jaszczak1970,
  title =	 {$^{12}\mathrm{C}(\ensuremath{\alpha},\ensuremath{\gamma})^{16}\mathrm{O}$ Capture Cross Section Below 3.2\,{MeV}},
  author =	 {Jaszczak, Ronald J. and Gibbons, John H. and Macklin, Richard L.},
  journal =	 PRC,
  volume =	 {2},
  issue =	 {1},
  pages =	 {63--69},
  numpages =	 {0},
  year =	 {1970},
  month =	 7,
  doi =		 {10.1103/PhysRevC.2.63}
}

@article{Jiang2007,
  title =	 {Expectations for $^{12}\mathrm{C}$ and $^{16}\mathrm{O}$ induced fusion cross sections at energies of astrophysical interest},
  author =	 {Jiang, C. L. and Rehm, K. E. and Back, B. B. and Janssens, R. V. F.},
  journal =	 PRC,
  volume =	 75,
  issue =	 1,
  pages =	 015803,
  numpages =	 11,
  year =	 2007,
  month =	 1,
  doi =		 {10.1103/PhysRevC.75.015803}
}

@article{Johnson1953,
  title =	 {Angular Aberrations in Sector Shaped Electromagnetic Lenses for Focusing Beams of Charged Particles},
  author =	 {Johnson, Edgar G. and Nier, Alfred O.},
  journal =	 PR,
  volume =	 91,
  issue =	 1,
  pages =	 {10--17},
  numpages =	 0,
  year =	 1953,
  month =	 7,
  doi =		 {10.1103/PhysRev.91.10}
}

@ARTICLE{Junker1998,
  author =	 {{Junker}, M. and others},
  OPTauthor =	 {{Junker}, M. and {D'alessandro}, A. and {Zavatarelli}, S. and {Arpesella}, C. and {Bellotti}, E. and {Broggini}, C. and {Corvisiero}, P. and {Fiorentini}, G. and {Fubini}, A. and {Gervino}, G. and {Greife}, U. and {Gustavino}, C. and {Lambert}, J. and {Prati}, P. and {Rodney}, W.~S. and {Rolfs}, C. and {Strieder}, F. and {Trautvetter}, H.~P. and {Zahnow}, D.},
  title =	 "{Cross section of $^{3}$He($^{3}$He,2p)$^{4}$He measured at solar energies}",
  journal =	 PRC,
  year =	 1998,
  month =	 5,
  volume =	 {57},
  number =	 {5},
  pages =	 {2700-2710},
  doi =		 {10.1103/PhysRevC.57.2700}
}

@ARTICLE{Kaeppeler1994,
  OPTauthor =	 {{Kaeppeler}, F. and {Wiescher}, M. and {Giesen}, U. and {Goerres}, J. and {Baraffe}, I. and {El Eid}, M. and {Raiteri}, C.~M. and {Busso}, M. and {Gallino}, R. and {Limongi}, M. and {Chieffi}, A.},
  author =	 {{Käppeler}, F. and others},
  title =	 {Reaction Rates for {$^{18}\mathrm{O}(\alpha,\gamma)^{22}\mathrm{Ne}$, $^{22}\mathrm{Ne}(\alpha,\gamma)^{26}\mathrm{Mg}$}, and {$^{22}\mathrm{Ne}\an^{25}\mathrm{Mg}$} in Stellar Helium Burning and $s$-Process Nucleosynthesis in Massive Stars},
  journal =	 APJ,
  year =	 1994,
  month =	 12,
  volume =	 437,
  pages =	 396,
  doi =		 {10.1086/175004}
}

@ARTICLE{Kaeppeler2011,
  author =	 {{K{\"a}ppeler}, F. and {Gallino}, R. and {Bisterzo}, S. and {Aoki}, Wako},
  title =	 {The {$s$} process: Nuclear physics, stellar models, and observations},
  journal =	 RMP,
  year =	 2011,
  month =	 1,
  volume =	 83,
  number =	 1,
  pages =	 {157-194},
  doi =		 {10.1103/RevModPhys.83.157}
}

@article{Kamimura81,
  author =	 {{Kamimura}, M.},
  title =	 {Transition densities between the {$0_{1}^{+}$, $2_{1}^{+}$, $4_{1}^{+}$, $0_{2}^{+}$, $2_{2}^{+}$, $1_{1}^{-}$} and {$3_{1}^{-}$} states in {$^{12}\mathrm{C}$} derived from the three-alpha resonating-group wave functions},
  journal =	 NuclPhysA,
  year =	 1981,
  month =	 1,
  volume =	 351,
  number =	 3,
  pages =	 {456-480},
  doi =		 {10.1016/0375-9474(81)90182-2}
}

@ARTICLE{Karakas2014,
  author =	 {{Karakas}, Amanda I. and {Lattanzio}, John C.},
  title =	 {The {Dawes} Review 2: Nucleosynthesis and Stellar Yields of Low- and Intermediate-Mass Single Stars},
  journal =	 pasa,
  year =	 2014,
  month =	 7,
  volume =	 31,
  eid =		 {e030},
  pages =	 {e030},
  doi =		 {10.1017/pasa.2014.21}
}

@article{Kibedi2020,
  OPTauthor =	 {{Kib{\'e}di}, T. and {Alshahrani}, B. and {Stuchbery}, A.~E. and {Larsen}, A.~C. and {G{\"o}rgen}, A. and {Siem}, S. and {Guttormsen}, M. and {Giacoppo}, F. and {Morales}, A.~I. and {Sahin}, E. and {Tveten}, G.~M. and {Garrote}, F.~L. Bello and {Campo}, L. Crespo and {Eriksen}, T.~K. and {Klintefjord}, M. and {Maharramova}, S. and {Nyhus}, H. -T. and {Tornyi}, T.~G. and {Renstr{\o}m}, T. and {Paulsen}, W.},
  author =	 {{Kib{\'e}di}, T. and others},
  title =	 {Radiative Width of the {Hoyle} State from {\ensuremath{\gamma}}-Ray Spectroscopy},
  journal =	 PRL,
  year =	 2020,
  month =	 10,
  volume =	 125,
  number =	 18,
  eid =		 182701,
  pages =	 182701,
  doi =		 {10.1103/PhysRevLett.125.182701}
}

@article{Kiener1991,
  title = {Measurements of the Coulomb dissociation cross section of {156\,MeV} $^{6}\mathrm{Li}$ projectiles at extremely low relative fragment energies of astrophysical interest},
  author = {Kiener, J. and others},
  OPTauthor = {Kiener, J. and Gils, H. J. and Rebel, H. and Zagromski, S. and Gsottschneider, G. and Heide, N. and Jelitto, H. and Wentz, J. and Baur, G.},
  journal = PRC,
  volume = 44,
  issue = 5,
  pages = {2195--2208},
  numpages = 0,
  year = 1991,
  month = 11,
  doi = {10.1103/PhysRevC.44.2195}
}

@article{Kirsebom12,
  OPTauthor =	 {{Kirsebom}, O.~S. and {Alcorta}, M. and {Borge}, M.~J.~G. and {Cubero}, M. and {Diget}, C. Aa. and {Fraile}, L.~M. and {Fulton}, B.~R. and {Fynbo}, H.~O.~U. and {Galaviz}, D. and {Jonson}, B. and {Madurga}, M. and {Nilsson}, T. and {Nyman}, G. and {Riisager}, K. and {Tengblad}, O. and {Turri{\'o}n}, M.},
  author =	 {{Kirsebom}, O.~S. and others},
  title =	 {Improved Limit on Direct {\ensuremath{\alpha}} Decay of the {Hoyle} State},
  journal =	 PRL,
  year =	 2012,
  month =	 5,
  volume =	 108,
  number =	 20,
  eid =		 202501,
  pages =	 202501,
  doi =		 {10.1103/PhysRevLett.108.202501}
}

@article{Kutschera2005,
  title =	 {Progress in isotope analysis at ultra-trace level by {AMS}},
  journal =	 IJMS,
  volume =	 242,
  number =	 2,
  pages =	 {145-160},
  year =	 2005,
  OPTnote =	 {Isotope Ratio Measurements SI},
  issn =	 {1387-3806},
  doi =		 {10.1016/j.ijms.2004.10.029},
  author =	 {Walter Kutschera}
}

@ARTICLE{Lacognata2013,
  author =	 {{La Cognata}, M. and others},
  OPTauthor =	 {{La Cognata}, M. and {Spitaleri}, C. and {Trippella}, O. and {Kiss}, G.~G. and {Rogachev}, G.~V. and {Mukhamedzhanov}, A.~M. and {Avila}, M. and {Guardo}, G.~L. and {Koshchiy}, E. and {Kuchera}, A. and {Lamia}, L. and {Puglia}, S.~M.~R. and {Romano}, S. and {Santiago}, D. and {Spart{\`a}}, R.},
  title =	 "{On the Measurement of the $^{13}$C({\ensuremath{\alpha}}, n)$^{16}$O S-factor at Negative Energies and its Influence on the s-process}",
  journal =	 APJ,
  year =	 2013,
  month =	 11,
  volume =	 {777},
  pages =	 {143},
  doi =		 {10.1088/0004-637X/777/2/143},
}

@ARTICLE{LaCognata2022,
  OPTauthor =	 {{La Cognata}, M. and {Palmerini}, S. and {Adsley}, P. and {Hammache}, F. and {Di Pietro}, A. and {Figuera}, P. and {Alba}, R. and {Cherubini}, S. and {Dell'Agli}, F. and {Guardo}, G.~L. and {Gulino}, M. and {Lamia}, L. and {Lattuada}, D. and {Maiolino}, C. and {Oliva}, A. and {Pizzone}, R.~G. and {Prajapati}, P.~M. and {Romano}, S. and {Santonocito}, D. and {Spart{\'a}}, R. and {Sergi}, M.~L. and {Tumino}, A.},
  author =	 {{La Cognata}, M. and others},
  title =	 {Exploring the astrophysical energy range of the {$^{27}\mathrm{Al}(\mathrm{p},\alpha){}^{24}\mathrm{Mg}$} reaction: A new recommended reaction rate},
  journal =	 PLB,
  year =	 2022,
  month =	 3,
  volume =	 826,
  eid =		 136917,
  pages =	 136917,
  doi =		 {10.1016/j.physletb.2022.136917}
}

@ARTICLE{Lamia2017,
  OPTauthor =	 {{Lamia}, L. and {Spitaleri}, C. and {Bertulani}, C.~A. and {Hou}, S.~Q. and {La Cognata}, M. and {Pizzone}, R.~G. and {Romano}, S. and {Sergi}, M.~L. and {Tumino}, A.},
  author =	 {{Lamia}, L. and others},
  title =	 {On the Determination of the {$^{7}\mathrm{Be}(\mathrm{n},\alpha)^{4}\mathrm{He}$} Reaction Cross Section at {BBN} Energies},
  journal =	 APJ,
  year =	 2017,
  month =	 12,
  volume =	 850,
  number =	 2,
  eid =		 175,
  pages =	 175,
  doi =		 {10.3847/1538-4357/aa965c}
}

@ARTICLE{Lamia2019,
  OPTauthor =	 {{Lamia}, L. and {Mazzocco}, M. and {Pizzone}, R.~G. and {Hayakawa}, S. and {La Cognata}, M. and {Spitaleri}, C. and {Bertulani}, C.~A. and {Boiano}, A. and {Boiano}, C. and {Broggini}, C. and {Caciolli}, A. and {Cherubini}, S. and {D'Agata}, G. and {da Silva}, H. and {Depalo}, R. and {Galtarossa}, F. and {Guardo}, G.~L. and {Gulino}, M. and {Indelicato}, I. and {La Commara}, M. and {La Rana}, G. and {Menegazzo}, R. and {Mrazek}, J. and {Pakou}, A. and {Parascandolo}, C. and {Piatti}, D. and {Pierroutsakou}, D. and {Puglia}, S.~M.~R. and {Romano}, S. and {Rapisarda}, G.~G. and {S{\'a}nchez-Ben{\'\i}tez}, A.~M. and {Sergi}, M.~L. and {Sgouros}, O. and {Soramel}, F. and {Soukeras}, V. and {Spart{\'a}}, R. and {Strano}, E. and {Torresi}, D. and {Tumino}, A. and {Yamaguchi}, H. and {Zhang}, G.~L.},
  author =	 {{Lamia}, L. and others},
  title =	 {Cross-section Measurement of the Cosmologically Relevant {$^{7}\mathrm{Be}(\mathrm{n},\alpha)^{4}\mathrm{He}$} Reaction over a Broad Energy Range in a Single Experiment},
  journal =	 APJ,
  year =	 2019,
  month =	 7,
  volume =	 879,
  number =	 1,
  eid =		 23,
  pages =	 23,
  doi =		 {10.3847/1538-4357/ab2234}
}

@ARTICLE{Laubenstein2020,
  author =	 {{Laubenstein}, Matthias and {Lawson}, Ian},
  title =	 "{Low Background Radiation Detection Techniques and Mitigation of Radioactive Backgrounds}",
  journal =	 FP,
  year =	 2020,
  month =	 11,
  volume =	 {8},
  eid =		 {506},
  pages =	 {506},
  doi =		 {10.3389/fphy.2020.577734}
}

@article{Leggett1980PTPS,
  author =	 {{Leggett}, A.~J.},
  title =	 {Macroscopic Quantum Systems and the Quantum Theory of Measurement},
  journal =	 PTPS,
  year =	 1980,
  month =	 1,
  volume =	 69,
  pages =	 {80-100},
  publisher =	 {Oxford Academic},
  doi =		 {10.1143/PTP.69.80}
}

@article{Li2016,
  title =	 {{Cross section measurement of $^{14}\mathrm{N}\pg^{15}\mathrm{O}$ in the CNO cycle}},
  OPTauthor =	 {Li, Q. and G{\"o}rres, J. and deBoer, R. J. and Imbriani, G. and Best, A. and Kontos, A. and LeBlanc, P. J. and Uberseder, E. and Wiescher, M.},
  author =	 {Li, Q. and others},
  journal =	 PRC,
  volume =	 93,
  issue =	 5,
  pages =	 055806,
  numpages =	 11,
  year =	 2016,
  month =	 5,
  doi =		 {10.1103/PhysRevC.93.055806}
}

@article{Li22,
  OPTauthor =	 {{Li}, K.~C.~W. and {Smit}, F.~D. and {Adsley}, P. and {Neveling}, R. and {Papka}, P. and {Nikolskii}, E. and {Br{\"u}mmer}, J.~W. and {Donaldson}, L.~M. and {Freer}, M. and {Harakeh}, M.~N. and {Nemulodi}, F. and {Pellegri}, L. and {Pesudo}, V. and {Wiedeking}, M. and {Buthelezi}, E.~Z. and {Chudoba}, V. and {F{\"o}rtsch}, S.~V. and {Jones}, P. and {Kamil}, M. and {Mira}, J.~P. and {O'Neill}, G.~G. and {Sideras-Haddad}, E. and {Singh}, B. and {Siem}, S. and {Steyn}, G.~F. and {Swartz}, J.~A. and {Usman}, I.~T. and {van Zyl}, J.~J.},
  author =	 {K.C.W. Li and others},
  title =	 {Investigating the predicted breathing-mode excitation of the {Hoyle} state},
  journal =	 PLB,
  year =	 2022,
  month =	 4,
  volume =	 827,
  eid =		 136928,
  pages =	 136928,
  doi =		 {10.1016/j.physletb.2022.136928}
}

@article{Limata2008,
  title =	 {{$^7\mathrm{Be}$} radioactive beam production at {CIRCE} and its utilization in basic and applied physics},
  journal =	 NIMB,
  volume =	 266,
  number =	 10,
  pages =	 {2117-2121},
  year =	 2008,
  OPTnote =	 {Accelerators in Applied Research and Technology},
  issn =	 {0168-583X},
  doi =		 {10.1016/j.nimb.2008.02.083},
  OPTauthor =	 {Benedicta Normanna Limata and Lucio Gialanella and Antonino {Di Leva} and Nicola De Cesare and Antonio D’Onofrio and G. Gyurky and Claus Rolfs and Mario Romano and Detlef Rogalla and Cesare Rossi and Michele Russo and Endre Somorjai and Filippo Terrasi},
  author =	 {Limata, Benedicta Normanna and others}
}

@ARTICLE{Limata2010,
  author =	 {{Limata}, B. N. and others},
  OPTauthor =	 {{Limata}, B. and {Strieder}, F. and {Formicola}, A. and {Imbriani}, G. and {Junker}, M. and {Becker}, H.~W. and {Bemmerer}, D. and {Best}, A. and {Bonetti}, R. and {Broggini}, C. and {Caciolli}, A. and {Corvisiero}, P. and {Costantini}, H. and {Di Leva}, A. and {Elekes}, Z. and {F{\"u}l{\"o}p}, Z. and {Gervino}, G. and {Guglielmetti}, A. and {Gustavino}, C. and {Gy{\"u}rky}, G. and {Lemut}, A. and {Marta}, M. and {Mazzocchi}, C. and {Menegazzo}, R. and {Prati}, P. and {Roca}, V. and {Rolfs}, C. and {Rossi Alvarez}, C. and {Salvo}, C. and {Somorjai}, E. and {Straniero}, O. and {Terrasi}, F. and {Trautvetter}, {H.-P.}},
  title =	 "{New experimental study of low-energy {$\pg$} resonances in magnesium isotopes}",
  journal =	 PRC,
  year =	 2010,
  month =	 7,
  volume =	 82,
  number =	 1,
  pages =	 {015801-+},
  doi =		 {10.1103/PhysRevC.82.015801}
}

@ARTICLE{Limongi2018,
  author =	 {{Limongi}, Marco and {Chieffi}, Alessandro},
  title =	 {Presupernova Evolution and Explosive Nucleosynthesis of Rotating Massive Stars in the Metallicity Range -3{\ensuremath{\leq}}{[Fe/H]} {\ensuremath{\leq}}0},
  journal =	 APJSS,
  year =	 2018,
  month =	 7,
  volume =	 237,
  number =	 1,
  eid =		 13,
  pages =	 13,
  doi =		 {10.3847/1538-4365/aacb24}
}

@ARTICLE{Lombardo2015,
  OPTauthor =	 {{Lombardo}, I. and {Dell'Aquila}, D. and {Di Leva}, A. and {Indelicato}, I. and {La Cognata}, M. and {La Commara}, M. and {Ordine}, A. and {Rigato}, V. and {Romoli}, M. and {Rosato}, E. and {Spadaccini}, G. and {Spitaleri}, C. and {Tumino}, A. and {Vigilante}, M.},
  author =	 {{Lombardo}, I. and others},
  title =	 {Toward a reassessment of the {${}^{19}\mathrm{F}(\mathrm{p},\alpha_{0}){}^{16}\mathrm{O}$} reaction rate at astrophysical temperatures},
  journal =	 PLB,
  year =	 2015,
  month =	 9,
  volume =	 748,
  pages =	 {178-182},
  doi =		 {10.1016/j.physletb.2015.06.073}
}

@article{Lombardo2019,
  title =	 {New analysis of $\mathrm{p}\!+\!{}^{19}\mathrm{F}$ reactions at low energies and the spectroscopy of natural-parity states in ${}^{20}\mathrm{Ne}$},
  OPTauthor =	 {Lombardo, Ivano and Dell'Aquila, Daniele and He, Jian-Jun and Spadaccini, Giulio and Vigilante, Mariano},
  author =	 {Lombardo, Ivano and others},
  journal =	 PRC,
  volume =	 100,
  issue =	 4,
  pages =	 044307,
  numpages =	 13,
  year =	 2019,
  month =	 10,
  doi =		 {10.1103/PhysRevC.100.044307}
}

@ARTICLE{Lugaro2017,
  OPTauthor =	 {{Lugaro}, M. and {Karakas}, A.~I. and {Bruno}, C.~G. and {Aliotta}, M. and {Nittler}, L.~R. and {Bemmerer}, D. and {Best}, A. and {Boeltzig}, A. and {Broggini}, C. and {Caciolli}, A. and {Cavanna}, F. and {Ciani}, G.~F. and {Corvisiero}, P. and {Davinson}, T. and {Depalo}, R. and {di Leva}, A. and {Elekes}, Z. and {Ferraro}, F. and {Formicola}, A. and {F{\"u}l{\"o}p}, Zs. and {Gervino}, G. and {Guglielmetti}, A. and {Gustavino}, C. and {Gy{\"u}rky}, Gy. and {Imbriani}, G. and {Junker}, M. and {Menegazzo}, R. and {Mossa}, V. and {Pantaleo}, F.~R. and {Piatti}, D. and {Prati}, P. and {Scott}, D.~A. and {Straniero}, O. and {Strieder}, F. and {Sz{\"u}cs}, T. and {Tak{\'a}cs}, M.~P. and {Trezzi}, D.},
  author =	 {{Lugaro}, M. and others},
  title =	 {Origin of meteoritic stardust unveiled by a revised proton-capture rate of {$^{17}\mathrm{O}$}},
  journal =	 NatureAstro,
  year =	 2017,
  month =	 1,
  volume =	 1,
  eid =		 0027,
  pages =	 0027,
  doi =		 {10.1038/s41550-016-0027}
}

@article{LulliInPreparation,
  author =	 {{Lulli}, Matteo and {Marciano}, Antonino and {Piscicchia}, Kristian},
  title =	 "{Stochastic Ricci Flow dynamics of the gravitationally induced wave-function collapse}",
  journal =	 {arXiv e-prints},
  year =	 2023,
  month =	 7,
  OPTeid =	 {arXiv:2307.10136},
  pages =	 {arXiv:2307.10136},
  doi =		 {10.48550/arXiv.2307.10136},
  OPTarchivePrefix ={arXiv},
  OPTeprint =	 {2307.10136},
  OPTprimaryClass ={gr-qc}
}

@article{Mak1975,
  OPTauthor =	 {{Mak}, H. -B. and {Evans}, H.~C. and {Ewan}, G.~T. and {McDonald}, A.~B. and {Alexander}, T.~K.},
  author =	 {{Mak}, H. -B. and others},
  title =	 {Radiative decay of the second excited state of {$^{12}\mathrm{C}$}},
  journal =	 PRC,
  year =	 1975,
  month =	 10,
  volume =	 12,
  number =	 4,
  pages =	 {1158-1166},
  doi =		 {10.1103/PhysRevC.12.1158}
}

@article{Makii2005,
  title =	 {Measurement system of the {$\gamma$}-ray angular distributions of the reactions of the {${}^{12}\mathrm{C}(\alpha,\gamma){}^{16}\mathrm{O}$} reaction},
  journal =	 NIMA,
  volume =	 547,
  number =	 "2-3",
  pages =	 "411 - 423",
  year =	 2005,
  issn =	 "0168-9002",
  doi =		 "10.1016/j.nima.2005.03.164",
  OPTauthor =	 "H. Makii and K. Mishima and M. Segawa and E. Sano and H. Ueda and T. Shima and Y. Nagai and M. Igashira and T. Ohsaki",
  author =	 "H. Makii and others"
}

@article{Makii2009,
  OPTauthor =	 {Makii, H. and Nagai, Y. and Shima, T. and Segawa, M. and Mishima, K. and Ueda, H. and Igashira, M. and Ohsaki, T.},
  Author =	 {Makii, H. and others},
  Doi =		 {10.1103/PhysRevC.80.065802},
  Issue =	 6,
  Journal =	 PRC,
  Month =	 12,
  Numpages =	 16,
  Pages =	 065802,
  Title =	 {{$E1$} and {$E2$} cross sections of the {${}^{12}\mathrm{C}(\alpha,\gamma){}^{16}\mathrm{O}$} reaction using pulsed $\alpha$ beams},
  Volume =	 80,
  Year =	 2009
}

@article{Manfredi12,
  OPTauthor =	 {{Manfredi}, J. and {Charity}, R.~J. and {Mercurio}, K. and {Shane}, R. and {Sobotka}, L.~G. and {Wuosmaa}, A.~H. and {Banu}, A. and {Trache}, L. and {Tribble}, R.~E.},
  author =	 {{Manfredi}, J. and others},
  title =	 {{\ensuremath{\alpha}} decay of the excited states in {$^{12}$C} at 7.65 and 9.64\,{MeV}},
  journal =	 PRC,
  year =	 2012,
  month =	 3,
  volume =	 85,
  number =	 3,
  eid =		 037603,
  pages =	 037603,
  doi =		 {10.1103/PhysRevC.85.037603}
}

@article{Marcucci2005,
  OPTauthor =	 {Marcucci, L. E. and Viviani, M. and Schiavilla, R. and Kievsky, A. and Rosati, S.},
  author =	 {Marcucci, L. E. and others},
  title =	 {Electromagnetic structure of {$A=2$} and {$3$} nuclei and the nuclear current operator},
  doi =		 "10.1103/PhysRevC.72.014001",
  journal =	 PRC,
  volume =	 72,
  pages =	 014001,
  year =	 2005
}

@article{Marcucci2016,
  author =	 {Marcucci, L. E. and Mangano, G. and Kievsky, A. and Viviani, M.},
  title =	 {Implication of the proton-deuteron radiative capture for {Big Bang Nucleosynthesis}},
  doi =		 "10.1103/PhysRevLett.116.102501",
  journal =	 PRL,
  volume =	 116,
  number =	 10,
  pages =	 102501,
  year =	 2016,
  note =	 "[Erratum: Phys. Rev. Lett. {\bf 117}, 049901 (2016)]"
}

@article{Marin-Lambarri14,
       OPTauthor = {{Mar{\'\i}n-L{\'a}mbarri}, D.~J. and {Bijker}, R. and {Freer}, M. and {Gai}, M. and {Kokalova}, Tz. and {Parker}, D.~J. and {Wheldon}, C.},
       author = {{Mar{\'\i}n-L{\'a}mbarri}, D.~J. and others},
        title = {Evidence for Triangular {D3h} Symmetry in {$^{12}\mathrm{C}$}},
      journal = PRL,
         year = 2014,
        month = 7,
       volume = {113},
       number = {1},
          eid = {012502},
        pages = {012502},
          doi = {10.1103/PhysRevLett.113.012502}
}

@article{Marta2010,
  OPTauthor =	 {{Marta}, M. and {Trompler}, E. and {Bemmerer}, D. and {Beyer}, R. and {Broggini}, C. and {Caciolli}, A. and {Erhard}, M. and {F{\"u}l{\"o}p}, Z. and {Grosse}, E. and {Gy{\"u}rky}, G. and {Hannaske}, R. and {Junghans}, A.~R. and {Menegazzo}, R. and {Nair}, C. and {Schwengner}, R. and {Sz{\"u}cs}, T. and {Vezz{\'u}}, S. and {Wagner}, A. and {Yakorev}, D.},
  author =	 {{Marta}, M. and others},
  doi =		 {10.1103/PhysRevC.81.055807},
  journal =	 PRC,
  pages =	 055807,
  title =	 {Resonance strengths in the {$^{14}\mathrm{N}\pg^{15}\mathrm{O}$} and {$^{15}\mathrm{N}\pag^{12}\mathrm{C}$} reactions},
  volume =	 81,
  year =	 2010
}

@article{Marzaioli2011a,
  title =	 {Forensic applications of $^{14}\mathrm{C}$ at {CIRCE}},
  journal =	 NIMB,
  volume =	 269,
  number =	 24,
  pages =	 {3171-3175},
  year =	 2011,
  OPTnote =	 {Proceedings of the 10th European Conference on Accelerators in Applied Research and Technology (ECAART10)},
  issn =	 {0168-583X},
  doi =		 {10.1016/j.nimb.2011.04.025},
  OPTauthor =	 {F. Marzaioli and V. Fiumano and M. Capano and I. Passariello and N. {De Cesare} and F. Terrasi},
  author =	 {F. Marzaioli and others}
}

@article{Marzaioli2011b,
  OPTauthor =	 {Marzaioli, Fabio and Lubritto, Carmine and Nonni, Sara and Passariello, Isabella and Capano, Manuela and Terrasi, Filippo},
  author =	 {Marzaioli, Fabio and others},
  title =	 {Mortar Radiocarbon Dating: Preliminary Accuracy Evaluation of a Novel Methodology},
  journal =	 AC,
  volume =	 83,
  number =	 6,
  pages =	 {2038-2045},
  year =	 2011,
  doi =		 {10.1021/ac1027462}
}

@ARTICLE{Matei2006,
  OPTauthor =	 {{Matei}, C. and {Buchmann}, L. and {Hannes}, W.~R. and {Hutcheon}, D.~A. and {Ruiz}, C. and {Brune}, C.~R. and {Caggiano}, J. and {Chen}, A.~A. and {D'Auria}, J. and {Laird}, A. and {Lamey}, M. and {Li}, Zh. and {Liu}, Wp. and {Olin}, A. and {Ottewell}, D. and {Pearson}, J. and {Ruprecht}, G. and {Trinczek}, M. and {Vockenhuber}, C. and {Wrede}, C.},
  author =	 {{Matei}, C. and others},
  title =	 {Measurement of the Cascade Transition via the First Excited State of {$^{16}\mathrm{O}$} in the $\cag$ Reaction, and Its {$S$} Factor in Stellar Helium Burning},
  journal =	 PRL,
  year =	 2006,
  month =	 12,
  volume =	 97,
  number =	 24,
  eid =		 242503,
  pages =	 242503,
  doi =		 {10.1103/PhysRevLett.97.242503}
}

@inproceedings{Mavromatos2018EPJ,
       author = {{Mavromatos}, Nick E.},
        title = {Models \& Searches of {CPT} Violation: a personal, very partial, list},
      journal = EPJWC,
         year = 2018,
       series = EPJWC,
       volume = {166},
        month = 1,
          eid = {00005},
        pages = {00005},
          doi = {10.1051/epjconf/201816600005}
}

@article{Messiah1964PR,
  doi =		 {10.1103/physrev.136.b248},
  year =	 1964,
  month =	 10,
  volume =	 136,
  number =	 {1B},
  pages =	 {B248--B267},
  author =	 {A. M. L. Messiah and O. W. Greenberg},
  title =	 {Symmetrization Postulate and Its Experimental Foundation},
  journal =	 PR
}

@misc{Milotti2007AX,
  doi =		 {10.48550/ARXIV.0705.1363},
  author =	 {Milotti, Edoardo},
  title =	 {Enrico Fermi's view of identical particles},
  publisher =	 {arXiv},
  year =	 2007,
  copyright =	 {Assumed arXiv.org perpetual, non-exclusive license to distribute this article for submissions made before January 2004}
}

@article{Milotti2018EN,
  doi =		 {10.3390/e20070515},
  year =	 2018,
  month =	 7,
  volume =	 20,
  number =	 7,
  pages =	 515,
  OPTauthor =	 {Edoardo Milotti and Sergio Bartalucci and Sergio Bertolucci and Massimiliano Bazzi and Mario Bragadireanu and Michael Cargnelli and Alberto Clozza and Catalina Curceanu and Luca De Paolis and Jean-Pierre Egger and Carlo Guaraldo and Mihail Iliescu and Matthias Laubenstein and Johann Marton and Marco Miliucci and Andreas Pichler and Dorel Pietreanu and Kristian Piscicchia and Alessandro Scordo and Hexi Shi and Diana Sirghi and Florin Sirghi and Laura Sperandio and Oton V{\'{a}}zquez Doce and Eberhard Widmann and Johann Zmeskal},
  author =	 {Edoardo Milotti and others},
  title =	 {On the Importance of Electron Diffusion in a Bulk-Matter Test of the {Pauli} Exclusion Principle},
  journal =	 {Entropy}
}

@article{Milotti2020SY,
  doi =		 {10.3390/sym13010006},
  year =	 2020,
  month =	 12,
  volume =	 13,
  number =	 1,
  pages =	 6,
  OPTauthor =	 {Edoardo Milotti and Sergio Bartalucci and Sergio Bertolucci and Massimiliano Bazzi and Mario Bragadireanu and Michael Cargnelli and Alberto Clozza and Catalina Curceanu and Luca De Paolis and Raffaele Del Grande and Carlo Guaraldo and Mihail Iliescu and Matthias Laubenstein and Johann Marton and Marco Miliucci and Fabrizio Napolitano and Kristian Piscicchia and Alessandro Scordo and Hexi Shi and Diana Laura Sirghi and Florin Sirghi and Laura Sperandio and Oton V{\'{a}}zquez Doce and Johann Zmeskal},
  author =	 {Edoardo Milotti and others},
  title =	 {Semi-Analytical Monte Carlo Method to Simulate the Signal of the {VIP}-2 Experiment},
  journal =	 {Symmetry}
}

@article{Miyake2012,
  author =	 {Miyake, Fusa and Nagaya, Kentaro and Masuda, Kimiaki and Nakamura, Toshio},
  date =	 {2012/06/01},
  doi =		 {10.1038/nature11123},
  journal =	 Nature,
  number =	 7402,
  pages =	 {240--242},
  title =	 {A signature of cosmic-ray increase in ad 774--775 from tree rings in {Japan}},
  volume =	 486,
  year =	 2012
}

@article{Miyake2013,
  author =	 {Miyake, Fusa and Masuda, Kimiaki and Nakamura, Toshio},
  date =	 {2013/04/23},
  doi =		 {10.1038/ncomms2783},
  isbn =	 {2041-1723},
  journal =	 NatureCom,
  number =	 1,
  pages =	 1748,
  title =	 {Another rapid event in the carbon-14 content of tree rings},
  volume =	 4,
  year =	 2013
}

@phdthesis{Morales-Gallegos2017,
  author =	 {Morales-Gallegos, Elia Lizeth},
  school =	 {The University of Edinburgh},
  title =	 {Carbon burning in stars: An experimental study of the {$^{12}\mathrm{C}(^{12}\mathrm{C,p})^{23}\mathrm{Na}$} reaction towards astrophysical energies},
  year =	 2017,
  url =		 {http://hdl.handle.net/1842/28967}
}

@ARTICLE{Morales2018,
  OPTauthor =	 {{Morales-Gallegos}, L. and {Aliotta}, M. and {Bruno}, C.~G. and {Buompane}, R. and {Davinson}, T. and {De Cesare}, M. and {Di Leva}, A. and {D'Onofrio}, A. and {Duarte}, J.~G. and {Gasques}, L.~R. and {Gialanella}, L. and {Imbriani}, G. and {Porzio}, G. and {Rapagnani}, D. and {Romoli}, M. and {Sch{\"u}rmann}, D. and {Terrasi}, F. and {Zhang}, L.~Y.},
  author =	 {{Morales-Gallegos}, L. and others},
  title =	 {{Reduction of deuterium content in carbon targets for ${}^{12}\mathrm{C} + {}^{12}\mathrm{C}$ reaction studies of astrophysical interest}},
  journal =	 EPJA,
  year =	 2018,
  month =	 8,
  volume =	 54,
  number =	 8,
  eid =		 132,
  pages =	 132,
  doi =		 {10.1140/epja/i2018-12564-8}
}

@article{Morinaga1956,
  title =	 {Interpretation of Some of the Excited States of {$4n$} Self-Conjugate Nuclei},
  author =	 {Morinaga, H.},
  journal =	 PR,
  volume =	 101,
  issue =	 1,
  pages =	 {254--258},
  numpages =	 0,
  year =	 1956,
  month =	 1,
  doi =		 {10.1103/PhysRev.101.254}
}

@article{Mossa2020,
  OPTauthor =	 {Mossa, V. and St\"ockel, K. and Cavanna, F. and Ferraro, F. and Aliotta, M. and Barile, F. and Bemmerer, D. and Best, A. and Boeltzig, A. and Broggini, C. and Bruno, C. G. and Caciolli, A. and Chillery, T. and Ciani, G. F. and Corvisiero, P. and Csedreki, L. and Davinson, T. and Depalo, R. and Di Leva, A. and Elekes, Z. and Fiore, E. M. and Formicola, A. and F\"ul\"op, Zs. and Gervino, G. and Guglielmetti, A. and Gustavino, C. and Gyurky, G. and Imbriani, G. and Junker, M. and Kievsky, A. and Kochanek, I. and Lugaro, M. and Marcucci, L. E. and Mangano, G. and Marigo, P. and Masha, E. and Menegazzo, R. and Pantaleo, F. R. and Paticchio, V. and Perrino, R. and Piatti, D. and Pisanti, O.  and Prati, P. and Schiavulli, L. and Straniero, O. and Sz\"ucs, T. and Tak\'acs, M. P. and Trezzi, D. and Viviani, M. and Zavatarelli, S.},
  Author =	 {Mossa, V. and others},
  Title =	 {The baryon density of the Universe from an improved rate of deuterium burning},
  Journal =	 Nature,
  Year =	 2020,
  Volume =	 587,
  Pages =	 {210+},
  DOI =		 {10.1038/s41586-020-2878-4}
}

@ARTICLE{Mossa2020-EPJA,
       author = {{Mossa}, V. and others},
       OPTauthor = {{Mossa}, V. and {St{\"o}ckel}, K. and {Cavanna}, F. and {Ferraro}, F. and {Aliotta}, M. and {Barile}, F. and {Bemmerer}, D. and {Best}, A. and {Boeltzig}, A. and {Broggini}, C. and {Bruno}, C.~G. and {Caciolli}, A. and {Csedreki}, L. and {Chillery}, T. and {Ciani}, G.~F. and {Corvisiero}, P. and {Davinson}, T. and {Depalo}, R. and {Di Leva}, A. and {Elekes}, Z. and {Fiore}, E.~M. and {Formicola}, A. and {F{\"u}l{\"o}p}, Zs. and {Gervino}, G. and {Guglielmetti}, A. and {Gustavino}, C. and {Gy{\"u}rky}, G. and {Imbriani}, G. and {Junker}, M. and {Kochanek}, I. and {Lugaro}, M. and {Marcucci}, L.~E. and {Marigo}, P. and {Masha}, E. and {Menegazzo}, R. and {Pantaleo}, F.~R. and {Paticchio}, V. and {Perrino}, R. and {Piatti}, D. and {Prati}, P. and {Schiavulli}, L. and {Straniero}, O. and {Sz{\"u}cs}, T. and {Tak{\'a}cs}, M.~P. and {Trezzi}, D. and {Zavatarelli}, S. and {Zorzi}, G.},
        title = {Setup commissioning for an improved measurement of the {$\mathrm{D}(\mathrm{p},\gamma){}^{3}\mathrm{He}$} cross section at {Big Bang Nucleosynthesis} energies},
      journal = EPJA,
         year = 2020,
       volume = {56},
          eid = {144},
        pages = {144},
          doi = {10.1140/epja/s10050-020-00149-1}
}

@article{Mugnaioli2022,
	author = {Mugnaioli, E. and others},
	OPTauthor = {Mugnaioli, Enrico and Zucchini, Azzurra and Comodi, Paola and Frondini, Francesco and Bartolucci, Luca and Di Michele, Alessandro and Sassi, Paola and Gemmi, Mauro},
	journal = {Mineral. Mag.},
	number = {2},
	pages = {272--281},
	title = {{3D} electron diffraction study of terrestrial iron oxide alteration in the {Mineo} pallasite},
	volume = {86},
	year = {2022},
doi = {10.1180/mgm.2022.20}
 }

@article{Muller1977,
  author =	 {Richard A. Muller },
  title =	 {Radioisotope Dating with a Cyclotron},
  journal =	 Science,
  volume =	 196,
  number =	 4289,
  pages =	 {489-494},
  year =	 1977,
  doi =		 {10.1126/science.196.4289.489}
}

@article{Napolitano2022SY,
  doi =		 {10.3390/sym14050893},
  year =	 2022,
  month =	 4,
  volume =	 14,
  number =	 5,
  pages =	 893,
  OPTauthor =	 {Fabrizio Napolitano and Sergio Bartalucci and Sergio Bertolucci and Massimiliano Bazzi and Mario Bragadireanu and Cesidio Capoccia and Michael Cargnelli and Alberto Clozza and Luca De Paolis and Raffaele Del Grande and Carlo Fiorini and Carlo Guaraldo and Mihail Iliescu and Matthias Laubenstein and Johann Marton and Marco Miliucci and Edoardo Milotti and Federico Nola and Kristian Piscicchia and Alessio Porcelli and Alessandro Scordo and Francesco Sgaramella and Hexi Shi and Diana Laura Sirghi and Florin Sirghi and Oton Vazquez Doce and Johann Zmeskal and Catalina Curceanu},
  author =	 {Fabrizio Napolitano and others},
  title =	 {Testing the {Pauli} Exclusion Principle with the {VIP}-2 Experiment},
  journal =	 {Symmetry}
}

@article{Nelson1966PR,
  title =	 {Derivation of the {Schr\"odinger} Equation from Newtonian Mechanics},
  author =	 {Nelson, Edward},
  journal =	 PR,
  volume =	 150,
  issue =	 4,
  pages =	 {1079--1085},
  numpages =	 0,
  year =	 1966,
  month =	 10,
  doi =		 {10.1103/PhysRev.150.1079},
}

@book{Nelson1967PUP,
  year =	 1967,
  publisher =	 {Princeton University Press},
  address =	 {US},
  author =	 {Edward Nelson},
  editor =	 {Department of Mathematics Princeton University},
  title =	 {Dynamical Theories of Brownian Motion},
  url =		 {https://web.math.princeton.edu/~nelson/books/bmotion.pdf}
}

@article{Okun1987JL,
       author = {{Okun'}, L.~B.},
        title = {Possible violation of the {Pauli} principle in atoms},
      journal = JETPL,
         year = 1987,
        month = 12,
       volume = 46,
        pages = 529,
          url = {http://www.jetpletters.ru/ps/1234/article_18632.shtml}
}

@article{Pearle1989PRA,
  doi =		 {10.1103/physreva.39.2277},
  year =	 1989,
  month =	 3,
  volume =	 39,
  number =	 5,
  pages =	 {2277--2289},
  author =	 {Philip Pearle},
  title =	 {Combining stochastic dynamical state-vector reduction with spontaneous localization},
  journal =	 PRA
}

@article{Penrose1996SPR,
       author = {{Penrose}, Roger},
        title = {On Gravity's role in Quantum State Reduction},
      journal = GRG,
         year = 1996,
        month = 5,
       volume = {28},
       number = {5},
        pages = {581-600},
          doi = {10.1007/BF02105068}
}

@article{Penrose2014SPR,
       author = {{Penrose}, Roger},
        title = {On the Gravitization of Quantum Mechanics 1: {Quantum} State Reduction},
      journal =	 FOUNDP,
         year = 2014,
        month = 5,
       volume = {44},
       number = {5},
        pages = {557-575},
          doi = {10.1007/s10701-013-9770-0}
}

@Article{Petraglia2022,
  OPTauthor =	 {Petraglia, Antonio and Sirignano, Carmina and Marzaioli, Fabio and Sabbarese, Carlo and D’Onofrio, Antonio and Porzio, Giuseppe and Buompane, Raffaele and Roca, Vincenzo and Stellato, Luisa and Esposito, Alfonso Maria and Mazziotta, Pietro and Terrasi, Filippo},
  AUTHOR =	 {Petraglia, Antonio and others},
  TITLE =	 {{Ultrasensitive Radionuclide Analysis in Water and Sediments for Environmental Radiological Assessment near the Decommissioning Garigliano Nuclear Power Plant (Italy)}},
  JOURNAL =	 APPSC,
  VOLUME =	 12,
  YEAR =	 2022,
  NUMBER =	 16,
  ARTICLE-NUMBER =8033,
  ISSN =	 {2076-3417},
  DOI =		 {10.3390/app12168033}
}

@ARTICLE{Piersanti2022,
  OPTauthor =	 {{Piersanti}, Luciano and {Bravo}, Eduardo and {Straniero}, Oscar and {Cristallo}, Sergio and {Dom{\'\i}nguez}, Inmaculada},
  author =	 {{Piersanti}, Luciano and others},
  title =	 {Pre-explosive Accretion and Simmering Phases of {SNe Ia}},
  journal =	 APJ,
  year =	 2022,
  month =	 2,
  volume =	 926,
  number =	 1,
  eid =		 103,
  pages =	 103,
  doi =		 {10.3847/1538-4357/ac403b}
}

@ARTICLE{Pignatari2008,
  OPTauthor =	 {{Pignatari}, M. and {Gallino}, R. and {Meynet}, G. and {Hirschi}, R. and {Herwig}, F. and {Wiescher}, M.},
  author =	 {{Pignatari}, M. and others},
  title =	 {The $s$-Process in Massive Stars at Low Metallicity: The Effect of Primary {$^{14}\mathrm{N}$} from Fast Rotating Stars},
  journal =	 APJ,
  year =	 2008,
  month =	 11,
  volume =	 687,
  pages =	 {L95-L98},
  doi =		 {10.1086/593350}
}

@ARTICLE{Pignatari2010,
  OPTauthor =	 {{Pignatari}, M. and {Gallino}, R. and {Heil}, M. and {Wiescher}, M. and {K{\"a}ppeler}, F. and {Herwig}, F. and {Bisterzo}, S.},
  author =	 {{Pignatari}, M. and others},
  title =	 {The Weak {$s$}-Process in Massive Stars and its Dependence on the Neutron Capture Cross Sections},
  journal =	 APJ,
  year =	 2010,
  month =	 2,
  volume =	 710,
  number =	 2,
  pages =	 {1557-1577},
  doi =		 {10.1088/0004-637X/710/2/1557}
}

@article{Piscicchia2017E,
  doi =		 {10.3390/e19070319},
  year =	 2017,
  month =	 6,
  volume =	 19,
  number =	 7,
  pages =	 319,
  OPTauthor =	 {Kristian Piscicchia and Angelo Bassi and Catalina Curceanu and Raffaele Grande and Sandro Donadi and Beatrix Hiesmayr and Andreas Pichler},
  author =	 {Kristian Piscicchia and others},
  title =	 {{CSL} Collapse Model Mapped with the Spontaneous Radiation},
  journal =	 {Entropy}
}

@article{Piscicchia2020EN,
  doi =		 {10.3390/e22111195},
  year =	 2020,
  month =	 10,
  volume =	 22,
  number =	 11,
  pages =	 1195,
  OPTauthor =	 {Kristian Piscicchia and Johann Marton and Sergio Bartalucci and Massimiliano Bazzi and Sergio Bertolucci and Mario Bragadireanu and Michael Cargnelli and Alberto Clozza and Raffaele Del Grande and Luca De Paolis and Carlo Fiorini and Carlo Guaraldo and Mihail Iliescu and Matthias Laubenstein and Marco Miliucci and Edoardo Milotti and Fabrizio Napolitano and Andreas Pichler and Alessandro Scordo and Hexi Shi and Diana Laura Sirghi and Florin Sirghi and Laura Sperandio and Oton Vazquez Doce and Johann Zmeskal and Catalina Curceanu},
  author =	 {Kristian Piscicchia and others},
  title =	 {{VIP}-2 {\textemdash}High-Sensitivity Tests on the {Pauli} Exclusion Principle for Electrons},
  journal =	 {Entropy}
}

@article{Piscicchia2020EPJC,
  doi =		 {10.1140/epjc/s10052-020-8040-5},
  year =	 2020,
  month =	 6,
  volume =	 80,
  number =	 6,
  OPTauthor =	 {Kristian Piscicchia and Edoardo Milotti and Aidin Amirkhani and Sergio Bartalucci and Sergio Bertolucci and Massimiliano Bazzi and Mario Bragadireanu and Michael Cargnelli and Alberto Clozza and Raffaele Del Grande and Luca De~Paolis and Jean-Pierre Egger and Carlo Fiorini and Carlo Guaraldo and Mihail Iliescu and Matthias Laubenstein and Johann Marton and Marco Miliucci and Andreas Pichler and Dorel Pietreanu and Alessandro Scordo and Hexi Shi and Diana Laura Sirghi and Florin Sirghi and Laura Sperandio and Oton Vazquez~Doce and Johann Zmeskal and Catalina Curceanu},
  author =	 {Kristian Piscicchia and others},
  title =	 {Search for a remnant violation of the {Pauli} exclusion principle in a Roman lead target},
  journal =	 EPJC
}

@article{Piscicchia2022APPA,
       OPTauthor = {{Piscicchia}, K. and {Bartalucci}, S. and {Bertolucci}, S. and {Bazzi}, M. and {Borghi}, G. and {Bragadireanu}, M. and {Capoccia}, C. and {Cargnelli}, M. and {Clozza}, A. and {Del Grande}, R. and {De Paolis}, L. and {Fiorini}, C. and {Guaraldo}, C. and {Iliescu}, M. and {Laubenstein}, M. and {Marton}, J. and {Miliucci}, M. and {Milotti}, E. and {Napolitano}, F. and {Porcelli}, A. and {Scordo}, A. and {Shi}, H. and {Sirghi}, D.~L. and {Sirghi}, F. and {Sgaramella}, F. and {Zorzi}, N. and {Zmeskal}, J. and {Curceanu}, C.},
       author =	 {Kristian Piscicchia and others},
        title = {High Sensitivity {Pauli Exclusion Principle} Tests by the {VIP} Experiment: Status and Perspectives},
      journal =	 APPA,
         year = 2022,
        month = 9,
       volume = {142},
       number = {3},
        pages = {361-366},
          doi = {10.12693/APhysPolA.142.361}
}

@article{Piscicchia2022PRL,
  doi =		 {10.1103/physrevlett.129.131301},
  year =	 2022,
  month =	 9,
  volume =	 129,
  number =	 13,
  OPTauthor =	 {Kristian Piscicchia and Andrea Addazi and Antonino Marcian{\`{o}} and Massimiliano Bazzi and Michael Cargnelli and Alberto Clozza and Luca De Paolis and Raffaele Del Grande and Carlo Guaraldo and Mihail Antoniu Iliescu and Matthias Laubenstein and Johann Marton and Marco Miliucci and Fabrizio Napolitano and Alessio Porcelli and Alessandro Scordo and Diana Laura Sirghi and Florin Sirghi and Oton Vazquez Doce and Johann Zmeskal and Catalina Curceanu},
  author =	 {Kristian Piscicchia and others},
  title =	 {Strongest Atomic Physics Bounds on Noncommutative Quantum Gravity Models},
  journal =	 PRL
}

@article{Piscicchia2023PRD,
  doi =		 {10.1103/physrevd.107.026002},
  year =	 2023,
  month =	 1,
  volume =	 107,
  number =	 2,
  OPTauthor =	 {Kristian Piscicchia and Andrea Addazi and Antonino Marcian{\`{o}} and Massimiliano Bazzi and Michael Cargnelli and Alberto Clozza and Luca De Paolis and Raffaele Del Grande and Carlo Guaraldo and Mihail Antoniu Iliescu and Matthias Laubenstein and Johann Marton and Marco Miliucci and Fabrizio Napolitano and Alessio Porcelli and Alessandro Scordo and Diana Laura Sirghi and Florin Sirghi and Oton Vazquez Doce and Johann Zmeskal and Catalina Curceanu},
  author =	 {Kristian Piscicchia and others},
  title =	 {Experimental test of noncommutative quantum gravity by {VIP}-2 Lead},
  journal =	 PRD
}

@article{Pitrou2018,
  title =	 {{Precision big bang nucleosynthesis with improved Helium-4 predictions}},
  journal =	 PRep,
  volume =	 754,
  pages =	 {1-66},
  year =	 2018,
  issn =	 {0370-1573},
  doi =		 {10.1016/j.physrep.2018.04.005},
  author =	 {Cyril Pitrou and Alain Coc and Jean-Philippe Uzan and Elisabeth Vangioni}
}

@article{Plag2012,
  OPTauthor =	 {Plag, R. and Reifarth, R. and Heil, M. and K\"appeler, F. and Rupp, G. and Voss, F. and Wisshak, K.},
  Author =	 {Plag, R. and others},
  Doi =		 {10.1103/PhysRevC.86.015805},
  Issue =	 1,
  Journal =	 PRC,
  Month =	 7,
  Numpages =	 9,
  Pages =	 015805,
  Title =	 {{$^{12}\mathrm{C}(\alpha,\gamma){}^{16}\mathrm{O}$} studied with the {Karlsruhe} {4$\pi$ BaF$_2$} detector},
  Volume =	 86,
  Year =	 2012
}

@ARTICLE{Prada2002,
  author =	 {{Prada Moroni}, Pier Giorgio and {Straniero}, Oscar},
  title =	 {Calibration of White Dwarf Cooling Sequences: Theoretical Uncertainty},
  journal =	 APJ,
  year =	 2002,
  month =	 12,
  volume =	 581,
  number =	 1,
  pages =	 {585-597},
  doi =		 {10.1086/344052}
}

@ARTICLE{Prantzos1990,
  author =	 {{Prantzos}, N. and {Hashimoto}, M. and {Nomoto}, K.},
  title =	 {The {$s$}-process in massive stars: yields as a function of stellar mass and metallicity.},
  journal =	 AAP,
  year =	 1990,
  month =	 8,
  volume =	 234,
  pages =	 211
}

@ARTICLE{Prantzos2020,
  OPTauthor =	 {{Prantzos}, N. and {Abia}, C. and {Cristallo}, S. and {Limongi}, M. and {Chieffi}, A.},
  author =	 {{Prantzos}, N. and others},
  title =	 {Chemical evolution with rotating massive star yields {II. A} new assessment of the solar {$s$}- and {$r$}-process components},
  journal =	 MNRAS,
  year =	 2020,
  month =	 1,
  volume =	 491,
  number =	 2,
  pages =	 {1832-1850},
  doi =		 {10.1093/mnras/stz3154}
}

@article{Psaltis2022a,
  title = {Direct Measurement of Resonances in $^{7}\mathrm{Be}(\ensuremath{\alpha},\ensuremath{\gamma})^{11}\mathrm{C}$ Relevant to $\ensuremath{\nu}p$-Process Nucleosynthesis},
  author = {Psaltis, A. and others},
  OPTauthor = {Psaltis, A. and Chen, A. A. and Longland, R. and Connolly, D. S. and Brune, C. R. and Davids, B. and Fallis, J. and Giri, R. and Greife, U. and Hutcheon, D. A. and Kroll, L. and Lennarz, A. and Liang, J. and Lovely, M. and Luo, M. and Marshall, C. and Paneru, S. N. and Parikh, A. and Ruiz, C. and Shotter, A. C. and Williams, M.},
  journal = PRL,
  volume = {129},
  issue = {16},
  pages = {162701},
  numpages = {6},
  year = {2022},
  month = 10,
  doi = {10.1103/PhysRevLett.129.162701}
}

@article{Psaltis2022b,
  title = {First inverse kinematics measurement of resonances in $^{7}\mathrm{Be}$($\ensuremath{\alpha},\ensuremath{\gamma})^{11}\mathrm{C}$ relevant to neutrino-driven wind nucleosynthesis using DRAGON},
  author = {Psaltis, A. and others},
  OPTauthor = {Psaltis, A. and Chen, A. A. and Longland, R. and Connolly, D. S. and Brune, C. R. and Davids, B. and Fallis, J. and Giri, R. and Greife, U. and Hutcheon, D. A. and Kroll, L. and Lennarz, A. and Liang, J. and Lovely, M. and Luo, M. and Marshall, C. and Paneru, S. N. and Parikh, A. and Ruiz, C. and Shotter, A. C. and Williams, M.},
  journal = PRC,
  volume = {106},
  issue = {4},
  pages = {045805},
  numpages = {14},
  year = {2022},
  month = 10,
  doi = {10.1103/PhysRevC.106.045805}
}

@article{Purser1977,
  OPTauthor =	 {{Purser}, K.H. and {Liebert}, R.B. and {Litherland}, A.E. and {Beukens}, R.P. and {Gove}, H.E. and {Bennett}, C.L. and {Clover}, M. R. and {Sondheim}, W.E.},
  author =	 {{Purser}, K.H. and others},
  title =	 {{An attempt to detect stable N- ions from a sputter ion source and some implications of the results for the design of tandems for ultra-sensitive carbon analysis}},
  DOI =		 "10.1051/rphysap:0197700120100148700",
  journal =	 RPA,
  year =	 1977,
  volume =	 12,
  number =	 10,
  pages =	 "1487-1492"
}

@article{Raduta11,
  OPTauthor =	 {{Raduta}, Ad. R. and {Borderie}, B. and {Geraci}, E. and {Le Neindre}, N. and {Napolitani}, P. and {Rivet}, M.~F. and {Alba}, R. and {Amorini}, F. and {Cardella}, G. and {Chatterjee}, M. and {De Filippo}, E. and {Guinet}, D. and {Lautesse}, P. and {La Guidara}, E. and {Lanzalone}, G. and {Lanzano}, G. and {Lombardo}, I. and {Lopez}, O. and {Maiolino}, C. and {Pagano}, A. and {Pirrone}, S. and {Politi}, G. and {Porto}, F. and {Rizzo}, F. and {Russotto}, P. and {Wieleczko}, J.~P.},
  author =	 {{Raduta}, Ad. R. and others},
  title =	 {Evidence for {\ensuremath{\alpha}}-particle condensation in nuclei from the {Hoyle} state deexcitation},
  journal =	 PLB,
  year =	 2011,
  month =	 11,
  volume =	 705,
  number =	 {1-2},
  pages =	 {65-70},
  doi =		 {10.1016/j.physletb.2011.10.008}
}

@article{Rahal1988PRA,
  doi =		 {10.1103/physreva.38.3728},
  year =	 1988,
  month =	 10,
  volume =	 38,
  number =	 7,
  pages =	 {3728--3731},
  author =	 {V. Rahal and A. Campa},
  title =	 {Thermodynamical implications of a violation of the {Pauli} principle},
  journal =	 PRA
}

@ARTICLE{Raiteri1991,
  OPTauthor =	 {{Raiteri}, C.~M. and {Busso}, M. and {Gallino}, R. and {Picchio}, G. and {Pulone}, L.},
  author =	 {{Raiteri}, C.~M. and others},
  title =	 {{$s$-Process} Nucleosynthesis in Massive Stars and the Weak Component. I. Evolution and Neutron Captures in a {25\,$M_{\odot}$} Star},
  journal =	 APJ,
  year =	 1991,
  month =	 1,
  volume =	 367,
  pages =	 228,
  doi =		 {10.1086/169622}
}

@article{Ramberg1990PLB,
  doi =		 {10.1016/0370-2693(90)91762-z},
  year =	 1990,
  month =	 4,
  publisher =	 {Elsevier {BV}},
  volume =	 238,
  number =	 {2-4},
  pages =	 {438--441},
  author =	 {Erik Ramberg and George A. Snow},
  title =	 {Experimental limit on a small violation of the {Pauli} principle},
  journal =	 PLB
}

@article{Rana13,
  OPTauthor =	 {{Rana}, T.~K. and {Bhattacharya}, S. and {Bhattacharya}, C. and {Kundu}, S. and {Banerjee}, K. and {Ghosh}, T.~K. and {Mukherjee}, G. and {Pandey}, R. and {Roy}, P. and {Srivastava}, V. and {Gohil}, M. and {Meena}, J.~K. and {Pai}, H. and {Saha}, A.~K. and {Sahoo}, J.~K. and {Saha}, R.~M.},
  author =	 {{Rana}, T.~K. and others},
  title =	 {Estimation of direct components of the decay of the {Hoyle} state},
  journal =	 PRC,
  year =	 2013,
  month =	 8,
  volume =	 88,
  number =	 2,
  eid =		 021601,
  pages =	 021601,
  doi =		 {10.1103/PhysRevC.88.021601}
}

@article{Rana19,
  OPTauthor =	 {{Rana}, T.~K. and {Bhattacharya}, S. and {Bhattacharya}, C. and {Manna}, S. and {Kundu}, Samir and {Banerjee}, K. and {Pandey}, R. and {Roy}, Pratap and {Dhal}, A. and {Mukherjee}, G. and {Srivastava}, V. and {Dey}, A. and {Chaudhuri}, A. and {Ghosh}, T.~K. and {Sen}, A. and {Asgar}, Md. A. and {Roy}, T. and {Sahoo}, J.~K. and {Meena}, J.~K. and {Saha}, A.~K. and {Saha}, R.~M. and {Sinha}, M. and {Roy}, Amit},
  author =	 {{Rana}, T.~K. and others},
  title =	 {New high precision study on the decay width of the {Hoyle} state in {$^{12}$C}},
  journal =	 PLB,
  year =	 2019,
  month =	 6,
  volume =	 793,
  pages =	 {130-133},
  doi =		 {10.1016/j.physletb.2019.04.028}
}

@article{Rapagnani2017,
  title =	 "A supersonic jet target for the cross section measurement of the {$\cag$} reaction with the recoil mass separator {ERNA}",
  journal =	 NIMB,
  volume =	 407,
  pages =	 "217-221",
  year =	 2017,
  issn =	 "0168-583X",
  doi =		 {10.1016/j.nimb.2017.07.003},
  OPTauthor =	 {D. Rapagnani and R. Buompane and A. {Di Leva} and L. Gialanella and M. Busso and M. {De Cesare} and G. {De Stefano} and J.G. Duarte and L.R. Gasques and L. Morales-Gallegos and S. Palmerini and M. Romoli and F. Tufariello},
  author =	 {D. Rapagnani and others}
}

@article{Rapagnani2020,
  title =	 {Ion Beam Analysis for recession determination and composition estimate of Aerospace Thermal Protection System materials},
  journal =	 NIMB,
  volume =	 {467},
  pages =	 {53-57},
  year =	 {2020},
  issn =	 {0168-583X},
  doi =		 {10.1016/j.nimb.2020.01.006},
  author =	 {D. Rapagnani and others},
  OPTauthor =	 {D. Rapagnani and M. {De Cesare} and D. Alfano and R. Buompane and S. Cantoni and M. {De Stefano Fumo} and A. {Del Vecchio} and A. D'Onofrio and G. Porzio and G.C. Rufolo and L. Gialanella}
}

@article{Rapagnani2021,
  author =	 {{Rapagnani}, D. and others},
  OPTauthor =	 {{Rapagnani}, D. and {De Cesare}, M. and {Buompane}, R. and {Del Vecchio}, A. and {Di Leva}, A. and {D'Onofrio}, A. and {Porzio}, G. and {Gialanella}, L.},
  title =	 "{Validation of a novel technique with radioactive implanted ions for recession rate estimate of aerospace material}",
  journal =	 JPhysD,
  year =	 2021,
  month =	 8,
  volume =	 {54},
  number =	 {32},
  eid =		 {32LT01},
  pages =	 {32LT01},
  doi =		 {10.1088/1361-6463/ac006e}
}

@article{Rapagnani2023,
  title =	 {Feasibility study of a compact and multi-gas supersonic plasma jet for nuclear astrophysics and space research},
  journal =	 NIMA,
  volume =	 {1056},
  pages =	 {168536},
  year =	 {2023},
  issn =	 {0168-9002},
  doi =		 {10.1016/j.nima.2023.168536},
  author =	 {D. Rapagnani and others},
  OPTauthor =	 {D. Rapagnani and L. Cutrone and G. Ranuzzi and M. {De Stefano Fumo} and L. Savino and M. {De Cesare} and R. Buompane and A. {Del Vecchio} and A. {Di Leva} and J.G. Duarte and L. Morales-Gallegos and M. Romoli and A. Schettino and L. Gialanella}
}

@ARTICLE{Renzini2015,
  author =	 {{Renzini}, Alvio},
  title =	 {Origin of multiple stellar populations in globular clusters and their helium enrichment},
  journal =	 MNRAS,
  year =	 2008,
  month =	 11,
  volume =	 391,
  number =	 1,
  pages =	 {354-362},
  doi =		 {10.1111/j.1365-2966.2008.13892.x}
}

@article{Ricci2022,
  OPTauthor =	 {Ricci, Giulia and Secco, Michele and Addis, Anna and Pistilli, Anna and Preto, Nereo and Brogiolo, Gian Pietro and Arnau, Alexandra Chavarria and Marzaioli, Fabio and Passariello, Isabella and Terrasi, Filippo and Artioli, Gilberto},
  author =	 {Ricci, Giulia and others},
  date =	 {2022/02/28},
  doi =		 {10.1038/s41598-022-07406-x},
  journal =	 SciRep,
  number =	 1,
  pages =	 3339,
  title =	 {Integrated multi-analytical screening approach for reliable radiocarbon dating of ancient mortars},
  volume =	 12,
  year =	 2022
}

@Article{RicciLisa2022,
  author =	 {Ricci, L. and others},
  OPTAUTHOR =	 {Ricci, Lisa and Petrelli, Maurizio and Frondini, Francesco and Zucchini, Azzurra and Comodi, Paola and Bisciotti, Andrea and Vescovi, Diego and Trippella, Oscar},
  title =	 {The Achievements of the {RockStar Group (Perugia)} on Astrophysical Modelling and {Pallasite} Geochemistry},
  journal =	 {Universe},
  volume =	 {8},
  year =	 {2022},
  number =	 {3},
  article-NUMBER ={156},
  issn =	 {2218-1997},
  doi =		 {10.3390/universe8030156}
}

@article{Rogalla1999,
  title =	 {Recoil separator {ERNA}: ion beam purification},
  journal =	 NIMA,
  volume =	 437,
  number =	 2,
  pages =	 {266-273},
  year =	 1999,
  issn =	 {0168-9002},
  doi =		 {10.1016/S0168-9002(99)00767-6},
  OPTauthor =	 {D Rogalla and S Theis and L Campajola and A D'Onofrio and L Gialanella and U Greife and G Imbriani and A Ordine and V Roca and C Rolfs and M Romano and C Sabbarese and F Schümann and F Strieder and F Terrasi and H.P Trautvetter},
  author =	 {D Rogalla and others}
}

@article{Rogalla2003,
  title =	 {Recoil separator {ERNA}: acceptances in angle and energy},
  journal =	 NIMA,
  volume =	 513,
  number =	 3,
  pages =	 {573-578},
  year =	 2003,
  issn =	 {0168-9002},
  doi =		 {10.1016/j.nima.2003.07.001},
  OPTauthor =	 {D. Rogalla and D. Schürmann and F. Strieder and M. Aliotta and N. DeCesare and A. DiLeva and C. Lubritto and A. D’Onofrio and L. Gialanella and G. Imbriani and J. Kluge and A. Ordine and V. Roca and H. Röcken and C. Rolfs and M. Romano and F. Schümann and F. Terrasi and H.P. Trautvetter},
  author =	 {D. Rogalla and others}
}

@book{Rolfs,
  author =	 {C.E. Rolfs and W.S. Rodney},
  title =	 {Cauldrons in the Cosmos},
  publisher =	 {The University of Chicago Press},
  address =	 {Chicago},
  year =	 1988
}

@Article{Romoli2018,
  OPTauthor =	 {{Romoli}, M. and {Morales-Gallegos}, L. and {Aliotta}, M. and {Bruno}, C. G. and {Buompane}, R. and {D'Onofrio}, A. and {Davinson}, T. and {De Cesare}, M. and {Di Leva}, A. and {Di Meo}, P. and {Duarte}, J. and {Gasques}, L. and {Gialanella}, L. and {Imbriani}, G. and {Porzio}, G. and {Rapagnani}, D. and {Vanzanella}, A.},
  author =	 {{Romoli}, M. and others},
  title =	 "Development of a two-stage detection array for low-energy light charged particles in nuclear astrophysics applications",
  journal =	 EPJA,
  year =	 2018,
  month =	 9,
  day =		 30,
  volume =	 54,
  number =	 8,
  pages =	 142,
  issn =	 "1434-601X",
  doi =		 "10.1140/epja/i2018-12575-5"
}

@mastersthesis{Santonastaso2019,
  author =	 {Santonastaso, C.},
  title =	 {Change in the $^7\mathrm{Be}$ half life in different environments},
  year =	 {2019},
  school =	 {Università degli Studi di Salerno},
  type =	 {Master Thesis},
  publication_stage ={Final},
  source =	 {},
}

@ARTICLE{Santonastaso2021,
  author =	 {Santonastaso, C. and others},
  OPTauthor =	 {Santonastaso, C. and Buompane, R. and Di Leva, A. and Morales-Gallegos, L. and Itaco, N. and Landi, G. and Neitzert, H.C. and Rapagnani, D. and Gialanella, L.},
  title =	 {Change in the {$^7\mathrm{Be}$} half-life in different environments},
  year =	 {2021},
  journal =	 {Il Nuovo Cimento C},
  volume =	 {44},
  number =	 {2-3},
  doi =		 {10.1393/ncc/i2021-21075-8}
}

@ARTICLE{Sbordone2010,
  OPTauthor =	 {{Sbordone}, L. and {Bonifacio}, P. and {Caffau}, E. and {Ludwig}, H. -G. and {Behara}, N.~T. and {Gonz{\'a}lez Hern{\'a}ndez}, J.~I. and {Steffen}, M. and {Cayrel}, R. and {Freytag}, B. and {van't Veer}, C. and {Molaro}, P. and {Plez}, B. and {Sivarani}, T. and {Spite}, M. and {Spite}, F. and {Beers}, T.~C. and {Christlieb}, N. and {Fran{\c{c}}ois}, P. and {Hill}, V.},
  author =	 {{Sbordone}, L. and others},
  title =	 {The metal-poor end of the Spite plateau. {I. Stellar} parameters, metallicities, and lithium abundances},
  journal =	 AAP,
  year =	 2010,
  month =	 11,
  volume =	 522,
  eid =		 {A26},
  pages =	 {A26},
  doi =		 {10.1051/0004-6361/200913282}
}

@article{schrodinger1935NW,
  title =	 {{Die gegenw{\"a}rtige Situation in der Quantenmechanik}},
  author =	 {Schr{\"o}dinger, Erwin},
  journal =	 {Naturwissenschaften},
  volume =	 23,
  number =	 49,
  pages =	 {823--828},
  year =	 1935,
  doi = {10.1007/BF01491914}
}

@article{Schurmann2004,
  title =	 {Recoil separator {ERNA}: charge state distribution, target density, beam heating, and longitudinal acceptance},
  journal =	 NIMA,
  volume =	 531,
  number =	 3,
  pages =	 {428-434},
  year =	 2004,
  issn =	 {0168-9002},
  doi =		 {10.1016/j.nima.2004.05.131},
  OPTauthor =	 {D. Schürmann and F. Strieder and A. {Di Leva} and L. Gialanella and N. {De Cesare} and A. D'Onofrio and G. Imbriani and J. Klug and C. Lubritto and A. Ordine and V. Roca and H. Röcken and C. Rolfs and D. Rogalla and M. Romano and F. Schümann and F. Terrasi and H.P. Trautvetter},
  author =	 {D. Schürmann and others}
}

@Article{Schurmann2005,
  OPTauthor =	 {Sch{\"u}rmann, D. and Di Leva, A. and Gialanella, L. and Rogalla, D. and Strieder, F. and De Cesare, N. and D'Onofrio, A. and Imbriani, G. and Kunz, R. and Lubritto, C. and Ordine, A. and Roca, V. and Rolfs, C. and Romano, M. and Sch{\"u}mann, F. and Terrasi, F. and Trautvetter, H.-P.},
  author =	 {Sch{\"u}rmann, D. and others},
  title =	 {First direct measurement of the total cross-section of $\cag$},
  journal =	 EPJA,
  year =	 2005,
  month =	 11,
  day =		 01,
  volume =	 26,
  number =	 2,
  pages =	 {301-305},
  issn =	 {1434-601X},
  doi =		 {10.1140/epja/i2005-10175-2}
}

@article{Schurmann2011,
  title =	 {Study of the 6.05\,{MeV} cascade transition in $\cag$},
  journal =	 PLB,
  volume =	 703,
  number =	 5,
  pages =	 {557-561},
  year =	 2011,
  issn =	 {0370-2693},
  doi =		 {10.1016/j.physletb.2011.08.061},
  OPTauthor =	 {D. Sch\"urmann and A. {Di Leva} and L. Gialanella and R. Kunz and F. Strieder and N. {De Cesare} and M. {De Cesare} and A. D'Onofrio and K. Fortak and G. Imbriani and D. Rogalla and M. Romano and F. Terrasi},
  author =	 {D. Sch\"urmann and others}
}

@article{Schurmann2012,
  Author =	 {D. Sch\"urmann and L. Gialanella and R. Kunz and F. Strieder},
  Doi =		 {10.1016/j.physletb.2012.03.064},
  Journal =	 PLB,
  Number =	 1,
  Pages =	 {35-40},
  Title =	 {{The astrophysical $S$ factor of $^{12}\mathrm{C}(\alpha,\gamma){}^{16}\mathrm{O}$ at stellar energy}},
  Volume =	 711,
  Year =	 2012
}

@ARTICLE{Schuermann2013,
  OPTauthor =	 {{Sch{\"u}rmann}, D. and {Di Leva}, A. and {Gialanella}, L. and {De Cesare}, M. and {De Cesare}, N. and {Imbriani}, G. and {D'Onofrio}, A. and {Romano}, M. and {Romoli}, M. and {Terrasi}, F.},
  author =	 {{Sch{\"u}rmann}, D. and others},
  title =	 {A windowless hydrogen gas target for the measurement of {$^{7}\mathrm{Be}(\mathrm{p},\gamma)^{8}\mathrm{B}$} with the recoil separator {ERNA}},
  journal =	 EPJA,
  year =	 2013,
  month =	 6,
  volume =	 49,
  pages =	 80,
  doi =		 {10.1140/epja/i2013-13080-1}
}

@article{Sen2019,
  title =	 "A high intensity, high stability 3.5\,{MV} {S}ingletron™ accelerator",
  journal =	 NIMB,
  volume =	 450,
  pages =	 "390 - 395",
  year =	 2019,
  OPTnote =	 "The 23rd International Conference on Ion Beam Analysis",
  issn =	 "0168-583X",
  doi =		 "10.1016/j.nimb.2018.09.016",
  OPTauthor =	 "A. Sen and G. Domínguez-Cañizares and N.C. Podaru and D.J.W. Mous and M. Junker and G. Imbriani and V. Rigato",
  author =	 "A. Sen and others"
}

@article{Shen2020,
  title =	 {Constraining the External Capture to the {$^{16}\mathrm{O}$} Ground State and the {$E2$ $S$} Factor of the {$^{12}\mathrm{C}(\alpha,\gamma){}^{16}\mathrm{O}$} Reaction},
  OPTauthor =	 {Shen, Y. P. and Guo, B. and deBoer, R. J. and Li, Z. H. and Li, Y. J. and Tang, X. D. and Pang, D. Y. and Adhikari, S. and Basu, C. and Su, J. and Yan, S. Q. and Fan, Q. W. and Liu, J. C. and Chen, C. and Han, Z. Y. and Li, X. Y. and Lian, G. and Ma, T. L. and Nan, W. and Nan, W. K. and Wang, Y. B. and Zeng, S. and Zhang, H. and Liu, W. P.},
  author =	 {Shen, Y. P. and others},
  journal =	 PRL,
  volume =	 124,
  issue =	 16,
  pages =	 162701,
  numpages =	 6,
  year =	 2020,
  month =	 4,
  doi =		 {10.1103/PhysRevLett.124.162701}
}

@article{Shi2018EPJC,
  doi =		 {10.1140/epjc/s10052-018-5802-4},
  year =	 2018,
  month =	 4,
  volume =	 78,
  number =	 4,
  OPTauthor =	 {H. Shi and E. Milotti and S. Bartalucci and M. Bazzi and S. Bertolucci and A. M. Bragadireanu and M. Cargnelli and A. Clozza and L. De Paolis and S. Di Matteo and J.-P. Egger and H. Elnaggar and C. Guaraldo and M. Iliescu and M. Laubenstein and J. Marton and M. Miliucci and A. Pichler and D. Pietreanu and K. Piscicchia and A. Scordo and D. L. Sirghi and F. Sirghi and L. Sperandio and O. Vazquez Doce and E. Widmann and J. Zmeskal and C. Curceanu},
  author =	 {H. Shi and others},
  title =	 {Experimental search for the violation of {Pauli} exclusion principle},
  journal =	 EPJC
}

@article{Sie2000,
  title =	 {{A fast bouncing system for the high-energy end of AMS}},
  journal =	 NIMB,
  volume =	 172,
  number =	 1,
  pages =	 {268-273},
  year =	 2000,
  OPTnote =	 {8th International Conference on Accelerator Mass Spectrometry},
  issn =	 {0168-583X},
  doi =		 {10.1016/S0168-583X(00)00126-9},
  author =	 {S.H Sie and D.A Sims and G.F Suter and T.R Niklaus}
}

@ARTICLE{Skowronski2023,
  author =	 {{Skowronski}, J. and others},
  OPTauthor =	 {{Skowronski}, J. and {Gesu{\`e}}, R.~M. and {Boeltzig}, A. and {Ciani}, G.~F. and {Piatti}, D. and {Rapagnani}, D. and {Aliotta}, M. and {Ananna}, C. and {Barile}, F. and {Bemmerer}, D. and {Best}, A. and {Broggini}, C. and {Bruno}, C.~G. and {Caciolli}, A. and {Campostrini}, M. and {Cavanna}, F. and {Colombetti}, P. and {Compagnucci}, A. and {Corvisiero}, P. and {Csedreki}, L. and {Davinson}, T. and {Depalo}, R. and {Di Leva}, A. and {Elekes}, Z. and {Ferraro}, F. and {Formicola}, A. and {F{\"u}l{\"o}p}, Zs and {Gervino}, G. and {Guglielmetti}, A. and {Gustavino}, C. and {Gy{\"u}rky}, Gy and {Imbriani}, G. and {Junker}, M. and {Lugaro}, M. and {Marigo}, P. and {Masha}, E. and {Menegazzo}, R. and {Paticchio}, V. and {Perrino}, R. and {Prati}, P. and {Rigato}, V. and {Schiavulli}, L. and {Sidhu}, R.~S. and {Straniero}, O. and {Sz{\"u}cs}, T. and {Zavatarelli}, S.},
  title =	 {Advances in radiative capture studies at {LUNA} with a segmented {BGO} detector},
  journal =	 JPhysG,
  year =	 2023,
  month =	 4,
  volume =	 {50},
  number =	 {4},
  eid =		 {045201},
  pages =	 {045201},
  doi =		 {10.1088/1361-6471/acb961}
}

@ARTICLE{Slemer2017,
  OPTauthor =	 {{Slemer}, A. and {Marigo}, P. and {Piatti}, D. and {Aliotta}, M. and {Bemmerer}, D. and {Best}, A. and {Boeltzig}, A. and {Bressan}, A. and {Broggini}, C. and {Bruno}, C.~G. and {Caciolli}, A. and {Cavanna}, F. and {Ciani}, G.~F. and {Corvisiero}, P. and {Davinson}, T. and {Depalo}, R. and {Di Leva}, A. and {Elekes}, Z. and {Ferraro}, F. and {Formicola}, A. and {F{\"u}l{\"o}p}, Zs. and {Gervino}, G. and {Guglielmetti}, A. and {Gustavino}, C. and {Gy{\"u}rky}, G. and {Imbriani}, G. and {Junker}, M. and {Menegazzo}, R. and {Mossa}, V. and {Pantaleo}, F.~R. and {Prati}, P. and {Straniero}, O. and {Sz{\"u}cs}, T. and {Tak{\'a}cs}, M.~P. and {Trezzi}, D.},
  author =	 {{Slemer}, A. and others},
  title =	 {{$^{22}\mathrm{Ne}$} and {$^{23}\mathrm{Na}$} ejecta from intermediate-mass stars: the impact of the new {LUNA} rate for {${}^{22}\mathrm{Ne}(\mathrm{p},\gamma){}^{23}\mathrm{Na}$}},
  journal =	 MNRAS,
  year =	 2017,
  month =	 3,
  volume =	 465,
  number =	 4,
  pages =	 {4817-4837},
  doi =		 {10.1093/mnras/stw3029}
}

@article{Smith2021,
  OPTauthor =	 {R. Smith and M. Gai and S. R. Stern and D. K. Schweitzer and M. W. Ahmed},
  author =	 {R. Smith and others},
  title =	 {{Precision measurements on oxygen formation in stellar helium burning with gamma-ray beams and a Time Projection Chamber}},
  journal =	 NatureCom,
  volume =	 12,
  pages =	 5920,
  year =	 2021,
  doi =		 {10.1038/s41467-021-26179-x}
}

@article{Smith2017,
  OPTauthor =	 {{Smith}, R. and {Kokalova}, Tz. and {Wheldon}, C. and {Bishop}, J.~E. and {Freer}, M. and {Curtis}, N. and {Parker}, D.~J.},
  author =	 {{Smith}, R. and others},
  title =	 {New Measurement of the Direct {3\ensuremath{\alpha}} Decay from the {$^{12}\mathrm{C}$} {Hoyle} State},
  journal =	 PRL,
  year =	 2017,
  month =	 9,
  volume =	 119,
  number =	 13,
  eid =		 132502,
  pages =	 132502,
  doi =		 {10.1103/PhysRevLett.119.132502}
}

@ARTICLE{Smolin2006Arxiv,
  author =	 {{Smolin}, Lee},
  title =	 {Could quantum mechanics be an approximation to another theory?},
  journal =	 {arXiv e-prints},
  year =	 2006,
  month =	 9,
  eid =		 {quant-ph/0609109},
  pages =	 {quant-ph/0609109},
  doi =		 {10.48550/arXiv.quant-ph/0609109}
}

@article{Soti2019,
  author =	 {{S\'oti}, Zsolt and {Magill}, Joseph and {Dreher}, Raymond},
  title =	 {{Karlsruhe Nuclide Chart - New 10th edition 2018}},
  DOI =		 {10.1051/epjn/2019004},
  journal =	 EPJN,
  year =	 2019,
  volume =	 5,
  pages =	 6
}

@phdthesis{sperandio2008TV,
  title =	 {New experimental limit on the {Pauli} Exclusion Principle violation by electrons from the {VIP} experiment},
  author =	 {Sperandio, Laura},
  school =	 {Università degli Studi di Roma ``Tor Vergata"},
  year =	 2008
}

@ARTICLE{Straniero1995,
  OPTauthor =	 {{Straniero}, O. and {Gallino}, R. and {Busso}, M. and {Chiefei}, A. and {Raiteri}, C.~M. and {Limongi}, M. and {Salaris}, M.},
  author =	 {{Straniero}, O. and others},
  title =	 {Radiative {$^{13}\mathrm{C}$} Burning in {Asymptotic Giant Branch} Stars and $s$-Processing},
  journal =	 APJL,
  year =	 1995,
  month =	 2,
  volume =	 440,
  pages =	 {L85},
  doi =		 {10.1086/187767}
}

@ARTICLE{Straniero2006,
  author =	 {{Straniero}, Oscar and {Gallino}, Roberto and {Cristallo}, Sergio},
  title =	 {{$s$} process in low-mass {Asymptotic Giant Branch} stars},
  journal =	 NuclPhysA,
  year =	 2006,
  month =	 10,
  volume =	 777,
  pages =	 {311-339},
  doi =		 {10.1016/j.nuclphysa.2005.01.011}
}

@INPROCEEDINGS{Straniero2019,
  author =	 {{Straniero}, Oscar and {Piersanti}, Luciano and {Dominguez}, Inma and {Tumino}, Aurora},
  title =	 {On the Mass of Supernova Progenitors: The Role of the {$^{12}\mathrm{C}$+$^{12}\mathrm{C}$} Reaction},
  booktitle =	 {Nuclei in the Cosmos XV},
  year =	 2019,
  volume =	 219,
  month =	 8,
  pages =	 {7-11},
  doi =		 {10.1007/978-3-030-13876-9_2}
}

@ARTICLE{Suhonen2017,
  author =	 {Suhonen, Jouni T.},
  title =	 {Value of the Axial-Vector Coupling Strength in {$\beta$} and {$\beta\beta$} Decays: A Review},
  journal =	 FP,
  volume =	 5,
  year =	 2017,
  doi =		 {10.3389/fphy.2017.00055},
  issn =	 {2296-424X}
}

@article{Suno15,
  author =	 {{Suno}, Hiroya and {Suzuki}, Yasuyuki and {Descouvemont}, Pierre},
  title =	 {Triple-{\ensuremath{\alpha}} continuum structure and {Hoyle} resonance of {$^{12}$C} using the hyperspherical slow variable discretization},
  journal =	 PRC,
  year =	 2015,
  month =	 1,
  volume =	 91,
  number =	 1,
  eid =		 014004,
  pages =	 014004,
  doi =		 {10.1103/PhysRevC.91.014004}
}

@article{Terrasi2007,
  title =	 {A new {AMS} facility in {Caserta/Italy}},
  journal =	 NIMB,
  volume =	 259,
  number =	 1,
  pages =	 {14 - 17},
  year =	 2007,
  OPTnote =	 {Accelerator Mass Spectrometry - Proceedings of the Tenth International Conference on Accelerator Mass Spectrometry},
  issn =	 {0168-583X},
  doi =		 {10.1016/j.nimb.2007.01.139},
  OPTauthor =	 {Terrasi, F. and Rogalla, D. and De Cesare, N. and D'Onofrio, A. and Lubritto, C. and Marzaioli, F. and Passariello, I. and Rubino, M. and Sabbarese, C. and Casa, G. and Palmieri, A. and Gialanella, L. and Imbriani, G. and Roca, V. and Romano, M. and Sundquist, M. and Loger, R.},
  author =	 {Terrasi, F. and others}
}

@article{Terrasi2020,
  title =	 {Can The {$^{14}\mathrm{C}$} Production in 1055\,{CE} be Affected by {SN1054}?},
  volume =	 62,
  DOI =		 {10.1017/RDC.2020.58},
  number =	 5,
  journal =	 {Radiocarbon},
  OPTauthor =	 {Terrasi, F and Marzaioli, F and Buompane, R and Passariello, I and Porzio, G and Capano, M and Helama, S and Oinonen, M and Nöjd, P and Uusitalo, J and et al.},
  author =	 {Terrasi, F and others},
  year =	 2020,
  pages =	 {1403–1418}
}

@article{Tohsaki01,
  title =	 {Alpha Cluster Condensation in {$^{12}\mathrm{C}$} and {$^{16}\mathrm{O}$}},
  author =	 {Tohsaki, A. and Horiuchi, H. and Schuck, P. and R\"opke, G.},
  journal =	 PRL,
  volume =	 87,
  issue =	 19,
  pages =	 192501,
  numpages =	 4,
  year =	 2001,
  month =	 10,
  doi =		 {10.1103/PhysRevLett.87.192501}
}

@article{Toro2018JPA,
  doi =		 {10.1088/1751-8121/aaabc6},
  year =	 2018,
  month =	 2,
  volume =	 51,
  number =	 11,
  pages =	 115302,
  author =	 {Marko Toro{\v{s}} and Angelo Bassi},
  title =	 {Bounds on quantum collapse models from matter-wave interferometry: calculational details},
  journal =	 JPhysA
}

@article{Trezzi2017,
  title =	 {{Big Bang} {$^6\mathrm{Li}$} nucleosynthesis studied deep underground},
  journal =	 APARTP,
  volume =	 89,
  pages =	 {57-65},
  year =	 2017,
  issn =	 {0927-6505},
  doi =		 {10.1016/j.astropartphys.2017.01.007},
  OPTauthor =	 {D. Trezzi and M. Anders and M. Aliotta and A. Bellini and D. Bemmerer and A. Boeltzig and C. Broggini and C.G. Bruno and A. Caciolli and F. Cavanna and P. Corvisiero and H. Costantini and T. Davinson and R. Depalo and Z. Elekes and M. Erhard and F. Ferraro and A. Formicola and Zs. F\"ul\"op and G. Gervino and A. Guglielmetti and C. Gustavino and Gy. GyÃ¼rky and M. Junker and A. Lemut and M. Marta and C. Mazzocchi and R. Menegazzo and V. Mossa and F. Pantaleo and P. Prati and C. {Rossi Alvarez} and D.A. Scott and E. Somorjai and O. Straniero and T. Sz\"ucs and M. Tak\'acs},
  author =	 {D. Trezzi and others}
}

@ARTICLE{Trippella2017,
  author =	 {{Trippella}, O. and {La Cognata}, M.},
  title =	 "{Concurrent Application of ANC and THM to assess the $^{13}$C({\ensuremath{\alpha}}, n)$^{16}$O Absolute Cross Section at Astrophysical Energies and Possible Consequences for Neutron Production in Low-mass AGB Stars}",
  journal =	 APJ,
  year =	 2017,
  volume =	 {837},
  pages =	 {41},
  doi =		 {10.3847/1538-4357/aa5eb5},
}

@article{Tumino2018,
  title =	 {An increase in the {$^{12}\mathrm{C}$+$^{12}\mathrm{C}$} fusion rate from resonances at astrophysical energies},
  OPTauthor =	 {Tumino, A. and Spitaleri, C. and La Cognata, M. and Cherubini, S. and Guardo, G.L. and Gulino, M. and Hayakawa, S. and Indelicato, I.  and Lamia, L. and Petrascu, H. and Pizzone, R.G. and Puglia, S.M.R. and Rapisarda, G.G. and Romano, S. and Sergi, M.L. and Spartá R. and Trache, L.},
  author =	 {Tumino, A. and others},
  journal =	 Nature,
  volume =	 557,
  pages =	 {687-690},
  year =	 2018,
  doi =		 {10.1038/s41586-018-0149-4}
}

@article{Tumino2021,
  OPTauthor =	 {Tumino, Aurora and Bertulani, Carlos A. and La Cognata, Marco and Lamia, Livio and Pizzone, Rosario Gianluca and Romano, Stefano and Typel, Stefan},
    author =	 {Tumino, Aurora and others},
  title =	 {The {Trojan Horse Method}: A Nuclear Physics Tool for Astrophysics},
  journal =	 ARNPS,
  volume =	 71,
  number =	 1,
  pages =	 {345-376},
  year =	 2021,
  doi =		 {10.1146/annurev-nucl-102419-033642}
}

@article{Uegaki77,
  author =	 {Uegaki, Eiji and Okabe, Shigetō and Abe, Yasuhisa and Tanaka, Hajime},
  title =	 {Structure of the Excited States in {$^{12}\mathrm{C}$}. {I}},
  journal =	 PTP,
  volume =	 57,
  number =	 4,
  pages =	 1262,
  year =	 1977,
  doi =		 {10.1143/PTP.57.1262}
}

@article{Uusitalo2018,
  OPTauthor =	 {Uusitalo, J. and Arppe, L. and Hackman, T. and Helama, S. and Kovaltsov, G. and Mielik{\"a}inen, K. and M{\"a}kinen, H. and N{\"o}jd, P. and Palonen, V. and Usoskin, I. and Oinonen, M.},
  author =	 {Uusitalo, J. and others},
  date =	 {2018/08/28},
  doi =		 {10.1038/s41467-018-05883-1},
  journal =	 NatureCom,
  number =	 1,
  pages =	 3495,
  title =	 {{Solar superstorm of AD 774 recorded subannually by Arctic tree rings}},
  volume =	 9,
  year =	 2018
}

@article{VandeKolk2022,
  title =	 {{Investigation of the $^{10}\mathrm{B}\pa^{7}\mathrm{Be}$ reaction from 0.8 to 2.0\,MeV}},
  OPTauthor =	 {Vande~Kolk, B. and Macon, K. T. and deBoer, R. J. and Anderson, T. and Boeltzig, A. and Brandenburg, K. and Brune, C. R. and Chen, Y. and Clark, A. M. and Danley, T. and Frentz, B. and Giri, R. and G\"orres, J. and Hall, M. and Henderson, S. L. and Holmbeck, E. and Howard, K. B. and Jacobs, D. and Lai, J. and Liu, Q. and Long, J. and Manukyan, K. and Massey, T. and Moran, M. and Morales, L. and Odell, D. and O'Malley, P. and Paneru, S. N. and Richard, A. and Schneider, D. and Skulski, M. and Sensharma, N. and Seymour, C. and Seymour, G. and Soltesz, D. and Strauss, S. and Voinov, A. and W\"ustrich, L. and Wiescher, M.},
  author =	 {Vande~Kolk, B. and others},
  journal =	 PRC,
  volume =	 105,
  issue =	 5,
  pages =	 055802,
  numpages =	 13,
  year =	 2022,
  month =	 5,
  doi =		 {10.1103/PhysRevC.105.055802},
}

@article{Vescovi2019,
  OPTauthor =	 {{Vescovi, D.} and {Piersanti, L.} and {Cristallo, S.} and {Busso, M.} and {Vissani, F.} and {Palmerini, S.} and {Simonucci, S.} and {Taioli, S.}},
  author =	 {{Vescovi, D.} and others},
  title =	 {Effects of a revised {$^7\mathrm{Be}$ $e$}-capture rate on solar neutrino fluxes},
  DOI =		 "10.1051/0004-6361/201834993",
  journal =	 AAP,
  year =	 2019,
  volume =	 623,
  pages =	 "A126"
}

@article{vinante2016PRL,
  title =	 {Upper Bounds on Spontaneous Wave-Function Collapse Models Using Millikelvin-Cooled Nanocantilevers},
  OPTauthor =	 {Vinante, A. and Bahrami, M. and Bassi, A. and Usenko, O. and Wijts, G. and Oosterkamp, T. H.},
  author =	 {Vinante, A. and others},
  journal =	 PRL,
  volume =	 116,
  issue =	 9,
  pages =	 090402,
  numpages =	 5,
  year =	 2016,
  month =	 3,
  doi =		 {10.1103/PhysRevLett.116.090402}
}

@article{Vinante2017PRL,
  title =	 {Improved Noninterferometric Test of Collapse Models Using Ultracold Cantilevers},
  OPTauthor =	 {Vinante, A. and Mezzena, R. and Falferi, P. and Carlesso, M. and Bassi, A.},
  author =	 {Vinante, A. and others},
  journal =	 PRL,
  volume =	 119,
  issue =	 11,
  pages =	 110401,
  numpages =	 5,
  year =	 2017,
  month =	 9,
  doi =		 {10.1103/PhysRevLett.119.110401}
}

@article{Vinante2020PRL,
  title =	 {Narrowing the Parameter Space of Collapse Models with Ultracold Layered Force Sensors},
  OPTauthor =	 {Vinante, A. and Carlesso, M. and Bassi, A. and Chiasera, A. and Varas, S. and Falferi, P. and Margesin, B. and Mezzena, R. and Ulbricht, H.},
  author =	 {Vinante, A. and others},
  journal =	 PRL,
  volume =	 125,
  issue =	 10,
  pages =	 100404,
  numpages =	 6,
  year =	 2020,
  month =	 9,
  doi =		 {10.1103/PhysRevLett.125.100404}
}

@article{Viviani2000,
  OPTauthor =	 {Viviani, M. and Kievsky, A. and Marcucci, L. E. and Rosati, S. and Schiavilla, R.},
  author =	 {Viviani, M. and others},
  title =	 {Photodisintegration and electrodisintegration of {${}^{3}\mathrm{He}$} at threshold and {$\mathrm{p}\mathrm{d}$} radiative capture},
  doi =		 "10.1103/PhysRevC.61.064001",
  journal =	 PRC,
  volume =	 61,
  pages =	 064001,
  year =	 2000
}

@article{VonOertzen97,
  author =	 {{von Oertzen}, W.},
  title =	 {Dimers based on the {$\alpha$+$\alpha$} potential and chain states of carbon isotopes},
  journal =	 ZFPA,
  year =	 1997,
  month =	 5,
  volume =	 357,
  number =	 4,
  pages =	 {355-365},
  doi =		 {10.1007/s002180050255}
}

@article{Wagner2018,
  title =	 {{Astrophysical $S$ factor of the $^{14}\mathrm{N}\pg^{15}\mathrm{O}$ reaction at 0.4-1.3\,{MeV}}},
  OPTauthor =	 {Wagner, L. and Akhmadaliev, S. and Anders, M. and Bemmerer, D. and Caciolli, A. and Gohl, St. and Grieger, M. and Junghans, A. and Marta, M. and Munnik, F. and Reinhardt, T. P. and Reinicke, S. and R{\"o}der, M. and Schmidt, K. and Schwengner, R. and Serfling, M. and Tak{\'a}cs, M. P. and Sz{\"u}cs, T. and Vomiero, A. and Wagner, A. and Zuber, K.},
  author =	 {Wagner, L. and others},
  journal =	 PRC,
  volume =	 97,
  issue =	 1,
  pages =	 015801,
  numpages =	 15,
  year =	 2018,
  month =	 1,
  doi =		 {10.1103/PhysRevC.97.015801}
}

@article{Wallner2005,
  OPTauthor =	 {Wallner, A.  and Golser, R.  and Kutschera, W. and Priller,A. and Steier, P. and Vockenhuber, C.  and Vonach,H.  and Faestermann,T.  and Knie,K.  and Korschinek,G. },
  author =	 {Wallner, A.  and others},
  title =	 {Potential of {AMS} for Quantifying Long‐Lived Reaction Products},
  journal =	 {AIP Conference Proceedings},
  volume =	 769,
  number =	 1,
  pages =	 {621-624},
  year =	 2005,
  doi =		 {10.1063/1.1945086}
}

@article{Wang2021,
  doi =		 {10.3847/1538-4357/ac2d90},
  year =	 2021,
  month =	 12,
  volume =	 923,
  number =	 2,
  pages =	 219,
  OPTauthor =	 {Xilu Wang and Adam M. Clark and John Ellis and Adrienne F. Ertel and Brian D. Fields and Brian J. Fry and Zhenghai Liu and Jesse A. Miller and Rebecca Surman},
  author =	 {Xilu Wang and others},
  title =	 {r-Process Radioisotopes from Near-{Earth} Supernovae and Kilonovae},
  journal =	 APJ
}

@article{Weinberg2012PRA,
  author =	 {{Weinberg}, Steven},
  title =	 {Collapse of the state vector},
  journal =	 PRA,
  year =	 2012,
  month =	 6,
  volume =	 85,
  number =	 6,
  eid =		 062116,
  pages =	 062116,
  doi =		 {10.1103/PhysRevA.85.062116}
}

@inbook{Weinberg2014OKML,
  author =	 {{Weinberg}, Steven},
  title =	 {Precision Tests of Quantum Mechanics},
  booktitle =	 {The Oskar Klein Memorial Lectures: 1988-1999},
  year =	 2014,
  editor =	 {{Ekspong}, Gosta},
  pages =	 {61-68},
  doi =		 {10.1142/9789814571616_0005},
  publisher =	 {World Scientific},
  address =	 {Singapore}
}

@book{White2023,
  year =	 2023,
  publisher =	 {John Wiley \& Sons},
  address =	 {US},
  author =	 {White, William M.},
  title =	 {Isotope Geochemistry}
}

@article{Wiescher2010,
  title =	 {The Cold and Hot {CNO} Cycles},
  OPTauthor =	 {Wiescher, M. and G\"{o}rres, J. and Uberseder, E. and Imbriani, G. and Pignatari, M.},
  author =	 {Wiescher, M. and others},
  journal =	 ARNP,
  volume =	 60,
  number =	 1,
  pages =	 {381-404},
  year =	 2010,
  doi =		 {10.1146/annurev.nucl.012809.104505}
}

@article{Xu2013,
  title =	 {{NACRE II: an update of the NACRE compilation of charged-particle-induced thermonuclear reaction rates for nuclei with mass number $A<$~16}},
  journal =	 NuclPhysA,
  volume =	 918,
  pages =	 {61-169},
  year =	 2013,
  issn =	 {0375-9474},
  doi =		 {10.1016/j.nuclphysa.2013.09.007},
  OPTauthor =	 {Y. Xu and K. Takahashi and S. Goriely and M. Arnould and M. Ohta and H. Utsunomiya},
  author =	 {Y. Xu and others}
}

@ARTICLE{Yong2006,
  author =	 {{Yong}, David and {Aoki}, Wako and {Lambert}, David L.},
  title =	 {{Mg} Isotope Ratios in Giant Stars of the Globular Clusters {M13} and {M71}},
  journal =	 APJ,
  year =	 2006,
  month =	 2,
  volume =	 638,
  number =	 2,
  pages =	 {1018-1027},
  doi =		 {10.1086/498974}
}

@article{Zhang2020,
  doi =		 {10.1088/1361-6471/ab6a71},
  year =	 2020,
  month =	 3,
  volume =	 47,
  number =	 5,
  pages =	 054002,
  author =	 {Xilin Zhang and Kenneth M Nollett and D R Phillips},
  title =	 {$S$-factor and scattering-parameter extractions from ${}^{3}\mathrm{He} + {}^{4}\mathrm{He}\to{}^{7}\mathrm{Be} + \gamma$},
  journal =	 JPhysG
}

@ARTICLE{Zhang2021,
  OPTauthor =	 {{Zhang}, L.~Y. and {Su}, J. and {He}, J.~J. and {Wiescher}, M. and {deBoer}, R.~J. and {Kahl}, D. and {Chen}, Y.~J. and {Li}, X.~Y. and {Wang}, J.~G. and {Zhang}, L. and {Cao}, F.~Q. and {Zhang}, H. and {Zhang}, Z.~C. and {Jiao}, T.~Y. and {Sheng}, Y.~D. and {Wang}, L.~H. and {Song}, L.~Y. and {Jiang}, X.~Z. and {Li}, Z.~M. and {Li}, E.~T. and {Wang}, S. and {Lian}, G. and {Li}, Z.~H. and {Tang}, X.~D. and {Zhao}, H.~W. and {Sun}, L.~T. and {Wu}, Q. and {Li}, J.~Q. and {Cui}, B.~Q. and {Chen}, L.~H. and {Ma}, R.~G. and {Guo}, B. and {Xu}, S.~W. and {Li}, J.~Y. and {Qi}, N.~C. and {Sun}, W.~L. and {Guo}, X.~Y. and {Zhang}, P. and {Chen}, Y.~H. and {Zhou}, Y. and {Zhou}, J.~F. and {He}, J.~R. and {Shang}, C.~S. and {Li}, M.~C. and {Zhou}, X.~H. and {Zhang}, Y.~H. and {Zhang}, F.~S. and {Hu}, Z.~G. and {Xu}, H.~S. and {Chen}, J.~P. and {Liu}, W.~P.},
  author =	 {{Zhang}, L.~Y. and others},
  title =	 {Direct Measurement of the Astrophysical {${}^{19}\mathrm{F}(\mathrm{p},\alpha\gamma){}^{16}\mathrm{O}$} Reaction in the Deepest Operational Underground Laboratory},
  journal =	 PRL,
  year =	 2021,
  month =	 10,
  volume =	 127,
  number =	 15,
  eid =		 152702,
  pages =	 152702,
  doi =		 {10.1103/PhysRevLett.127.152702}
}

@ARTICLE{Zhang2022,
  author =	 {{Zhang}, Liyong and others},
  OPTauthor =	 {{Zhang}, Liyong and {He}, Jianjun and {deBoer}, Richard J. and {Wiescher}, Michael and {Heger}, Alexander and {Kahl}, Daid and {Su}, Jun and {Odell}, Daniel and {Chen}, Yinji and {Li}, Xinyue and {Wang}, Jianguo and {Zhang}, Long and {Cao}, Fuqiang and {Zhang}, Hao and {Zhang}, Zhicheng and {Jiang}, Xinzhi and {Wang}, Luohuan and {Li}, Ziming and {Song}, Luyang and {Zhao}, Hongwei and {Sun}, Liangting and {Wu}, Qi and {Li}, Jiaqing and {Cui}, Baoqun and {Chen}, Lihua and {Ma}, Ruigang and {Li}, Ertao and {Lian}, Gang and {Sheng}, Yaode and {Li}, Zhihong and {Guo}, Bing and {Zhou}, Xiaohong and {Zhang}, Yuhu and {Xu}, Hushan and {Cheng}, Jianping and {Liu}, Weiping},
  title =	 {Measurement of {$^{19}\mathrm{F}(p,\gamma)^{20}\mathrm{Ne}$} reaction suggests {CNO} breakout in first stars},
  journal =	 Nature,
  year =	 2022,
  month =	 10,
  volume =	 {610},
  number =	 {7933},
  pages =	 {656-660},
  doi =		 {10.1038/s41586-022-05230-x}
}

@article{Zickefoose2018,
  title =	 {{Measurement of the ${}^{12}\mathrm{C}({}^{12}\mathrm{C},\mathrm{p}){}^{23}\mathrm{Na}$ cross section near the Gamow energy}},
  OPTauthor =	 {Zickefoose, J. and Di Leva, A. and Strieder, F. and Gialanella, L. and Imbriani, G. and De Cesare, N. and Rolfs, C. and Schweitzer, J. and Spillane, T. and Straniero, O. and Terrasi, F.},
  author =	 {Zickefoose, J. and others},
  journal =	 PRC,
  volume =	 97,
  issue =	 6,
  pages =	 065806,
  numpages =	 9,
  year =	 2018,
  month =	 6,
  doi =		 {10.1103/PhysRevC.97.065806}
}

@article{Zucchini2018,
  author =	 {A. Zucchini and others},
  OPTauthor =	 {A. Zucchini and M. Petrelli and F. Frondini and C. M. Petrone and P. Sassi and A. Di Michele and S. Palmerini and O. Trippella and M. Busso},
  journal =	 MPS,
  number =	 {2},
  pages =	 {268-283},
  title =	 {Chemical and mineralogical characterization of the {Mineo (Sicily, Italy)} pallasite: A unique sample},
  volume =	 {53},
  year =	 {2018},
  doi =		 {10.1111/maps.13002}
}

\end{document}